\documentclass[twocolumn,english]{article}
\usepackage{ae,aecompl}
\usepackage[T1]{fontenc}
\usepackage[latin9]{inputenc}
\usepackage[a4paper]{geometry}
\usepackage{color}
\definecolor{shadecolor}{rgb}{0.957031, 0.921875, 0.894531}
\usepackage{babel}
\usepackage{array}
\usepackage{rotating}
\usepackage{float}
\usepackage{framed}
\usepackage{textcomp}
\usepackage{multirow}
\usepackage{amsmath}
\usepackage{amsthm}
\usepackage{amssymb}
\usepackage{graphicx}
\usepackage{setspace}
\usepackage[unicode=true]
 {hyperref}

\makeatletter

\providecommand{\tabularnewline}{\\}

\numberwithin{equation}{section}
\numberwithin{figure}{section}
\newcommand{\lyxaddress}[1]{
	\par {\raggedright #1
	\vspace{1.4em}
	\noindent\par}
}

\usepackage[usenames,dvipsnames,svgnames,table]{xcolor}
\hypersetup{
     colorlinks   = true,
     citecolor    = blue
}
\definecolor{myred}{rgb}{0.66, 0.15, 0.15}
\usepackage{flushend}
\usepackage{cuted}
\usepackage{widetext}
\usepackage{sectsty} 

\makeatother

\begin{document}
\begin{strip}

\textsf{\textbf{\Huge{}Efimov Physics: a review}}{\Huge\par}

\bigskip{}

{\Large{}Pascal Naidon$^{1}$ and Shimpei Endo$^{2}$}\\
\\

\lyxaddress{$^{1}$RIKEN Nishina Centre, RIKEN, Wako, 351-0198 Japan.\hypersetup{urlcolor=myred}
\emph{\href{mailto:pascal@riken.jp}{pascal@riken.jp}}}

\lyxaddress{$^{2}$School of Physics and Astronomy, Monash University, Clayton,
VIC, 3800, Australia.\hypersetup{urlcolor=myred}\emph{\href{mailto:shimpei.endo@monash.edu}{shimpei.endo@monash.edu}}}
\begin{flushleft}
\noindent\begin{minipage}[t]{1\paperwidth}%
\begin{flushleft}
Published on March 28$^{\text{th}}$, 2017 in \emph{Rep. Prog. Phys.
80, 056001}, at the invitation of Professor Gordon Baym.\\
Revised on October 3rd, 2022 (see list of changes at the end).
\par\end{flushleft}%
\end{minipage}
\par\end{flushleft}

\begin{center}
\begin{minipage}[t]{0.965\textwidth}%
\begin{shaded}%
\begin{center}
\begin{minipage}[t]{0.99\textwidth}%
\begin{onehalfspace}
\vspace{0cm}
\textbf{\Large{}Abstract}\vspace{0.2cm}

\end{onehalfspace}

This article reviews theoretical and experimental advances in Efimov
physics, an array of quantum few-body and many-body phenomena arising
for particles interacting via short-range resonant interactions, that
is based on the appearance of a scale-invariant three-body attraction
theoretically discovered by Vitaly Efimov in 1970. This three-body
effect was originally proposed to explain the binding of nuclei such
as the triton and the Hoyle state of carbon-12, and later considered
as a simple explanation for the existence of some halo nuclei. It
was subsequently evidenced in trapped ultra-cold atomic clouds and
in diffracted molecular beams of gaseous helium. These experiments
revealed that the previously undetermined three-body parameter introduced
in the Efimov theory to stabilise the three-body attraction typically
scales with the range of atomic interactions. The few- and many-body
consequences of the Efimov attraction have been since investigated
theoretically, and are expected to be observed in a broader spectrum
of physical systems.%
\end{minipage}
\par\end{center}\end{shaded}%
\end{minipage}
\par\end{center}

\smallskip{}
\end{strip}

\hypersetup{linkcolor=myred}

\tableofcontents{}

\partfont{\color{black}}
\sectionfont{\color{myred}\raggedright}
\subsectionfont{\color{myred}}
\subsubsectionfont{\color{myred}}
\paragraphfont{\color{myred}}
\subparagraphfont{\color{myred}}

\clearpage{}

\part{Introduction\label{part:Introduction}}

\section{What is Efimov physics?\label{sec:What-is-Efimov}}

In 1970, Vitaly Efimov found a remarkable effect in the quantum spectrum
of three particles~\cite{Efimov1970a,Efimov1970b}. He considered
particles interacting through short-range attractive interactions
that are nearly resonant. By \emph{short range}, one means interactions
decaying faster than $1/r^{3}$ where $r$ is the interparticle distance,
and by \emph{nearly resonant}, one means attractive interactions that
can almost or just barely support a weakly two-body bound state. The
fact that, in quantum mechanics, an attractive interaction may be
too weak to bind two particles is due the quantum fluctuations of
the kinetic energy (also known as the zero-point energy) that competes
with the attractive interaction. When the interaction is just strong
enough to cancel the repulsive effect of the kinetic energy, the interaction
is said to be resonant because two particles scattering at low energy
are very close to binding during their collision: they spend a long
time together (they ``resonate'') before separating, which is characterised
in scattering theory by an $s$-wave scattering length that is much
larger than the range of the interactions.

Under these conditions, Vitaly Efimov found that an effective long-range
three-body attraction arises, and this attraction may support an infinite
family of three-body bound states (called Efimov states or Efimov
trimers), in which the three particles are bound at larger and larger
distances, beyond the range of the interactions. The Efimov effect,
as it became known, is striking in several aspects:

\paragraph*{Induced long-range interaction}

Even though the interactions are short-ranged, the three particles
feel a long-range three-body attraction. This seemingly counter-intuitive
situation can be explained by the fact that an effective interaction
is mediated between two particles by the third particle moving back
and forth between the two. It is thus possible for the three particles
to feel their influence at distances much larger than the range of
interactions, typically up to distances on the order of the scattering
length.

\paragraph*{Discrete scale invariance}

Right at the resonance, the scattering length is infinite and the
effective attraction extends to infinite distances. Being of kinetic
origin (the exchange of a particle between two others), the attraction
scales like the kinetic energy of the particles and brings no characteristic
length scale. As a result, the three-body system is scale invariant.
Quantisation in this attractive potential gives an infinite series
of bound states, the Efimov trimers, whose properties such as size
and energy are related to each others' by a scale transformation with
a universal scaling factor. The energy spectrum, for instance, forms
a geometric series with an accumulation at the zero energy threshold,
corresponding to infinitely weakly bound states. This situation is
referred to as the ``discrete scale invariance'' of Efimov states.
Efimov states thus look like a infinite family of matryoshka, the
Russian wooden dolls that can be nested inside each other. This image
was originally given to describe renormalisation-group limit cycles~\cite{LeClair2004,Braaten2006},
which constitute a possibility among the general classification of
renormalisation-group limits, originally proposed by Kenneth G. Wilson~\cite{Wilson1971}.
This possibility, which exhibits discrete scale invariance, is indeed
realised in systems exhibiting the Efimov effect~\cite{Bedaque1999a}\footnote{Other examples of systems exhibiting the renormalisation-group limit
cycle are systems with $1/r^{2}$ two-body interactions \cite{Glazek2002,LeClair2004}
such as an electron scattering off an excited hydrogen atom. A more
general discussion on discrete scale invariance is given in reference~\cite{Sornette1998}. }.

\paragraph*{Borromean binding}

When the interaction is not strong enough to support a two-body bound
state, it may nonetheless support one, up to infinitely many, Efimov
trimers. This possibility of binding $N$ particles, while the $N-1$
subsystems are unbound is called ``Borromean'' binding\footnote{Some authors~\cite{Yamashita2011,Baas2014} reserve the term ``Borromean\textquotedbl{}
for $N=3$ and use the term ``Brunnian\textquotedbl{} for larger
$N$. }. This denomination derives from the ancient symbol of intricated
circles called ``Borromean rings\textquotedbl , which have been
used, among others, by the Borromeo family in their coat of arms.
Borromean rings are arranged in such a way that they cannot be separated,
although cutting one of them sets the others free. They therefore
constitute a classical example of Borromean binding. In this case,
the binding is due to their specific topology. In the case of quantum
particles, however, Borromean binding is possible even if the interparticle
interaction is isotropic and does not enjoy such topological properties.
Although it is counterintuitive from a classical point of view, it
may be understood by considering that the number of degrees of freedom
providing a zero-point kinetic energy scales like $N$ whereas the
number of pairwise interactions scales like $N^{2}$, making the interactions
win for sufficiently large $N$. Efimov trimers are an example of
this phenomenon for $N=3$.

More difficult to interpret is the fact that when the interaction
is strong enough to support a two-body bound state, further increasing
the interaction reduces the binding of the three-body bound state
with respect to that of the two-body bound state. \\

In recent years, it has been realised that the Efimov effect gives
rise to a broad class of phenomena that have been referred to as \emph{Efimov
physics}. Consequences and extensions of the Efimov effect have indeed
been found in systems of various kinds of particles, from three to
many particles, with various kinds of interactions and in various
mixtures of dimensions. The denomination ``Efimov physics'' is however
not clearly defined and somewhat subjective. Even the notion of what
constitutes an Efimov state has been debated and considerably extended
by some authors - see section~\ref{subsec:What-is-an-Efimov-state}.
In the strictest sense, ``Efimov physics'' designates physics that
is a direct consequence of the occurence of the Efimov effect. More
generally speaking, one may say that a system exhibits Efimov (or
Efimov-like) physics when a three-or-more-body attraction emerges
from short-range interactions and possibly exhibits some kind of discrete
scale invariance.

The purpose of this review is to cover the recent theoretical and
experimental advances in Efimov physics, taken in its broadest sense.

\section{Why is it important? For which systems?\label{sec:Why-is-it-important}}

Efimov physics is not only remarkable for its distinctive properties,
it is also part of what is often referred to as the \emph{universality}
of low-energy physics. When a physical system with short-range interaction
has a sufficiently low energy, its wave function is so delocalised
that many microscopic details of the interactions become irrelevant,
and most of its properties can be effectively described by a few parameters.
Such physics is universal as it can be applied to many different systems,
regardless of their microscopic details. Efimov physics is an example
of such a situation, as it involves states in which the particles
are on average at larger separations than the range of their interactions.
For instance the discrete scale invariance of Efimov states is a universal
feature that depends only on a few general properties such as the
particles' masses and quantum statistics. As a result of this universality,
Efimov physics applies to virtually any field of quantum physics,
be it atomic and molecular physics~\cite{Braaten2007,Kunitski2015},
nuclear physics~\cite{Jensen2004}, condensed matter~\cite{Nishida2013}
or even high-energy physics~\cite{Hammer2008,Hammer2010}. Interestingly,
thanks to the formal connection between quantum theory and statistical
physics, it may also apply to the thermal equilibrium of classical
systems, such as three-stranded DNA~\cite{Maji2010}. 

The universality of Efimov physics does not mean that it occurs in
any system. It means that any system meeting the conditions for its
appearance exhibits the same universal features. These conditions
turn out to be quite restrictive, which is why it has taken around
forty years since the original theoretical prediction of the Efimov
effect to obtain convincing experimental confirmations. Generally
speaking, the Efimov effect requires resonant short-range interactions.
Such interactions are rare, because they require a bound or virtual
state to exist accidentally just below the scattering threshold of
two particles. This situation turns out to be common in nuclear physics,
but most nuclear particles obey Fermi statistics, and the Pauli exclusion
between fermions overcomes the Efimov attraction in most cases, preventing
the Efimov effect from occurring. On the other hand, bosonic particles
or excitations are common in various fields of physics, but their
interaction is rarely resonant. Nevertheless, there are now a significant
number of physical systems where Efimov physics has been observed
or is expected to be observed. In particular, with the advent of controllable
Feshbach resonances in ultra-cold atomic gases it has become possible
to fulfill at will the conditions for the occurence of Efimov physics,
and study it extensively.

Since there have been many theoretical developments in Efimov physics
recently, this review is organised from the theoretical point of view
in terms of physical situations leading to Efimov or Efimov-like physics.
For each situation, the current state of experimental observation
in different fields of physics is presented. Although this choice
of presentation requires the reader to read different sections to
know about the experimental achievements in a particular field, it
should give a comprehensive overview of what has, and what has not
yet, been observed in Efimov physics. The sections are relatively
independent, so that the reader can jump directly to the situation
of their interest. As for the readers who desire to grasp the bare
essentials of the Efimov effect, we have included a concise derivation
of Efimov theory in section~\ref{subsec:Efimov-theory} and discussed
the main features of Efimov states in section~\ref{subsec:What-is-an-Efimov-state}.
In addition, we give in the following section a short history of the
development of Efimov physics underlining the landmarks contributions.

\section{A short history of Efimov physics}

\begin{figure*}
\hfill{}\includegraphics[scale=0.7]{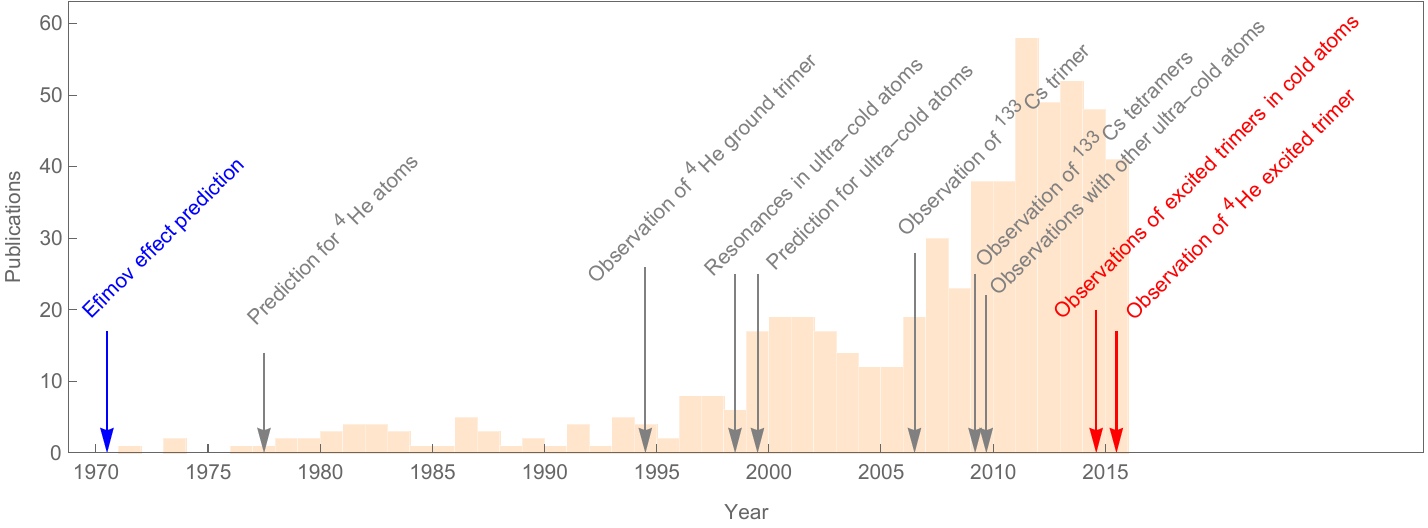}\hfill{}

\caption{\label{fig:History-of-Efimov}History of Efimov physics from the original
theoretical prediction by Vitaly Efimov to the latest experimental
observations, along with the number of related publications (source:
Web of Knowledge).}
\end{figure*}
In 1970, Vitaly Efimov was working as a junior researcher at the Ioffe
Institute in Leningrad, where he had completed his doctoral thesis
four years earlier. Following the seminal work of Llewellyn H. Thomas
in 1935~~\cite{Thomas1935} and later works by G.~V.~Skorniakov
and Karen~A~Ter-Martirosian~\cite{Skorniakov1957}, he was interested
in the three-body problem in quantum mechanics to describe nuclear
systems such as the triton (the nucleus of tritium, made of one proton
and two neutrons). Thomas had shown that three particles with a symmetric
wave function, unlike two particles, can be bound with arbitrarily
large binding energy for sufficiently small range of the interparticle
attractive force. This finding, referred to as the ``Thomas collapse''
or ``fall to the centre'' seemed somewhat peculiar, but allowed Thomas
to estimate a lower bound for the range of nuclear forces from the
measured energy of tritium, before it was confirmed by neutron-proton
scattering experiments.

Using the hyper-spherical coordinates, Efimov found that when two
of the particles can nearly bind, the three particles actually admit
an infinite series of bound states of ever-increasing sizes, instead
of just one as previously anticipated. This was due to an effective
three-body attractive force, which gave a simple interpretation for
the Thomas collapse. He published his result in both the Soviet literature~\cite{Efimov1970a,Efimov1972}
and Western journals~\cite{Efimov1970b,Efimov1973} where it became
known as the ``Efimov effect''. The first publication in English did
not provide the derivation and the effect was thus met with scepticism.
However, it prompted some theorists to look into the problem and soon
after, the validity of Efimov's result was confirmed both analytically
and numerically by R. D. Amado and J. V. Noble~\cite{Amado1972}.
For a long time, however, the Efimov effect was regarded by many as
a theoretical peculiarity of the formal three-body problem that would
have little to virtually no observable consequences on real physical
systems. On the other hand, some people took the effect seriously
and tried to find physical systems where it could be observed.

Vitaly Efimov proposed in his original papers that the Efimov effect
could describe nuclear systems such as the triton and the famous Hoyle
state of carbon-12. Subsequently, it was suggested that the Efimov
effect may be revealed in some hypernuclei by T. K. Lim in 1986~\cite{Lim1986},
and in halo nuclei by Dmitri V. Fedorov, Aksel S. Jensen, and Karsten
Riisager in 1994~\cite{Fedorov1994}. The proposed nuclear systems
indeed feature resonant two-body subsystems, which is a requirement
for the Efimov effect to occur. The closer to resonance the two-body
subsystems are, the larger the number of three-body bound states.
However, having more than one three-body bound state requires a very
close tuning near the resonance, something that happens only accidentally
in nature. As a result, the proposed nuclear systems allow only one
three-body bound state to exist, and do not reveal the infinity of
other states predicted by the Efimov theory closer to resonance. Moreover,
it is difficult to show that such a single three-body bound state
originates from the Efimov effect for two reasons. First, Efimov's
theory relies on an unknown three-body parameter to describe the three-body
states, and is thus not quite predictive for the properties of a single
three-body state, whereas it makes definite and universal predictions
(independent of the three-body parameter) for the relative properties
of two three-body states. Second, the first three-body state is the
smallest of the Efimov series and is significantly affected by the
details of the interparticle forces, to the point that it is debatable
to call it Efimov state. Because of these ambiguities, and despite
the experimental observations of the proposed nuclear systems, it
has been difficult to prove or disprove that they are indeed Efimov
states.

To obtain better experimental evidence of the Efimov effect, researchers
turned to other kinds of particles for which the two-body resonance
condition could be more closely fulfilled. Seven years after Efimov's
theory, T. K. Lim already pointed out the particular case of helium-4
atoms \cite{Lim1977}, whose interatomic interaction is close enough
to resonance to admit two three-body bound states, as was checked
subsequently by many few-body theorists. This prompted a decade-long
experimental search for these two helium-4 trimer states by the group
of Jan Peter Toennies in G\"ottingen, by analysing diffracted beams
of helium-4 clusters~\cite{Schoellkopf1994,Bruehl2005}. While the
ground-state trimer could be observed, the excited trimer state, which
is regarded as a true Efimov trimer and an evidence of the Efimov
effect, could not be observed.

The breakthrough that established Efimov physics came from the field
of ultra-cold atoms. In the 1990s, it was predicted~\cite{Tiesinga1993}
and demonstrated experimentally~\cite{Inouye1998,Courteille1998}
that the interactions between atoms could be controlled and brought
to resonance by applying a magnetic field. This led to the proposal
by Brett D. Esry, Chris H. Greene, and James P. Burke Jr~\cite{Esry1999}
to observe the signatures of Efimov states in such systems. Such experimental
signature of a three-body state near the two-body resonance of caesium-133
atoms was obtained in 2002 in the group of Hanns-Christoph N\"agerl
and Rudolf Grimm in Innsbruck, and after careful analysis reported
in 2006~\cite{Kraemer2006}. Although it revealed only one trimer,
as in the nuclear systems or the previous experiments on helium, its
Efimovian nature appeared more convincing due to its Borromean nature
(the trimer exists in a region where two-body subsystems are known
to be unbound). This landmark experiment opened the way for a systematic
investigation of Efimov physics, because the interaction could now
be controlled. This led to many similar experimental results from
various laboratories around the world using other species of ultra-cold
atoms (in particular during the year 2009) as well as an intense theoretical
activity to understand and explore various aspects of Efimov physics
in ultra-cold atoms. In the same year, universal four-body bound states
tied to Efimov states were evidenced in the caesium experiment in
Innsbruck by Francesca Ferlaino and co-workers~\cite{Ferlaino2009},
just after being predicted by theorists~\cite{Hammer2007,Stecher2009}.
The year 2009 culminated with the ITAMP workshop in Rome entitled
``Efimov 2009\textquotedbl , where the wealth of new experimental
and theoretical results was presented.

With the accumulation of experimental results in ultra-cold atoms,
the theoretically unknown three-body parameter of the Efimov theory
could be obtained from experimental measurements for many different
Efimov states. In particular, it could be compared for different two-body
resonances in the lithium-7 experiments by the group of Lev Khaykovich
at Bar-Ilan University~\cite{Gross2011}, and in the caesium experiments
by the group of Innsbruck~\cite{Berninger2011}. To everyone's surprise,
the three-body parameter was found to be nearly the same for all the
resonances of a given atomic species. It even appeared to be universally
correlated to the van der Waals length of the atoms, while it was
thought to depend on many other microscopic details. This so-called
``van der Waals universality\textquotedbl{} of the three-body parameter
was later explained by theoretical works~\cite{Wang2012,Naidon2014a},
which showed that a sudden deformation of the trimer configuration
prevents the three atoms from reaching separations smaller than the
van der Waals length, making the trimers insensitive to more microscopic
features of the interatomic interaction.

While different measurements of three-body recombination and atom-dimer
relaxation provided several experimental points in the three-body
spectrum confirming the ``scenario''' obtained by Vitaly Efimov,
the most striking aspect of this scenario, namely the discrete scale
invariance leading to the geometric series of three-body bound states
was not confirmed clearly since the experiments did not reveal consecutive
three-body bound states. Observing consecutive Efimov states is an
experimental challenge since each new state is by definition much
larger in size with a much weaker binding energy. This endeavour was
ultimately successful in 2014, when the experimenters in Innsbruck
managed to observe a second Efimov state of caesium atoms by pushing
the limits of their experiment~\cite{Huang2014}, while the groups
of Cheng Chin at the University of Chicago~\cite{Tung2014} and Matthias
Weidem\"uller at the University of Heidelberg~\cite{Pires2014}
independently observed up to three Efimov states of two caesium and
one lithium atoms, whose energy levels were predicted to be closer
to each other due to the large mass imbalance between these two atomic
species. The same year, outside the ultra-cold atom community, the
group of Reinhard D\"orner in Frankfurt could finally observe the
long-sought second trimer of helium-4 by the Coulomb explosion imaging
technique, a result published the following year \cite{Kunitski2015}.
Not only this brought further experimental confirmation of the Efimov
effect, it also provided the first spatial imaging of an Efimov state.
One may say that the year 2014 concluded a 44-year-long search for
a full confirmation of the Efimov effect. The history of this search
is summarised in figure~\ref{fig:History-of-Efimov} where landmark
contributions are indicated.

\newpage{}

\part{Three particles\label{part:Three-particles}}

\section{Three identical bosons\label{sec:Three-identical-bosons}}

The simplest situation for which Efimov physics occurs corresponds
to three identical bosons interacting via resonant short-range interactions.
In section~\ref{subsec:Efimov-theory}, we briefly present the corresponding
theory originally proposed by Vitaly Efimov and its various extensions
in sections~\ref{subsec:Finite-range-interactions}-\ref{subsec:Relativistic-case}.
In section~\ref{subsec:What-is-an-Efimov-state}, we look into the
question of what constitutes an Efimov state, before reviewing in
sections~\ref{subsec:Observations-in-nuclear}-\ref{subsec:Prospects-for-observation}
the experimental observations and prospects for observations of bosonic
Efimov states in nuclear, atomic, and condensed matter systems.

\subsection{The Efimov universal theory\label{subsec:Efimov-theory}}

We consider identical bosonic particles of mass $m$, with no internal
degree of freedom, interacting via short-range two-body (and possibly
three-body) interactions. Here, \emph{short-range} interactions means
that the interaction potentials decay faster than $1/r^{3}$, where
$r$ is the separation between two particles. In this situation, there
exists a separation $b$, called the range of the interaction, beyond
which the relative motion of two particles is almost free. It is in
this asymptotically free region where the particles' energy is purely
kinetic that the Efimov effect takes its roots, and that is why it
is universal.

Although the relative motion of two particles is free in this region,
each angular partial wave of the wave function $\psi(\vec{r})$ describing
the two-body relative motion has a phase shift $\delta_{\ell}$ with
respect to the non-interacting wave function, as a result of the particles
interacting at shorter separation. Namely, in the partial-wave expansion
of $\psi(\vec{r})$, 
\begin{equation}
\psi(\vec{r})=\psi(r,\theta,\phi)=\sum_{\ell=0}^{\infty}\frac{f_{\ell}(r)}{r}P_{\ell}(\cos\theta),\label{eq:PartialWaveExpansion}
\end{equation}
where $P_{\ell}$ are the Legendre polynomials, the partial wave component
$f_{\ell}(r)$ has the form, 
\begin{equation}
f_{\ell}(r)=\begin{cases}
\mbox{complicated} & \mbox{for }r\lesssim b\mbox{ (interaction)}\\
\propto\sin(kr-\ell\frac{\pi}{2}+\delta_{\ell}) & \mbox{for }r\gg b\mbox{ (free region)}
\end{cases}\label{eq:PartialWave}
\end{equation}
where $k$ is the relative wave number between the two particles.
In the absence of interaction, the phase shift $\delta_{\ell}=0$
(no scattering occurs). On the opposite, the strongest dephasing the
interaction can induce is $\delta_{\ell}=\pi/2$ (modulo $\pi$),
in which case the interaction is said to be \emph{resonant} in that
partial wave. 

Efimov physics arises when the two-body interaction is near-resonant
in the $s$-wave partial wave ($\ell=0$), which means that the phase
shift $\delta_{0}$ of the $s$ wave is close to $\pi/2$ (modulo
$\pi$).

\paragraph{The scattering length}

It is well-known from scattering theory~\cite{Mott1965} that at
low scattering energy ($k\ll b^{-1}$), only the $s$ wave is scattered,
i.e. has a non-zero phase shift. Moreover, the phase shift can be
written as 
\begin{equation}
\delta_{0}\sim-\arctan(ka)\quad\mbox{ for }k\ll b^{-1},\label{eq:SWavePhaseShift}
\end{equation}
where $a$ defines the \emph{scattering length}\footnote{Despite its name, the scattering length can be positive or negative.}.
Therefore, for the two-body interaction to be resonant at low energy,
the scattering length $a$ has to be much larger than $b$:
\begin{equation}
\vert a\vert\gg b\label{eq:ResonanceCondition}
\end{equation}
In particular, the limit $a\to\pm\infty$ is sometimes called the
\emph{unitary limit} or \emph{unitarity}, because in this limit the
factor $\sin^{2}\delta_{0}$ in the expression of the scattering cross
section $\sigma=\frac{4\pi}{k^{2}}\sin^{2}\delta_{0}$, approaches
its maximal value $\sin^{2}\delta_{0}=1$. This maximum of the scattering
cross section is the consequence of a fundamental property of quantum
mechanics, the unitarity of the S-matrix. It can be reached precisely
for resonant interactions. 

Near unitarity, the scattering length $a$ is the only parameter that
controls the physics of two particles at low energy, either positive
or negative: it determines the cross section for scattering states
(positive energy), and the binding energy of a weakly bound state
below the break-up threshold (negative energy). This bound state,
also called \emph{dimer}, exists only for a positive scattering length
and its binding energy is close to 
\begin{equation}
\frac{\hbar^{2}}{ma^{2}},\label{eq:UniversalDimerEnergy}
\end{equation}
where $m$ is the mass of the particles and $\hbar$ is the reduced
Planck constant. The resonance of the interaction is therefore related
to the appearance of the two-body bound state from below the scattering
threshold exactly at the unitary limit $a\to\pm\infty$. It is represented
by a black line in figure~\ref{fig:EfimovPlot}. 

\paragraph{Zero-range theory}

Short-range near-resonant interactions at low energy constitute a
limit that can be treated by the zero-range theory. This theory assumes
that the short-range region where the interaction directly affects
the wave function can be neglected and only the asymptotically free
region that is parameterised by the scattering length is relevant.
This amounts to saying that the range $b$ of the interaction is vanishingly
small compared to the scattering length $a$ or wave length $k^{-1}$
of the particles. This can be implemented in various ways.

A first way is to consider a simple interaction potential with a finite
range $b$, calculate observables, and take the limit $b\to0$ for
a fixed scattering length $a$. Another way is to consider a zero-range
pseudopotential, such as a contact interaction represented by a Dirac
delta function potential, sometimes referred to as a ``Fermi pseudopotential''~\cite{Fermi1936}.
This introduces ultraviolet divergences in exact calculations which
need to be renormalised to obtain observables~\cite{Weinberg1995}.
Renormalisation can be implemented by introducing a cut-off in momentum
space~\cite{Brueckner1957,Fetter2003,Bedaque1999a}, or regularising
the delta function using the Lee-Huang-Yang pseudopotential~\cite{Huang1987}:
\begin{equation}
\hat{V}(r)=\frac{4\pi\hbar^{2}a}{m}\delta^{3}(\vec{r})\frac{\partial}{\partial r}(r\cdot)\label{eq:LeeHuangYangPotential}
\end{equation}

Yet another way is to consider the system as free (no interaction)
and impose the so-called Bethe-Peierls boundary condition~\cite{Bethe1935a}
on the many-particle wave function $\Psi$ when any two particles
separated by $r$ come in contact:
\begin{equation}
-\frac{1}{r\Psi}\frac{\partial}{\partial r}(r\Psi)\xrightarrow[r\to0]{}\frac{1}{a}\label{eq:BethePeierls}
\end{equation}

The essence of all these methods is to correctly reproduce the form
of the two-body wave function in the region $b\ll r\ll k^{-1}$,
\begin{equation}
\psi(\vec{r})\propto\frac{1}{r}-\frac{1}{a},\label{eq:TwoBodyWaveFunction}
\end{equation}
which can be obtained from equations~(\ref{eq:PartialWaveExpansion}),
(\ref{eq:PartialWave}) and (\ref{eq:SWavePhaseShift}). The zero-range
methods make the simplification that this form remains true down to
$r=0$, although this is unphysical for $r\lesssim b$.

Such zero-range methods can be directly implemented in the Schr\"odinger
equation describing the three-boson system~\cite{Fedorov1993}, or
alternative formalisms such as integral equations~\cite{Skorniakov1957},
functional renormalisation equations~\cite{Wetterich1993,Moroz2009},
and effective field theory~\cite{Bedaque1999,Bedaque1999a}. Here,
we will make use of the Schr\"odinger equation along with the Bethe-Peierls
boundary condition~(\ref{eq:BethePeierls}).

We should note that Vitaly Efimov's original derivation~\cite{Efimov1970a}
did not invoke explicitly a zero-range interaction, but instead considered
equation~(\ref{eq:TwoBodyWaveFunction}), i.e. the effect of the
resonant interaction outside its range $b$, without requiring $b\to0$.
As we shall see, the zero-range theory for three particles is in fact
ill-defined. The more physical approach of Efimov avoids this difficulty
and naturally introduces the three-body parameter. It should thus
be referred to as a universal theory, instead of a zero-range theory.
Nevertheless, it is essentially equivalent to the zero-range theory
cured by a three-body boundary condition. For the sake of simplicity,
we will take this path, which formally follows very closely Efimov's
original derivation. 

\paragraph{Derivation of the Efimov attraction}

\begin{figure*}
\hfill{}\includegraphics[scale=0.6]{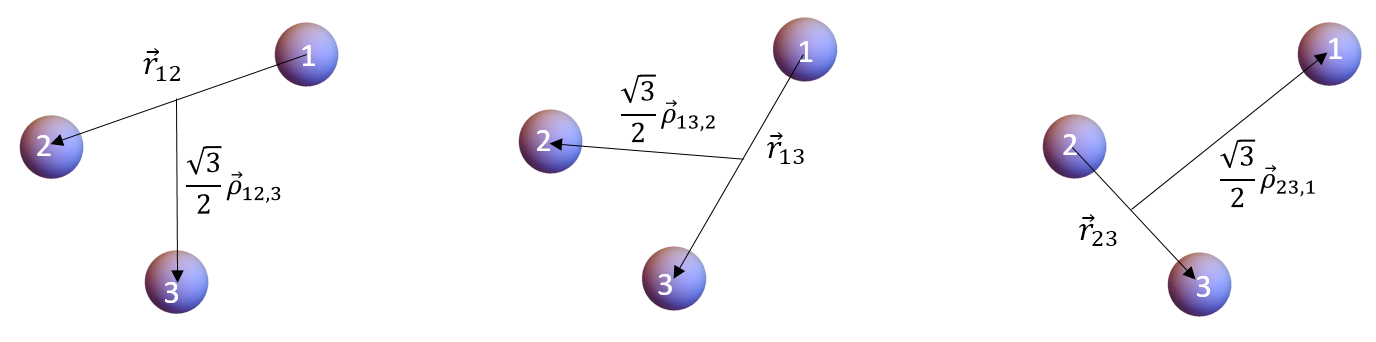}\hfill{}

\caption{\label{fig:Jacobi}The three sets of Jacobi coordinates describing
the relative positions of three identical particles.}

\end{figure*}

\begin{figure}[H]
\hfill{}\includegraphics[scale=0.68]{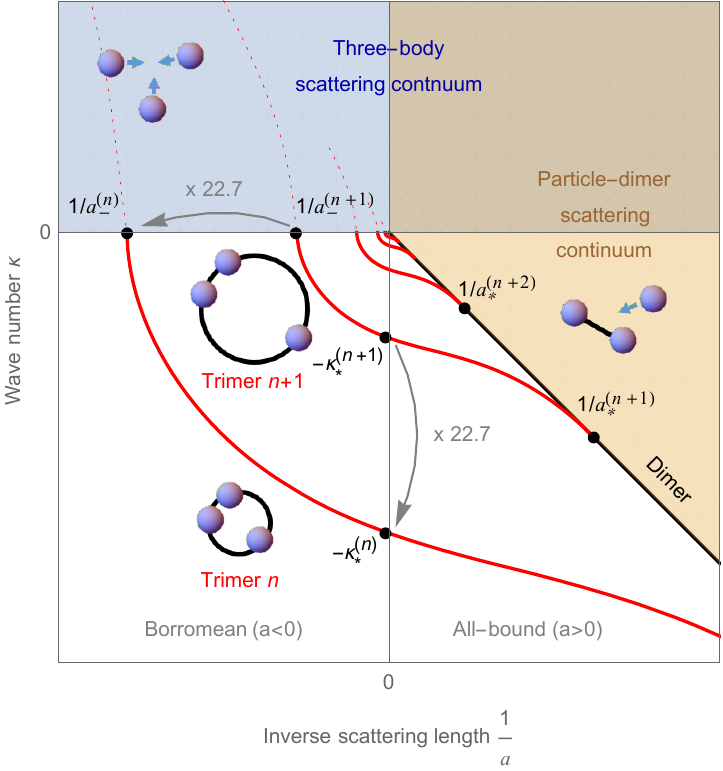}\hfill{}

\caption{\label{fig:EfimovPlot}Schematic representation of the so-called ``Efimov
plot'' or ``Efimov scenario'' showing the discrete scale invariance
of the three-body spectrum for identical bosons in the zero-range
theory. The wave number $\kappa=E\sqrt{m/(\hbar^{2}\vert E\vert)}$
associated with the energy $E$ of the dimer (black) and trimers (red)
is plotted against the inverse scattering length $1/a$. The blue
and orange filled regions represent the three-body scattering continuum
and the particle-dimer scattering continuum, respectively. Note that
these continua overlap for $1/a>0$ and $E>0$. Special values of
$\kappa$ and $1/a$ are indicated by the dots: a trimer appears from
the three-body scattering threshold at $1/a_{-}$, has a binding wave
number $\kappa_{*}$ at unitarity, and disappears below the particle-dimer
scattering threshold at $1/a_{*}$. Trimer resonances in the three-body
continuum are indicated by dotted curves. The discrete scale invariance
of the spectrum is indicated by the grey arrows showing the universal
scaling ratio between consecutive levels. For clarity, the value of
the strength $s_{0}$ has been artificially set to 3, instead of 1.00624,
thus reducing the spacings between the trimer levels to make them
more visible. }
\end{figure}

For three bosons located at $\vec{x}_{1}$, $\vec{x}_{2}$ and $\vec{x}_{3}$,
one can eliminate the centre of mass, and the system can be described
by two vectors, called \emph{Jacobi coordinates}:
\begin{eqnarray}
\vec{r}_{ij} & = & \vec{x}_{j}-\vec{x}_{i}\label{eq:Jacobi-r}\\
\vec{\rho}_{ij,k} & = & \frac{2}{\sqrt{3}}\left(\vec{x}_{k}-\frac{\vec{x}_{i}+\vec{x}_{j}}{2}\right)\label{eq:Jacobi-rho}
\end{eqnarray}
where $(i,j,k)$ are to be chosen among (1,2,3). There are thus three
possible Jacobi coordinate sets, shown in figure~\ref{fig:Jacobi},
which are related as follows:
\begin{eqnarray}
\vec{r}_{23} & = & -\frac{1}{2}\vec{r}_{12}+\frac{\sqrt{3}}{2}\vec{\rho}_{12,3}\label{eq:Jacobi-r23}\\
\vec{\rho}_{23,1} & = & -\frac{\sqrt{3}}{2}\vec{r}_{12}-\frac{1}{2}\vec{\rho}_{12,3}\label{eq:Jacobi-rho231}
\end{eqnarray}
\begin{eqnarray}
\vec{r}_{31} & = & -\frac{1}{2}\vec{r}_{12}-\frac{\sqrt{3}}{2}\vec{\rho}_{12,3}\label{eq:Jacobi-r31}\\
\vec{\rho}_{31,2} & = & \frac{\sqrt{3}}{2}\vec{r}_{12}-\frac{1}{2}\vec{\rho}_{12,3}\label{eq:Jacobi-rho312}
\end{eqnarray}
Choosing one set of Jacobi coordinates, the time-independent three-body
wave function satisfies the free Schr\"odinger equation at total
energy $E=\hbar^{2}k^{2}/m$:
\begin{equation}
(-\nabla_{r_{12}}^{2}-\nabla_{\rho_{12,3}}^{2}-k^{2})\Psi=0\label{eq:FreeSchrodingerEqPsi}
\end{equation}
along with the Bethe-Peierls boundary condition~(\ref{eq:BethePeierls})
for all pairs of bosons. Because of the bosonic exchange symmetry,
the wave function $\Psi$ can be decomposed as follows:
\begin{equation}
\Psi=\chi(\vec{r}_{12},\vec{\rho}_{12,3})+\chi(\vec{r}_{23},\vec{\rho}_{23,1})+\chi(\vec{r}_{31},\vec{\rho}_{31,2})\label{eq:FaddeevDecomposition}
\end{equation}
where the function $\chi$ (known as Faddeev component~\cite{Faddeev1961,Fedorov1993})
satisfies the equation: 
\begin{equation}
(-\nabla_{r}^{2}-\nabla_{\rho}^{2}-k^{2})\chi(\vec{r},\vec{\rho})=0\label{eq:FreeSchrodingerEqChi}
\end{equation}

Applying the Bethe-Peierls boundary condition~(\ref{eq:BethePeierls})
for the pair (1,2) to equation~(\ref{eq:FaddeevDecomposition}),
one obtains:
\begin{multline}
\left[\!\frac{\partial}{\partial r}\left(r\chi(\vec{r},\vec{\rho})\right)\!\right]_{r\to0}+\chi\!\left(\!\!\!\begin{array}{c}
\frac{\sqrt{3}}{2}\vec{\rho},-\frac{1}{2}\vec{\rho}\end{array}\!\!\!\right)+\chi\!\left(\!\!\!\begin{array}{c}
-\frac{\sqrt{3}}{2}\vec{\rho},-\frac{1}{2}\vec{\rho}\end{array}\!\!\!\right)\\
=\left[\!-\frac{r}{a}\left(\chi(\vec{r},\vec{\rho})+\chi\!\left(\!\!\!\begin{array}{c}
\frac{\sqrt{3}}{2}\vec{\rho},-\frac{1}{2}\vec{\rho}\end{array}\!\!\!\right)+\chi\!\left(\!\!\!\begin{array}{c}
-\frac{\sqrt{3}}{2}\vec{\rho},-\frac{1}{2}\vec{\rho}\end{array}\!\!\!\right)\right)\!\right]_{r\to0}\label{eq:BethePeierlsJacobi}
\end{multline}
where $\vec{r}\equiv\vec{r}_{12}$ and $\vec{\rho}\equiv\vec{\rho}_{12,3}$.
From the bosonic exchange symmetry, the same equation is obtained
by applying the Bethe-Peierls boundary condition for the other two
pairs. In the right-hand side of equation~(\ref{eq:BethePeierlsJacobi}),
only the first term remains when $r\to0$, because $\chi(\vec{r},\vec{\rho})$
diverges for $r\to0$ but is finite elsewhere. The function $\chi$
can be expanded in partial waves, which can be shown to be independent
in the zero-range theory. The Efimov effect for bosons occurs in the
partial-wave channel with total angular momentum $L=0$ . In this
channel, $\chi$ is independent of the directions of $\vec{r}$ and
$\vec{\rho}$ and can be written as
\begin{equation}
\chi(\vec{r},\vec{\rho})=\frac{\chi_{0}(r,\rho)}{r\rho}.\label{eq:SWaveChi}
\end{equation}
$\chi_{0}$ is finite for $r\to0$, consistent with the divergence
of $\chi$, but must satisfy:
\begin{equation}
\chi_{0}(r,\rho)\xrightarrow[\rho\to0]{}0\label{eq:Condition1}
\end{equation}
to keep $\chi$ finite in this limit. Inserting equation~(\ref{eq:SWaveChi})
into equations~(\ref{eq:FreeSchrodingerEqChi}) and (\ref{eq:BethePeierlsJacobi})
yields the equation
\[
\left(-\frac{\partial^{2}}{\partial r^{2}}-\frac{\partial^{2}}{\partial\rho^{2}}-k^{2}\right)\chi_{0}(r,\rho)=0
\]
and the boundary condition for $r\to0$:
\begin{equation}
\left[\!\frac{\partial}{\partial r}\left(\chi_{0}(r,\rho)\right)\!\right]_{r\to0}+2\frac{1}{\frac{\sqrt{3}}{4}\rho}\chi_{0}\!\left(\!\!\!\begin{array}{c}
\frac{\sqrt{3}}{2}\rho,\frac{1}{2}\rho\end{array}\!\!\!\right)=\!-\frac{1}{a}\chi_{0}(0,\rho)\!\label{eq:BethePeierlsJacobi-1}
\end{equation}
One can finally perform a transformation of the coordinates ($r,\rho$)
to the polar coordinates $(R,\alpha)$ known as hyper-spherical coordinates~\cite{Delves1958,Fedorov1993}:
\begin{eqnarray}
r & = & R\sin\alpha\label{eq:HypersphericalCoordinate1}\\
\rho & = & R\cos\alpha\label{eq:HypersphericalCoordinate2}
\end{eqnarray}
where $R$ is the hyper-radius satisfying
\begin{equation}
R^{2}=r^{2}+\rho^{2}=\frac{2}{3}\left(r_{12}^{2}+r_{23}^{2}+r_{31}^{2}\right)\label{eq:Hyper-radius}
\end{equation}
and $\alpha$ is the Delves hyper-angle. In these coordinates, one
obtains the equation:
\begin{equation}
\left(-\frac{\partial^{2}}{\partial R^{2}}-\frac{1}{R}\frac{\partial}{\partial R}-\frac{1}{R^{2}}\frac{\partial^{2}}{\partial\alpha^{2}}-k^{2}\right)\chi_{0}(R,\alpha)=0\label{eq:HypersphericalEquation}
\end{equation}
with the boundary condition for $\alpha\to0$:
\begin{equation}
\left[\!\frac{\partial}{\partial\alpha}\left(\chi_{0}(R,\alpha)\right)\!\right]_{\alpha\to0}+\frac{8}{\sqrt{3}}\chi_{0}\!\left(\!\!\!\begin{array}{c}
R,\frac{\pi}{3}\end{array}\!\!\!\right)=\!-\frac{R}{a}\chi_{0}(R,0)\label{eq:HypersphericalCondition}
\end{equation}
The problem then becomes separable in $R$ and $\alpha$, for the
case $a\to\pm\infty$ corresponding to the unitary limit. Indeed,
in this limit the right-hand side of equation~(\ref{eq:HypersphericalCondition})
vanishes and one is left with a boundary condition at $\alpha=0$
that is independent of $R$. On the other hand, the other boundary
condition~(\ref{eq:Condition1}) corresponds to $\chi_{0}(R,\frac{\pi}{2})=0$,
which is a boundary condition at $\alpha=\frac{\pi}{2}$ that is also
independent of $R$. One can thus find a solution of equation~(\ref{eq:HypersphericalEquation})
in the form:
\begin{equation}
\chi_{0}(R,\alpha)=F(R)\phi(\alpha)\label{eq:SeparableSolution}
\end{equation}
where $\phi$ satisfies $-\frac{d^{2}}{d\alpha^{2}}\phi(\alpha)=s_{n}^{2}\phi(\alpha)$
with the boundary conditions at $\alpha=0$ and $\alpha=\pi/2$. This
gives the following solutions:
\begin{equation}
\phi_{n}(\alpha)=\sin\left(s_{n}(\frac{\pi}{2}-\alpha)\right)\label{eq:HyperangularSolution}
\end{equation}
where $s_{n}$ is a solution of the equation:
\begin{equation}
-s_{n}\cos\Big(s_{n}\frac{\pi}{2}\Big)+\frac{8}{\sqrt{3}}\sin\Big(s_{n}\frac{\pi}{6}\Big)=0.\label{eq:TranscendentalEquation}
\end{equation}
Each solution labelled by $n$ constitutes a \emph{channel} for the
hyper-radial motion. That is to say, for each solution $\phi_{n}$
there is a corresponding hyper-radial function $F_{n}(R)$ such that
$F_{n}(R)\phi_{n}(\alpha)$ is a solution of equation~(\ref{eq:HypersphericalEquation}).
It satisfies the equation:
\begin{equation}
\left(-\frac{\partial^{2}}{\partial R^{2}}-\frac{1}{R}\frac{\partial}{\partial R}+\frac{s_{n}^{2}}{R^{2}}-k^{2}\right)F_{n}(R)=0\label{eq:HyperradialEquation}
\end{equation}
which can be written as a one-dimensional Schr\"odinger equation:
\begin{equation}
\left(-\frac{\partial^{2}}{\partial R^{2}}+V_{n}(R)-k^{2}\right)\sqrt{R}F_{n}(R)=0\label{eq:HyperradialEquation2}
\end{equation}
with the hyper-radial potential, 
\begin{equation}
V_{n}(R)=\frac{s_{n}^{2}-1/4}{R^{2}}\label{eq:HyperradialPotential}
\end{equation}

All solutions of equation~(\ref{eq:TranscendentalEquation}) are
real, except one denoted as $s_{0}\approx\pm1.00624i$ which is purely
imaginary. As a result, the effective $\propto R^{-2}$ potential
in equation~(\ref{eq:HyperradialEquation}) is attractive for the
channel $n=0$. This is in contrast with the non-interacting three-body
problem, where the boundary condition~(\ref{eq:HypersphericalCondition})
is replaced by $\chi_{0}(R,\alpha)\xrightarrow[r\to0]{}0$, leading
to equation~(\ref{eq:HyperangularSolution}) with eigenvalues $s_{n}=2(n+1)$
that are all real. In this case, the effective $\propto R^{-2}$ potential
of equation~(\ref{eq:HyperradialPotential}) is repulsive for all
$n$. This repulsion is interpreted as a generalised centrifugal barrier
due to the free motion of deformation of the three-body system. In
the interacting problem at unitarity however, the channel $n=0$ leads
to an effective three-body attraction 
\begin{equation}
\boxed{V_{0}(R)=-\frac{\vert s_{0}\vert^{2}+1/4}{R^{2}}}.\label{eq:EfimovAttraction}
\end{equation}
This unexpected attraction is the basis for Efimov physics and is
referred to as the \emph{Efimov attraction}. It can be interpreted
as the result of a mediated attraction between two particles by exchange
of the third particle.

The existence of this attraction shows that the zero-range theory
for three bosons is not well defined. Indeed, equation~(\ref{eq:HyperradialEquation})
for $n=0$ is a Schr\"odinger equation for an attractive $1/R^{2}$
potential, which is scale invariant since a $\propto1/R^{2}$ potential
scales as the kinetic energy $\propto d^{2}/dR^{2}$ under a scaling
transformation $R\to\lambda R$. It is known that such an equation
admits a solution at any energy, and its spectrum is therefore not
bounded from below~\cite{Newton2013,Frank1971}. Indeed, if the equation
admits a solution at energy $E<0$, making the scaling transformation
$R\to\lambda R$ with an arbitrary scaling factor $\lambda$ gives
another solution at energy $\lambda^{2}E<0$. This means that under
the Efimov attraction the three-boson system collapses on itself,
a phenomenon discovered long ago by Llewellyn H. Thomas~~\cite{Thomas1935}
and referred to as the ``Thomas collapse'' or ``fall of the particles
to the centre''. The same problem was found~\cite{Minlos1961} in
the formulation of the zero-range theory for three particles by an
integral equation, known as the Skorniakov and Ter-Martorisian equation~\cite{Skorniakov1957}.
This is of course a shortcoming of the zero-range theory, since the
finite-range effects of the interaction can no longer be neglected
when the distance between the three bosons becomes comparable with
the finite range of interactions\footnote{Throughout this article and much of the cited literature, the expression
``finite range\textquotedbl{} means a range that is not zero.}. 

A practical solution to this problem, originally suggested by Vladimir~N.~Gribov
and demonstrated by G.~S.~Danilov~\cite{Danilov1961}, consists
in imposing a condition on the solutions of the three-body equation,
or a momentum cut-off on the equation~\cite{Kharchenko1972}, in
order to reproduce a known three-body observable, such as a three-body
bound state energy or particle-dimer scattering property. For instance,
fixing the triton energy to the observed value, and solving the three-body
equation with that condition enables the prediction of the neutron-deuteron
scattering length~\cite{Danilov1963}.

In Vitaly Efimov's formulation of the three-body problem in terms
of equation~(\ref{eq:HyperradialEquation}), a similar procedure
can be achieved by imposing a boundary condition below some arbitrarily
small hyper-radius $R_{0}$. Thus, in addition to the Bethe-Peierls
two-body boundary condition~(\ref{eq:BethePeierls}), the three-body
problem in the zero-range theory requires an extra three-body boundary
condition. This boundary condition can be implemented in various ways,
for example setting a hard wall at the hyper-radius $R_{0}$ where
$F_{0}$ has to vanish, or imposing the value of the logarithmic derivative
of $F_{0}$ at $R_{0}$, by analogy to the Bethe-Peierls condition~(\ref{eq:BethePeierls}).
Note that these two implementations are not strictly equivalent: a
hard wall prevents the fall to the centre and sets a ground-state
energy, whereas a logarithmic derivative condition only makes the
spectrum discrete but still unbounded from below (states below the
physically relevant energy are therefore unphysical features). In
any case, both implementations introduce a new length scale in the
problem, which is referred to as the \emph{three-body parameter}.
It is this parameter that fixes the three-body observables. The necessity
to introduce this parameter may be regarded as a quantum anomaly in
the scaling symmetry of the system~\cite{Ananos2003}. Physically,
the three-body parameter encapsulates the effects of the two-body
(and possibly three-body) interactions at short distance.

To see how the three-body parameter arises, let us consider the solutions
of equation~(\ref{eq:HyperradialEquation}) for $n=0$. Near the
small hyper-radius $R_{0}$, any solution with sufficiently small
energy $\vert k^{2}\vert\ll\vert s_{0}\vert^{2}/R_{0}^{2}$ is of
the form:
\begin{equation}
F_{0}(R)=\alpha R^{i\vert s_{0}\vert}+\beta R^{-i\vert s_{0}\vert}\quad\mbox{for }R\gtrsim R_{0}\label{eq:F0}
\end{equation}
Imposing a boundary condition at $R_{0}$ imposes a specific ratio
$\beta/\alpha$. From dimensional analysis, this ratio has units of
inverse length $\Lambda$ to the power $-2i\vert s_{0}\vert$. Thus,
we can write $F_{0}(R)=\alpha(R^{i\vert s_{0}\vert}+\Lambda^{-2i\vert s_{0}\vert}R^{-i\vert s_{0}\vert})$,
which can be further expressed as
\begin{equation}
F_{0}(R)\underset{R\gtrsim R_{0}}{\propto}e^{i\vert s_{0}\vert\ln\Lambda R}+e^{-i\vert s_{0}\vert\ln\Lambda R}=\cos(\vert s_{0}\vert\ln\Lambda R).\label{eq:F0bis}
\end{equation}
The three-body wave function therefore shows log-periodic ocillations
in the hyper-radius, and the phase of these oscillations
\begin{equation}
\Phi=\vert s_{0}\vert\ln(\Lambda/\Lambda_{0})\label{eq:ThreeBodyPhase}
\end{equation}
is given by the new scale $\Lambda$ (expressed in some previously
fixed unit $\Lambda_{0}$), which is a possible representation of
the three-body parameter.

One of the fundamentally new findings of Vitaly Efimov is that the
three-body problem with the three-body boundary condition does not
only yield just one three-body bound state, as previously thought,
but infinitely many bound states. This is a simple consequence of
the effective attractive $1/R^{2}$ potential in equation~(\ref{eq:HyperradialEquation}).
Indeed, although the boundary condition~(\ref{eq:F0bis}) breaks
the scale invariance of the system under arbitrary scale transformations,
one can easily check that equation~(\ref{eq:F0bis}) is still invariant
under a discrete set of scale transformations $R\to\lambda_{0}^{n}R$,
with scaling factors that are integral powers of $\lambda_{0}=e^{\pi/\vert s_{0}\vert}\approx22.7$.
Thus, if the boundary condition gives a solution at some energy $E<0$,
it also gives solutions with energies $E/\lambda_{0}^{2n}<0$. There
is therefore an infinite number of bound states, forming a geometric
series of energies accumulating at zero energy, with scaling factor
$\lambda_{0}^{2}\approx515$. This situation is referred to as the
\emph{discrete scale invariance}.

Remarkably, Vitaly Efimov has shown that this discrete scale invariance
not only holds at unitarity ($a\to\pm\infty$) but also at finite
scattering length $a$, when one considers the spectrum in the polar
coordinates ($h,\xi$) of the inverse scattering length $a^{-1}$
and the wave number $\kappa=E\sqrt{m/(\hbar^{2}\vert E\vert)}$:
\begin{eqnarray}
a^{-1} & = & h\cos\xi,\label{eq:PolarCoordinates1}\\
\kappa & = & h\sin\xi.\label{eq:PolarCoordinates2}
\end{eqnarray}

The spectrum along a line at a fixed angle $\xi$ has the discrete
scale invariance with the scaling factor $\lambda_{0}=e^{\pi/\vert s_{0}\vert}\approx22.7$.
This property can be checked by scaling $k$, $a^{-1}$, and $R^{-1}$
by $\lambda_{0}$ in equations~(\ref{eq:HypersphericalEquation}-\ref{eq:HypersphericalCondition}).
As a result, all the three-body bound states show the same trajectory
in the $(a^{-1},\kappa)$ plane up to a scale transformation, as shown
in figure~\ref{fig:EfimovPlot}. The infinite series of bound-states
energies $E^{(n)}$ can therefore be described by the discrete scaling
of a single universal function $\Delta(\xi)\equiv2\vert s_{0}\vert\ln h(\xi)$
through the formula
\begin{equation}
\vert E^{(n)}\vert+\frac{\hbar^{2}}{ma^{2}}=\frac{\hbar^{2}\kappa_{*}^{2}}{m}e^{-2\pi n/\vert s_{0}\vert}e^{\Delta(\xi)/\vert s_{0}\vert},\label{eq:UniversalFormula}
\end{equation}
where $n$ is an integer labelling the states, and $\kappa_{*}$ is
the binding wave number at unitarity of the state $n=0$. The value
of $\kappa_{*}$ is set by the three-body boundary condition, so that
it may be regarded as a representation of the three-body parameter.
A change in the value of $\kappa_{*}$ simply scales the curves in
figure~\ref{fig:EfimovPlot} inwards or outwards from the accumulation
point ($a^{-1}=0$, $E=0$).

The universal function $\Delta(\xi)$ has been determined numerically
and approximated by analytical expressions in reference~\cite{Braaten2003}.
This numerical approximation has inaccuracies on the order of 3\%
for $\xi$ close to $-\pi$, and we give here an improved version:
\begin{equation}
\Delta(\xi)=\begin{cases}
-0.825-0.05z-0.77z^{2} & \mbox{for }\xi\in\big[-\pi,-\frac{5}{8}\pi\big]\\
\qquad\quad+1.26z^{3}-0.37z^{4}\\
2.11y+1.96y^{2}+1.38y^{3} & \mbox{for }\xi\in\big[-\frac{5}{8}\pi,-\frac{3}{8}\pi\big]\\
6.027-9.64x+3.14x^{2} & \mbox{for }\xi\in\big[-\frac{3}{8}\pi,-\frac{\pi}{4}\big]
\end{cases}\label{eq:AnalyticApproximation}
\end{equation}
where $z=\xi+\pi$, $y=\xi+\pi/2$, and $x=\sqrt{-\xi-\pi/4}$.

The discrete scale invariance not only holds for the three-body bound
states, but for the whole three-body spectrum including the scattering
continua~\cite{Wang2010,Wang2011c}. In the three-body scattering
continuum for $a<0$, resonances that arise at threshold from the
three-body bound states and persist up to energies $\sim2\hbar^{2}/(ma^{2})$
also exhibit a discrete-scale-invariant pattern with the scaling factor
$\lambda_{0}$~\cite{Wang2011c}. These Efimov resonances have been
used to evidence Efimov states in experiments with ultra-cold atoms,
as discussed in section~\ref{subsec:Ultracold-atoms}. On the other
hand, for $a>0$, the disappearance of the three-body bound states
below the particle-dimer threshold does not lead to trimer resonances
but trimer virtual states, similarly to the disappearance of the two-body
bound state below the two-body threshold. 

\subsection{Finite-range interactions\label{subsec:Finite-range-interactions}}

The Efimov effect has been confirmed by many calculations using finite-range
interactions~\cite{Thogersen2008,Grinyuk2003}. In particular, a
series of trimer states is obtained in these calculations. One important
aspect of systems with finite-range interactions is that the discrete
scale invariance is necessarily broken below some distance comparable
with the range of interactions. As a result, the spectrum is bounded
from below, as it should be physically: the series of trimer states
starts from a ground state.

\begin{figure*}[t]
\hfill{}\includegraphics[scale=0.75]{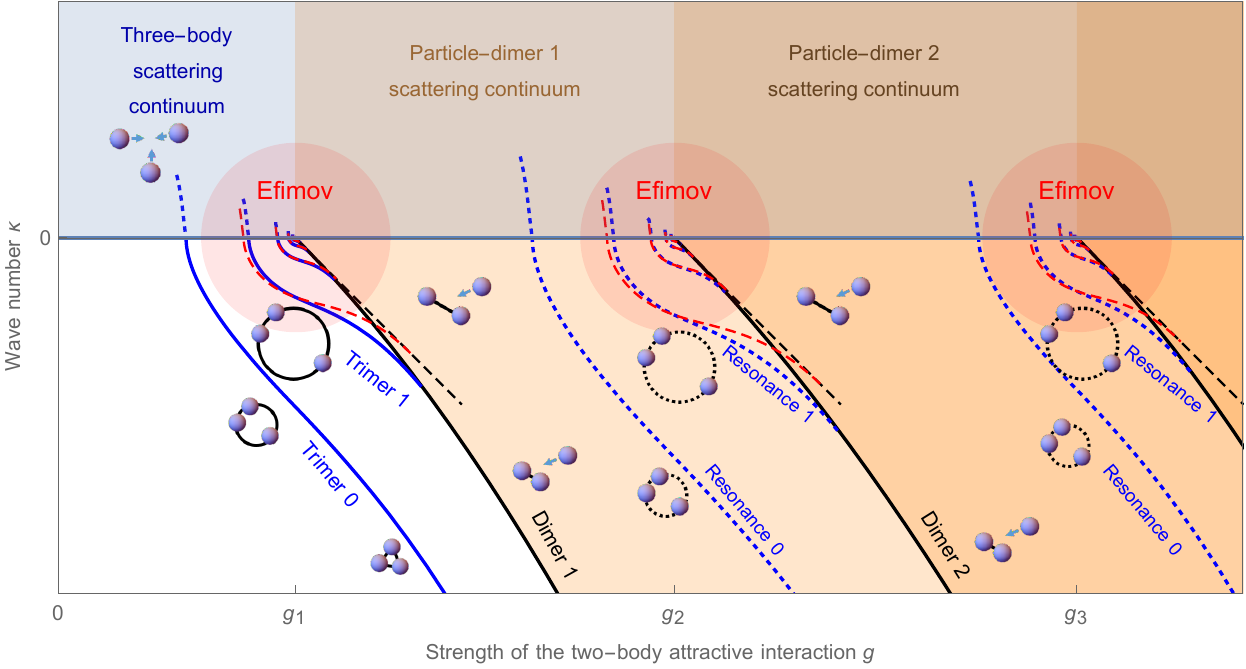}\hfill{}

\caption{\label{fig:EfimovPlotFiniteRange}Schematic plot of the three-body
spectrum of three identical bosons of mass $m$, as the strength $g$
of their two-body attractive interaction $gV(r)$ is increased. For
clarity, the wave number $\kappa=E\sqrt{m/(\hbar\vert E\vert)}$ is
represented instead of the energy $E$, and only states with zero
angular momentum are shown. At a certain interaction strength $g_{1}$,
an $s$-wave two-body bound state (solid black curve labelled as dimer~1)
appears, whose binding energy increases with increasing strength.
At larger strengths $g_{2}$, $g_{3}$, etc., a second, third, etc.,
$s$-wave two-body bound states (solid black curves labelled as dimer~2,
etc.) appear. Just before the appearance of the first two-body bound
state, an infinite set of three-body bound states emerge, indicated
by solid blue curves. Just before the appearance of the other two-body
bound states, a similar set of three-body states appear, indicated
by dotted curves. These states are not true bound states, but resonant
states embedded in the continua of scattering states between a particle
and a deeper two-body bound state. These continua are indicated by
the shaded areas above the curves corresponding to the two-body bound
states. Around each appearance of a two-body bound state (the ``two-body
resonances''), the inverse scattering length $1/a$ is proportional
to $g-g_{i}$. As a result, the three-body states follow the Efimov
spectrum of figure~\ref{fig:EfimovPlot}, here shown in dashed red
curves. These regions of good agreement with the zero-range theory
are called ``Efimov windows of universality'' and indicated by red
discs. Away from these regions, the two-body and three-body bound
states significantly deviate from the ideal Efimov spectrum. }
\end{figure*}

Finite-range calculations show that in the universal window where
the scattering length $a$ is much larger than the range $b$ of interactions,
and the wave number $k$ associated with the three-body energy $\hbar^{2}k^{2}/m$
is much smaller than the inverse range $b^{-1}$, the dimer and trimers
follow closely the zero-range theory predictions, in particular the
energy spectrum follows the discrete scale-invariant Efimov spectrum
given by equation~(\ref{eq:UniversalFormula}). Outside this window,
however, the spectrum deviates from the universal predictions of the
zero-range theory. Typically, the ground-state trimer shows marked
deviations from the universal spectrum and does not meet the particle-dimer
threshold, which can be understood from a variational argument~\cite{Bruch1973}.
The first excited state may also show some deviations near the particle-dimer
threshold, which it approaches closely, following the Efimov scenario,
but does not necessarily meet~\cite{Mestrom2016}. These features
are summarised in figure~\ref{fig:EfimovPlotFiniteRange} for a three-body
system whose two-body attractive potential $V(r)$ is gradually scaled
by a strength factor $g$, enabling to change the scattering length
and make it resonant for certain values of $g$.

\paragraph{Finite-range corrections}

Experimental observations often lie on the border of the universal
window, where the zero-range theory may not be accurate enough. Some
efforts have therefore been devoted to understanding the finite-range
corrections to the zero-range theory.

A first line of approach is based on the effective range theory~\cite{Bethe1949}.
At the two-body level, deviations from the zero-range theory involve
the \emph{effective range} $r_{e}$, which appears in the low-energy
expansion of the $s$-wave phase shift~\cite{Bethe1949}:
\begin{equation}
\frac{k}{\tan\delta_{0}(k)}=-\frac{1}{a}+\frac{1}{2}r_{e}k^{2}+o(k^{2})\label{eq:LowEnergyPhaseShift}
\end{equation}

The effective range $r_{e}$ is typically, but not always, on the
order of the range $b$ of the interactions. It is relatively straightforward
to take into account this two-body range correction into the three-body
problem. In the three-body Schr\"odinger formalism, one can apply
a generalised Bethe-Peierls boundary condition that replaces the scattering
length by the energy-dependent quantity $-k/\tan\delta_{0}(k)$. This
quantity also appears explicitly in the integrated Schr\"odinger
equation of the zero-range three-body problem, known as the Skorniakov-
Ter-Martirosian equation~\cite{Skorniakov1957}. One can therefore
use the low-energy expression of the phase shift given by equation~(\ref{eq:LowEnergyPhaseShift})
into these formalisms~\cite{Efimov1991,Efimov1993,Thogersen2009,Sorensen2013}.
This brings out a correction $\propto r_{e}/R^{3}$ to the Efimov
attraction of equation~(\ref{eq:EfimovAttraction})~\cite{Efimov1991,Efimov1993,Thogersen2009}.
One can use more elaborate expressions describing the energy dependence
of the phase shift over a wider range of energy~\cite{Nakajima2010,Massignan2008}.

It would be tempting to think that such a procedure regularises the
Thomas collapse problem of the zero-range theory and sets the three-body
parameter through the new length scale given by $r_{e}$~\cite{Massignan2008}.
It is indeed the case for a large and negative effective range, a
situation that arises in the case of narrow Feshbach resonances~\cite{Petrov2004}
- see section~\ref{subsec:Narrow-Feshbach-resonances}. However,
in general the procedure does not regularise the equations, and one
still has to impose a regularisation of the equations that introduces
a three-body parameter. Such an approach with a fixed three-body parameter
has not been quite successful in reproducing experimental data and
theoretical calculations with finite-range interactions; an energy
dependence of the three-body parameter is needed to reproduce these
results~\cite{Nakajima2010,Nakajima2011,Naidon2011}. A likely reason
is that equation~(\ref{eq:LowEnergyPhaseShift}) only accounts for
the range corrections of the phase shift, i.e. the on-the-energy-shell
scattering properties, which correspond to asymptotic properties of
two-body systems, but not the off-the-energy-shell properties which
correspond to their short-range correlations. In this respect, separable
potentials~\cite{Yamaguchi1954} are useful tools to account for
finite-range effects, since they can reproduce both on- and off-the-energy-shell
finite-range effetcs, while keeping the simplicity of the zero-range
theory~\cite{Naidon2012a,Naidon2014} - see Appendix for details.

An alternative and more systematic approach to range corrections is
based on the effective field theory~\cite{Platter2009,Ji2010}. Effective
field theory~\cite{Bedaque1999} is the effective theory that one
can write at low energy respecting the basic symmetries of the systems.
In this framework, the ratio $b/\vert a\vert$ of the range of interaction
over the scattering length can be treated as an expansion parameter.
The leading order in this expansion reproduces the zero-range theory~\cite{Bedaque1999}.
Calculations to the next-to-leading order have been performed in Refs.~\cite{Platter2006,Platter2006a,Ji2010,Ji2012}
and show the necessity to introduce a second three-body parameter
to renormalise the equation at this order.

A more recent approach~\cite{Kievsky2013,Garrido2013,Kievsky2014}
based on numerical calculations with model potentials has provided
an empirical way to reproduce range corrections to the zero-range
theory. These works show that finite-range deviations from universal
formulas such as equation~(\ref{eq:UniversalFormula}) can be accounted
for to a good accuracy over a wide range of scattering length and
energy by simply replacing the scattering length $a$ by a length
$a_{B}$, and shifting the three-body parameter by a quantity inversely
proportional to $a$. The length $a_{B}$ is defined as the value
$\kappa^{-1}$ that is the solution of $\tan\delta_{0}(i\kappa)=-i$,
corresponding to the pole of the scattering amplitude $f(k)=(k/\tan\delta_{0}(k)-ik)^{-1}$,
provided that an analytic continuation to imaginary $k$ is possible.
For $a>0$, the energy $-\frac{\hbar^{2}}{ma_{B}^{2}}$ therefore
coincides with the two-body bound-state energy, while for $a<0$ it
corresponds to the energy of a virtual bound state, since there is
no physical bound state. This procedure has been used to fit theoretical
results obtained with finite-range interactions, as well as experimental
data obtained for lithium-7~\cite{Kievsky2014,Gross2011}. According
to this procedure, the universal formula~(\ref{eq:UniversalFormula})
for the trimer energy is modified as follows (changes are emphasised
in red),
\begin{equation}
\vert E^{(n)}\vert+\frac{\hbar^{2}}{m{\color{red}a_{B}^{{\normalcolor 2}}}}=\frac{\hbar^{2}\left(\kappa_{*}{\color{red}+\Gamma_{n}/a}\right)^{2}}{m}e^{-2\pi n/\vert s_{0}\vert}e^{\Delta(\xi)/\vert s_{0}\vert}.\label{eq:ModifiedUniversalFormula}
\end{equation}
Equivalently, the finite-range energy curve can be mapped to the original
Efimov curve by plotting the renormalised energy $E^{\prime(n)}=\lambda_{n}^{2}E^{(n)}$
(or wave number $\kappa^{\prime(n)}=\lambda_{n}\kappa^{(n)}$) as
a function of the renormalised inverse scattering length $a^{\prime-1}=\lambda_{n}a_{B}^{-1}$
with the $a$-dependent renormalisation coefficient $\lambda_{n}=\left(1+\Gamma_{n}/(\kappa_{*}a)\right)^{-1}$.
An example of such mapping will be shown in the case of helium-4 in
section~\ref{subsec:Helium-4}. 

The replacement $a\to a_{B}$ is related to the two-body range correction
given by equation~(\ref{eq:LowEnergyPhaseShift}). Indeed, according
to the definition of $a_{B}$ and to equation~(\ref{eq:LowEnergyPhaseShift}),
one has:
\begin{eqnarray}
\frac{1}{a_{B}} & \approx & r_{e}^{-1}\left(1-\sqrt{1-2r_{e}/a}\right)\label{eq:aB}\\
 & \approx & \frac{1}{a}\left(1+\frac{1}{2}\frac{r_{e}}{a}+\dots\right)
\end{eqnarray}
In contrast, the shift $\Gamma_{n}/a$ is a range correction to the
three-body parameter,
\[
\kappa_{*}^{\prime}=\kappa_{*}\left(1+\frac{\Gamma_{n}/\kappa_{*}}{a}+\dots\right)
\]
that is likely associated with two- and three-body short-range correlations.
The form of this shift was recently justified from effective-field
theory~\cite{Ji2015}, but the value of $\Gamma_{n}$ has so far
been determined only numerically for each value of $n$ to reproduce
finite-range calculations. These results suggest that, with the introduction
of the parameters $r_{e}$ and $\Gamma_{n}$ characterising finite-range
corrections, the universality of Efimov physics may be extended beyond
the window of validity of the zero-range theory.

\subsection{Other interactions\label{subsec:Long-range-interactions}}

\subsubsection{Coulomb interactions\label{subsec:Coulomb-interactions}}

Electrically charged particles are subjected to the Coulomb interaction.
It is a long-range interaction, whose potential decays as $1/r$,
thus more slowly than $1/r^{3}$. For such interactions, there is
no range beyond which the particles effectively cease to interact.
Therefore, there is no Efimov physics associated with Coulomb interactions
themselves. However, particles interacting with short-range interactions
may also interact with additional Coulomb interactions due to their
electric charge. Such is the case of protons or nuclei, which interact
through the short-range nuclear forces as well as the repulsive Coulomb
interaction. If the short-range interactions are resonant, there is
an expected interplay between the $1/R^{2}$ Efimov attraction~(\ref{eq:EfimovAttraction})
and Coulomb forces. 

To our knowledge, this interplay has not been studied explicitly,
due to the technical difficulties in solving the three-body problem
with Coulomb interactions~\cite{Deltuva2005}. Nevertheless, some
simple considerations can be made, as discussed by Vitaly Efimov in
his original paper~\cite{Efimov1970a}. Since the Coulomb potential
decays as $1/r$ and the Efimov attraction decays as $1/R^{2}$, the
latter dominates at short distances and the former dominates at large
distances, breaking the scaling invariance. The distance where this
transition occurs is given by the Bohr radius 
\begin{equation}
a_{C}=\frac{\hbar^{2}}{mk_{e}q^{2}},\label{eq:BohrRadius}
\end{equation}
where $m$ is the mass of particles, $q$ is their electric charge,
and $k_{e}=1/4\pi\epsilon_{0}$ is Coulomb's constant. On the other
hand, the Efimov attraction exists only beyond the range $b$ of the
short-range resonant interaction. Therefore, the window of existence
for the Efimov attraction is delimited by the range $b$ of the short-range
forces and the Bohr radius $a_{C}$. Within this window, Efimov states
can be bound by the Efimov attraction, and their number scales as
$\ln(a_{C}/b)$. A necessary condition for the existence of Efimov
states in these systems is thus
\begin{equation}
b<a_{C}.\label{eq:ConditionForEfimovInCoulombSystems}
\end{equation}

The single-particle problem in a sum of $1/r^{2}$ and $1/r$ potentials
was treated quantitavely in reference~\cite{Hammer2008a}, and corroborates
these qualitative considerations.

\subsubsection{Dipolar interactions\label{subsec:Dipolar-interactions}}

\hspace{0.5cm}Even if particles are electrically neutral, they may
possess an electric or magnetic dipole moment that creates a dipole-dipole
interaction between them. In the ultra-cold atom research community,
there has been a growing interest in studying particles interacting
via dipole-dipole interactions. Prime examples are atoms with a large
magnetic dipole moment $\mu$, such as $^{52}$Cr ($\mu=6\mu_{B}$,
where $\mu_{B}\approx9.274\times10^{-24}\mbox{J}.\mbox{T}^{-1}$ is
the Bohr magneton)~\cite{Griesmaier2005,Stuhler2005}, $^{164}$Dy
($\mu=10\mu_{B}$)~\cite{Lu2011}, and $^{168}$Er ($\mu=7\mu_{B}$)~\cite{Aikawa2012,Frisch2014}.
Systems with dipolar interactions can also be realised with polar
molecules that possesses a permanent electric dipole moment $d$~\cite{Doyle1995,Ni2010,Neyenhuis2012}.
For these atoms and molecules, the strength and the polarisation direction
of the dipole interaction can be controlled by external magnetic or
electric fields, aligning the dipoles in certain directions.

For two particles with dipole moments induced by an external field
and aligned along the vertical direction, the dipole-dipole interaction
potential at large distance has the form
\begin{equation}
V(r,\theta)=d^{2}\frac{1-3\cos^{2}\theta}{r^{3}},\label{eq:DipoleDipoleInteraction}
\end{equation}
where $d$ is the dipole moment (expressed in units of $\text{m}^{5/2}\cdot\text{kg}^{1/2}\cdot\text{s}^{-1}$),
$r$ is the distance, and $\theta$ is the polar angle between the
two particles. It is thus an anisotropic interaction with a long-range
tail. While it is seemingly more complicated than isotropic short-range
interactions discussed in sections~\ref{subsec:Efimov-theory} and
\ref{subsec:Finite-range-interactions}, the two-body physics of dipoles
turns out to exhibit the same universal behaviour as that of short-range
interactions around the threshold regime at which two dipoles are
about to form an $s$-wave dominated bound state~\cite{Marinescu1998,Roudnev2009,Ticknor2010}.
Close to such $s$-wave dominated resonances, the coupling between
different partial waves induced by the dipole interaction occurs at
a distance much smaller than the spatial extent of the bound state
since the coupling decays as $1/r^{6}$. One can therefore essentially
consider a single channel scattering in the $s$-wave channel, in
which the dipole interaction averages out to zero and one is left
with a short-range interaction in this channel. The calculation of
this interaction in second-order perturbation theory (through the
coupling of the $s$ wave to the $d$ wave) shows that it decays as
$-C_{4}/r^{4}$ with $C_{4}=4\hbar^{2}\ell_{4}^{2}/m$ and $\ell_{4}\approx0.365\ell_{d}$~\cite{Bohn2009}\footnote{The value of $\ell_{4}\approx0.365\ell_{d}$ is derived from equation~(34)
in reference~\cite{Bohn2009}, as we could not reproduce the value
$2\ell_{4}=1.09\ell_{d}$ given after equation~(34) of the same reference. }, where $\ell_{d}$ is the dipole length defined as 
\begin{equation}
\ell_{d}=\frac{md^{2}}{2\hbar^{2}}.\label{eq:DipoleLength}
\end{equation}

Because of this similarity of the dipole-dipole interaction with a
$1/r^{4}$ short-range attraction in the $s$ channel, the two- and
three-body physics near an $s$-wave dominated two-body resonance
shows the same universal behaviour as that of short-range interactions.
Namely, near an $s$-wave dominated resonance, the dimer energy scales
with the scattering length according to the universal formula~(\ref{eq:UniversalDimerEnergy})~\cite{Ticknor2010},
and the Efimov effect occurs~\cite{Wang2011a}. In reference~\cite{Wang2011a},
Yujun Wang and co-workers have considered the three-body problem of
identical bosons with the dipole-dipole interaction in the proximity
of the $s$-wave dominated resonance. They have found the appearance
of Efimov states with the same universal scaling factor $22.7$ as
that of systems with short-range interactions. One notable feature
of the dipolar Efimov states is that their three-body parameter is
universally set by the dipole length,
\begin{equation}
\kappa_{*}^{(0)}=0.173(2)\,\ell_{d}^{-1},\label{eq:DipoleThreeBodyParameter}
\end{equation}
in the absence of other forces at distances comparable with the dipole
length. The insensitivity of the three-body parameter to forces at
shorter distances than the dipole length is due to the strong repulsion
created by the partial wave couplings induced by the dipole interaction.
This repulsion appears at a distance on the order of the dipole length.
It prevents the three dipoles from getting closer than the dipole
length and renders the three-body parameter universal. 

As we shall see in sections~\ref{subsec:Van-der-Waals-Universality}
and \ref{subsec:Other-types-of-interactions}, a three-body repulsion
also appears in systems with isotropic short-range interactions, in
particular power-law decaying interactions, and makes in some limit
the three-body parameter universally determined by the effective range.
It would be therefore tempting to think that the three-body universality
of the dipole -dipole interaction is related to the three-body universality
of its effective $1/r^{4}$ interaction in the $s$ wave channel.
However, the universal three-body parameter for such an $1/r^{4}$
interaction has been estimated to $\kappa_{*}^{(0)}\approx0.174\,r_{4}^{-1}\approx0.48\,\ell_{d}^{-1}$~\cite{Naidon2014},
which is almost a factor of three different from the value in equation~(\ref{eq:DipoleThreeBodyParameter}).
This indicates that the dipole-dipole interaction belongs to a different
class of three-body universality that involves the explicit partial
wave coupling at short distance. We will discuss the universality
classes of the three-body parameter for isotropic short-range interactions
in more detail in section~\ref{subsec:Other-types-of-interactions}.

The dipolar three-body physics leads to an even more interesting behaviour
when the particles are identical fermions. While identical fermions
do not exhibit Efimov physics (see section~\ref{sec:Three-identical-fermions}),
Yujun Wang and co-workers have found in reference~\cite{Wang2011b}
that there exists a new type of three-body bound state, which is universally
described by the dipole interaction. The size of this bound state
is of the order of the dipole length, and it has a shape of an obtuse
isosceles triangle, whose longer side is parallel to the polarisation
axis of the dipoles. This particular shape originates from a competition
between the dipole interaction and the Pauli exclusion principle,
maximising the attraction between the dipoles by aligning them in
parallel while preserving the antisymmetrisation condition by having
nodes between them.

\subsubsection{Inverse-square interactions and generalised Efimov effect\label{subsec:Inverse-square-interactions}}

The scale invariance of Efimov physics comes from the $1/r^{2}$ dependence
of the Efimov attraction, which scales like the kinetic energy. This
long-range three-body attraction is remarkable because it originates
from short-range two-body interactions. If the two-body interactions
are not short-ranged but have an attractive $1/r^{2}$ dependence
themselves, then the three-body system is also expected to feature
an effective scale-invariant three-body attraction. In such systems,
however, the long-range nature of the forces is set by construction,
and does not emerge from short-range forces, as in the Efimov effect.
In this sense, they may not be considered to be related to the Efimov
effect. On the other hand, an interesting question about these systems
is whether they exhibit a discrete scale invariance that requires
an extra length scale, as in the Efimov effect. Concretely, the question
is whether the $-\alpha/r^{2}$ two-body potentials are strong enough
to support two-body and three-body bound states. For two-body systems,
$\alpha$ must be larger than $\alpha_{2}=\hbar^{2}/(8\mu)$, where
$\mu$ is the particles\textquoteright{} reduced mass. In this case,
there is an infinity of two-body bound states whose energy spectrum
forms a geometric series as in the Efimov effect. This breaks the
scaling invariance into a discrete scaling invariance and requires
the knowledge of a microscopic length scale\footnote{As in the Efimov effect, this length scale originates from the short-range
details of the interactions, which inevitably depart from the inverse
square form assumed at larger distance, since the purely inverse-square
potential has no ground state in this case and is therefore unphysical.}. Likewise, the three-body spectrum features an infinite number of
bound states or resonances below each two-body bound state. The problem
becomes particularly interesting for $\alpha<\alpha_{2}$, for which
the attraction is not strong enough to bind two particles, but may
be strong enough to bind three particles. In this case, the energy
spectrum of these systems resembles that of Efimov systems, in the
sense that an infinity of discrete-scale-invariant three-body states
may exist despite the absence of two-body bound states. For this reason,
Sergej Moroz, José D'Incao and Dmitry Petrov~\cite{Moroz2015} have
advocated a generalised definition of the Efimov effect, as ``the
emergence of discrete scaling symmetry in a three-body problem if
the particles attract each other via a two-body scale-invariant potential'',
regardless of the short- or long-range nature of this potential. Situation
in 3D The situation of three identical particles interacting via inverse-square
potentials was studied by Nicolais L. Guevara, Yujun Wang, and Brett
D. Esry \cite{Guevara2012}. For identical bosons, they found that
the generalised Efimov effect can occur slightly below the critical
strength required for binding two particles, namely for $\alpha_{3}<\lesssim\alpha<\alpha_{2}$,
with $\alpha_{3}=0.97\alpha_{2}$. In this range, an infinite family
of three-body bound states exists. Their energies form an almost geometric
series, as the effective three-body potential turns out to be almost
but not exactly an inverse-square potential. One must note that these
states are extremely weakly bound, and their scaling ratio is typically
on the order of $10^{10}$. For $\alpha>\alpha_{2}$, an infinite
number of two-body bound states arise, and the three-body bound states
remain below the lowest two-body bound state, while new families of
three-body resonances exist below each two-body bound state. Interestingly,
similar results were obtained for three identical fermions in the
$1^{+}$ symmetry, with $\alpha_{2}=9\hbar^{2}/(8\mu)$ and $\alpha_{3}=0.82\alpha_{2}$,
whereas the standard Efimov effect with short-range interactions does
not apply to identical fermions (see section~\ref{sec:Three-identical-fermions}).

\paragraph{Situation in 1D}

In one dimension, the problem of three identical particles interacting
via attractive inverse square potentials was solved analytically by
F. Calogero~\cite{Calogero1969} who found that the system remains
scale-invariant for $\alpha<\alpha_{2}$, i.e. there is no generalised
Efimov effect. However, the situation changes if one considers a particle
interacting via an inverse square two-body potential with two identical
particles. In this case, Sergej Moroz, José D'Incao and Dmitry Petrov~\cite{Moroz2015}
found that for any value of $\alpha_{2}/2<\alpha<\alpha_{2}$, the
generalised Efimov effect occurs for a sufficiently large mass ratio.
It can also occur for smaller values of $\alpha$ by a fine tuning
of the short-range details of the two-body interactions. As a possible
physical realisation, the authors have proposed a system of two polar
molecules interacting with an electron, all confined along a line.

\paragraph{Connection with the standard Efimov effect}

In addition to the generalised definition of the Efimov effect, systems
with $1/r^{2}$ interactions may also be interesting from the point
of view of the standard Efimov effect. Indeed, $N$-body systems with
$1/r^{2}$ pairwise interactions could constitute in some limit an
approximation of systems of $N$ particles undergoing a $1/r^{2}$
Efimov attraction that is induced by their resonant (short-range)
interaction with a lighter particle (see section~\ref{subsec:2-Identical-particles+1particle}).
This limit requires that the $N+1$ system may be treated in the Born-Oppenheimer
approximation (the light particle being much lighter than the $N$
particles) and that the resulting Born-Oppenheimer potential between
the $N$ particles may be approximated by a sum of pairwise $1/r^{2}$
attractive potentials. This idea was introduced in reference~\cite{Guevara2012},
where the authors argued that the four-body Efimov states found in
a 3 heavy fermions + 1 particle system (see reference~\cite{Castin2010}
and section~\ref{subsec:3-fermions-+1particle}) may be described
by the $1/r^{2}$ interaction model of three identical fermions. As
mentioned above, this model indeed predicts the existence of a geometric
series of three-body bound states in the $1^{+}$ symmetry above a
critical strength $\alpha_{3}$, which would correspond to a mass
ratio of 11.58 in the 3+1 system. This is qualitatively similar to
the appearance of four-body bound states in the 3+1 system above the
critical mass ratio 13.384~\cite{Castin2010}. The authors thus interpret
the four-body Efimov effect found in reference~\cite{Castin2010}
as a three-body generalised Efimov effect for inverse-square interactions,
that originate themselves from the three-body Efimov effect. It is
however unclear to what extent this appealing picture is valid, since
the Born-Oppenheimer potential between the three heavy particles is,
strictly speaking, different from the sum of pair-wise $1/r^{2}$
interaction. In particular, applying the same approximation to the
3 heavy bosons + 1 particle system, one expects an infinite number
of four-body bound states tied to each Efimov trimer state, whereas
numerical studies have so far found at most two four-body bound states
(see section~\ref{subsec:4-body-states-associated} and figure~\ref{fig:Three+One}). 

\subsection{Relativistic case\label{subsec:Relativistic-case}}

The Efimov effect, seen as the infinite accumulation of three-body
bound states with smaller and smaller binding energies in the three-body
spectrum, is by definition a low-energy phenomenon. It is thus not
directly affected by relativity. Nevertheless, if the range of interactions
between particles is smaller than their Compton wave length, relativistic
corrections may affect the most deeply bound Efimov states and the
three-body parameter. In particular, when the range of interactions
is so small that they can be approximated by contact interactions,
there is still a length scale in the relativistic theory, the Compton
wave length, that may prevent the Thomas collapse and set the three-body
parameter, instead of the interaction itself.

The first authors to look at the Efimov effect in a relativistic framework
were James V. Lindesay and H. Pierre Noyer in the 1980s~\cite{Lindesey1980,Lindesey1986}.
They considered three bosons of rest mass $M$, interacting with attractive
contact interactions such that the total energy $M_{2}$ of two particles
may be less than $2M$. They obtained the following integral equation\footnote{Note that in both references \cite{Lindesey1980,Lindesey1986}, the
equation has the wrong factor $4\pi$ instead of $1/\pi$. In the
first paper \cite{Lindesey1980}, the numerical calculations were
performed with this wrong factor and are therefore incorrect for identical
bosons.} for the three-body energy $M_{3}$: 

\begin{eqnarray}
W(k) & \!\!\!\!\!\!=\!\!\!\!\!\! & -\frac{1}{\pi}\int_{0}^{\frac{M_{3}^{2}-M^{2}}{2M_{3}}}\frac{dk^{\prime}}{\varepsilon^{\prime}}\frac{k^{\prime}}{k}\frac{\sqrt{s^{\prime}}}{\frac{1}{a}-\sqrt{M^{2}-\frac{s^{\prime}}{4}}}\nonumber \\
 &  & \times\ln\left(\frac{\sqrt{M^{2}+(k+k^{\prime})^{2}}+\varepsilon+\varepsilon^{\prime}-M_{3}}{\sqrt{M^{2}+(k-k^{\prime})^{2}}+\varepsilon+\varepsilon^{\prime}-M_{3}}\right)W(k^{\prime})\nonumber \\
\label{eq:Lindesay-Noyer}
\end{eqnarray}
where $\varepsilon=\sqrt{k^{2}+M^{2}}$, $\varepsilon^{\prime}=\sqrt{k^{\prime2}+M^{2}}$,
$s^{\prime}=M_{3}^{2}+M^{2}-2M_{3}\varepsilon^{\prime}$, and the
scattering length $a$ is given by $1/a=\pm\sqrt{\vert M^{2}-M_{2}^{2}/4\vert}$
where $\pm$ is the sign of $2M-M_{2}$. For $k/M\ll1$ and $\vert3M-M_{3}\vert\ll M$,
one retrieves the non-relativistic integral equation. However, unlike
the nonrelativistic equation, the integral above has a finite upper
limit of integration, which comes from the relativistic kinematics.
This prevents the Thomas collapse and the three-body energy is set
by the rest mass $M$ (or equivalently the Compton wavelength $h/Mc$).
From this equation, the authors concluded that the Efimov scenario
is qualitatively unchanged. The ground-state trimer appears for $M_{2}\approx2.006M$
and its energy at unitarity ($M_{2}=2M$) is $\approx2.988M$, i.e.
it is bound by an energy $\approx0.0122M$ with respect to the three-body
threshold $3M$. It disappears below the particle-dimer threshold
at $M_{2}\approx1.03M$. The first excited trimer is bound by $2.41\times10^{-5}M$
at unitarity, which is about a factor 507 from the ground state, relatively
close to the non-relativistic scaling ratio $e^{2\pi/\vert s_{0}\vert}\approx515$
for excited states.

\begin{figure*}[t]
\includegraphics[scale=0.8]{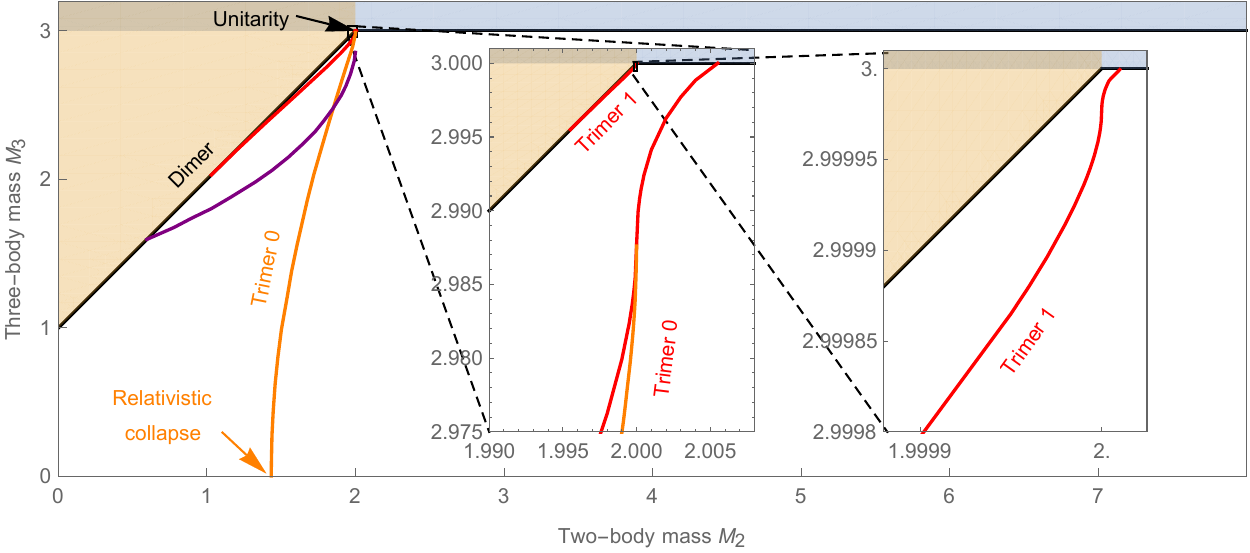}

\caption{\label{fig:Relativistic}Relativistic Efimov spectrum: mass $M_{3}$
of the three-body state as a function of the mass $M_{2}$ of the
two-body state in units of the particles' mass $M$. The two-body
state is bound for $M_{2}<2M$. Note that the direction of the horizontal
axis is inverted with respect to that of Figs.~\ref{fig:EfimovPlot}
and \ref{fig:EfimovPlotFiniteRange}. The horizontal line shows the
three-body threshold at $M_{3}=M+M+M$ and the diagonal line shows
the dimer-particle threshold $M_{3}=M_{2}+M$. The three-body states
are bound below these two thresholds. At their intersection is the
unitary point ($M_{2}=2M,\,M_{3}=3M$) below which the Efimov states
accumulate, as shown in the insets. The curve in red shows the results
of reference~\cite{Lindesey1986}, based on equation~(\ref{eq:Lindesay-Noyer}).
The curve in purple shows the results of reference~\cite{Frederico1992},
based on equation~(\ref{eq:Frederico}) with $x_{\tiny\mbox{min}}=M^{2}/M_{3}^{2}$
and $q_{\tiny\mbox{max}}=\sqrt{(1-x^{\prime})(M_{3}^{2}x^{\prime}-M^{2})}$.
The curve in orange shows the result of reference~\cite{Carbonell2003},
based on equation~(\ref{eq:Frederico}) with $x_{\tiny\mbox{min}}=0$
and $q_{\tiny\mbox{max}}=\infty$. In the insets, only the excited
trimer obtained from equation~(\ref{eq:Lindesay-Noyer}) is shown.}
\end{figure*}
The same problem was independently addressed a few years later by
Tobias Frederico~\cite{Frederico1992} in the light-front dynamics
formalism~\cite{Frederico2011}. The author derived the following
integral equation,
\begin{eqnarray}
\Gamma(q,x) & = & F(M_{12})\frac{1}{(2\pi)^{3}}\int_{x_{\mbox{\tiny min}}}^{1-x}\frac{dx^{\prime}}{x^{\prime}(1-x-x^{\prime})}\nonumber \\
 &  & \times\int_{0}^{q_{\mbox{\tiny max}}}\frac{d^{2}q^{\prime}}{\mathcal{M}^{2}-M_{3}^{2}}\Gamma(q^{\prime},x^{\prime})\label{eq:Frederico}
\end{eqnarray}
where $F(M_{12})=8\pi^{2}\left(\frac{\arctan y_{M_{12}}}{y_{M_{12}}}-\frac{\arctan y_{M_{2}}}{y_{M_{2}}}\right)^{-1}$
with $y_{m}=\frac{m}{\sqrt{4M^{2}-m^{2}}}$ and $M_{12}^{2}=(1-x)M_{3}^{2}-\frac{q^{2}+(1-x)M^{2}}{x}$,
and $\mathcal{M}^{2}=\frac{q^{\prime2}+M^{2}}{x^{\prime}}+\frac{q^{2}+M^{2}}{x}+\frac{(\vec{q}^{\prime}+\vec{q})^{2}+M^{2}}{1-x-x^{\prime}}$.
Here, the integral boundaries are set to $x_{\tiny\mbox{min}}=M^{2}/M_{3}^{2}$
and $q_{\tiny\mbox{max}}=\sqrt{(1-x^{\prime})(M_{3}^{2}x^{\prime}-M^{2})}$.
Like equation~(\ref{eq:Lindesay-Noyer}), this equation also reduces
to the non-relativistic integral equation in the non-relativistic
limit. Solving this equation, the author reached conclusions similar
to those of Refs.~\cite{Lindesey1980,Lindesey1986}, with relatively
different numerical results\footnote{Note that the numerical results of reference~\cite{Frederico1992}
are not converged, as mentioned in reference~\cite{Carbonell2003}.
We give here the numerical values from reference~\cite{Carbonell2003}}. In particular, the ground-state trimer is bound by about $\approx0.2M$
at unitarity and disappears below the particle-dimer threshold at
$M_{2}\approx0.6M$.

In a more recent work using a similar formalism~\cite{Carbonell2003},
Jaume Carbonell and V. A. Karmanov argued that for zero-range interactions
the boundaries of the integrals assumed in the previous work \cite{Frederico1992}
should be changed to $x_{\tiny\mbox{min}}=0$ and $q_{\tiny\mbox{max}}=\infty$.
This results in a drastically smaller binding energy at unitariy $\approx0.012M$
for the ground state, which is remarkably close to the results of
Lindesay and Pierre Noyer in reference~\cite{Lindesey1986}. In addition,
the ground-state trimer does not disappear below the particle-dimer
threshold. Instead, its energy vanishes at $M_{2}\approx1.43M$, which
the authors called the ``relativistic Thomas collapse''. For a smaller
mass $M_{2}$ than this critical value (i.e. a stronger two-body attraction),
the three-body energy $M_{3}$ is formally imaginary, making it unphysical.

The results of these works are summarised in figure~\ref{fig:Relativistic}.

\subsection{What is an Efimov state?\label{subsec:What-is-an-Efimov-state}}

Before we address the observations of Efimov states, we have to ask
ourselves what is an Efimov state, and what constitutes an experimental
evidence of such a state. The answer to these questions varies somewhat
from one person to the other. The major issue is whether the ground-state
trimer, which is the most likely to exist and most easily observable,
should be included or not in the series of Efimov states.

\subsubsection{Energy\label{subsec:EfimovEnergy}}

Let us review some proposed definitions of Efimov states based on
their energy spectrum.
\begin{enumerate}
\item In the strictest sense, Efimov states are evidence of the Efimov effect,
i.e. an accumulation point in the three-body spectrum at zero energy
as shown in figure~\ref{fig:EfimovPlot}. In this sense, one cannot
evidence a single Efimov state, one needs to exhibit several (at least
two) of these states and show that they follow the predicted discrete
scale invariant pattern. In this definition, the observation of the
ground-state trimer is not an evidence of an Efimov state. 
\item In another definition, one may allow a single state to qualify as
an Efimov state, if it can be shown that its variation with scattering
length follows qualitatively the universal curve obtained in the zero-range
theory and shown in figure~\ref{fig:EfimovPlot}, sometimes referred
to as the ``Efimov scenario''. Namely, the trimer has to appear from
the three-body threshold at some negative scattering length, and disappear
in (or approach closely) the particle+dimer threshold at some positive
scattering length. The ground-state trimer often remains far below
the particle+dimer threshold~\cite{Bruch1973,Esry1996,Giannakeas2016},
and thus does not qualify as an Efimov state in this definition~\cite{Lee2007}.
\item In their review article~\cite{Braaten2006}, Eric Braaten and Hans-Werner
Hammer advocated a broader definition: ``a trimer is defined to be
an Efimov state if a deformation that tunes the scattering length
to \textpm{} \ensuremath{\infty} moves its binding energy along the
universal curve''. In this definition, the trimer does not have to
meet the particle+dimer threshold on the positive scattering length
side. The ground state trimer is therefore usually an Efimov state
according to this definition.
\item The results of Refs.~\cite{Kievsky2013,Garrido2013} suggest a somewhat
related definition: a trimer is defined to be an Efimov state if the
trimer energy as a function of $a_{B}$ can be fitted by the modified
universal formula~(\ref{eq:ModifiedUniversalFormula}). The authors
of Refs.~\cite{Kievsky2013,Garrido2013} have presented numerical
evidence that close to unitarity the ground state is expected to be
an Efimov state in this definition. 
\end{enumerate}
The first two definitions are qualitative: a given state is either
an Efimov state or not. The last two definitions are less restrictive
and can be made quantitative: by comparing the energy with the universal
or modified universal formulas, one may quantify how much of an Efimov
state a given trimer is. The disadvantage of definitions 2, 3, and
4 is that they require the variation of the two-body scattering length,
which is not always possible experimentally if the interactions cannot
be controlled but are just set by nature.

From a physical point of view, it would be natural to say that a trimer
is an Efimov state if the Efimov attraction is present and necessary
to ensure its binding, although this point is difficult to characterise
experimentally, and even theoretically for real systems. 

We should mention that the notion of Efimov trimer is not restricted
to bound states and can be extended to resonant states. When the two-body
interaction potential supports several two-body bound states and one
of them has zero angular momentum and is very close to threshold,
it leads to a situation similar to what the Efimov theory predicts
for just one two-body bound state, except that the Efimov trimers
are resonant states that can dissociate into a particle and a deeper
two-body bound state. Such resonant states are shown as dotted curves
in figure~\ref{fig:EfimovPlotFiniteRange}. As long as this dissociation
is weak, the trimer resnonances are narrow and follow the Efimov scenario.
They can therefore be qualified as Efimov states. In fact, it is the
situation encountered in ultra-cold atomic gases - see section~\ref{subsec:Ultracold-atoms}.
For very strong losses by dissociation, the trimer resonances are
broad and a theoretical study indicates that the Efimov spectrum is
rotated in the complex energy plane~\cite{Werner2009}, where the
imaginary part of the energy correspond to the resonance width.

\begin{figure*}[t]
\hfill{}\includegraphics[viewport=0bp 80bp 1045bp 852bp,clip,scale=0.45]{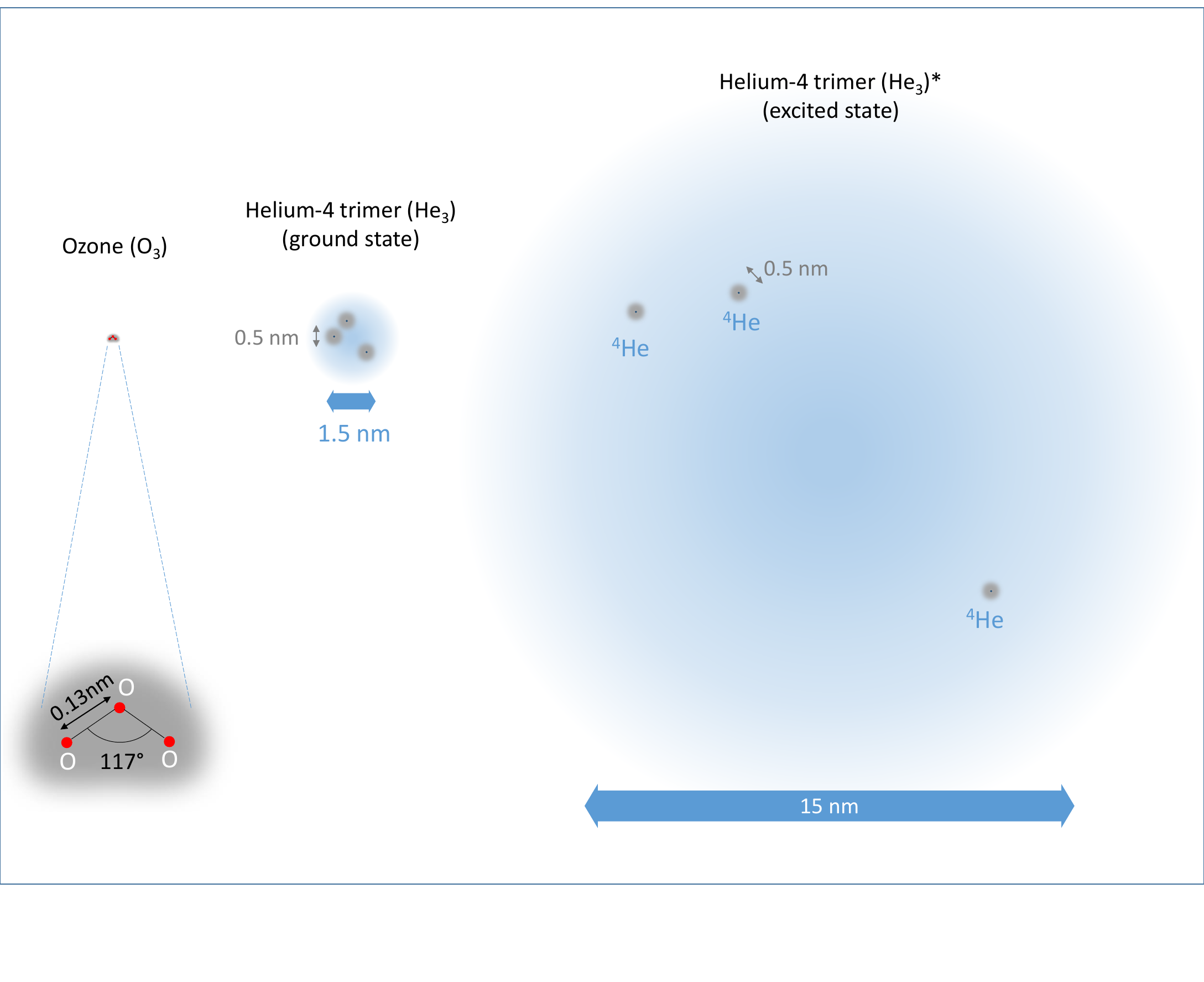}\hfill{}

\caption{\label{fig:Molecules}Ozone molecule (O$_{3}$) compared to Efimov
states of helium-4 atoms (He$_{3}$) on the same scale. Each molecule
is represented schematically as a typical snapshot of its geometry.
The ozone molecule is enlarged to show its structure (readers of the
electronic version are invited to zoom in). The nuclei are represented
by small dots (red for oxygen, blue for helium), and the electronic
cloud is shown as a grey halo. For the helium trimers, the atomic
cloud formed by the motion of the helium atoms is shown as a light
blue halo. The structure of the excited helium trimer is closest to
that of an ideal Efimov state, which is typically described by an
elongated triangle configuration. The structural properties of the
ground helium trimer do not conform fully to those of Efimov states,
but are nonetheless very different from more compact molecules like
O$_{3}$.}
\end{figure*}

\subsubsection{Structure\label{subsec:EfimovStructure}}

Although the peculiar energy spectrum of Efimov states is often presented
as their defining characteristic, their spatial structure is also
worth considering, as it makes them very different from other three-body
bound states such as water or ozone molecules. Ozone molecules, for
instance, have a relatively well defined geometry, with an O-O length
of about 0.127~nm and an O-O-O angle of about 117°. Although the
electrons are delocalised around the oxygen nuclei, the relative positions
of the nuclei are quite localised, as a result of the strong binding
interaction provided by the bonding electrons. Efimov states, on the
other hand, are very diffuse objects without a well defined geometry.

In the intuitive picture of the Efimov attraction, the particles keep
moving back and forth between one another, thereby inducing the Efimov
attraction that keeps them together. Thus, for an Efimov state made
of atoms, one may say that the atoms themselves play the role of bonding
electrons by performing an exchange motion. The electrons play a role
only when the atoms come in contact, within the radius of their electronic
cloud, which is on the order of a few tenths of nanometre for light
atoms. The motion of the atoms, on the other hand, occurs at distances
larger than the size of their electronic cloud, around three times
for the ground state and much more for excited states. As result,
Efimov states of atoms are much larger and diffuse than usual molecules.
This is shown in figure~\ref{fig:Molecules} for the case of helium-4
trimers, whose excited state is typically fifty times or more larger
than ozone molecules.

In addition to the broad distribution of sizes, there is also a broad
distribution of geometries for three particles forming an Efimov state.
The most probable geometries in this distribution correspond to elongated
triangles, where two particles are relatively close and the third
one is farther away. This can be checked at unitarity from the hyper-angular
wave function in equation~(\ref{eq:HyperangularSolution}), which
peaks at $\alpha=0$, corresponding to two particles in contact with
the third particle away.

These distinctive structural properties can be used to experimentally
characterise a trimer as an Efimov state, when structural properties
can be measured. This was recently demonstrated in the Coulomb explosion
imaging of helium-4 trimers~\cite{Voigtsberger2014,Kunitski2015}
- see section~\ref{subsec:Helium-4}. This experiment revealed that
the distribution of geometries for the excited trimer state of helium-4
conforms to the distribution of an Efimov state favouring elongated
triangles, whereas the distribution for the ground state is broader
and does not seem to favour any particular geometry. This is due to
the fact that finite-range effects are more important for the ground
state and tend to push the system to a more equilateral configuration
(see a discussion of this effect in section~\ref{par:Connection-with-two-body-physics}).
For this reason, ground-state Efimov trimers are in general not expected
to exhibit the distribution of geometries of an ideal Efimov state.
Nevertheless, the absence of a well-defined geometry make even the
ground-state Efimov trimers very distinct from conventional molecules,
as figure~\ref{fig:Molecules} illustrates.

\begin{figure*}
\hfill{}\includegraphics[scale=0.85]{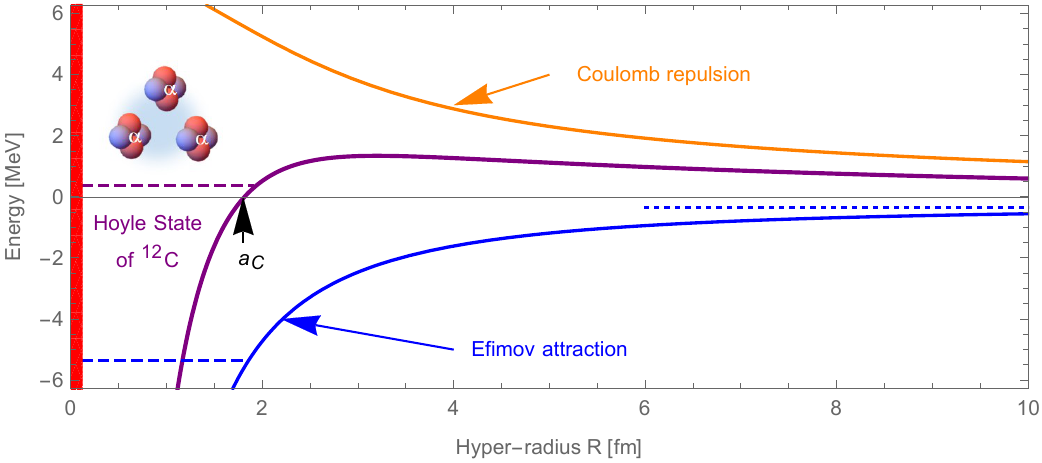}\hfill{}

\caption{\label{fig:HoyleEfimov}Schematic picture of the conjecture that the
Hoyle state of carbon-12 is bound by the Efimov attraction. In this
picture, the Hoyle state appears as a resonant state supported by
the sum of two potentials as a function of the hyper-radius between
the three alpha particles. The first potential is a Coulomb repulsive
potential, here taken to be $-\alpha(2e)^{2}k_{e}\mbox{erf}(R/\beta)/R$
where $\alpha=2$ and $\beta=1.66$~fm, and represented in orange.
The second potential is an Efimov attractive potential shown in blue,
asymptoting at small hyper-radius to equation~(\ref{eq:EfimovAttraction})
and at large hyper-radius to the dimer energy $-\hbar^{2}/ma^{2}$
shown by the dotted blue line), where the alpha-alpha scattering length
$a=5.4$~fm. The sum of the two potential is shown in purple. A three-body
repulsive hard wall shown in red is set at $R_{0}=0.122$~fm to fix
the three-body parameter and reproduce the energy of the Hoyle state
at 0.38~MeV, shown by the dashed purple line. The energy of that
state without Coulomb repulsion is shown by the dashed blue line.
Note that the value of $R_{0}$ is much smaller than the range of
the alpha-alpha interaction, and thus quite unlikely.}
\end{figure*}

\subsection{Observations in nuclear physics\label{subsec:Observations-in-nuclear}}

Bosonic particles in nuclear physics are compounds of fermions. In
the case of nucleons, since neutrons do not bind, bosonic clusters
have to involve protons. This introduces Coulomb interactions which
do not qualify as short-range interactions. Nevertheless, as discussed
in section~\ref{subsec:Coulomb-interactions}, Efimov physics can
survive in this context: if the short-range nuclear interactions are
resonant, they create a three-body Efimov attraction that competes
with the Coulomb repulsion to form bound states. Although these states
do not necessary follow the definitions of Efimov states put forward
in section~\ref{subsec:EfimovEnergy}, they could still be regarded
as Efimov states in the sense of being bound by the Efimov attraction.
A necessary condition for the Efimov attraction to survive is given
by equation~(\ref{eq:ConditionForEfimovInCoulombSystems}). This
condition can only be satisfied by light nuclei.

\subsubsection{The Hoyle state of carbon-12}

The Hoyle state is an excited resonant state of carbon-12 predicted
by Fred Hoyle in 1954. It plays a crucial role in the stellar nucleosynthesis
of carbon. In his original papers~\cite{Efimov1970a,Efimov1970b},
Vitaly~Efimov suggested that the Hoyle state may be viewed as a trimer
of alpha particles (i.e. helium nuclei, which are bosons) bound by
the Efimov attraction. The works of Renato Higa and Hans-Werner Hammer~\cite{Higa2008,Hammer2008a}
based on Effective-Field Theory looks into the effect of the Coulomb
interactions on alpha systems close to unitarity. They conjectured
that the Hoyle state is indeed a remnant of the Efimov spectrum broken
by the Coulomb interaction, surviving as a resonance above the three-alpha
scattering threshold. The corresponding picture of the Hoyle state
would be a resonant state resulting from the balance between the Efimov
attraction and the Coulomb repulsion. This picture is shown in figure~\ref{fig:HoyleEfimov}\@.
\begin{figure*}[t]
\hfill{}\includegraphics{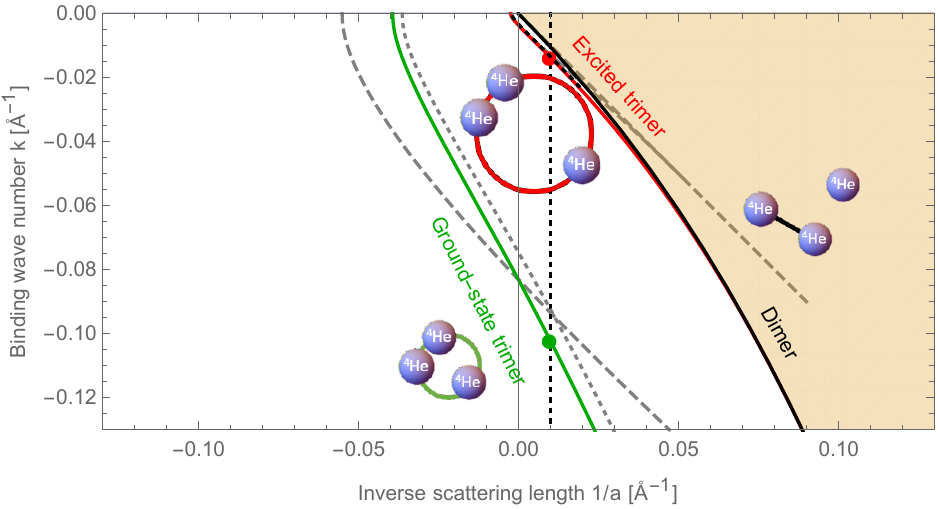}\hfill{}

\caption{\label{fig:Helium1}Efimov plot for three helium-4 atoms, showing
the wave number $k=E\big(\frac{m}{\vert E\vert\hbar^{2}}\big)^{1/2}$
as a function of the inverse scattering length $1/a$ artificially
varied by scaling the interaction potential for two helium-4 atoms
(the potential used is the so-called LM2M2 potential of reference~\cite{Aziz1991}).
The dimer state is shown in solid black, while two trimer states are
shown in green (ground) and red (excited). The vertical dotted line
shows the value of the scattering length for the unscaled potential,
i.e. the physical value for the true helium-4 system, and the corresponding
trimer energies are shown by the dots. They roughly follow the structure
of the zero-range Efimov spectrum, shown by the dashed curves. The
dotted curves show the trimer energies obtained for a separable interaction
reproducing exactly the two-body wave function at zero energy of the
scaled helium-4 potential (see Appendix).}
\end{figure*}
\begin{figure*}
\hfill{}\includegraphics{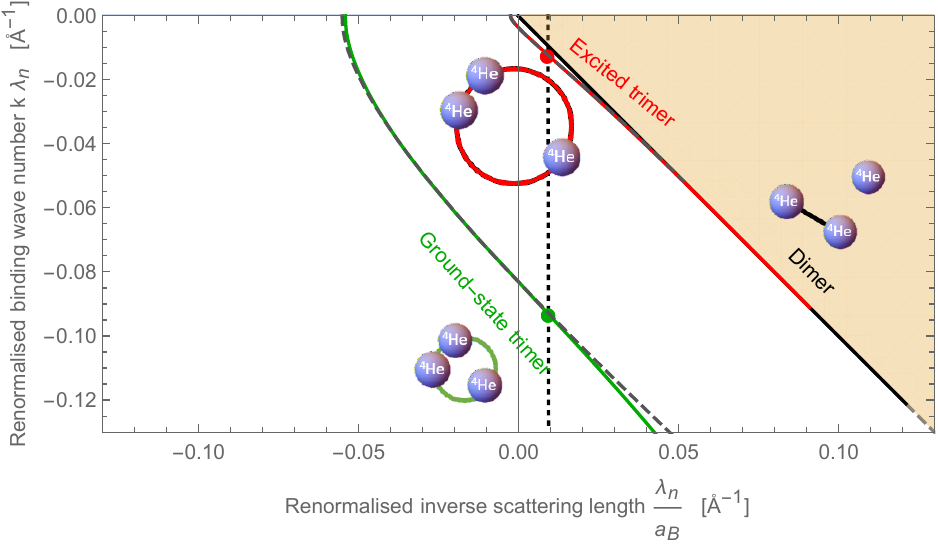}\hfill{}

\caption{\label{fig:Helium2}Same figure as figure~\ref{fig:Helium1}, where
the scattering $a$ is length is replaced by the length $a_{B}$,
and both axes are renormalised by the $a$-dependent factor $\lambda_{n}=\left(1+\frac{\Gamma_{n}}{\kappa_{*}a}\right)^{-1}$.
The values of $\Gamma_{n}$ are adjusted to $\Gamma_{0}=0.79$ and
$\Gamma_{1}=0.99$ to best reproduce the Efimov spectrum curves (shown
by the dashed curves).}
\end{figure*}

Although this picture of the Hoyle state is quite appealing, there
are two points that make it questionable. First, excluding the Coulomb
repulsion, the nuclear force between two alpha particles does not
seem to be resonant, as the scattering length of the model potentials~\cite{Ali1966}
for the alpha-alpha interaction is about 5~fm, which is similar to
the range $b$ and effective range $r_{e}\approx3.4$~fm of these
potentials. The resonance condition~(\ref{eq:ResonanceCondition})
may therefore not be satisfied. This would suggest that the attraction
between alpha particles is directly due to the nuclear force rather
than the Efimov attraction. Second, even if the alpha-alpha interaction
is resonant, the Efimov attraction seems too weak to overcome the
Coulomb repulsion and support a resonant state at distances larger
than the range $b$. Indeed, the value of the Bohr radius given by
equation~(\ref{eq:BohrRadius}) for alpha particles is $a_{C}\approx1.8$~fm.
The condition $b<a_{C}$ of equation~(\ref{eq:ConditionForEfimovInCoulombSystems})
therefore does not appear to be satisfied.

These conclusions rely on rough estimates, and only a full treatment
of the three-body problem with short-range and Coulomb interactions
can give a definite answer. The three-body model calculation of the
Hoyle state by Hiroya Suno, Yasuyuki Suzuki, and Pierre Descouvemont~\cite{Suno2015}
gives a preliminary answer. In their work, they show the contributions
from the Coulomb, nuclear and centrifugal (kinetic) energies as a
function of the hyper-radius. Although an attractive well (presumably
due to the Efimov attraction) can be seen in the centrifugal energy,
it appears that it is not enough to overcome the Coulomb repulsion,
and it is the nuclear force that is responsible for the stability
of the Hoyle state in this model. It is therefore likely that the
Hoyle state may not be considered as an Efimov state.

\subsection{Observations with atoms\label{subsec:Observations-with-atoms}}

Most atomic species have bosonic isotopes, and interactions between
neutral atoms is of the short-range type, decaying with a $1/r^{6}$
van der Waals tail. They therefore appear as ideal systems for the
observation of the Efimov physics of identical bosons described previously.
However, the remaining condition that the interaction be resonant
is more difficult to satisfy.

In most cases, the scattering length is on the order of the range
of the atomic interaction. Only in accidental cases does the scattering
length happen to be much larger than the range of the interaction.
Such is the case of helium-4 atoms discussed below. For a long time,
helium-4 appeared as the only promising atomic species for observing
Efimov states.

The situation changed drastically with the implementation of magnetic
Feshbach resonances in ultra-cold atoms experiments~\cite{Chin2010}.
Thanks to these resonances, it has been possible to change at will
the scattering length between various species of atoms by simply varying
the strength of an applied magnetic field. This has enabled a detailed
exploration of Efimov physics in these atomic systems.

\subsubsection{Helium-4\label{subsec:Helium-4}}

\paragraph{Theoretical predictions}

Because of its simplicity, the interaction potential of atomic helium
has been widely studied. The range of this potential can be characterised
by its van der Waals length, which is about 0.27 nm and its scattering
length turns out to be about 20 times larger than this range, around
10 nm. This fact makes helium-4 systems close to the resonance condition~(\ref{eq:ResonanceCondition})
for Efimov physics. It was first noted by T. K. Lim and co-workers
\cite{Lim1977} who suggested that helium-4 atoms may exhibit the
first example of Efimov trimer in nature, based on approximate three-body
calculations. This triggered many subsequent three-body calculations
using refined methods and helium potentials~\cite{Cornelius1986,Esry1996,Kolganova2011}
(see also references in \cite{Kolganova2011}). These calculations
predicted the existence of two helium-4 trimers. The one with lowest
binding energy (the excited trimer) was identified as an Efimov trimer,
in accordance with the definitions of section~\ref{subsec:What-is-an-Efimov-state}.
The ground-state one has not been regarded as an Efimov trimer, because
its energy does not fit well in the discrete scale-invariant pattern
of the Efimov universal theory. According to definitions 3 and 4 of
section~\ref{subsec:What-is-an-Efimov-state}, however, the ground-state
trimer is also an Efimov state. 

To appreciate how the two trimer states of helium-4 fit in the Efimov
picture, one can change the scattering length between two helium atoms
by scaling their interaction potential by some variable factor, and
plot the trimer energies as a function of the varied scattering length.
This is represented in Fig\@.~\ref{fig:Helium1}. When the scattering
length is tuned to infinity, the Efimov effect occurs and an infinite
number of trimer states exist. At the physical value of the scattering
length, only two trimer states remain, below the atom-dimer threshold.
As the scattering length is varied, these two trimer states roughly
follow the Efimov scenario, although none of them dissociate into
the atom-dimer threshold for the helium potential used in figure~\ref{fig:Helium1}.
As expected, the ground-state trimer shows the most marked deviations.
The excited trimer energy comes very close but always remains below
the dimer energy, which cannot be seen in figure~\ref{fig:Helium1}.

The significant deviations of the dimer and trimer energies with respect
to the ideal Efimov spectrum can be accounted for by finite-range
corrections. One can for instance replace the zero-range interaction
by a separable interaction parameterised to reproduce exactly the
scattering length and effective range of the helium potential (see
Appendix for details). This interaction indeed reproduces the scattering
length dependence of the deviations, although the trimer energies
are off by a small shift that can be removed by adding a three-body
force~\cite{Naidon2012a}. Alternatively, one can use the finite-range
corrected universal formula~(\ref{eq:ModifiedUniversalFormula}).
This formula allows one to map the helium trimer energies to the ideal
Efimov spectrum by plotting the renormalised wave number $\kappa^{\prime}=\lambda_{n}\kappa$
as a function of the renormalised inverse scattering length $1/a^{\prime}=\lambda_{n}/a_{B}$,
where $\lambda_{n}=(1+\Gamma_{n}/(\kappa_{*}a))^{-1}$. The resulting
plot is shown in figure~\ref{fig:Helium2}.

\paragraph{Experimental observations}

The first experimental investigation was carried out by the group
of Jan Peter Toennies in G\"ottingen \cite{Bruehl2005}. A beam of
helium clusters was generated by cryogenic expansion and diffracted
through a nanostructured transmission grating. The mass and spatial
extent of diffracted clusters were measured, revealing the existence
of the ground-state helium trimer with spatial extent of $\langle r\rangle=1.1_{-0.5}^{+0.4}$
nm, in agreement with theoretical predictions. The excited trimer
was not observed in that experiment\emph{. }Both trimers were eventually
observed in the group of Reinhard D\"orner in Frankfurt\emph{~}\cite{Voigtsberger2014,Kunitski2015}
by Coulomb explosion imaging. Although the trimer energies have not
been measured, the Coulomb explosion imaging technique enables to
measure structural properties of the trimers and to some extent reconstruct
their wave functions, in very good agreement with the theoretical
calculations. From these observations, the authors concluded that
the ground trimer's structural properties are very different from
those of an ideal Efimov trimer, whereas the excited trimer conforms
relatively well to the structure of an Efimov trimer.

\subsubsection{ultra-cold atoms under Feshbach resonances\label{subsec:Ultracold-atoms}}

The field of ultra-cold atomic gases has developed from laser-cooling
experiments in the 1980s. It consists in magnetically or optically
trapping inside a vacuum chamber a cloud of atoms cooled to ultra-low
temperatures, from microkelvins down to a few nanokelvins. In this
setting, it was realised that the scattering length between the atoms
could be changed through a Feshbach resonance by applying a magnetic
field to the cloud~\cite{Tiesinga1993,Chin2010}.

A Feshbach resonance~\cite{Feshbach1958,Feshbach1962,Fano1961} is
a general resonance phenomenon of particles with different possible
internal states (for instance, hyperfine states in the case of atoms).
The various pair combinations of internal states constitute different
two-body scattering channels. These channels are in general coupled
at short distance by some interactions (for example, the hyperfine
interaction in the case of atoms). If the scattering threshold of
some incoming channel (the ``open'' channel) is close to the energy
level of a two-body bound state in another or several other channels
(the ''closed'' channel), the coupling between these two channels
makes the scattering in the open channel resonant, i.e. the scattering
length can become much larger than the range of the interaction potentials
in these various channels. If the spacing between the threshold and
bound-state energy can be controlled (for instance by applying a magnetic
field and shifting by the Zeeman effect the threshold and bound-state
energy by different amounts), the resonance condition and therefore
the scattering length can be tuned. This technique has been extremely
successful for controlling interactions in ultra-cold atomic gases,
and studying Efimov physics in particular.

\subparagraph{Observation through loss}

Atomic Efimov trimers arise when the scattering length of the interatomic
potentials is tuned to a large value. This corresponds to the presence
of an $s$-wave two-body bound state or virtual state near the two-body
scattering threshold. However, for atoms commonly used in ultra-cold
gases, these potentials also support many other two-body bound states
that are more deeply bound, i.e. diatomic molecular states of various
rotational symmetries. This introduces two complications with respect
to the ideal Efimov scenario.

The first one is that the many diatomic molecular levels experience
different Zeeman shifts as the magnetic field is varied to tune the
scattering length. This can result in a second molecular state reaching
the threshold and creating a Feshbach resonance overlaping with the
resonance of interest, thus complicating the relationship between
the magnetic field and the scattering length. This situation is illustrated
by the many overlapping Feshbach resonances of caesium-133 ground-state
atoms~\cite{Ferlaino2011}. However, when the two resonances have
very different widths, the Efimov physics associated with one resonance
may not be significantly affected by the presence of the other. A
second consequence of the different Zeeman shifts of the molecular
levels is that there can be many avoided crossings between these levels.
Sometimes such a crossing may strongly affect the $s$-wave two-body
bound state associated with the Feshbach resonance, resulting in a
quick departure from the zero-range picture. This is the case, for
instance, for the caesium dimer associated with the Feshbach resonance
near 800~G, which remains very close to the threshold due to an avoided
crossing with a weakly bound molecular level of the open channel~\cite{Lee2007}.
Nevertheless, the Efimov effect still occurs close to resonance. 

The second complication is that Efimov trimers are not true bound
states, since they have a much smaller binding energy than the many
diatomic molecular states. Instead, these trimers exist as resonances
embedded in the continua formed by the scattering of one atom and
one diatomic molecule, as shown by the dotted curves in figure~\ref{fig:EfimovPlotFiniteRange}.
As a result, these trimers have a finite lifetime as they can decay
in these continua by dissociating into an atom and diatomic molecule.
This situation is different from that of helium-4, for which the potential
supports only one two-body bound state and Efimov trimers are true
bound states that are infinitely long lived. Nevertheless, the dissociation
processes are weak enough to allow to resolve the trimer resonances,
which were found to conform to the general Efimov scenario. It is
for this reason that these resonances are thought to be a good approximation
of Efimov states, even though they are not true bound states.

Ultra-cold atom experiments typically start with a gas of unbound
atoms, dimers, or atom-dimer mixtures. Three-body bound states can
be formed through inelastic collisions~\cite{Wang2009} but these
states are difficult to observe in the standard setups. Nevertheless,
these collisions tend to deplete the gas of atoms and this loss can
be monitored by imaging the gas. This can be used to indirectly observe
the three-body bound states through their influence on inelastic collisions.
For example, when three atoms collide in an ultra-cold gas, two of
them may recombine into a diatomic molecule. The energy gained by
this binding is then redistributed as kinetic energy between the formed
molecule and third atom. This inelastic process can be strongly enhanced
at the low collisional energy of ultra-cold atoms by Efimov resonances
that occur whenever an Efimov trimer lies just below the three-body
scattering threshold. As one varies the scattering length through
the values $a_{-}$ shown in figure~\ref{fig:EfimovPlot}, one therefore
expects to see peaks in the loss rate. The existence of these peaks
were first predicted in 1999 by Brett D. Esry, Chris H. Greene, and
James P. Burke Jr in reference~\cite{Esry1999}. Eric Braaten and
Hans-Werner Hammer subsequently derived from the zero-range theory
a simple analytical formula for the three-body recombination loss
rate coefficient $L_{3}$ at zero temperature~\cite{Braaten2001,Braaten2004}:
\begin{equation}
L_{3}=\frac{C\sinh(2\eta)}{\sin^{2}[\vert s_{0}\vert\ln(a/a_{-})]+\sinh^{2}\eta}\frac{\hbar a^{4}}{m},\qquad(\mbox{for }a<0)\label{eq:RecombinationRate}
\end{equation}
as a function of the scattering length $a$, the three-body parameter
characterised by $a_{-}$, and an inelasticity parameter $\eta$ describing
the decay to diatomic molecules. $C\approx4590$ is a numerical constant.
This formula shows that $L_{3}$ has a local maximum when the $\sin^{2}$
term vanishes, corresponding to scattering lengths that are multiples
of $a_{-}$ with the Efimov scaling factor, i.e. $a=e^{\pi n/\vert s_{0}\vert}a_{-}\approx(22.7)^{n}a_{-}$. 

Such a peak was seen in the early experiments on Feshbach resonances
and subsequently interpreted as an Efimov resonance. The first observation
was made in 2002 and the final results were reported in 2006~\cite{Kraemer2006}
by the group led by Hanns-Christoph N\"agerl and Rudolf Grimm in
Innsbruck for an ultra-cold gas of caesium-133 atoms in their hyperfine
ground state $\vert F=3,m_{F}=3\rangle$, where $F$ and $m_{F}$
designate the hyperfine quantum numbers. The Feshbach resonance used
to tune the scattering length in this experiment is a bit particular,
because it occurs at a ``negative value of the magnetic field'' $B\approx-10$~G,
which physically corresponds to the excited hyperfine state $\vert F=3,m_{F}=-3\rangle$
of caesium-133. This excited state is not stable against two-body
decay, so the experiment was limited to the positive values of $B$.
Although the dimer causing the resonance exists only for $B\le-10$~G,
the resonance makes the scattering length very large and negative
for $0\le B\lesssim10$~G, making it possible to observe Efimov physics.
In this range, however, only the ground Efimov state exists~\cite{Lee2007},
and was revealed by a peak in the measured three-body loss rate.

This allowed not only to prove the existence of an Efimov trimer,
but also measure the value $a_{-}$ which is related to Efimov's three-body
parameter. Since the ground-state Efimov trimer does not completely
follow the Efimov universal scenario, some objections on the terminology
of ``Efimov state'' were raised~\cite{Lee2007}. Indeed, the caesium
trimer evidenced in the experiment, when followed to the negative
side of magnetic field values where the dimer appears from the two-body
threshold, is not expected to approach the atom-dimer threshold, and
therefore does not conform to definition 2 given in section~\ref{subsec:What-is-an-Efimov-state}.
However, it is expected to be an Efimov state according to definition
3 and 4. Moreover, unlike the ground-state trimer of helium-4, which
occurs in the region of positive scattering length where deviations
from the Efimov scenario are significant, the loss peak observed for
caesium corresponds to the region of negative scattering length, where
the ground-state trimer is borromean and follows more closely the
Efimov scenario. For these reasons, this experiment may be thought
to constitute the first experimental demonstration of a borromean
Efimov state.

In the following years, many similar observations of Efimov resonances
were made with various kinds of atomic species. For positive scattering
lengths, the trimer energy is below the dimer energy. At some value
$a_{*}$ of the scattering length, the trimer energy can reach, or
approach closely, the atom-dimer threshold, creating a low-energy
resonance in atom-dimer scattering, which can be seen as a peak in
the rate of losses associated with inelastic atom-dimer scattering.
Efimov physics can also be seen in the three-body inelastic scattering
for positive scattering length. In this case, recombination to dimers
can take two paths, and as a result of quantum interference between
the two paths, a minimum in the three-body recombination loss rate
is expected for a value $a_{+}$ of the scattering length\footnote{Such mininum was already seen in the original experiment of reference~\cite{Kraemer2006},
although it is located at higher magnetic fields, accross a zero of
the scattering length; it therefore corresponds to an adjacent Feshbach
resonance, i.e. a different window of universality than that of the
observed peak}. The zero-range theory for these Efimov features in the loss measurement
at zero temperature were done in several theoretical contributions
\cite{Braaten2001,Braaten2004,Braaten2007} (see references in \cite{Braaten2007}).
The influence of temperature has appeared to play an important role
for the identification and interpretation of the resonance position,
and was treated theoretically in Refs.~\cite{Braaten2008,Rem2013}.

Studying the inelastic collisions between weakly bound dimers and
unbound atoms of caesium-133~\cite{Knoop2009}, the Innsbruck group
observed an atom-dimer resonance as predicted by the Efimov scenario
and measured the value of $a_{*}$ of figure~\ref{fig:EfimovPlot}.
The group of Giovanni Modugno and Massimo Inguscio in Florence \cite{Zaccanti2009}
observed both a local maximum and local minima in the three-body recombination
rate of potassium-39, thus determining the values of $a_{-}$ and
$a_{+}$, as well as inferring the atom-dimer resonance position $a_{*}$
from a small peak of the three-body loss rate appearing on the positive
scattering length side. A similar experiment was performed for lithium-7
atoms by the group of Randall G.~Hulet at Rice University~\cite{Pollack2009}.
From these experiments, the following measured ratios were reported:
$a_{-}/a_{+}=-6.7\pm0.6\;[^{39}\text{K}],\ -2.5\pm0.2\;[^{7}\text{Li}]$
and $a_{-}/a_{*}=-50\pm3\;[^{39}\text{K}],10.4\pm1.5\;[^{7}\text{Li}]$.
These results were later modified in 2013 in Refs.~\cite{Roy2013}
and \cite{Dyke2013} due to a misassignment of the three-body loss
peak\footnote{In reference\cite{Zaccanti2009}, two peaks were found at $a=-1500a_{0},-650a_{0}$,
which were interpreted as signatures of an Efimov trimer ($a_{-}=-1500a_{0}$)
and an associated tetramer. Later however, the same group could not
find the peak at $a=-1500a_{0}$ \cite{Roy2013}, and reassigned the
other peak to a trimer, with the updated value $a_{-}=-690(40)a_{0}$. } and to a recalibration of the Feshbach resonance parameters, respectively.
The updated values $a_{-}/a_{+}=-3.1\pm0.3\;[^{39}\text{K}],\ -2.8\pm0.3\;[^{7}\text{Li}]$
and $a_{-}/a_{*}=-22\pm2\;[^{7}\text{Li}]$ are in fair agreement
with the values predicted by the universal Efimov theory $a_{-}/a_{+}\approx-4.9$
and $a_{-}/a_{*}\approx-22.0$~\cite{Braaten2006}. The group of
Lev Khaykovich \cite{Gross2009} performed a similar experiment using
lithium-7 atoms, and measured $a_{-}$ and $a_{*}$. The ratio $a_{*}/a_{-}=-1.01(15)$
obtained in the experiment agrees with the theoretical value above
if $a_{-}$ and $a_{*}$ are interpreted as corresponding to the ground
and first-excited Efimov trimers and the universal scaling factor
22.7 is multiplied. The groups from Florence \cite{Zaccanti2009}
and Rice University \cite{Pollack2009} have both reported the observation
of two dips in the loss rate for positive scattering lengths, i.e.
two values of $a_{+}$ corresponding to a ground-state and an excited
Efimov trimers. They found the ratio between these two values to be
25(4) and 22.5(3) respectively, roughly confirming the discrete scale
invariance of Efimov trimers with the universal ratio 22.7. The universal
scaling was reported to be observed for negative scattering lengths
as well in reference~\cite{Pollack2009}, although a subsequent recalibration
of the scattering length with respect to the magnetic field dismissed
the second value of \emph{$a_{-}$ }and corrected the values of $a_{+}$,
giving a updated ratio 16(2) \cite{Dyke2013}.

\begin{figure}[H]
\hfill{}\includegraphics[scale=0.7]{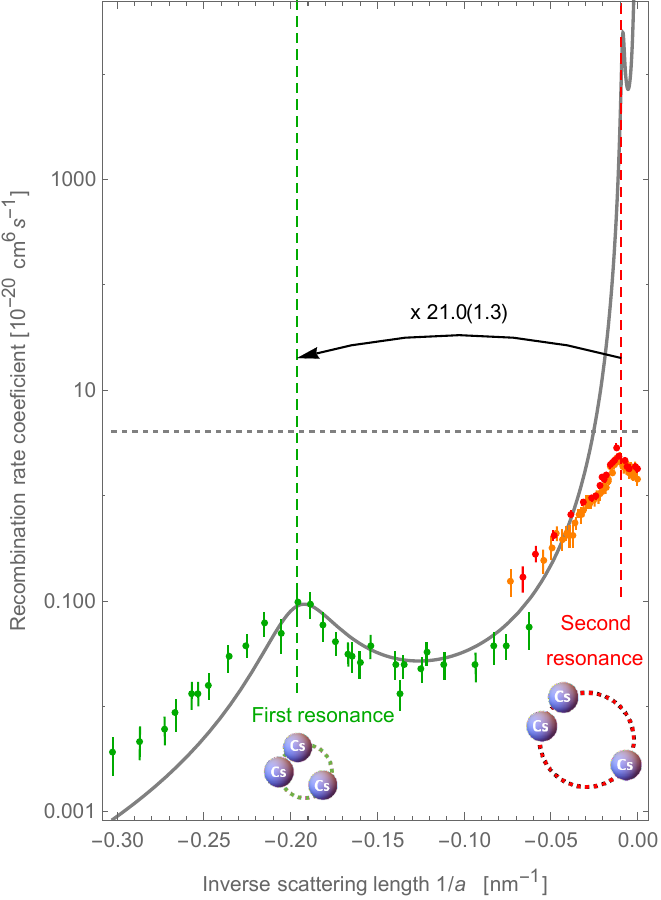}\hfill{}

\caption{\label{fig:Cesium}Efimov resonances in an ultra-cold gas of caesium-133
atoms in their lowest hyperfine state, whose scattering length is
varied by applying a magnetic field on the order of 800 gauss (adapted
from reference~\cite{Huang2014}). The resonances appear as peaks
in the three-body recombination rate as a function of scattering length.
The experimental data of reference~\cite{Huang2014} is shown with
different colours corresponding to different data sets, all taken
at a temperature of about 10~nK. The data around the first resonance
is fitted by the zero-temperature universal formula~(\ref{eq:RecombinationRate})
with $a_{-}=-51.0(0.6)$~nm and $\eta=0.10(1)$. The formula predicts
a second resonance at $a=22.7\times-51=-1158$~nm. The second resonance
was observed at $a=-1068(63)$~nm, corresponding to a factor $21(1.3)$
indicated by the arrow. The height of the peak is much smaller than
the zero-temperature prediction, due to the saturation effect of finite
temperature~\cite{Braaten2008,Rem2013}, indicated by the horizontal
dotted line for 10~nK.}
\end{figure}

More recently, the second Efimov trimer could be observed for negative
scattering lengths with caesium-133 atoms near a 800~G Feshbash resonance~\cite{Huang2014},
as originally proposed in reference~\cite{Lee2007}. The scaling
factor between the ground state Efimov trimer and second Efimov trimer
was found to be 21.0(1.3), close to the universal ratio 22.7. The
experimental data for the two Efimov resonances are shown in figure~\ref{fig:Cesium}.
This is so far the most convincing experimental evidence of Efimov
states of identical bosons, according to definition 1 of section~\ref{subsec:What-is-an-Efimov-state}.

\subparagraph{Bound-state spectroscopy}

Before the evidence of Efimov states through three-body and two-body
losses, the association of three colliding atoms in an Efimov trimer
was theoretically proposed~\cite{Stoll2005}. Such an association,
although different from the original proposal, could be achieved experimentally
and allowed the direct spectroscopy of Efimov trimers. Although the
first demonstration was done with three distinguishable atoms (see
section~\ref{subsec:Observations-with-atoms-Multi-component}), the
association of three bosons into an Efimov trimer was later achieved
with lithium-7 atoms in the group of Lev Khaykovich~\cite{Machtey2012}.
The experiment consists in driving a transition between three colliding
atoms and the underlying Efimov trimer state by applying a radio-frequency
(rf) modulation of the magnetic field. When the frequency of the modulation
matches the energy difference between the energy of the colliding
atoms and the trimer energy, the three atoms are associated in trimers,
resulting in a loss of the atoms from the imaged cloud. The resonance
could be seen as a small dip in the number of atoms as the frequency
of the modulation is varied, on the shoulder of a broader dip due
to the association of two atoms into a dimer. 

\subsection{Prospects for observation in condensed matter\label{subsec:Prospects-for-observation}}

Most studies of Efimov physics have been done in the fields of nuclear
physics and atomic physics, but recently there have been works seeking
the possibility of Efimov physics in other physical systems.  Here,
we present the current prospects for quantum spin systems \cite{Nishida2013a,Nishida2013b}
and excitons \cite{Omachi2013}.

\subsubsection{Efimov states in quantum spin systems}

Quantum spin systems have constituted an important and active field
of research in condensed matter physics for more than 70 years. They
have been been used with some success to explain and predict various
magnetic phenomena in insulating solids. A quantum spin system is
a system of spins fixed on the sites of a lattice, which can interact
with each other through an exchange interaction. Depending on the
geometry of the lattice, spatial dimension, and the sign and the spatial
extent of the exchange interaction, quantum spin systems can exhibit
various magnetic phases.

When the ground state of a quantum spin system shows a non-trivial
magnetic phase, such as the ferromagnetic or anti-ferromagnetic phases,
the symmetry of the Hamiltonian is spontaneously broken. The Nambu-Goldstone
theorem dictates that there must exist gapless low-energy excitations,
called magnons, in such broken-symmetry systems. For a ferromagnet
system, the magnons have the dispersion relation $\varepsilon(k)\propto k^{2}$
~\cite{Watanabe2012}. This dispersion is similar to that of a non-relativistic
particle in the vacuum.

From this similarity between the magnons and the non-relativistic
particles, Yusuke~Nishida and co-workers \cite{Nishida2013} have
sought the possibility of finding Efimov states in a ferromagetic
quantum spin system. To achieve the resonant-interaction condition~(\ref{eq:ResonanceCondition})
necessary for the appearance of the Efimov states, they considered
the Heisenberg model in three spatial dimensions with an anisotropic
exchange coupling and the single-ion anisotropy. These terms originate
from the anisotropy in the lattice and the crystal field effect that
are present in real materials described by quantum spin systems. In
the presence of these terms, the magnons no longer behave as independent
quasi-particles, and start to interact. This interacting problem is
greatly simplified from the fact that the number of magnons is a conserved
quantity in the anisotropic Heisenberg model. It is therefore justified
to consider the few-body problem for magnons with a well-defined number
of magnons. Y.~Nishida and co-workers have shown by analytically
solving the two-body problem of magnons that the magnons can form
a bound state when the anisotropies are increased. As a result, the
scattering length between the magnons diverges at some critical values
of the anisotropies at which the bound state of magnons appears, a
necessary condition for the onset of Efimov physics. Y.~Nishida and
co-workers have solved numerically the three-magnon problem at the
resonant point. For different values of the spin quantum number and
critical anisotropies, the three-body energy spectrum of magnons is
found to be in good agreement with the universal Efimov spectrum,
and to exhibit a discrete scale invariance with a scaling factor close
to 22.7

It is important to note that the following features of the anisotropic
Heisenberg model are the same as the resonantly interacting non-relativistic
particles V. Efimov considers in his original argument: (a) magnons
have a low-energy dispersion $\varepsilon(k)\propto k^{2}$, (b) the
number of magnons is conserved, (c) the interaction between magnons
is short-ranged. These sufficient conditions seem to be a useful guideline
for searching Efimov states in other systems.

To achieve the Efimov states of magnons, one needs a fine tuning of
the parameters of the system close to the resonantly interacting regime.
In the above case, one needs to increase the anisotropy of the exchange
interaction or the single-ion anisotropy term, such that the bound
state of two magnons is about to appear. Although such a fine tuning
is rather challenging, it is known that in organic crystals, the exchange
coupling can be modified by applying an external pressure. This may
be one route to realise Efimov states in quantum spin systems. Once
one finds a suitable system, Y.~Nishida and co-workers \cite{Nishida2013}
have argued that the appearance of Efimov states can be tested experimentally
from the electron spin resonance signal \cite{Nishida2013a}.

\subsubsection{Universal few-body physics with excitons}

\hspace{0.5cm}An exciton is a bound state of an electron and a hole
which appears in a semiconductor excited by a laser. Because it is
an association of two fermions, it behaves as a boson. While there
has been a long history of exciton studies since its first theoretical
prediction~\cite{Frenkel1931,Wannier1937}, there has been a revival
of interest recently by the realisation of the Bose-Einstein condensates
of the excitons~\cite{Yoshioka2011}, which had been one of the holy
grails in the research on excitons.

Recently, a possible signature of Efimov physics in an excitonic system
has been claimed to be observed~\cite{Omachi2013}. In reference~\cite{Omachi2013},
$N$-body bound states of excitons, called poly-excitons, have been
observed up to $N=6$ by a photoluminescence measurement in a diamond
crystal. The binding energies of these poly-exciton states were measured,
and compared with those found in other crystals~\cite{Steele1987,Thewalt1996}.
They have been found to agree excellently between different crystals
if normalized by the binding energy of an exciton, suggesting that
the poly-exciton states behave universally, i.e., they are independent
of the details of the crystal and the energy scale of the system.
The binding energies are reminiscent of the series of universal binding
energies for the $N$-body clusters of resonantly interacting bosons
discussed in section~\ref{subsec:Universal-clusters}, although it
is not clear whether they are related since the exciton-exciton interaction
is typically non-resonant.

\section{Three identical fermions\label{sec:Three-identical-fermions}}

Identical fermions (i.e., fermions in the same internal state) cannot
interact in the $s$-wave channel due to the antisymmetrisation, so
that the conventional Efimov physics appearing for the $s$-wave resonance
(see section~\ref{sec:Three-identical-bosons}) does not occur for
such Fermi systems. Two identical fermions can only scatter in the
odd angular momentum channels, and among them, the most dominant channel
at low energy is the $p$-wave channel. Therefore, the scattering
amplitude at low energy is written as 
\begin{equation}
f(k,\theta)=\frac{k^{2}\cos\theta}{-1/a_{p}+\frac{1}{2}r_{p}k^{2}-ik^{3}},\label{eq:p-wave_scat}
\end{equation}
where $a_{p}$ is the $p$-wave scattering volume, $r_{p}$ is the
$p$-wave inverse effective range. Note that, although the notations
are similar to those of equation~(\ref{eq:LowEnergyPhaseShift}),
$a_{p}$ is a volume and $r_{p}$ is the inverse of a length. A natural
step is to investigate universal three-body physics of identical fermions
in the vicinity of resonant $p$-wave interaction, i.e., when the
moduli of $p$-wave scattering volume becomes divergently large. Although
such $p$-wave resonances are experimentally challenging because they
are much narrower than the $s$-wave ones and subject to larger atomic
number losses, they have been successfully realised in ultra-cold
atom experiments with Feshbach resonances~\cite{Regal2003a,Zhang2004,Schunck2005,Nakasuji2013}.
In this section, we review the recent theoretical progress on the
few-fermion physics in the vicinity of a $p$-wave resonance in three
dimensions (see \ref{subsec:Resonant-p-wave-interactions}), and two
dimensions (see \ref{subsec:Super-Efimov-effect}).

\subsection{Resonant $p$-wave interactions\label{subsec:Resonant-p-wave-interactions}}

In three dimensions, the possibility of Efimov trimers has been investigated
for three fermions with $p$-wave resonant interactions~\cite{Macek2006,Braaten2012,Nishida2012}.
If one takes $1/a_{p}=0$ and $r_{p}=0$ simultaneously in equation~(\ref{eq:p-wave_scat}),
the moduli of the scattering amplitude takes its maximum value allowed
by the unitary bound and the system becomes scale invariant, similarly
to the unitary limit in the $s$-wave case: $1/a=0$ and $r_{e}=0$.
In this unitary limit $1/a_{p}=0$ and $r_{p}=0$, Macek and Sternberg
have shown, using a pseudo-potential method, that the Efimov effect
occurs for the identical fermions for a spin 1/2 system (i.e., among
three identical fermions, two are in the same internal state, while
the other is in another internal state)~\cite{Macek2006}. The same
conclusion was obtained in reference~\cite{Braaten2012}.

However, this unitary limit is unphysical, violating the positivity
of probability~\cite{Braaten2012,Nishida2012,Jona-Lasinio2008,Pricoupenko2006}.
Indeed, if one believes the $p$-wave form of the scattering amplitude
in equation~(\ref{eq:p-wave_scat}) for any $k<\Lambda$, where $\Lambda$
is a cutoff momentum $\Lambda\gg|a_{p}|^{-1/3},|r_{p}|$, then the
probability of the bound-state wave function outside the range $b$
of the potential exceeds unity, suggesting a negative probability
at short distance~\cite{Braaten2012,Nishida2012,Jona-Lasinio2008,Pricoupenko2006}.
The positivity of the probability thus dictates the inverse range
$r_{p}$ to have a finite negative value, satisfying the Wigner bound~\cite{Hammer2009,Hammer2010a}.
\begin{equation}
r_{p}<-\frac{2}{b}
\end{equation}
Equivalently, the scattering amplitude equation~(\ref{eq:p-wave_scat})
is valid only for $k\ll|r_{p}|$ at the $p$-wave resonance, with
$r_{p}<0$ satisfying the Wigner bound.

In Refs.~\cite{Jona-Lasinio2008,Levinsen2007}, spin-polarised three
identical fermions around the $p$-wave resonance have been studied,
keeping the Wigner bound. Three-body bound states with threefold rotational
degeneracy have been found, one state for each channel in $L=1^{-}$
channel~\cite{Jona-Lasinio2008,Levinsen2007} and in $L=1^{+}$ channel~\cite{Jona-Lasinio2008}.
Both three-body bound states have larger binding energies than that
of the $p$-wave dimer around the $p$-wave resonance. This suggests
that the $p$-wave molecular BEC phase predicted in the studies of
the $p$-wave resonant two-component Fermi system~\cite{Gurarie2005,Cheng2005,Gurarie2007}
is not a genuine ground state, but can be subject to trimer formation
instability. The recombination rate to the trimers in a gas of $p$-wave
dimers has been estimated in Refs.~\cite{Levinsen2007,Levinsen2008},
while the three-body recombination to the $p$-wave dimer state in
a gas of identical fermions has been studied in Refs.~\cite{Jona-Lasinio2008,Suno2003}.
The trimer states found here are Borromean states~\cite{Jona-Lasinio2008}:
they have finite binding energies at the $p$-wave resonance $1/a_{p}=0$,
and they persist for $1/a_{p}<0$, where no $p$-wave dimer exists.
As one moves further away from the resonance towards the negative
scattering volume side, the trimers finally dissociate into three
fermions at three-body continuum $E=0$. On the positive scattering
volume side, the trimer energy becomes equal to the dimer energy,
so that the trimers dissociate into a particle plus a $p$-wave dimer.

\subsection{The super-Efimov effect\label{subsec:Super-Efimov-effect}}

In two dimensions, more exotic few-body states emerge. In reference~\cite{Nishida2013b},
Yusuke Nishida and co-workers have found that there exists an infinite
series of three-body bound states in $\ell=\pm1$ channels with discrete
scale invariance. These states, unlike the Efimov states, show a \emph{double
exponential scaling} of their energies 
\begin{equation}
E^{(n)}=E_{*}\exp\left[-2\exp\left(\frac{3\pi n}{4}+\theta\right)\right].\label{eq:supEfimov_descale}
\end{equation}
For this reason, they have been dubbed ``super-Efimov states\textquotedbl .
Here, $\theta$ is a parameter determined by the short-range part
of the interaction (see the final paragraph of this section for more
details). The super-Efimov states, in addition to a non-super-Efimovian
$\ell=0$ ground state, have been found to be Borromean~\cite{Volosniev2014,Gao2014a},
i.e., they remain bound even in the absence of a two-body $p$-wave
dimer. We note that the three-fermion problem in two dimensions at
the $p$-wave resonance was also solved in reference~\cite{Levinsen2008}
before the super-Efimov states have been found by Yusuke Nishida and
co-workers in reference~\cite{Nishida2013b}, and the on-shell $T$-matrix
at the threshold energy was found to show the same double exponential
scaling behavior as equation~(\ref{eq:supEfimov_descale}).

While Yusuke Nishida and co-workers have found the super-Efimov states
with the renormalization group analysis, as well as by solving a momentum-space
integral equation for a separable potential~\cite{Nishida2013b},
they can also be found by other formalisms~\cite{Volosniev2014,Gao2014a,Gridnev2014,Gao2014}.
A rigorous mathematical study based on a spectral analysis has proved
the presence of the super-Efimov effect~\cite{Gridnev2014}. The
super-Efimov states can also be demonstrated by the hyper-spherical
formalism, but some remarks are in order. In the hyper-spherical formalism,
the diagonal adiabatic potential in the super-Efimovian channel $\ell=\pm1$
is found to be ($V_{n}$ here has the same notation as in equation~(\ref{eq:HyperradialEquation2}))~\cite{Volosniev2014,Gao2014}
\begin{eqnarray}
V_{0}(R) & = & -\frac{1}{4R^{2}}-\frac{Y}{R^{2}\ln(R/R_{0})}-\frac{16/9}{R^{2}\ln^{2}(R/R_{0})}\nonumber \\
 &  & +O\left(\frac{1}{R^{2}\ln^{3}(R/R_{0})}\right)\label{eq:supEfimov_adpot}
\end{eqnarray}
where $R_{0}$ is the range of the interaction, and $Y$ is a parameter
dependent on the short-range part of the interaction. If $Y=0$ and
one can neglect the second term, one arrives at the super-Efimov states
with a double exponential scaling ${\displaystyle E^{(n)}\propto\exp\left[-2\exp\left(\pi n\Biggr/\sqrt{\frac{16}{9}-\frac{1}{4}}\right)\right]}$,
which is slightly different from equation~(\ref{eq:supEfimov_descale})~\cite{Volosniev2014}.
However, $Y$ turns out to be positive definite~\cite{Volosniev2014,Gao2014a},
and the second term is relevant at large $R$. Due to the second term,
equation~(\ref{eq:supEfimov_adpot}) leads to a different scaling
behavior ${\displaystyle E^{(n)}\propto\exp\left(-\frac{\pi^{2}n^{2}}{2Y}\right)}$
at low energy~\cite{Gao2014}. The discrepancy between these results
can be ascribed to the non-adiabatic term $Q_{nm}(R)$~\cite{Volosniev2014,Gao2014a}.
The diagonal non-adiabatic term in the super-Efimovian channel $Q_{00}(R)$
has been found to behave in the leading order as~\cite{Volosniev2014}
\begin{equation}
Q_{00}(R)=\frac{Y}{R^{2}\ln(R/R_{0})}+O\left(\frac{1}{R^{2}\ln^{2}(R/R_{0})}\right).
\end{equation}
One can see that the first term cancels the second term in equation~(\ref{eq:supEfimov_adpot}).
In reference~\cite{Gao2014a}, Chao Gao and co-authors have studied
this system numerically for various classes of potentials, and the
next leading order of $Q_{00}(R)$ has been found to be consistent
with ${\displaystyle -\frac{1}{4R^{2}\ln^{2}(R/R_{0})}}$. One thus
obtains a correct hyper-radial potential in the super-Efimovian channel
as 
\begin{eqnarray}
V_{0}(R)+Q_{00}(R) & = & -\frac{1}{4R^{2}}-\frac{16/9+1/4}{R^{2}\ln^{2}(R/R_{0})}\nonumber \\
 &  & +O\left(\frac{1}{R^{2}\ln^{3}(R/R_{0})}\right),
\end{eqnarray}
which reproduces the double exponential scaling of equation~(\ref{eq:supEfimov_descale})
found in reference~\cite{Nishida2013b}.

While the double exponential scaling $\ln\vert E^{(n+1)}\vert/\ln\vert E^{(n)}\vert=\exp\left(\frac{3\pi}{4}\right)\approx10.6$
seems too large to be observed in a realistic system, it can be reduced
significantly for mass-imbalanced two-component systems~\cite{Moroz2014}.
It has been found that the super-Efimov effect also occurs in a 2
identical fermions + 1 particle system or 2 identical bosons + 1 particle
system when the inter-species $p$-wave interaction is resonant. The
scaling factor of the super-Efimov states decreases as the inter-species
mass ratio is increased towards a 2 heavy + 1 light configuration.
For instance, for a mixture of $^{6}$Li and $^{133}$Cs atoms, corresponding
to the mass ratio 22.1, it reaches a reasonable value $\ln\vert E^{(n+1)}\vert/\ln\vert E^{(n)}\vert\approx1.3$.
Note that the adiabatic approximation, i.e., the Born-Oppenheimer
approximation, also fails for the mass-imbalanced system even for
large mass ratios~\cite{Moroz2014}, leading to a similarly wrong
energy spectrum as mentioned above~\cite{Efremov2014}.

Interestingly, Chao Gao and co-authors have found that the three-body
parameters $E_{*}$ and $\theta$ are universal for the class of van
der Waals potentials~\cite{Gao2014a}, in similarity with the van
der Waals universality of the three-body parameter found for the Efimov
states of three identical bosons (see section~\ref{subsec:Van-der-Waals-Universality}).
This similarity is reinforced by the fact that four-body bound states
($\ell=\pm2$ channels) exist and are tied to each super-Efimov states,
showing the same double exponential scaling, in the same way four-body
bound states are associated with the Efimov trimers (see section~\ref{sec:Many-Identical-bosons}).
Although the super-Efimov trimers have yet to be observed in experiments,
it is of interest to investigate the universality of the super-Efimov
states in the $N$-body sector ($N\ge4$) and understand differences
and similarities with the conventional Efimov physics for three bosons
in three dimensions. We also note that a Borromean three-body bound
state, possibly related to the super-Efimov states, has also been
found in a two dimensional Hubbard model in the $S=3/2$ channel,
which corresponds to a spin-polarised three-fermion system~\cite{Kornilovitch2014}.
It is an interesting avenue to search for super-Efimov physics in
such condensed-matter systems.

\section{Multi-component systems\label{sec:Multi-component-systems}}

Systems with different kinds of particles, or particles with different
internal states (either referred to as ``components'') exhibit an
even richer Efimov physics than systems of identical particles. These
systems have more parameters: the different kinds of particles may
have different masses, quantum statistics, and different interactions
between them. This situation introduces a few general facts:
\begin{itemize}
\item For a given three-particle system, there are three inter-particle
interactions. At least two of these interactions should be resonant
for the Efimov effect to occur. This can be understood simply from
the picture of mediated interaction: in order for one particle to
mediate an effective long-range interaction between two other particles,
it must interact resonantly with these two particles. If it interacts
resonantly with only one particle, then the mediation to another particle
is not possible.
\item Generally speaking, bosonic particles are favourable to the Efimov
effect, whereas fermionic particles tend to prevent the Efimov effect,
since their Pauli exclusion may overcome the Efimov attraction.
\item The lighter a particle is, the better it mediates interaction between
other particles. Thus, mass-imbalanced systems tend to enhance the
Efimov attraction, and enable the Efimov effect in fermionic systems.
\end{itemize}
In the following, we review various situations.

\subsection{Three distinguishable particles\label{subsec:3-distinguishable-particles}}

The general treatment of three different kinds of particles with different
masses and interacting with different scattering lengths was first
addressed by Amado and Noble~\cite{Amado1972} and Efimov~\cite{Efimov1972,Efimov1973}.
Here, we consider the simpler case of three different particles with
equal masses but different scattering lengths $a_{12}$, $a_{23}$,
$a_{31}$. A concrete example of this situation is given by particles
of the same kind, polarised in three different internal states - see
section~\ref{subsec:Polarised-systems}.

The general form of the Faddeev decomposition used in equation~(\ref{eq:FaddeevDecomposition})
involves three different Faddeev components:
\begin{equation}
\Psi(1,2,3)=\chi^{(1)}(\vec{r}_{23},\vec{\rho}_{23,1})+\chi^{(2)}(\vec{r}_{31},\vec{\rho}_{31,2})+\chi^{(3)}(\vec{r}_{12},\vec{\rho}_{12,3})\label{eq:FaddeevGeneral}
\end{equation}

Following the derivation of section~\ref{subsec:Efimov-theory},
we apply the Bethe-Peierls condition~(\ref{eq:BethePeierls}) for
each pair and obtain:{\small{}
\[
\frac{\partial\chi_{0}^{(i)}}{\partial\alpha}(R,0)+\frac{4}{\sqrt{3}}\left(\chi_{0}^{(j)}(R,{\textstyle \frac{\pi}{3}})+\chi_{0}^{(k)}(R,{\textstyle \frac{\pi}{3}})\right)=-\frac{R}{a_{jk}}\chi_{0}^{(i)}(R,0)
\]
}Using the expansion $\chi_{0}^{(i)}(R,\alpha)=\sum_{n}F_{n}^{(i)}(R)\phi_{n}^{(i)}(\alpha;R)$
where $\phi_{n}^{(i)}(\alpha;R)$ has the form:
\[
\phi_{n}^{(i)}(\alpha;R)=\sin\Big(s_{n}(R)\big({\textstyle \frac{\pi}{2}}-\alpha\big)\Big),
\]
 the conditions can be written in a matrix form~\cite{Braaten2006}:{\small{}
\begin{align}
\Bigg[-\cos(s_{n}{\textstyle \frac{\pi}{2}})\left(\begin{array}{ccc}
\!1\! & \!0\! & \!0\!\\
\!0\! & \!1\! & \!0\!\\
\!0\! & \!0\! & \!1\!
\end{array}\right)+\frac{4}{\sqrt{3}}\frac{\sin(s_{n}\frac{\pi}{6})}{s_{n}}\left(\begin{array}{ccc}
\!0\! & \!1\! & \!1\!\\
\!1\! & \!0\! & \!1\!\\
\!1\! & \!1\! & \!0\!
\end{array}\right)\nonumber \\
+\frac{\sin(s_{n}\frac{\pi}{2})}{s_{n}}R\left(\begin{array}{ccc}
\!a_{23}^{-1}\! & \!0\! & \!0\!\\
\!0\! & \!a_{31}^{-1}\! & \!0\!\\
\!0\! & \!0\! & \!a_{12}^{-1}\!
\end{array}\right)\Bigg]\cdot\left(\!\begin{array}{c}
\!F_{n}^{(1)}\!\\
F_{n}^{(2)}\!\\
\!F_{n}^{(3)}\!
\end{array}\right)=0\label{eq:ThreeDistinguishableTwoBodyCondition}
\end{align}
}To obtain a non-trivial solution $F_{n}^{(i)}\ne0$, the determinant
of the matrix should be zero.

\subsubsection{Three resonantly interacting pairs}

For three resonantly large scattering lengths $\vert a_{ij}\vert\gg b$
and hyper-radius $b\ll R\ll\vert a_{ij}\vert$, the terms $Ra_{ij}^{-1}$
can be neglected and the determinant becomes:
\[
\!\left(\!\cos(s_{n}{\textstyle \frac{\pi}{2}})+\frac{4}{\sqrt{3}}\frac{\sin(s_{n}\frac{\pi}{6})}{s_{n}}\!\right)^{2}\!\left(\!\cos(s_{n}{\textstyle \frac{\pi}{2}})-\frac{8}{\sqrt{3}}\frac{\sin(s_{n}\frac{\pi}{6})}{s_{n}}\!\right)\!,
\]
whose second factor admits the same imaginary root $s_{0}$ as that
of equation~(\ref{eq:TranscendentalEquation}) for three identical
bosons. This could be anticipated from the fact that the form of the
wave function in equation~(\ref{eq:FaddeevGeneral}) can reduce to
the bosonic case of equation~(\ref{eq:FaddeevDecomposition}) with
$\chi^{(i)}=\chi$, when all scattering lengths are equally large.
Therefore, the same Efimov effect that occurs for three bosons also
occurs for three distinguishable particles. In particular, this systems
exhibits discrete scaling invariance with the same scaling ratio $e^{\pi/\vert s_{0}\vert}\approx22.7$
at unitarity $a_{12}=a_{23}=a_{31}=\pm\infty$. As in the bosonic
case, the discrete scaling invariance persists away from unitarity.
However, it requires the simultaneous scaling of all scattering lengths.
It may therefore not be apparent if only one or two scattering lengths
are scaled.

\subsubsection{Two resonantly interacting pairs}

We now consider only two resonantly large scattering length, say $\vert a_{23}\vert,\vert a_{31}\vert\gg b$,
and one non-resonant scattering length $a_{12}$ on the order of the
interaction range $b$. In the region of hyper-radius $b\ll R\ll\vert a_{23}\vert,\vert a_{31}\vert$,
the terms $Ra_{23}^{-1}$, $Ra_{31}^{-1}\sim0$ in equation~(\ref{eq:ThreeDistinguishableTwoBodyCondition})
can be neglected, whereas the term $Ra_{12}^{-1}$ is large and imposes
$F_{n}^{(3)}\approx0$. We are thus left with the equations for $F_{n}^{(1)}$
and $F_{n}^{(2)}$. The determinant of the corresponding matrix is
\[
\!\left(\!\cos(s_{n}{\textstyle \frac{\pi}{2}})+\frac{4}{\sqrt{3}}\frac{\sin(s_{n}\frac{\pi}{6})}{s_{n}}\!\right)\left(\!\cos(s_{n}{\textstyle \frac{\pi}{2}})-\frac{4}{\sqrt{3}}\frac{\sin(s_{n}\frac{\pi}{6})}{s_{n}}\!\right)\!,
\]
whose second factor admits an imaginary root $s_{0}\approx i0.4137$.
The Efimov effect therefore occurs in this case as well, although
it is weaker. The corresponding discrete scaling invariance ratio
is $e^{\pi/\vert s_{0}\vert}\approx1986.12$, implying a very sparse
Efimov spectrum. 

\subsubsection{One resonantly interacting pair}

Finally, if there is only one resonantly large scattering length,
say $a_{23}$, only $F_{n}^{(1)}$ contributes in equation~(\ref{eq:ThreeDistinguishableTwoBodyCondition})
and one is left with:
\[
\cos(s_{n}{\textstyle \frac{\pi}{2}})=0,
\]
which admits only real roots $s_{n}$. There is therefore no Efimov
effect in this case. This is consistent with the physical picture
that the Efimov effect is due to the exchange of a particle resonantly
interacting with two other particles, which is not possible if only
one pair is resonantly interacting.

\subsection{2 Identical particles + 1 particle\label{subsec:2-Identical-particles+1particle}}

We now consider the case when two of the three particles are identical,
either identical bosons or identical fermions. Here, ``identical particles''
means that they are in the same internal states (for a treatment of
identical particles with different internal states, see section~\ref{subsec:Particles-with-spin}).
The mass $M$ of these two particles is in general different from
the mass $m$ of the third particle.

\begin{figure*}
\hfill{}\includegraphics[scale=0.6]{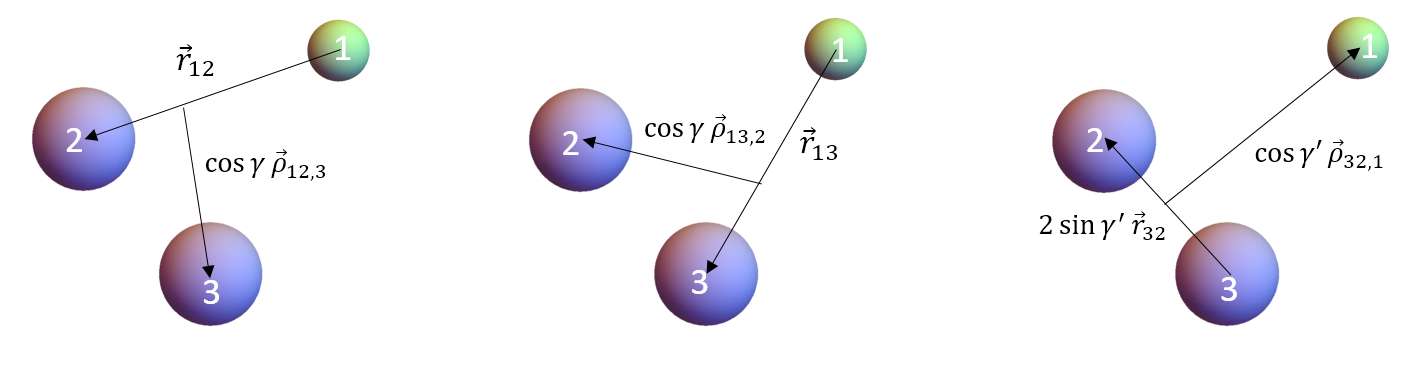}\hfill{}

\caption{\label{fig:Jacobi2+1}The three sets of Jacobi coordinates for two
identical particles (dark) plus one particle (pale).}

\end{figure*}

The two identical particles are located at $\vec{x}_{2}$ and $\vec{x}_{3}$
and the light particle at $\vec{x}_{1}$. The three sets of Jacobi
coordinates for this system read:
\begin{eqnarray*}
\vec{r}_{1i} & = & \vec{x}_{i}-\vec{x}_{1}\\
\vec{\rho}_{1i,j} & = & \frac{1}{\cos\gamma}\left(\vec{x}_{j}-\frac{M\vec{x}_{i}+m\vec{x}_{1}}{M+m}\right)
\end{eqnarray*}
for $\{i,j\}=\{1,2\}$, and
\begin{eqnarray*}
\vec{r}_{23} & = & \frac{1}{2\sin\gamma^{\prime}}\left(\vec{x}_{3}-\vec{x}_{2}\right)\\
\vec{\rho}_{23,1} & = & \frac{1}{\cos\gamma^{\prime}}\left(\vec{x}_{1}-\frac{\vec{x}_{2}+\vec{x}_{3}}{2}\right),
\end{eqnarray*}
where the angles $\gamma\in[0,\pi/2]$ and $\gamma^{\prime}\in[0,\pi/4]$
are defined by:
\begin{eqnarray*}
\gamma & = & \arcsin\frac{M}{M+m},\\
\gamma^{\prime} & = & \arcsin\sqrt{\frac{m}{2(M+m)}}.
\end{eqnarray*}
The three sets are shown in figure~\ref{fig:Jacobi2+1}. They all
satisfy $r_{ij}^{2}+\rho_{ij,k}^{2}=R^{2}$, where $R$ is the hyper-radius,
and they are related to each other by the following rotation transformations:
\begin{eqnarray*}
\vec{r}_{13} & = & \sin\gamma\;\vec{r}_{12}+\cos\gamma\;\vec{\rho}_{12,3}\\
\vec{\rho}_{13,2} & = & \cos\gamma\;\vec{r}_{12}-\sin\gamma\;\vec{\rho}_{12,3}
\end{eqnarray*}

\begin{eqnarray*}
\vec{r}_{32} & = & \sin\gamma^{\prime}\;\vec{r}_{12}-\cos\gamma^{\prime}\;\vec{\rho}_{12,3}\\
\vec{\rho}_{32,1} & = & -\cos\gamma^{\prime}\;\vec{r}_{12}-\sin\gamma^{\prime}\;\vec{\rho}_{12,3}
\end{eqnarray*}

 Using the general Faddeev decomposition of equation~(\ref{eq:FaddeevGeneral}),
and taking into account the symmetry or antisymmetry of the wave function
under the exchange of the identical particles 2 and 3, one obtains
the following form of the wave function:
\begin{align*}
\Psi(1,2,3)= & \chi^{(1)}(\vec{r}_{32},\vec{\rho}_{32,1})\\
 & +\chi^{(2)}(\vec{r}_{13},\vec{\rho}_{13,2})\pm\chi^{(2)}(\vec{r}_{12},\vec{\rho}_{12,3})
\end{align*}
The sign $\pm$ corresponds to a plus sign in the case of identical
bosons, and to a minus sign in the case of identical fermions.

Following the derivation of section~\ref{subsec:Efimov-theory},
we apply the Bethe-Peierls boundary condition~(\ref{eq:BethePeierls})
for pair 12 and 32, and obtain:{\small{}
\begin{align*}
\left[\left(\frac{\partial}{\partial r}+\frac{1}{a_{12}}\right)\chi_{0}^{(2)}(\vec{r},\vec{\rho})\right]_{r\to0}\pm2\frac{\chi_{0}^{(2)}(\cos\gamma\vec{\rho},-\sin\gamma\vec{\rho})}{\sin2\gamma\rho}\\
+2\frac{\chi_{0}^{(1)}(-\cos\gamma^{\prime}\vec{\rho},-\sin\gamma^{\prime}\vec{\rho})}{\sin2\gamma^{\prime}\rho}=0
\end{align*}
\begin{align}
\left[\left(\frac{\partial}{\partial r}+\frac{1}{a_{32}}\right)\chi_{0}^{(1)}(\vec{r},\vec{\rho})\right]_{r\to0}+2\frac{\chi_{0}^{(2)}(-\cos\gamma^{\prime}\vec{\rho},-\sin\gamma^{\prime}\vec{\rho})}{\sin2\gamma^{\prime}\rho}\nonumber \\
\pm2\frac{\chi_{0}^{(2)}(-\cos\gamma^{\prime}\vec{\rho},-\sin\gamma^{\prime}\vec{\rho})}{\sin2\gamma^{\prime}\rho}=0\label{eq:BethePeierls2+1}
\end{align}
}where $\chi_{0}^{(k)}(\vec{r},\vec{\rho})=\chi^{(k)}(\vec{r},\vec{\rho})/(r\rho)$.
The Faddeev components can be expressed as a function of the hyper-spherical
coordinates,
\[
R=\sqrt{r_{ij}^{2}+\rho_{ij,k}^{2}},
\]
\[
\alpha_{k}=\arctan\frac{r_{ij}}{\rho_{ij,k}},
\]
and the orientations $\hat{r}_{ij}$ and $\hat{\rho}_{ij,k}$ of $\vec{r}_{ij}$
and $\vec{\rho}_{ij,k}$. Since the particles interact in the $s$
wave, we consider the case when $\chi_{0}^{(k)}$ is independent of
the orientation $\hat{r}_{ij}$.  In these new coordinates, the Bethe-Peierls
boundary conditions (\ref{eq:BethePeierls2+1}) become:
\begin{align*}
\left[\left(\frac{\partial}{\partial\alpha}+\frac{R}{a_{12}}\right)\chi_{0}^{(2)}(R,\alpha,\hat{\rho})\right]_{\alpha\to0}\pm2\frac{\chi_{0}^{(2)}(R,\frac{\pi}{2}-\gamma,-\hat{\rho})}{\sin2\gamma}\\
+2\frac{\chi_{0}^{(1)}(R,\frac{\pi}{2}-\gamma^{\prime},-\hat{\rho})}{\sin2\gamma^{\prime}}=0
\end{align*}
\begin{align}
\left[\left(\frac{\partial}{\partial\alpha}+\frac{R}{a_{32}}\right)\chi_{0}^{(1)}(R,\alpha,\hat{\rho})\right]_{\alpha\to0}+2\frac{\chi_{0}^{(2)}(R,\frac{\pi}{2}-\gamma^{\prime},-\hat{\rho})}{\sin2\gamma^{\prime}}\nonumber \\
\pm2\frac{\chi^{(2)}(R,\frac{\pi}{2}-\gamma^{\prime},-\hat{\rho})}{\sin2\gamma^{\prime}}=0\label{eq:BetherPeierls2+1b}
\end{align}

One can then expand $\chi_{0}^{(k)}$ as follows:
\[
\chi_{0}^{(k)}(R,\alpha,\hat{\rho})=\sum_{n,\ell,m}F_{n,\ell,m}^{(k)}(R)\phi_{n,\ell}^{(k)}(\alpha;R)Y_{\ell m}(\hat{\rho})
\]
where $Y_{\ell m}$ are the spherical harmonics, and $\phi_{n,\ell}$
are the solutions of the eigenvector equation:
\begin{equation}
\left(-\frac{\partial^{2}}{\partial\alpha^{2}}+\frac{\ell(\ell+1)}{\cos^{2}\alpha}\right)\phi_{n,\ell}^{(k)}(\alpha;R)=s_{n,\ell}^{2}(R)\phi_{n,\ell}^{(k)}(\alpha;R)\label{eq:EquationForPhi2+1}
\end{equation}
with the boundary conditions $\phi_{n,\ell}^{(i)}(\frac{\pi}{2})=0$
and the Bethe-Peierls boundary conditions from equation~(\ref{eq:BetherPeierls2+1b}).
The eigenvalues $s_{n,\ell}^{(k)}(R)$ and eigenvectors $\phi_{n,\ell}(\alpha;R)$
determine a set coupled of coupled equations satisfied by $F_{n,\ell,m}^{(k)}(R)$,
from which the three-body problem can be solved. 

\begin{figure}[t]
\hfill{}\includegraphics[scale=0.65]{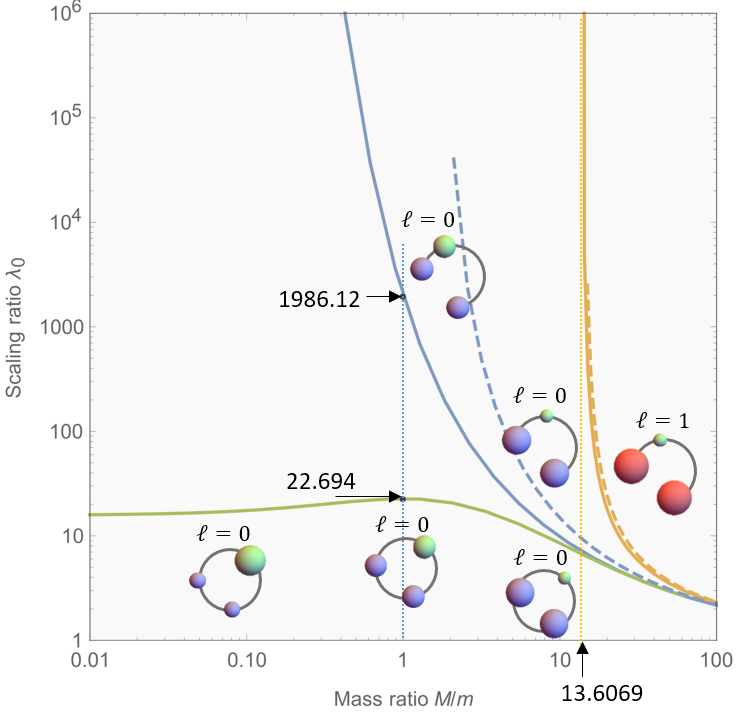}\hfill{}

\caption{\label{fig:2+1}Efimov scaling ratio $\lambda_{0}=e^{\pi/\vert s_{0}\vert}$
as a function of the mass ratio for Efimov states of two identical
particles and another particle. The lower curve (green) corresponds
to two identical bosons and one particle resonantly interacting with
each other. The middle curve (blue) corresponds to two identical bosons
interacting resonantly only with the other particle. The upper curve
(orange) corresponds to two identical fermions resonantly interacting
with another particle. The dashed curves show the results obtained
from the Born-Oppenheimer approximation for large mass ratios.}
\end{figure}

\subsubsection{2 bosons + 1 particle with $\ell=0$\label{subsec:2-bosons+1particle}}

If the two identical particles are bosonic, then $\pm=+$. For $\ell=0$,
the solutions of equation~(\ref{eq:EquationForPhi2+1}) are given
by $\phi_{n,0}^{(k)}(\alpha;R)=\sin\left(s_{n}\left(\frac{\pi}{2}-\alpha\right)\right)$.
The conditions resulting from equation~(\ref{eq:BetherPeierls2+1b})
can be written in a matrix form 
\begin{align}
\Bigg[-\cos(s_{n}{\textstyle \frac{\pi}{2}})\left(\begin{array}{cc}
\!1\! & \!0\!\\
\!0\! & \!1\!
\end{array}\right)+\frac{2}{s_{n}}\left(\begin{array}{cc}
\!\frac{\sin(s_{n}\gamma)}{\sin2\gamma}\! & \!\frac{\sin(s_{n}\gamma^{\prime})}{\sin2\gamma^{\prime}}\!\\
\!2\frac{\sin(s_{n}\gamma^{\prime})}{\sin2\gamma^{\prime}}\! & \!0\!
\end{array}\right)\nonumber \\
+\frac{\sin(s_{n}\frac{\pi}{2})}{s_{n}}R\left(\begin{array}{cc}
\!a_{12}^{-1}\! & \!0\!\\
\!0\! & \!a_{32}^{-1}\!
\end{array}\right)\Bigg]\cdot\left(\!\begin{array}{c}
\!F_{n}^{(2)}\!\\
F_{n}^{(1)}\!
\end{array}\right)=0\label{eq:BethePeierls2Boson+1}
\end{align}

\paragraph*{Three resonantly-interacting pairs}

We first consider the case when all scattering lengths are resonant,
$\vert a_{12}\vert,\vert a_{32}\vert\gg b$. In the region of hyper-radius
$b\ll R\ll\vert a_{12}\vert,\vert a_{32}\vert$, the terms $Ra_{12}^{-1}$,
$Ra_{32}^{-1}\sim0$ in equation~(\ref{eq:BethePeierls2Boson+1})
can be neglected, and the determinant of the matrix in that equation
is: 
\begin{equation}
\left(\cos(s_{n}{\textstyle \frac{\pi}{2}})-\frac{2}{s_{n}}\frac{\sin(s_{n}\gamma)}{\sin2\gamma}\right)\cos(s_{n}{\textstyle \frac{\pi}{2}})-2\left(\frac{2}{s_{n}}\frac{\sin(s_{n}\gamma^{\prime})}{\sin2\gamma^{\prime}}\right)^{2}.\label{eq:ThreeResonantPairs}
\end{equation}

\vspace{0.2cm}

As in the case of three identical particles, this determinant admits
an imaginary root $s_{0}$. Hence, the Efimov effect occurs in this
case, and the scaling ratio between Efimov states at unitarity ($a_{12},a_{32}\to\pm\infty$)
is shown in figure~\ref{fig:2+1}. In the case of equal mass $M=m$,
one retrieves the scaling ratio $e^{\pi/\vert s_{0}\vert}\approx22.7$
obtained for three identical bosons. This ratio is decreased by either
decreasing the mass ratio $M/m$ (it tends to $15.74$ in the limit
$M/m\to0$) or increasing the mass ratio (it tends to 1 in the limit
$M/m\to\infty$).

\paragraph{Two resonantly-interacting pairs}

We now consider the case when only the interaction between the particles
of mass $M$ and mass $m$ is resonant, $\vert a_{12}\vert\gg b$,
while the interaction between identical particles is non-resonant,
$\vert a_{32}\vert\sim b$. In the region of hyper-radius $b\ll R\ll\vert a_{12}\vert$,
the term $Ra_{12}^{-1}$ in equation~(\ref{eq:BethePeierls2Boson+1})
may be neglected, whereas the term $Ra_{32}^{-1}$ is very large and
imposes $F_{n}^{(1)}\approx0$~\footnote{The corrections due to the non-zero scattering length $a_{32}$ and
the effective range between the non-identical particles were recently
addressed in the framework of effective-field theory by Bijaya Acharya
and co-workers~\cite{Acharya2016}. It was shown that both corrections
can be accounted for by a single additional three-body parameter.}. The remaining condition reads: 
\begin{equation}
-\cos(s_{n}{\textstyle \frac{\pi}{2}})+\frac{2}{s_{n}}\frac{\sin(s_{n}\gamma)}{\sin2\gamma}=0,\label{eq:TwoResonantPairs}
\end{equation}
which admits one imaginary solution $s_{0}$. The Efimov effect occurs
in this case too, and the scaling ratio between Efimov states at unitarity
($a_{12}\to\pm\infty$) is shown in figure~\ref{fig:2+1}. Since
there are only two resonant pairs, the Efimov attraction is weaker
than for three resonant pairs, and for equal masses $M=m$, one retrieves
the scaling ratio $e^{\pi/\vert s_{0}\vert}\approx1986.12$ obtained
in section~\ref{subsec:3-distinguishable-particles}.

For large mass ratios, the scaling ratio of trimer energies becomes
the same as in the case of three-resonantly interacting pairs. This
large-mass-ratio limit is interesting because the trimer spectrum
is denser than that of identical bosons, allowing to more easily observe
several Efimov trimers. These trimers may be evidenced from the change
in particle-dimer scattering length and relaxation rate, or three-body
recombination rate~\cite{Helfrich2010,Petrov2015,Mikkelsen2015}.

\subsubsection{2 fermions + 1 particle with $\ell=1$\label{subsec:2-fermions-1particle}}

If the two identical particles are fermionic ($\pm=-$), there cannot
be any $s$-wave interaction between the two, which can be seen from
the boundary condition~(\ref{eq:BethePeierls2+1}), which imposes
$\phi^{(1)}=0$. For $\ell=0$, the solutions of equation~(\ref{eq:EquationForPhi2+1})
are given by $\phi_{n,0}^{(2)}(\alpha;R)=\sin\left(s_{n}\left(\frac{\pi}{2}-\alpha\right)\right)$,
with the following condition resulting from equation~(\ref{eq:BetherPeierls2+1b}),
\[
-\cos(s_{n}{\textstyle \frac{\pi}{2}})-\frac{2}{s_{n}}\frac{\sin(s_{n}\gamma)}{\sin2\gamma}=0.
\]

This equation only admits real solutions, and thus there is no Efimov
attraction in this case.

For $\ell=1$, however, solutions of equation~(\ref{eq:EquationForPhi2+1})
are given by~\cite{Kartavtsev2007} 
\[
\phi_{n,0}^{(2)}(\alpha;R)=s_{n}\cos\left(s_{n}\left({\textstyle \frac{\pi}{2}}-\alpha\right)\right)-\tan\alpha\sin\left(s_{n}({\textstyle \frac{\pi}{2}}-\alpha)\right),
\]

with the following condition resulting from equation~(\ref{eq:BetherPeierls2+1b})
\[
\frac{1-s_{n}^{2}}{s_{n}}\tan(s_{n}{\textstyle \frac{\pi}{2}})-\frac{2\cos\left(s_{n}\gamma\right)}{\sin2\gamma\cos\left(s_{n}\frac{\pi}{2}\right)}-\frac{\sin\left(s_{n}\gamma\right)/s_{n}}{\sin^{2}\gamma\cos\left(s_{n}\frac{\pi}{2}\right)}
\]

This condition admits one imaginary solution $s_{0}$ for a mass ratio
$\frac{M}{m}>\kappa_{c}$, where the critical mass ratio $\kappa_{c}\approx13.6069657$.
Although there is no Efimov effect for mass ratios smaller than $\kappa_{c}$
(in particular no Efimov effect in the equal-mass case), it occurs
for mass ratios larger than the critical mass ratio $\kappa_{c}$.
The corresponding scaling ratio at unitarity ($a_{12}=a_{13}\to\pm\infty$)
is shown in figure~\ref{fig:2+1}. It is infinitely large at the
critical mass ratio and rapidly decreases to approach the scaling
ratio for two bosons and one particle as the mass ratio is increased.

\subsubsection{Trimers with higher-angular momenta\label{subsec:Trimers-with-higher-angular-momentum}}

From the preceding discussion, it appears that there is in general
a competition between the Efimov attraction and the centrifugal repulsion
due to the angular momentum $\ell$. For large enough mass ratios,
the Efimov attraction can overcome the centrifugal repulsion, gradually
allowing the binding of Efimov trimers with higher angular momentum~\cite{Efimov1973,Nielsen2001,DIncao2006,Endo2011,Helfrich2011}.
Using the same approach as in the preceding discussion, one can determine
from equation~(\ref{eq:BetherPeierls2+1b}) the critical mass ratios
for the appearance of Efimov trimers of increasing angular momenta.
In the case of bosons, trimers with even angular momentum appear at
the following critical mass ratios~\cite{Kartavtsev2008,Endo2011}:

\begin{eqnarray*}
\ell & = & 2,\quad\kappa>38.630...\\
\ell & = & 4,\quad\kappa>125.765...\\
 & \dots
\end{eqnarray*}

In the case of fermions, trimers with odd angular momentum appear
at the following mass ratios:
\begin{eqnarray*}
\ell & = & 1,\quad\kappa>13.607...\\
\ell & = & 3,\quad\kappa>75.994...\\
 & \dots
\end{eqnarray*}

\subsubsection{The Born-Oppenheimer picture\label{subsec:The-Born-Oppenheimer-picture}}

In all three cases presented above and shown in figure~\ref{fig:2+1},
the Efimov effect is strengthened as the mass ratio $M/m$ is increased.
This can be simply understood from the Born-Oppenheimer approximation~\cite{Fonseca1979}.
This approximation exploits the fact that $M$ is much larger than
$m$ in this limit: the heavy particles of mass $M$ can thus be treated
as slow particles, and the particle of mass $m$ as a fast particle.

The Born-Oppenheimer approximation consists in first solving the motion
of the fast particle, for fixed positions of the heavy particles.
Let us call $\vec{R}$ the relative position between the two heavy
particles, and $\vec{r}$ the relative position between their centre
of mass and the light particle. The three-body wave function is approximated
by the form:
\[
\Psi(\vec{R},\vec{r})=F(\vec{R})\phi(\vec{r};\vec{R})
\]
where $\phi(\vec{r};\vec{R})$ is a solution of the problem for the
light particle in presence of the heavy particles at a fixed separation
$\vec{R}$. In the zero-range theory, this solution is a free wave
satisfying
\begin{equation}
-\frac{\hbar^{2}}{2m}\nabla_{r}^{2}\phi(\vec{r};\vec{R})=\epsilon\phi(\vec{r};\vec{R})\label{eq:FreeWaveLightParticle}
\end{equation}
with the Bethe-Peierls boundary condition~(\ref{eq:BethePeierls})
between the heavy and light particles. The solution with the lowest
energy $\epsilon<0$ is obtained with the following linear combination
of two free waves originating from two sources at the locations of
the heavy particles: 
\begin{equation}
\phi(\vec{r};\vec{R})=\frac{\exp(-\kappa\vert\vec{r}-\vec{R}/2\vert)}{\vert\vec{r}-\vec{R}/2\vert}+\frac{\exp(-\kappa\vert\vec{r}+\vec{R}/2\vert)}{\vert\vec{r}+\vec{R}/2\vert}\label{eq:BondingOrbital}
\end{equation}
where $\kappa=\frac{1}{\hbar}\sqrt{-2m\epsilon}$. Applying the Bethe-Peierls
boundary condition~(\ref{eq:BethePeierls}) to equation~(\ref{eq:BondingOrbital}),
one gets an equation for $\kappa$:
\begin{equation}
\kappa-\frac{e^{-\kappa R}}{R}=\frac{1}{a}.\label{eq:BondingPotentialEquation}
\end{equation}

\begin{figure}
\hfill{}\includegraphics[scale=0.6]{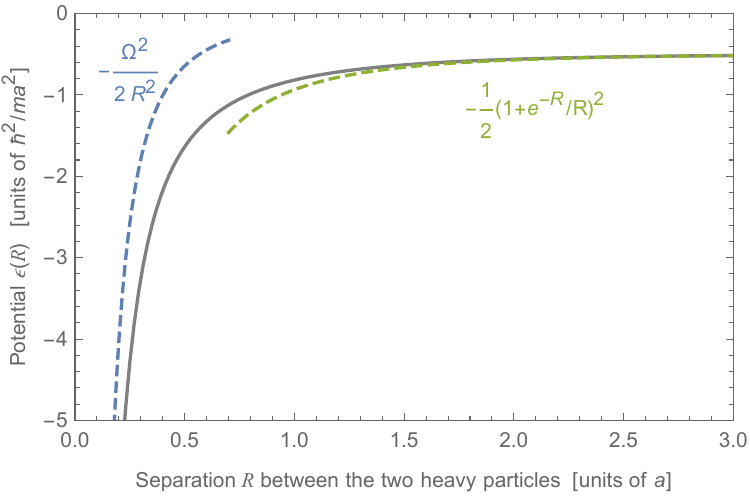}\hfill{}

\caption{\label{fig:Bonding-potential}Bonding potential $\epsilon(R)$ in
the Born-Oppenheimer approximation between two heavy particles separated
by $R$ and both resonantly interacting with a light particle, with
scattering length $a>0$. At separation much smaller than the scattering
length $a$ between the heavy and light particles, the heavy particles
experience an Efimov $R^{-2}$ attraction.}

\end{figure}

The wave number $\kappa$, and thus the energy $\epsilon$, are therefore
functions of the separation $R=\vert\vec{R}\vert$ between the two
heavy particles. The $R$-dependent energy $\epsilon(R)$ constitutes
a potential energy for the relative motion of the two heavy particles,
which obeys the following Schr\"odinger equation,
\begin{equation}
\left(-\frac{\hbar^{2}}{M}\nabla_{R}^{2}+\epsilon(R)\right)F(\vec{R})=E\,F(\vec{R}).\label{eq:TwoHeavyParticlesEquation}
\end{equation}
In addition to the potential $\epsilon(R)$ induced by the light particle,
there should also be a short-range interaction potential between the
two heavy particles, but it is neglected here. The potential energy
$\epsilon(R)$ can be calculated analytically from equation~(\ref{eq:BondingPotentialEquation})
using the Lambert function and is shown in figure~\ref{fig:Bonding-potential}
for a positive scattering length. It is an attractive potential, whereas
the potentials obtained for larger eigenvalues $\epsilon$ of equation~(\ref{eq:FreeWaveLightParticle})
are repulsive. For this reason, the lowest-eigenvalue solution given
by equation~(\ref{eq:BondingOrbital}) is called a \emph{bonding
orbital}, to reflect the fact that the light particle in such a state
acts as a glue between the two heavy particles. 

For large $R\gg a$, one finds from equation~(\ref{eq:BondingPotentialEquation})
that 
\[
\kappa\sim\frac{1}{a}+\frac{e^{-R/a}}{R},
\]
showing the bonding potential $\epsilon(R)$ asymptotes to the energy
$-\frac{\hbar^{2}}{2ma^{2}}$ corresponding to a two-body bound state
of light and heavy particles scattering with a free heavy particle
at zero energy. Moreover, the tail of the potential is as an attractive
Yukawa potential whose range is the scattering length $a$. 

For small $R\ll a$, on the other hand, one finds from equation~(\ref{eq:BondingPotentialEquation})
that $\kappa R$ approaches the Omega constant $\Omega\approx0.567143$,
solution of the equation $\Omega=e^{-\Omega}$. The bonding potential
is therefore $\epsilon(R)\sim-\frac{\hbar^{2}}{2m}\frac{\Omega^{2}}{R^{2}}$.
This reproduces the $1/R^{2}$ Efimov attraction. Hence, the Born-Oppenheimer
approximation shows that the Efimov attraction can indeed be interpreted
as resulting from the exchange of the light particle between the two
heavy particles. Moreover, since $\epsilon(R)$ is proportional to
$1/m$ and the relative kinetic energy of the two heavy particles
is proportional to $1/M$, one can see from equation~(\ref{eq:TwoHeavyParticlesEquation})
that the Efimov attraction is more effective for large mass ratios
$M/m$.

More precisely, for a given partial wave $F_{LM}(\vec{R})=R^{-1}f_{L}(R)Y_{LM}(\hat{R})$
with angular quantum number $L$, this equation can be rewritten in
a form similar to equation~(\ref{eq:HyperradialEquation2}),
\[
\left(-\frac{d^{2}}{dR^{2}}+V(R)-\frac{ME}{\hbar^{2}}\right)f_{L}(R)=0,
\]
with the potential 
\begin{equation}
V(R)=\frac{L(L+1)}{R^{2}}+\frac{M}{\hbar^{2}}\epsilon(R).\label{eq:HeavyHeavyPotential}
\end{equation}
Identification of this potential at small $R\ll a$ with the form
of equation~(\ref{eq:EfimovAttraction}) gives
\begin{equation}
\vert s_{0}\vert^{2}=\frac{M}{2m}\Omega^{2}-L(L+1)-\frac{1}{4}\label{eq:EffectiveEfimovCoefficient}
\end{equation}
from which one can calculate the scaling factor $\lambda_{0}=e^{\pi/\vert s_{0}\vert}$
associated with the $1/R^{2}$ attraction.

\begin{figure}[t]

\hfill{}\includegraphics[scale=0.6]{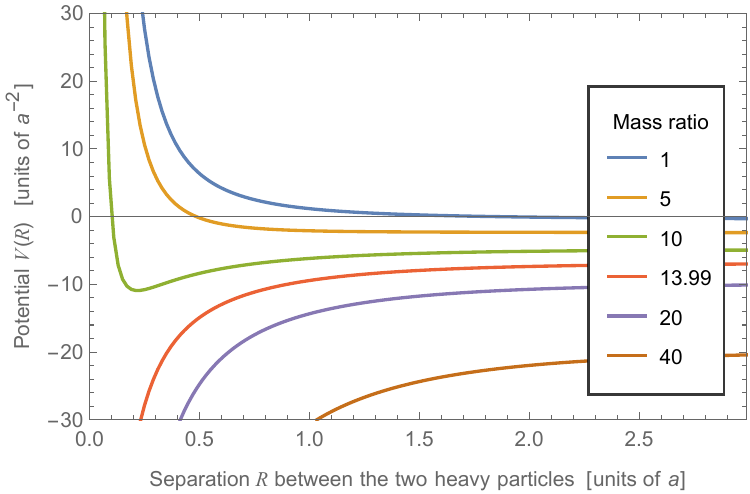}\hfill{}

\caption{\label{fig:Heavy-Heavy-potential}Effective potential $V(R)$ in the
Born-Oppenheimer approximation between two identical fermions separated
by $R$ and resonantly interacting (with a scattering length $a>0$)
with a light particle, for various values of their mass ratio. This
potential results from the competition between the bonding potential
$\epsilon(R)$ and the centrifugal repulsion with one unit of angular
momentum - see equation~(\ref{eq:HeavyHeavyPotential}). At short
separations $R\ll a$, the potential exhibits the $\propto R^{-2}$
Efimov attraction for a mass ratio larger than the critical value
$\kappa_{c}\approx13.99$ (corresponding to the red curve), while
it is repulsive for smaller mass ratios.}

\end{figure}
If the two heavy particles are distinguishable or identical bosons,
one can take $L=0$. The corresponding potential $V(R)\propto\epsilon(R)$
is purely attractive, as shown in figure~\ref{fig:Bonding-potential}.
If the two heavy particles are identical fermions, there should be
at least one unit of angular momentum between the two heavy particles
to respect the antisymmetry of their wave function. There is therefore
a competition between the Efimov attraction and the centrifugal repulsion.
The resulting potential $V(R)$ is represented in figure~\ref{fig:Heavy-Heavy-potential}
for different values of the mass ratio. According to equation~(\ref{eq:EffectiveEfimovCoefficient}),
the Efimov attraction wins for large enough mass ratio $M/m$. For
$\vert s_{0}\vert=0$, one obtains the critical mass ratio $\kappa_{c}\approx13.990296$
at which the potential becomes purely attractive, which is very close
to the exact result $\kappa_{c}\approx13.6069657$ presented in section~\ref{subsec:2-fermions-1particle}.
Above this critical mass ratio, the potential is dominated by the
$R^{-2}$ Efimov attraction for $R\ll a$, which leads to the Efimov
effect.

The Efimov scaling factor $\lambda_{0}$ obtained from equation~(\ref{eq:EffectiveEfimovCoefficient})
is shown by dashed curves in figure~\ref{fig:2+1} for both the bosonic
$L=0$ and fermionic $L=1$ cases, and is in good agreement with the
exact results for large mass ratios. The Born-Oppenheimer approximation
thus gives a simple account of the Efimov effect for 2+1 particles.

\subsubsection{Kartavtsev-Malykh universal trimers\label{subsec:Kartavtsev-Malykh-universal-trimers}}

As we saw in the preceding sections, a system of two identical fermions
resonantly interacting with a light particle can be bound by the Efimov
effect when the mass ratio between the fermions and the light particle
is larger than a critical value $\kappa_{c}$. Nevertheless, even
below the critical mass ratio, it is possible for the system to form
universal three-body bound states with the same quantum numbers for
$a>0$. This fact was pointed out by Oleg I. Kartavtsev and Anastasia
V. Malykh~\cite{Kartavtsev2007}. It can be seen from the Born-Oppenheimer
potential $V(R)$ between the two fermions shown in figure~\ref{fig:Heavy-Heavy-potential}.
For a mass ratio smaller than $\kappa_{c}$, even though the centrifugal
repulsion wins over the Efimov attraction, making the potential repulsive
at short separation, there can nonetheless be an attractive part at
larger distances, thereby creating a potential well. This potential
well can be seen in the curve corresponding to a mass ratio of 10
in figure~\ref{fig:Heavy-Heavy-potential}. The potential well deepens
as the mass ratio increases, until it becomes purely attractive at
the critical mass ratio. Kartavtsev and Malykh have shown by solving
the three-body problem exactly (using the hyperspherical formalism
sketched at the beginning of this section) that a ground three-body
bound state appears at the critical mass ratio $\kappa_{1}\approx8.17260$,
and an excited one appears at the mass ratio $\kappa_{2}\approx12.91743$.
Since there is no Efimov attraction at short distance for these states,
there is no need to introduce a three-body boundary condition, and
thus the states are universally determined by the scattering length
$a>0$ between a fermion and the light particle. Unlike Efimov trimers,
the energy of these trimer simply scales with the universal dimer
energy. For this reason, they are called ``universal trimers''. Similar
states appear for higher angular momenta at some critical mass ratios~\cite{Kartavtsev2008,Endo2011},
connecting to the non-zero angular-momentum Efimov trimers of section~\ref{subsec:Trimers-with-higher-angular-momentum}
appearing at larger mass ratios.

\begin{figure*}[t]
\hfill{}\includegraphics[scale=0.45]{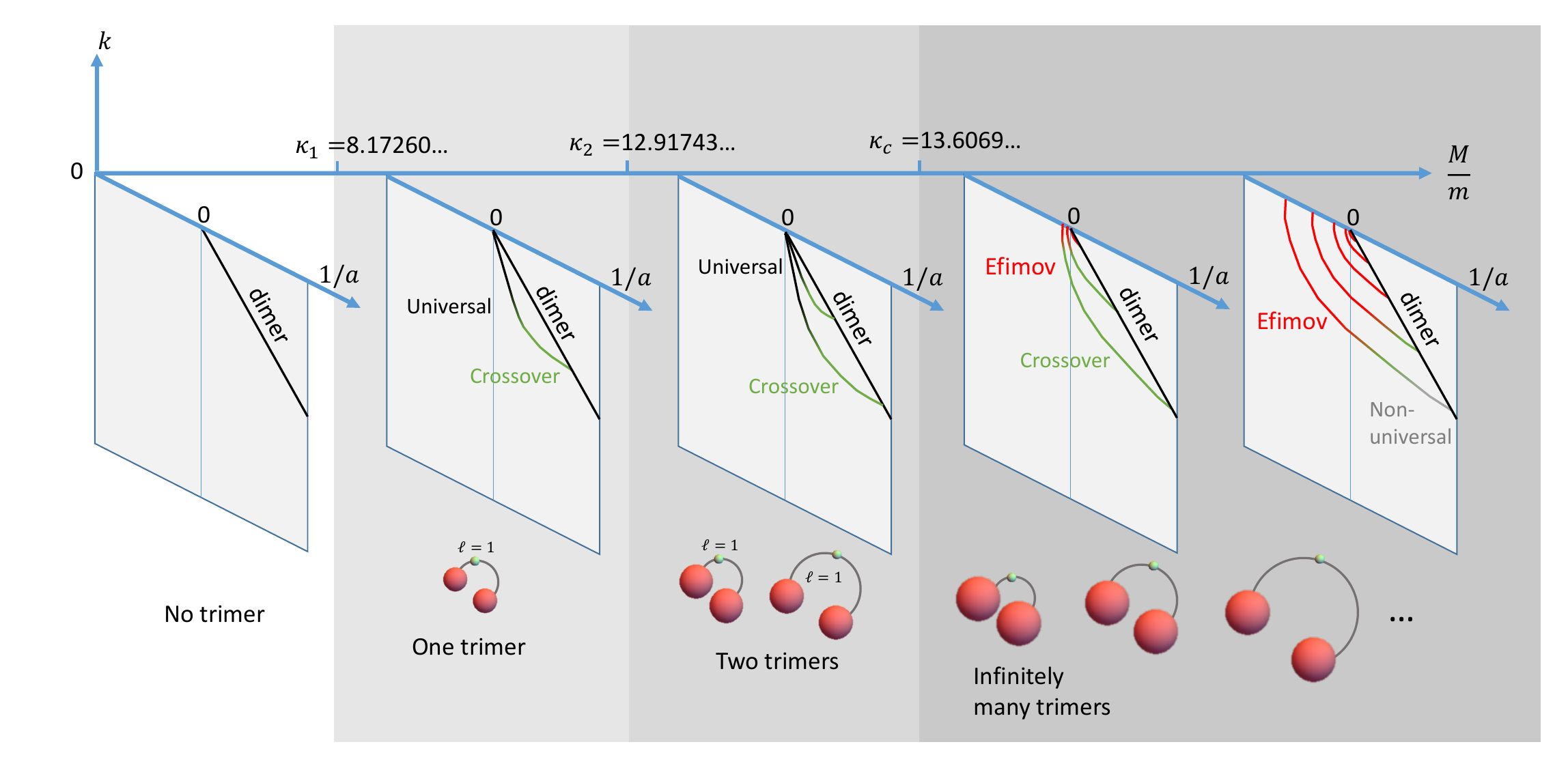}\hfill{}

\caption{\label{fig:Mass-imbalanced-fermions}Schematic three-body energy spectrum
(scaled as a wave number $k$) of two identical heavy fermions resonantly
interacting with a light particle, as a function of inverse scattering
length $1/a$ between the heavy and light particles, and mass ratio
$M/m$ between heavy and light particles. The three-body boundary
condition is set by a cutoff in momentum. A trimer appears above the
critical mass ratio $\kappa_{1}=8.17260...$, and a second trimer
appears above the critical mass ratio $\kappa_{2}=12.91743...$. The
wave number $k$ of these trimers universally scales with the inverse
scattering length $1/a$ near the unitarity limit $1/a\to0$ (as shown
in black) but deviates from this ideal behaviour for larger $1/a$
(as shown in green). Above the critial mass ratio $\kappa_{c}=13.6069...$,
the Efimov occurs and there is an infinite number of trimers. These
trimers exhibit the Efimov discrete scale invariance in the region
where the trimer curves are red. The trimer curves shown in green
break the discrete scale invariance but are characterised only by
the scattering length and a three-body parameter. They are referred
to as ``crossover trimers\textquotedbl{} because they dominate at
the critical mass ratio $\kappa_{c}$ and make a smooth connection
between the universal and Efimov trimers. The trimer curves shown
in grey depend on other microscopic details and are referred to as
non-universal.}

\end{figure*}

The existence of these states have important consequences for inelastic
three-body collisions by recombination into dimers~\cite{Petrov2003}
and the scattering of dimers of fermions with another fermion~\cite{Levinsen2011},
even at mass ratios slightly lower than $\kappa_{1}$, as was confirmed
experimentally~\cite{Jag2014}. However these states have not been
directly observed yet. It should be noted that the universality of
these states is in practice limited to very large scattering lengths~\cite{Endo2012,Safavi-Naini2013}.
Although there is formally no need to introduce a short-range three-body
boundary condition for the universal states, such condition exists
physically at a separation of the particles on the order of the interaction
range $b$. Even when the scattering length is more than ten times
that range, the trimers may be significantly affected by the three-body
boundary condition.

In reference~\cite{Endo2012} the 2+1 fermions problem was solved
with a three-body boundary condition implemented by imposing a cutoff
at some momentum $\sim b^{-1}$. For mass ratio $\kappa_{1}<\kappa<\kappa_{2}$,
a trimer was found to exist on the positive scattering length side
and shows the universal features predicted by Kartavtsev and Malykh
when the scattering length is very large $a/b\gg1$ (see figure~\ref{fig:Mass-imbalanced-fermions}).
As the scattering length is tuned away from unitarity, however, the
binding energy of the trimer gets smaller than that of the universal
trimer, due to the three-body boundary condition. A similar behaviour
is found for mass ratios $\kappa_{2}<\kappa<\kappa_{c}$, where two
universal trimers appear near unitarity (see the central column of
figure~\ref{fig:Mass-imbalanced-fermions}) but gradually turn into
what the authors of reference~\cite{Endo2012} called ``crossover
trimers'', which end up dissociating into a fermion and a fermion-light-particle
dimer as the scattering length is varied further. The crossover trimers
depend on the three-body parameter set by the three-body bounary condition
and smoothly connect the universal trimers to the Efimov trimers appearing
at larger mass ratio $\kappa\ge\kappa_{c}$: as the mass ratio is
increased from below with a fixed positive scattering length, the
universal trimers become more and more sensitive to the three-body
parameter, turning into the crossover trimer states, and then at $\kappa\ge\kappa_{c}$
turn into the ground and first excited states of the Efimov trimers
(the second and higher excited Efimov trimers start to appear at $\kappa>\kappa_{c}$
). This scenario illustrated in figure~\ref{fig:Mass-imbalanced-fermions}
describes how the two universal trimers for $\kappa<\kappa_{c}$ smoothly
connect to an infinite series of Efimov states for $\kappa>\kappa_{c}$. 

The 2+1 fermions problem was subsequently studied by Arghavan Safavi-Naini
and co-workers in reference~\cite{Safavi-Naini2013} with a general
three-body boundary condition implemented by setting the logarithmic
derivative of the hyper-radial wave function at distance on the order
of $b$. While this work confirmed that the trimers behave universally
when $a/b\gg1$ and that they tend to become more sensitive to the
three-body parameter as the mass ratio is increased towards $\kappa_{c}$,
there are some notable differences compared to the above work~\cite{Endo2012}.
In addition to the universal trimers, an additional ``non-universal''
trimer may exist. Here, non-universal means that it strongly depends
on the three-body boundary condition in addition to the scattering
length $a$. As the value of the three-body boundary condition is
varied, the non-universal trimer makes avoided crossings with the
universal trimers, and shifts their energies away from the universal
predictions. In contrast to the momentum cutoff method of reference~\cite{Endo2012},
the shift can be either positive or negative depending on binding
energy of the non-universal state. When it is bound deeper than the
universal trimers, it pushes up their energies. On the contrary, it
pushes down the energies of the universal trimers when it lies above
them.

In the limit where the scattering length $a$ is much larger than
$b$, the three-body boundary condition can be implemented as a zero-range
boundary condition parameterised by a three-body parameter $\Lambda$
(as in equation~(\ref{eq:F0bis})) for mass ratios larger than $\kappa_{r}\approx8.619$,
because both the regular solution (which vanishes at small hyper-radius)
and the irregular solution (which diverges at small hyper-radius)
are square-integrable in this case~\cite{Nishida2008b,Safavi-Naini2013,Kartavtsev2016}.
The precise energy spectrum in this limit for all possible values
of $\Lambda\vert a\vert$ and all mass ratios $\kappa_{r}<\kappa<\kappa_{c}$
has recently been calculated by Kartavtsev and Malykh~\cite{Kartavtsev2016}
for both $a>0$ and $a<0$. In some range of $\Lambda\vert a\vert$,
the trimers may appear, like Efimov states, at a negative scattering
length instead of being restricted to $a>0$. For $\Lambda^{-1}\ll a$,
one retrieves the universal results. 

To answer which kind of three-body boundary condition should be taken
for a given physical system, one needs further knowledge on the microscopic
details of the system in question. Indeed, recent works on the microscopic
origin of the three-body parameter suggest that pairwise interactions
inducing a significant drop of probability at short pair separation
create an effective repulsive barrier that serves as a three-body
boundary condition. This subtle effect is discussed in detail in section~\ref{sec:What-sets-the-3BP}.
Whether it is also relevant here is an open question.

\subsection{Particles with spin\label{subsec:Particles-with-spin}}

In many systems occurring in nature, particles have an internal spin
that complicates somewhat their description.

\subsubsection{Rotationally invariant systems\label{subsec:Rotationally-invariant-systems}}

In many cases, the interaction between two particles 1 and 2 is rotationally
invariant, and thus does not depend on the spin projection numbers
but only on the spin quantum number $s$ for the total spin of the
two particles. For example, for particles with spin 1/2, there are
two interaction potentials: the singlet potential (for $s=0$) and
the triplet potential (for $s=1$). A general treatment of the resonantly-interacting
three-body problem with spin has been given by Bulgac and Efimov~\cite{Bulgac1976}.
In general, for a given total spin $S$ of the three-body system,
one can generalise the Faddeev decomposition of equation~(\ref{eq:FaddeevDecomposition})
as follows~\cite{Braaten2006}: 
\begin{eqnarray}
\Psi_{S}(1,2,3) & = & \sum_{s}\chi_{s}^{(1)}(\vec{r}_{23},\vec{\rho}_{23,1})\vert s\rangle_{23}\label{eq:FaddeevDecompositionWithSpin}\\
 &  & +\sum_{s}\chi_{s}^{(2)}(\vec{r}_{31},\vec{\rho}_{31,2})\vert s\rangle_{31}\nonumber \\
 &  & +\sum_{s}\chi_{s}^{(3)}(\vec{r}_{12},\vec{\rho}_{12,3})\vert s\rangle_{12}.\nonumber 
\end{eqnarray}
The states $\vert s\rangle_{ij}$ denote three-spin states of total
spin quantum number $S$ and spin quantum number $s$ for the pair
$ij$. For a given pair $ij$, these states form a basis of the three-spin
space of total spin $S$. One can thus express the states for the
other pairs as a linear combination of the states for that particular
pair. In section~\ref{subsec:Triton}, we illustrate this by treating
the specific example of the triton.

\subsubsection{Polarised systems\label{subsec:Polarised-systems}}

In some other cases, such as in the presence of a magnetic field that
strongly breaks the rotational invariance, the pairwise interaction
can be assumed to depend only on the spin projections (magnetic quantum
number $m$) of the two particles. Then, each spin projection can
be regarded as a different kind of particle, and the situation is
equivalent to the cases discussed in sections~\ref{subsec:3-distinguishable-particles},
\ref{subsec:2-bosons+1particle}, and \ref{subsec:2-fermions-1particle}.
For instance, in the experiments discussed in section~\ref{subsec:Observations-in-nuclear-Multi-component}
of lithium-6 atoms in a magnetic field, polarised in three different
hyperfine states $\vert a\rangle$, $\vert b\rangle$, and $\vert c\rangle$,
the pairwise interaction just depends on the hyperfine states of the
pair~\cite{Naidon2011}. Although the atoms are fermionic, the fully
antisymmetrised wave function for three atoms in three different spin
states reads:
\begin{eqnarray*}
\Psi(1,2,3) & = & \phantom{+\,}\phi(1,2,3)\vert a,b,c\rangle-\phi(1,3,2)\vert a,c,b\rangle\\
 &  & +\,\phi(2,3,1)\vert b,c,a\rangle-\phi(2,1,3)\vert b,a,c\rangle\\
 &  & +\,\phi(3,1,2)\vert c,a,b\rangle-\phi(3,2,1)\vert c,b,a\rangle,
\end{eqnarray*}
and therefore is equivalent to a single wave function $\phi$ describing
a system of three distinguishable particles, with identical masses
but three different scattering lengths $a_{12},$ $a_{23}$, and $a_{31}$
- see section~\ref{subsec:3-distinguishable-particles} for the derivation
of the Efimov effect in this system.

\subsubsection{Spin-orbit interaction\label{subsec:Spin-orbit-interaction}}

Spin-orbit interaction is the coupling between a particle's spin and
its motion. It occurs for charged particles such as electrons in atoms
and solids, as well as nucleons inside a nucleus. In the recent years,
it has also been possible to create artificial spin-orbit interaction
for neutral atoms using laser techniques~\cite{Galitski2013}. The
influence of the spin-orbit interaction on Efimov physics was addressed
by Zhe-Yu Shi, Xiaoling Cui, and Hui Zhai~\cite{Shi2014,Shi2015}.
They considered a system of two heavy fermions resonantly interacting
with one light particle, in which the light particle is a spin-1/2
particle subject to an isotropic spin-orbit interaction of the form
$\lambda\vec{p}\cdot\vec{\sigma}$, where $\vec{p}$ is the particle's
momentum and $\vec{\sigma}$ its spin. From their calculations, some
general conclusions can de drawn. 

First of all, the spin-orbit interaction lowers the energy of the
heavy-light dimer and heavy-heavy-light trimers. However, the spin-orbit
interaction competes with the Efimov attraction and reduces the binding
energy of Efimov trimers with respect to the particle-dimer threshold.
Since the spin-orbit interaction only affects large distances, while
the Efimov attraction persists at shorter distances, the critical
mass ratio $\kappa_{c}$ for the onset of Efimov states, the discrete
scale invariance and scaling ratio $\lambda_{0}$ remain unchanged.
The discrete scale invariance is however broken below a certain energy
scale associated with the coupling strength $\lambda$, making the
number of trimers finite, as the excited trimers are pushed into the
particle-dimer threshold.

On the other hand, below the critical mass ratio, the spin-orbit interaction
favours the appearance of Kartavtsev-Malykh-like universal trimers
in a broader range of scattering lengths and mass ratios. In the presence
of spin-orbit coupling, these universal trimers not only exist for
positive scattering lengths but in a range of negative scattering
lengths as well, and the ground-state trimer appears at mass ratio
2.68 (for the state with total angular momentum $J=1/2$) and 5.92
(for $J=3/2$).

\subsection{Observations in nuclear physics\label{subsec:Observations-in-nuclear-Multi-component}}

So far, all known multicomponent systems in nuclear physics related
to Efimov physics involve two neutrons as two of the three particles.
Neutrons are favourable for the following reasons:
\begin{itemize}
\item First of all, the interaction between two neutrons is resonant, the
basic requirement for Efimov physics, since the neutrons can almost
form a two-body bound state. 
\item Second, they carry a spin 1/2, and can therefore be in two distinguishable
states, or more precisely they can form a spin singlet state. This
particular configuration is free of the centrifugal repulsion that
is otherwise present for fermions in a triplet state (such as two
identical fermions) and would suppress the Efimov attraction. 
\item Third, neutrons having no electrical charge, unlike protons, they
do not have any Coulomb repulsion between themselves and other particles,
that would compete with the Efimov attraction.
\end{itemize}

\subsubsection{Triton\label{subsec:Triton}}

The simplest case is the third particle being another nucleon. It
cannot be a neutron, since it would have to be in the same spin state
as one of the two others, and the resulting Fermi repulsion would
suppress the Efimov attraction. As seen in section~\ref{subsec:2-fermions-1particle},
the Efimov attraction for two identical fermions and one particle
of equal mass is not strong enough to overcome the Fermi repulsion.
As a matter of fact, there is no bound state of three neutrons. The
third nucleon therefore has to be a proton, and the three-body system
corresponds to the triton, the nucleus of tritium $^{3}\mbox{H}$.

The triton was considered in Vitaly Efimov's original work as a possible
candidate for Efimov state. Here, we show how the Efimov attraction
explains the binding of the triton and roughly reproduces its binding
energy. We first derive the Efimov attraction in the zero-range theory
of two neutrons and one proton, and show that is the same as for three
identical bosons. Then, we take into account the finite-range corrections,
and show how the triton fits in the Efimov spectrum.

\paragraph{Efimov attraction in the triton}

Systems of nucleons are often described as identical particles with
an internal property called isospin, which distinguishes between the
neutron state and proton state as different projections of a formal
spin 1/2. This is possible because protons and neutrons have nearly
the same mass, and nuclear interactions are approximately the same
for protons and neutrons, and thus nearly isospin-symmetric. The isospin
symmetry has been further confirmed to originate from the symmetries
of quantum chromodynamics (QCD) describing nucleons as made of quarks.
In the zero-range theory, however, the isospin formalism is not essential,
as the nuclear interactions turn out to depend only on spin at this
simple level of description. For the sake of readers unfamiliar with
isospin, we shall simply describe the triton as two identical fermions
(two neutrons) and a distinguishable fermion (proton), and assume
that the pairwise nuclear interaction depends only on the total spin
quantum number of two nucleons. We also assume the neutron and the
proton to have the same mass $m$. For a treatment with isospin, we
refer the readers to Refs.~\cite{Skorniakov1957,Sitenko1963}.

Each nucleon carries a spin 1/2, whose projection on a fixed axis
can be either up $\vert\uparrow\rangle$ or down $\vert\downarrow\rangle$.
The triton is characterised by a total spin $S$ equal to $1/2$,
and a projection $S_{z}=1/2$. Therefore, its wave function can be
expressed on the spin basis states,
\begin{eqnarray}
\vert0\rangle_{23} & = & \vert\uparrow\rangle_{1}\frac{1}{\sqrt{2}}\left(\vert\uparrow\rangle_{2}\vert\downarrow\rangle_{3}-\vert\downarrow\rangle_{2}\vert\uparrow\rangle_{3}\right)\label{eq:SpinSingletState}\\
\vert1\rangle_{23} & = & \frac{1}{\sqrt{3}}\vert\uparrow\rangle_{1}\frac{1}{\sqrt{2}}\left(\vert\uparrow\rangle_{2}\vert\downarrow\rangle_{3}+\vert\downarrow\rangle_{2}\vert\uparrow\rangle_{3}\right)\nonumber \\
 &  & -\sqrt{\frac{2}{3}}\vert\downarrow\rangle_{1}\vert\uparrow\rangle_{2}\vert\uparrow\rangle_{3}\label{eq:SpinTripletState}
\end{eqnarray}
which are obtained by standard summation of spins using Clebsh-Gordan
coefficients. Here, we assume that particle 1 is the proton, and particles
2 and 3 are the two neutrons. As can be seen from the above expressions,
$\vert0\rangle_{23}$ corresponds to a spin singlet state of the two
neutrons, whereas $\vert1\rangle_{23}$ corresponds to a spin triplet
state. The total wave function of the triton is thus:
\[
\Psi(123)=\psi_{S}\vert0\rangle_{23}+\psi_{A}\vert1\rangle_{23}
\]
and has to be antisymmetric under the exchange of the two neutrons,
particles 2 and 3. Since the singlet and triplet states are respectively
antisymmetric and symmetric under such exchange, as can be checked
from equations~(\ref{eq:SpinSingletState}-\ref{eq:SpinTripletState}),
$\psi_{S}$ and $\psi_{A}$ have to be symmetric and antisymmetric,
respectively. The Faddeev decompositions of $\psi_{S}$ and $\psi_{A}$
that preserve their symmetries are: 

\begin{eqnarray}
\!\!\!\!\!\!\!\!\!\!\!\!\psi_{S} & \!\!\!\!=\!\!\!\! & \chi(\vec{r}_{23},\vec{\rho}_{23,1})+\phi(\vec{r}_{12},\vec{\rho}_{12,3})+\phi(\vec{r}_{13},\vec{\rho}_{13,2})\quad\label{eq:TritonPsiS}\\
\!\!\!\!\!\!\!\!\!\!\!\!\psi_{A} & \!\!\!\!=\!\!\!\! & \xi(\vec{r}_{23},\vec{\rho}_{23,1})+\zeta(\vec{r}_{12},\vec{\rho}_{12,3})-\zeta(\vec{r}_{13},\vec{\rho}_{13,2})\quad\label{eq:TritonPsiA}
\end{eqnarray}
where $\chi$ and $\xi$ are respectively even and odd functions of
$\vec{r}_{23}$. One can see that the spin singlet configuration allows
the spatial configuration $\psi_{S}$ to have the bosonic exchange
symmetry when $\phi=\chi$, as in equation~(\ref{eq:FaddeevDecomposition}),
which in turn allows the Efimov effect to occur.

In the zero-range theory, $\psi_{S}$ and $\psi_{A}$ each satisfy
the free Schr\"odinger equation - see equation~(\ref{eq:FreeSchrodingerEqPsi})
- with the Bethe-Peierls boundary conditions for the contact of two
nucleons, either in the singlet or triplet spin state. In the case
of the neutron-neutron interaction, these conditions are readily expressed
as:
\begin{eqnarray}
\left(\frac{\partial}{\partial r_{23}}r_{23}\psi_{S}\right)_{r_{23}\to0} & = & -\frac{1}{a_{s}}\left(r_{23}\psi_{S}\right)_{r_{23}\to0}\quad\quad\quad\label{eq:NeutronNeutronSingletCondition}\\
\left(\frac{\partial}{\partial r_{23}}r_{23}\psi_{A}\right)_{r_{23}\to0} & = & -\frac{1}{a_{t}}\left(r_{23}\psi_{A}\right)_{r_{23}\to0}\quad\quad\quad\label{eq:NeutronNeutronTripletCondition}
\end{eqnarray}
where $a_{s}$ and $a_{t}$ are the nucleon singlet and triplet $s$-wave
scattering lengths. One can then proceed as in section \ref{subsec:Efimov-theory},
retaining only the zero-angular momentum contribution of the Faddeev
components of $\psi_{S}$ and $\psi_{A}$, i.e. assuming the form
of equation~(\ref{eq:SWaveChi}), and using the hyper-spherical coordinates
of equations~(\ref{eq:HypersphericalCoordinate1}-\ref{eq:HypersphericalCoordinate2}).
This gives a boundary condition analogous to equation~(\ref{eq:HypersphericalCondition})
for the singlet interaction,
\begin{equation}
\left[\!\frac{\partial}{\partial\alpha}\left(\chi_{0}(R,\alpha)\right)\!\right]_{\alpha\to0}+\frac{8}{\sqrt{3}}\phi_{0}\!\left(\!\!\!\begin{array}{c}
R,\frac{\pi}{3}\end{array}\!\!\!\right)=\!-\frac{R}{a_{s}}\chi_{0}(R,0)\label{eq:NeutronNeutronHypersphericalCondition}
\end{equation}
For the triplet interaction, equation~(\ref{eq:NeutronNeutronTripletCondition})
gives no constraint on $\psi_{A}$ because it is antisymmetric under
the exchange of the two neutrons, i.e. $\psi_{A}\to0$ when $r_{23}\to0$.
This expresses the fact that for zero-range interactions neutrons
interact only in the singlet state. For this reason, we can simply
set $\xi=0$.

In the case of proton-neutron interactions, we first have to rotate
the spin basis to obtain states that are singlet and triplet states
of the proton-neutron subsystem (say, particles 1 and 2):
\[
\left(\begin{array}{c}
\vert0\rangle_{23}\\
\vert1\rangle_{23}
\end{array}\right)=\left(\begin{array}{cc}
-1/2 & -\sqrt{3}/2\\
\sqrt{3}/2 & -1/2
\end{array}\right)\left(\begin{array}{c}
\vert0\rangle_{12}\\
\vert1\rangle_{12}
\end{array}\right)
\]
This gives
\[
\Psi=\psi_{0}\vert0\rangle_{12}+\psi_{1}\vert1\rangle_{12}
\]
with
\[
\psi_{0}=-\frac{1}{2}\psi_{S}+\frac{\sqrt{3}}{2}\psi_{A}\quad;\quad\psi_{1}=-\frac{\sqrt{3}}{2}\psi_{S}-\frac{1}{2}\psi_{A}.
\]
\begin{figure*}[t]
\includegraphics[scale=0.58]{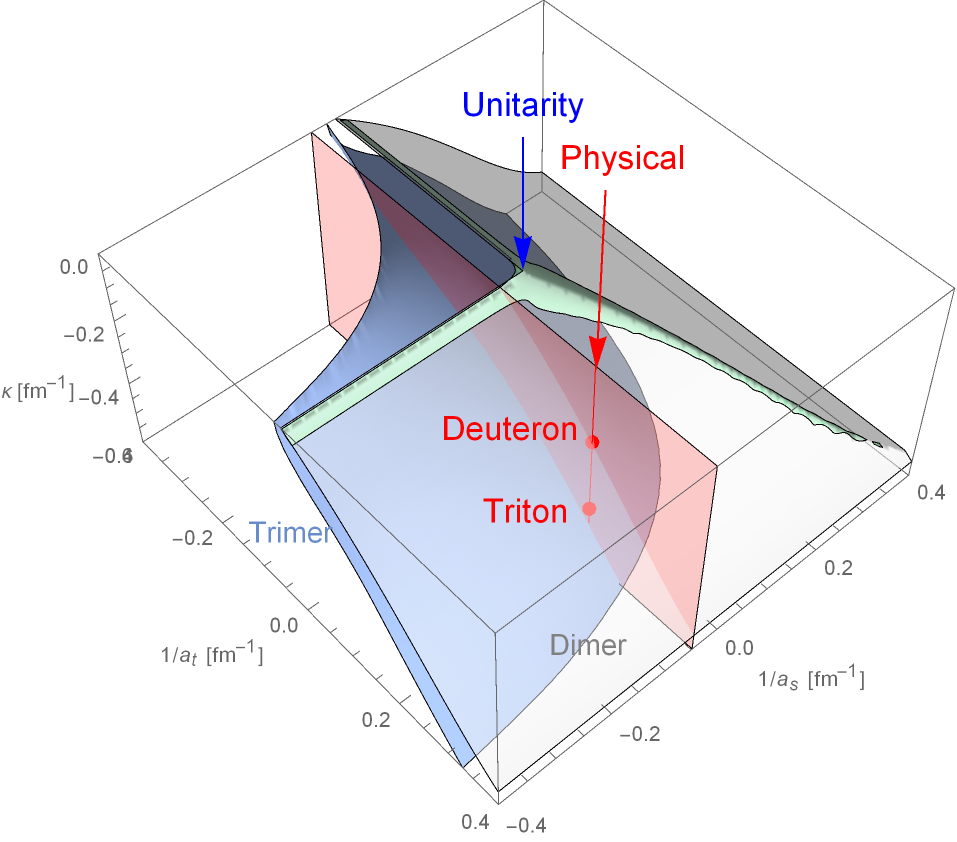}\includegraphics[scale=0.6]{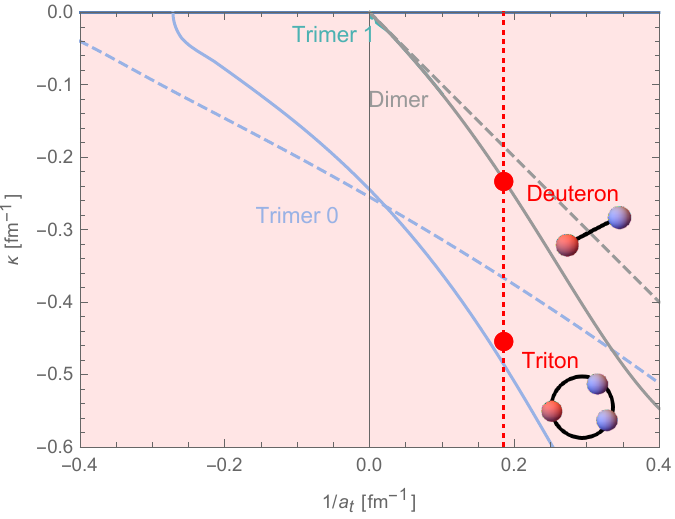}\caption{\label{fig:Triton}(Left) Energy spectrum of two neutrons and one
proton, as a function of the inverses of the singlet and triplet scattering
lengths $1/a_{s}$ and $1/a_{t}$. The energy $E$ (vertical axis)
is scaled as a binding momentum $\kappa=-\sqrt{m\vert E\vert}/\hbar$.
The lowest dimer energy (white transparent surface), the ground-state
trimer energy (blue surface), and the first excited trimer energy
(green surface) are obtained from a separable model (See Appendix
for details). The blue arrow indicates the unitarity point $1/a_{s}=1/a_{t}=\kappa=0$
where the Efimov series of trimers accumulate. The physical value
of $a_{s}$ is indicated by the red plane, and the physical value
of $a_{t}$ is indicated within that plane by the red line. The red
dots on that line indicate the experimental values for the energies
of the deuteron and the triton.\protect \\
 (Right) Section of left figure in the red plane. The excited trimer
is barely visible. The dashed curves show the results for the zero-range
theory with a three-body parameter set to match the highly-excited
states.}
\end{figure*}
The boundary conditions are then expressed as:
\begin{eqnarray*}
\left(\frac{\partial}{\partial r_{12}}r_{12}\psi_{0}\right)_{r_{12}\to0} & = & -\frac{1}{a_{s}}\left(r_{12}\psi_{0}\right)_{r_{12}\to0}\\
\left(\frac{\partial}{\partial r_{12}}r_{12}\psi_{1}\right)_{r_{12}\to0} & = & -\frac{1}{a_{t}}\left(r_{12}\psi_{1}\right)_{r_{12}\to0}
\end{eqnarray*}
This gives{\small{}
\begin{align}
\left[\frac{\partial}{\partial\alpha}\left(\phi_{0}-\sqrt{3}\zeta_{0}\right)\right]_{\alpha\to0}\!\!\!+\frac{4}{\sqrt{3}}\left(\chi_{0}+\phi_{0}+\sqrt{3}\zeta_{0}\right)(R,{\textstyle \frac{\pi}{3}})\nonumber \\
=-\frac{R}{a_{s}}\left(\phi_{0}-\sqrt{3}\zeta_{0}\right)(R,0),\label{eq:ProtonNeutronHypersphericalCondition1}
\end{align}
\begin{align}
\left[\frac{\partial}{\partial\alpha}\left(\phi_{0}+\frac{\zeta_{0}}{\sqrt{3}}\right)\right]_{\alpha\to0}\!\!\!+\frac{4}{\sqrt{3}}\left(\chi_{0}+\phi_{0}-\frac{\zeta_{0}}{\sqrt{3}}\right)(R,{\textstyle \frac{\pi}{3}})\nonumber \\
=-\frac{R}{a_{t}}\left(\phi_{0}+\frac{\zeta_{0}}{\sqrt{3}}\right)(R,0).\label{eq:ProtonNeutronHypersphericalCondition2}
\end{align}
} Combining equations~(\ref{eq:NeutronNeutronHypersphericalCondition})
and (\ref{eq:ProtonNeutronHypersphericalCondition1}), one obtains
a closed condition
\[
\left[\!\frac{\partial}{\partial\alpha}\left(\tilde{\chi}_{0}(R,\alpha)\right)\!\right]_{\alpha\to0}-\frac{4}{\sqrt{3}}\tilde{\chi}_{0}\!\left(\!\!\!\begin{array}{c}
R,\frac{\pi}{3}\end{array}\!\!\!\right)=\!-\frac{R}{a_{s}}\tilde{\chi}_{0}(R,0)
\]
for the quantity $\tilde{\chi}_{0}=\chi_{0}-\phi_{0}+\sqrt{3}\zeta_{0}$.
Unlike equation~(\ref{eq:HypersphericalCondition}) for the bosonic
case, this condition itself only yields real eigenvalues $s_{n}$
which do not lead to the Efimov effect. However, unlike the bosonic
case, this condition admits the extra solution $\tilde{\chi}_{0}=0$,
which does not set the total wave function to zero. In the language
of isospin symmetry, this corresponds to considering states with total
isospin $T=1/2$. Making this choice, the two remaining conditions
are: 
\[
\left[\frac{\partial}{\partial\alpha}\chi_{0}\right]_{\alpha\to0}+\frac{8}{\sqrt{3}}\phi_{0}(R,{\textstyle \frac{\pi}{3}})=-\frac{R}{a_{s}}\chi_{0}(R,0),
\]
\begin{align*}
\left[\frac{\partial}{\partial\alpha}\left(-4\phi_{0}+\chi_{0}\right)\right]_{\alpha\to0}+\frac{4}{\sqrt{3}}\left(-4\chi_{0}-2\phi_{0}\right)(R,{\textstyle \frac{\pi}{3}})\\
=-\frac{R}{a_{t}}\left(-4\phi_{0}+\chi_{0}\right)(R,0).
\end{align*}
Setting $\chi_{0}=f-\varphi$ and $\phi_{0}=f+\frac{1}{2}\varphi$,
one obtains
\begin{equation}
\frac{\partial f}{\partial\alpha}(R,0)+\frac{8}{\sqrt{3}}f(R,{\textstyle \frac{\pi}{3}})=-R\left(\frac{1}{a_{+}}f+\frac{1}{a_{-}}\varphi\right)(R,0)\label{eq:TritonCondition1}
\end{equation}
\begin{equation}
\frac{\partial\varphi}{\partial\alpha}(R,0)-\frac{4}{\sqrt{3}}\varphi(R,{\textstyle \frac{\pi}{3}})=-R\left(\frac{1}{a_{-}}f+\frac{1}{a_{+}}\varphi\right)(R,0)\label{eq:TritonCondition2}
\end{equation}
where $a_{\pm}^{-1}=\frac{1}{2}(a_{t}^{-1}\pm a_{s}^{-1})$. These
equations were first presented in Efimov's original work~\cite{Efimov1970a},
although their integral version in momentum space had been derived
thirteen years earlier by Skorniakov and Ter-Martirosian~\cite{Skorniakov1957}.
As shown in section~\ref{subsec:Efimov-theory}, the advantage of
this form is that one can immediately conclude that the system features
the Efimov attraction. Indeed, as anticipated from the symmetry of
equation~(\ref{eq:TritonPsiS}), when both the singlet and triplet
scattering lengths are infinite, equation~(\ref{eq:TritonCondition1})
reduces to equation~(\ref{eq:HypersphericalCondition}) for bosons
at unitarity. Therefore, \emph{there occurs an Efimov attraction of
the same strength as that of identical bosons, leading to the same
scaling ratio $e^{\pi/\vert s_{0}\vert}\approx22.7$ in the energy
spectrum}. 

\paragraph{Finite-range corrections}

The above zero-range theory predicts the existence of a two-body bound
state of neutron and proton (the deuteron), since $a_{t}>0$, as well
as possible Efimov three-body bound states of two neutrons and one
proton. Using the deuteron energy $E_{d}\approx2.224$~MeV and neutron-deuteron
spin-doublet scattering length $a_{2}\approx0.65$~fm as experimental
inputs to determine respectively the values of $a_{t}$ and the three-body
parameter in the zero-range theory, one finds one Efimov trimer, and
its energy 8.1~MeV \cite{Efimov1985,Efimov1988,Bedaque2003} is within
5\% that of the triton, $E_{t}\approx8.48$~MeV. However, this theory
is only qualitative. For instance, using the value $a_{t}=5.4112$~fm~\cite{Hackenburg2006},
the predicted binding energy from equation~(\ref{eq:UniversalDimerEnergy})
for the deuteron is 
\[
\frac{\hbar^{2}}{ma_{t}^{2}}\approx1.416\,\mbox{MeV},
\]
which significantly differs from the experimental value 2.224~MeV.
This means that non-zero range corrections are important.

The effective range theory is more quantitative. From equation~(\ref{eq:LowEnergyPhaseShift}),
with the triplet effective range $r_{e,t}=1.7436$~fm~\cite{Hackenburg2006},
one obtains the binding energy in the effective range approximation
\begin{equation}
\frac{\hbar^{2}}{m}\left(\frac{1-\sqrt{1-2r_{e,t}/a_{t}}}{r_{e,t}}\right)^{2}\approx2.223\,\mbox{MeV},\label{eq:EffectiveRangeDeuteron}
\end{equation}
which is remarkably close to the experimental value. Taking the effective
range correction to first order only, one finds:

\begin{equation}
\frac{\hbar^{2}}{ma_{t}^{2}}\left(1+\frac{r_{e,t}}{a_{t}}\right)\approx1.909\,\mbox{MeV}\label{eq:FirstOrderEffectiveRangeDeuteron}
\end{equation}
which differs by 15\% from the experimental value.

As discussed in section~\ref{subsec:Finite-range-interactions},
the first-order effective-range correction may also be taken into
account in the three-body calculations~\cite{Efimov1991}. Using
$a_{2}$ as the three-body input, and $a_{s}=-23.7$~fm~\cite{Hackenburg2006}
and $r_{e,s}$ as two-body inputs, V. Efimov and E. G. Tkachenko~\cite{Efimov1988}
obtained 8.8~MeV for the energy of the trimer (using $r_{e,t}=1.75$~fm
and $r_{e,s}=2.67$~fm), while P.~.F.~Bedaque and co-workers~\cite{Bedaque2003}
obtained 8.3~MeV (using $r_{e,t}=1.764$~fm and $r_{e,s}=2.73\,\mbox{fm}$)
in the framework of effective field theory. Thus, once the effective-range
corrections are taken into account, the predicted trimer energy remains
just a few percent away from the triton energy. That is because the
main effect of range corrections is to shift both the dimer and trimer
energies by a similar amount. \emph{These results show that the binding
of the triton is consistent with the Efimov scenario}.

\paragraph{Explanation of the Phillips line}

The zero-range theory, with or without finite-range corrections, requires
a three-body input to set the three-body energy, such as the neutron-deuteron
spin-doublet scattering length $a_{2}$. It therefore implies a correlation
between the trimer energy and that scattering length. Such correlation
was observed numerically for three-nucleon systems and known as the
Phillips line~\cite{Phillips1968}. If one plots the results for
the triton energy $E_{t}$ and neutron-deuteron scattering length
$a_{2}$ obtained from various nucleon-nucleon potential models, one
finds that the points $(E_{t},a_{2})$ tend to form a line. V. Efimov
and E. G. Tkachenko~\cite{Efimov1985,Efimov1988} pointed out that
the zero-range theory gives a natural explanation for this fact. Since
different nucleon-nucleon potentials lead to slightly different scattering
lenths and three-body parameters, their results sample a small portion
of the curve relating $E_{t}$ to $a_{2}$ in the zero-range theory,
thus forming a small line. The fact that they do so shows the relevance
of the zero-range picture for these systems. 

\paragraph{Triton and deuteron in the Efimov spectrum}

To visualise how the triton and deuteron fit into the Efimov spectrum,
we have represented in figure~\ref{fig:Triton} the energy spectrum
of two neutrons and one proton as a function of $1/a_{s}$ and $1/a_{t}$.
To take into account the effect of the effective range, the energy
is calculated using a separable interaction model~\cite{Kharchenko1967}
(see Appendix for details), which is parameterised to reproduce the
values~\cite{Hackenburg2006} of the effective ranges $r_{e,t}=1.7436$~fm
and $r_{e,s}=2.750$~fm at the physical values of $a_{t}$ and $a_{s}$.
Such a model accurately reproduces the effective-range approximation
of dimer energy given by equation~(\ref{eq:EffectiveRangeDeuteron}).
Moreover, it determines the trimer energy without requiring to set
a three-body parameter from a three-body input. The trimer energy
depends on the particular form of the separable interaction~\cite{Kharchenko1967},
but as we shall see in section~\ref{subsec:Other-types-of-interactions},
it is roughly determined by the effective ranges only.

The energy surfaces shown in figure~\ref{fig:Triton} are the ground
and excited trimers and the singlet and triplet dimers. Near the unitarity
point $(1/a_{s},1/a_{t})=(0,0)$, the spectrum is discrete-scale-invariant
by simultaneous scaling of $a_{s}$ and $a_{t}$ by a factor of $e^{\pi/\vert s_{0}\vert}\approx22.7$.
The physical values of $a_{s}$ and $a_{t}$ correspond to a vertical
line in that figure, where the experimental energies of the deuteron
and the triton are indicated. From this figure, one can conclude that
\emph{the deuteron is the triplet dimer, and the triton connects to
the ground state of an Efimov series of trimers} which accumulates
at $(1/a_{s},1/a_{t},\kappa)=(0,0,0)$.

To reproduce the experimental properties of the deuteron and the triton
with a higher accuracy, it is of course necessary to construct more
sophisticated models building in the details of nuclear interactions
(three-body force in particular), coupling of partial waves, the difference
of mass between the proton and the neutron, etc. Nonetheless, the
Efimov scenario gives a simple understanding of the deuteron and the
triton.

\begin{figure*}[t]
\hfill{}\includegraphics[viewport=0bp 80bp 720bp 510bp,clip,scale=0.65]{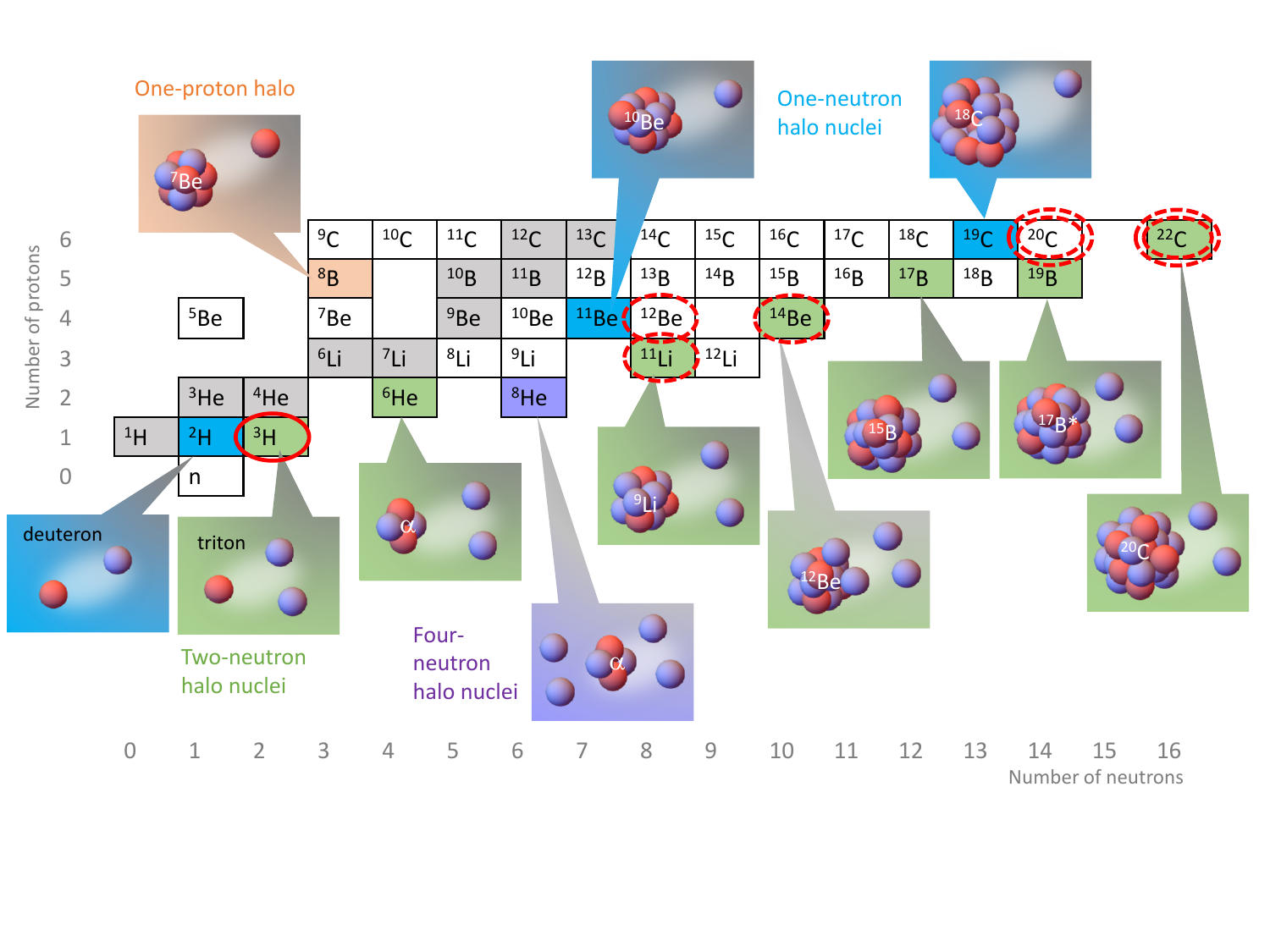}\hfill{}

\caption{\label{fig:Table-of-nuclides}Table of nuclides, as a function of
the number of protons and number of neutrons, showing stable elements
(grey) and observed light halo nuclei: one-proton halo (orange), one-neutron
halo (blue), two-neutron halo (green), and four-neutron halo (purple).
For each halo nuclei, a schematic representation of their structure
is given, where red spheres represent protons and blue spheres represent
neutrons. Note that $^{14}$Be has been argued to be a four-neutron
halo nucleus~\cite{Jensen2004}. Note also that the $^{17}$B core
of the two-neutron halo of $^{19}$B does not have the same structure
as the halo nucleus of $^{17}$B. The red circles indicate possible
candidates for ground-state Efimov trimers. The double circle indicates
the candidates for excited Efimov trimers. The solid circle denotes
a rather established one, while the dashed circles are more conjectural.}
\end{figure*}

\subsubsection{Two-neutron halo nuclei}

Other candidates for Efimov physics in nuclear systems are halo nuclei~\cite{Jensen2004,Hammer2008,Hammer2010,Frederico2014}.
Halo nuclei are exotic nuclei discovered from the 1980s, which have
an anomalously large mean radius and small binding energy. They usually
correspond to neutron-rich or proton-rich nuclei, or excited states
of normal nuclei, which means that their lifetime is usually short,
on the order of a few up to hundreds of milliseconds. Light halo nuclei
are shown in figure~\ref{fig:Table-of-nuclides}. Numerous experiments
and analyses have determined that they can be described as a compact
core nucleus surrounded by one or few loosely bound nucleons forming
a diffuse halo around the core~\cite{Tanihata2013}. Examples of
one-neutron halos are beryllium-11 and carbon-19. Examples of two-neutron
halo nuclei are helium-6, lithium-11, boron-17, boron-19 and carbon-22.
Helium-8 is considered to be a four-neutron halo nucleus.

At first sight, halo nuclei look very similar to the situation described
by the zero-range theory of section \ref{subsec:Efimov-theory}. The
one-neutron halos can be recognised as the universal dimer supported
by a resonant short-range interaction, assuming that the effective
interaction between the neutron and the core is resonant in the $s$
wave. It is thus tempting to identify the two-neutron halo nuclei
with Efimov states composed of a core and two neutrons, in accordance
with the discussion at the beginning of section~\ref{subsec:Observations-in-nuclear-Multi-component}.
This identification would naturally explain a number of features such
as the large extent of halo nuclei and the borromean nature of some
of them.

However, there may also be significant differences preventing the
identification of halo nuclei with Efimov states. For instance, the
two-neutron halo nucleus of helium-6 is known to feature a $p$-wave
resonance between the core and one neutron, in addition to their $s$-wave
interaction. As seen in section~\ref{subsec:Resonant-p-wave-interactions},
$p$-wave resonances do not lead to the discrete scale invariance
of Efimov physics. Therefore, halo nuclei such as helium-6 cannot
be considered as Efimov states, although they may be described by
universal models~\cite{Ji2014}. Another difficulty is that one would
expect to find Efimov states by adding one neutron to the one-neutron
halo nuclei $^{11}$Be and $^{19}$C, but $^{12}$Be and $^{20}$C
are not currently recognised as two-neutron halo nuclei. It is usually
thought that for these systems, the second neutron forms a Cooper
pair with the first neutron, which condenses with the other pairs
of the core, making the nucleus more compact. However, it is not excluded
that $^{12}$Be and $^{20}$C may constitute Efimov ground states
nonetheless. Indeed, the experimental characterisation of halo nuclei
relies on the measurement of anomalously large cross sections and
narrow momentum distributions of neutrons. While such features are
expected for a ground Efimov state close to the two-body or three-body
thresholds where its size becomes very large, the ground Efimov state
for interactions near unitarity is relatively compact, with a typical
size only marginally larger than the range of interactions. This situation
could explain why the nuclei obtained by adding one neutron to a one-neutron
halo nucleus are more compact and not recognised as halo states. The
assumption that the core remains inert in such compact states is however
questionable.

On the other hand, there exist two-neutron halo nuclei, such as $^{11}$Li
and $^{22}$C, with no corresponding one-neutron halo nucleus. These
nuclei could therefore correspond to borromean Efimov states of a
core and two neutrons, for which the core-neutron subsystems are unbound
but have a virtual state close to threshold (such as $^{10}$Li which
is observed as a resonance). Such Efimov states are typically much
larger in size than the range of interactions, which would explain
the halo structure of these nuclei.

In summary, although not all two-neutron halo nuclei are manifestations
of Efimov states ($^{6}$He is not), some halo nuclei (such as $^{11}$Li
and $^{22}$C) are possibly ground-state Efimov trimers in the borromean
regime, and other nuclei which are not experimentally regarded as
halo nuclei (such as $^{12}$Be and $^{20}$C) could be ground-state
Efimov trimers close to the unitary limit. For these nuclei, all interactions
(core-neutron and neutron-neutron) are assumed to be resonant, and
the neutrons can be in a symmetric state (spin singlet). If the core
has spin zero, the situation is equivalent to that of two identical
bosons and one particle with three resonantly-interacting pairs, as
described in section~\ref{subsec:2-bosons+1particle} and illustrated
in Fig\@.~\ref{fig:2+1}. Since the core is at least nine times
heavier than the neutron, the Efimov scaling ratio $\lambda$ when
both the neutron-neutron and core-neutron scattering lengths are infinite
should be close to the limiting value $\approx15.74$ for small mass
ratios, which is marginally smaller than the scaling ratio for identical
bosons $\approx22.7$. As discussed in section~\ref{subsec:What-is-an-Efimov-state},
the ground state of an Efimov series, though more easily observable,
is difficult to characterise as an Efimov state due to important finite-range
corrections. The excited states, on the other hand, conform more clearly
to the universal properties of Efimov states, but are less likely
to exist and more difficult to observe.

\paragraph{Efimov models of nuclei}

The first theorists to consider the possibility of Efimov nuclei made
of a core plus two neutrons under these assumptions are Dmitri V.
Fedorov, Aksel S. Jensen, and Karsten Riisager~\cite{Fedorov1994}
in 1994. They solved the Faddeev equations for such three-body system,
assuming Gaussian potentials for the neutron-neutron and neutron-core
interacions, and concluded that $^{14}$Be, $^{18}$C and $^{20}$C
are possible candidates for Efimov states.  The Efimov character
of nuclei such as $^{11}$Li, $^{14}$Be, and $^{22}$C was also investigated
using three-body models with separable interactions~\cite{Dasgupta1994,Mazumdar1997,Mazumdar2000,Mazumdar2006}.

In 1997, A. E. A. Amorim, Tobias Frederico, and Lauro Tomio~\cite{Amorim1997}
solved the same three-body problem for zero-range interactions, and
plotted the region of existence of the $(N+1)$th Efimov state, which
is universally determined by the core-neutron and neutron-neutron
scattering lengths normalised by the $N$th Efimov state's energy.
Assuming that nuclei such as $^{12}$Be, $^{18}$C, $^{20}$C are
ground-state Efimov trimers ($N=0$), and using the measured energies
of core-neutron bound states (or virtual states) and neutron-neutron
scattering length, the authors could represent the locations of these
nuclei in this plot. Only one nucleus, carbon-20, enters the region
of existence of the first excited Efimov state ($N=1$). Carbon-20
lies in fact just at the region boundary, and the large uncertainty
on the core-neutron bound state energy (the one-neutron halo nucleus
of carbon-19) makes it possible for an excited Efimov trimer state
to exist either as a bound state or a virtual state~\cite{Yamashita2008a,Yamashita2008b}. 

In 2008, David L. Canham and Hans-Werner Hammer~\cite{Canham2008}
refined the zero-range calculations of Amorim and co-workers in the
framework of effective-field theory and estimated the errors due to
finite-range corrections. The same authors included the finite-range
corrections explicitly in a later work~\cite{Canham2010}. They reached
the same conclusion that among the nuclei $^{11}$Li, $^{12}$Be,
$^{14}$Be, $^{18}$C, and $^{20}$C, carbon-20 is the only candidate
for the existence of an excited Efimov state. As mentioned before,
this analysis assumes that the ground state of these nuclei is a ground
Efimov state, \emph{i.e.} the three-body parameter of the zero-range
theory is adjusted to reproduce the ground-state energy of these nuclei\footnote{or more precisely, the energy required to separate two neutrons, since
the core is considered inert in the three-body model}. Even if the ground state of $^{20}$C turns out to be a compact
nucleus that has no connection with the Efimov attraction, it could
be that an excited state features two neutrons sufficiently far from
the core to conform to the structure an Efimov excited state. In this
case, however, the ground-state energy of $^{20}$C cannot be used
to predict the energy of that excited state.

In 2010, carbon-22 was experimentally identified as a two-neutron
halo nucleus, exhibiting a root-mean-square matter radius of $5.4\pm0.9$~fm~\cite{Tanaka2010}
- recently refined to $3.44\pm0.08$~fm~\cite{Togano2016}. Although
its two-neutron separation energy has not been measured, the radius
value could be used to parameterise three-body models \cite{Yamashita2011a,Acharya2013},
constraining the two-neutron separation energy to be smaller than
100~keV (400 keV with the latest radius value~\cite{Acharya2016a}).
This small two-neutron separation energy imposes the core-neutron
scattering length to be very large (corresponding to a virtual state
energy well below 1~keV) in order to allow an excited Efimov state.
Experimental data indicates that this is not the case~\cite{Mosby2013},
and it is thus likely that carbon-22, like all other borromean two-neutron
halo nuclei, does not admit an excited Efimov state. 

\paragraph{Comparison with experiments}

The fitting of a single observable such as the energy or the radius
of the nucleus does not of course demonstrate the validity of the
three-body models. To confirm the assumption that the ground state
of the considered nuclei is an Efimov trimer, the authors of the above-mentioned
works have calculated various structural properties from their three-body
model, in order to compare them with experimental data for these nuclei.
In particular, they considered the root mean square radii $r_{n}$
and $r_{c}$ of the neutrons and core from the centre of mass, the
root mean square $r_{nn}$ and $r_{nc}$ of the neutron-neutron and
neutron-core distances, as well as the opening angle $\theta_{nn}$
between the two neutrons~\cite{Yamashita2004,Canham2010}. For $^{11}$Li,
the three-body models (with and without range corrections, as they
turn out to be small) give $r_{n}\approx6.5$~fm, in fair agreement
with the experimental values $r_{n}=6.6\pm1.5$~fm and $r_{n}=6.1$~fm
reported in Refs.~\cite{Marques2000,Moriguchi2013}.

These results show that the basic properties of the candidate halo
nuclei look indeed consistent with Efimov states. The question whether
these nuclei can be considered as Efimov states will be answered more
clearly in the coming years as more experimental data becomes available,
and more comparison between three-body models and more microscopic
models like shell models are developed. 

\subsection{Observations with atoms\label{subsec:Observations-with-atoms-Multi-component}}

\subsubsection{Two-component trimers}

\paragraph{Potassium-Rubidium mixtures}

Efimov states of a mass-imbalanced system have been observed as atom-dimer
resonances in ultra-cold mixtures of potassium and rubidium atoms~\cite{Bloom2013,Hu2014,Kato2015,Kato2016},
in which the external magnetic field is tuned close to a Feshbach
resonance between the potassium and rubidium atoms. If both potassium
and rubidium atoms are bosonic, as is the case for $^{39}\mathrm{K}$,
$^{41}\mathrm{{K}}$, and $^{87}\mathrm{{Rb}}$ atoms, two kinds of
Efimov trimers are possible: KRbRb and KKRb, corresponding respectively
to two heavy bosons plus one light particle, and two light bosons
plus one heavy particle. Therefore, the ground-state trimers of these
two Efimov series should appear at the three-body threshold at some
negative scattering length $a_{-}^{KRbRb}$ and $a_{-}^{KKRb}$. As
the inverse of the potassium-rubidium scattering length is continuously
varied from negative to positive values, these trimers may break up
into an atom and a dimer at some positive scattering lengths $a_{*}^{KRbRb}$
and $a_{*}^{KKRb}$. This implies that atom losses in the mixture
due to recombination into deeper states are enhanced when the scattering
length is tuned around one of these scattering lengths, i.e. one expects
to observe peaks in three-body recombination loss rate at $a_{-}^{KRbRb}$
and $a_{-}^{KKRb}$ for the atomic mixture, and peaks in the atom-dimer
relaxation loss rate at $a_{*}^{KRbRb}$ ($a_{*}^{KKRb}$) for a mixture
of rubidium (potassium) atoms and KRb dimers.

A three-body calculation using potentials with a van der Waals tail~\cite{Wang2012a}
has shown that $|a_{-}^{KRbRb}|>3\times10^{4}\,a_{0}$ and $|a_{-}^{KKRb}|>1\times10^{6}\,a_{0}$
(where $a_{0}$ is the Bohr radius). It is currently not possible
to observe the three-body loss peaks associated with these negative
scattering lengths\footnote{In 2009, the group of Massimo Inguscio and Francesco Minardi in Florence~\cite{Barontini2009}
claimed to observe the first signatures of mass-imbalanced Efimov
trimers in a mixture of potassium-41 and rubidium-87 atoms. The group
found two peaks in the atom loss measurement at the scattering lengths
$-246(14)\,a_{0}$ and $-22(\begin{array}{c}
+4\\
-6
\end{array})\times10^{3}\,a_{0}$ which they assigned to $a_{-}^{KRbRb}$ and $a_{-}^{KKRb}$, respectively.
These values, however, turned out to be too small in magnitude compared
to the theoretical values. Furthermore, no three-body loss peak has
been observed in any potassium ($^{39}\mathrm{K}$,$^{40}\mathrm{K}$,$^{41}\mathrm{{K}}$)
rubidium-87 mixture by any of the subsequent experiments in the groups
of Deborah Jin at JILA~\cite{Bloom2013,Zirbel2008,Hu2014}, Shin
Inouye at the University of Tokyo~\cite{Kato2015,Kato2016}, and
Jan Arlt at Aarhus University~\cite{Wacker2016}. The work of reference~\cite{Wacker2016}
actually reported a three-body loss peak, but attributed it to a $p$-wave
two-body resonance (as predicted by theory) rather than an Efimov
resonance. }, since it is experimentally difficult to tune the magnetic field
to such large values of the scattering length and reach a temperature
that is low enough to resolve the peaks~\cite{Wacker2016}. On the
other hand, the positive scattering lengths $a_{*}^{KRbRb}$ and $a_{*}^{KKRb}$
take experimentally accessible values. 

In 2013, the group of Deborah Jin at JILA measured the three-body
loss rate and the atom-dimer relaxation rate in a mixture of potassium-40
and rubidium-87 atoms. Note that potassium-40 is fermionic, so that
the Efimov effect only occurs for KRbRb in this system. While they
could not find any evidence of three-body loss peaks for $-3000a_{0}<a<-200a_{0}$~\cite{Bloom2013,Zirbel2008,Hu2014},
they observed a peak in the atom ($^{87}\mathrm{{Rb}}$) - dimer ($^{40}\mathrm{{K}}{}^{87}\mathrm{{Rb}}$)
relaxation rate at the positive scattering length $a_{*}=230(30)\,a_{0}$
\cite{Bloom2013,Hu2014}. More recently, the group of Shin Inouye
at the University of Tokyo has also found a peak in the atom-dimer
relaxation rate in a $^{41}\mathrm{{K}}-^{87}\mathrm{{Rb}}$ mixture
at a similar position $a_{*}=348(49)\,a_{0}$~\cite{Kato2015,Kato2016}.
This agreement can be naturally explained if we interpret these peaks
as induced by the Efimov states: since the atoms in both experiments
are of the same atomic species with nearly the same mass, and the
interatomic interactions should have almost the same van der Waals
tail, one naturally expects on the basis of van der Waals universality
(see section~\ref{subsec:Van-der-Waals-Universality} and reference~\cite{Wang2012a})
that both experiments would yield essentially the same results for
the KRbRb Efimov physics. Indeed, both peak positions are found to
be consistent with $a_{*}^{KRbRb}$ obtained by a multi-channel three-body
calculation using a van der Waals type of potential~\cite{Wang2012a,Kato2016}.
The Efimov states of RbKK, corresponding to the ``Efimov-unfavoured''
case of one heavy and two light particles, have yet to be observed.

\paragraph{Lithium-caesium mixtures}

In 2014, in two independent experiments, one led by Cheng Chin at
the University of Chicago~\cite{Tung2014} and the other led by Matthias
Weidem\"uller at the University of Heidelberg~\cite{Pires2014},
experimentalists investigated mixtures of caesium-133 and lithium-6,
taking advantage of the large mass ratio to more easily observe Efimov
states since the scaling ratio is only 4.88 (compared to 22.7 for
bosons of equal mass). Both groups were able to observe three peaks
in the atom losses, corresponding to three Efimov states, when the
intensity of the external magnetic field is tuned near a Feshbach
resonance at $B=842.75$ G, although the authors of reference~\cite{Pires2014}
could not confirm the validity of the third peak after a careful analysis
of the three-body loss rate. The measured ratio of scattering lengths
between the first and second peak was found to be 5.1(2) in reference~\cite{Tung2014}
and 5.07(6)(13)(2) in reference~\cite{Pires2014}, close to the theoretical
value 4.88. The ratio for the second and third peaks in reference~\cite{Tung2014}
was found to be 4.8(7). This is so far the most relevant experimental
demonstration of the Efimov effect in its strictest sense (definition
1 of section \ref{subsec:What-is-an-Efimov-state}), which is the
geometric accumulation of trimer states below the three-body threshold.

\paragraph{Lithium-Rubidum mixtures}

In 2015, the group of C. Zimmermann in T\"ubingen \cite{Maier2015}
reported the signature of a heavy-heavy-light ground-state Efimov
trimer in an ultra-cold mixture of lithium-7 and rubidium-87. The
loss peak associated with that trimer was found at $a_{-}=-1870(121)\,a_{0}$,
which is consistent with the value expected from van der Waals models,
providing further experimental support for the van der Waals universality
of the three-body parameter discussed in section~\ref{subsec:Van-der-Waals-Universality}.

\subsubsection{Three-component trimers}

Three experimental groups working on ultra-cold lithium-6 have independently
evidenced the presence of Efimov states made of three distinguishable
atoms near broad Feshbach resonances: the group of Selim Jochim in
Heidelberg~\cite{Ottenstein2008,Lompe2010,Lompe2010a}, the group
of Kenneth M. O'Hara at the Pennsylvania State University~\cite{Huckans2009,Williams2009},
and the group of Takashi Mukaiyama at the University of Tokyo~\cite{Nakajima2010,Nakajima2011}.
In all experiments, the atoms were prepared in three different hyperfine
sublevels, making it possible for three distinguishable particles
to form Efimov trimers, as detailed in section~\ref{subsec:3-distinguishable-particles}.
This also makes the system more unstable by three-body recombination,
since nothing prevents three distinguishable particles from approaching
each other, unlike the two-component Fermi systems where two of three
particles are in the same state and therefore undergo Fermi exclusion
that limits the three-recombination processes. The experimentalists
thus found the first indications of Efimov physics in the strong variation
of the three-body losses with external magnetic field. As the intensity
of the magnetic field is varied, the scattering lengths between each
pair of states is strongly altered, due to the presence of a broad
Feshbach resonance for each scattering length. These broad resonances
overlap in the range $600-1200$ gauss (region~1), where all three
scattering lengths are much larger than the range of atomic interactions,
but also make the three scattering lengths relatively large in the
range $100-500$ gauss (region~2). Efimov physics is therefore possible
in these two regions. If the three scattering lengths diverged at
the same intensity of the magnetic field, one could expect an infinity
of Efimov trimers in principle, as in the case of identical bosons.
However, since the scattering lengths diverge at different intensities
of the magnetic field, only a finite number of trimers is expected.
As it turns out, the scattering lengths in region 1 are large enough
to support two Efimov trimers, while region 2 exhibits only a ground-state
Efimov trimer.

\paragraph{Observations through three-body losses}

The experimentalists first reported the observation of a plateau of
strong three-body losses in region~2 (larger than the background
losses by three orders of magntitude), delimited by a loss peak at
around 130 gauss and a much softer peak around 500 gauss. They suggested
that Efimov physics could be behind these strong losses. Using the
zero-range theory and fitting the three-body parameter to the experimental
data, theorists could indeed reproduce qualitatively the observed
plateau~\cite{Braaten2009,Naidon2009} and attributed it to the presence
of a single Efimov trimer appearing from the three-body scattering
threshold at around 130 gauss and dissociating back into that threshold
at 500 gauss~\cite{Naidon2009}. The fitted three-body parameter
also made it possible to predict a loss peak in region~1, at around
1160 gauss~\cite{Braaten2009}. Such a peak was indeed observed very
soon afterwards at lower temperatures by the group of K. O'Hara, but
it was located at a smaller intensity of 900 gauss. The experimentalists
identified this peak with the appearance of an excited Efimov trimer,
and showed that its location, along with those of the peaks previously
observed in region~2, were consistent with roughly the same three-body
parameter. This was the first experimental report of an excited Efimov
trimer.

\paragraph{Observations through atom-dimer losses}

Subsequent experiments in Heidelberg and Tokyo~\cite{Lompe2010,Lompe2010a,Nakajima2010,Nakajima2011}
confirmed this interpretation by first preparing a gas of dimers of
atoms in two different hyperfine states, by sweeping the magnetic
field near one of the Feshbach resonances. These large dimers could
then relax into deeper states by inelastic collisions with the remaining
atoms in the third hyperfine state. By measuring the relaxation rate
as a function of the external magnetic fields, the experimentalists
found two loss peaks around 600 gauss and 680 gauss. The zero-range
theory~\cite{Braaten2010,Naidon2011} showed that, qualitatively,
these peaks correspond to a ground and excited Efimov trimers dissociating
in the atom-dimer scattering threshold. This confirmed the existence
of the excited trimer previously observed at the three-body threshold.

\paragraph{Bound-state spectroscopy}

Finally, both groups in Heidelberg and Tokyo were able to perform
a radio-frequency spectroscopy of the excited trimer below the atom-dimer
threshold~\cite{Lompe2010a,Nakajima2011}. This was the first bound-state
spectroscopy of an Efimov trimer, i.e. the measurement of its binding
energy below a scattering threshold rather than its effects at the
scattering threshold.

Although the zero-range theory could unambigously interpret the features
observed in lithium-6, it could reproduce them only semi-quantitatively.
To reproduce all features quantitatively the three-body parameter
has to vary with energy \cite{Nakajima2010,Nakajima2011,Naidon2011}
by about 10\%, and the inleasticity parameter $\eta$ describing recombination
to deep dimers (see section~\ref{subsec:3BP-in-systems-with-loss})
has to be magnetic-field dependent~\cite{Wenz2009} or energy-dependent~\cite{Rittenhouse2010}.
To avoid these extra fitting assumptions, the work of \cite{Naidon2011}
incorporated two-body range corrections and the presence of deeper
dimers via a two-channel model with Gaussian separable potentials.
While the overall agreement with experiment is good, the model could
not reproduce the precise locations of all features with a single
set of parameters. Bo Huang and co-workers~\cite{Huang2014a} recently
used an updated two-body model of the Feshbach resonances in lithium-6~\cite{Zuern2013}
to extract more accurately the three-body parameter from the data,
using a three-body zero-range model. The three-body parameters obtained
from regions 1 and 2 still differ by 5\%, and deviate by 20\% from
the van der Waals universal value observed in other species (see section~\ref{subsec:Van-der-Waals-Universality}).
It would be worthwhile to revisit this problem with three-body models
that properly incorporate van der Waals physics and the more accurate
description of the Feshbach resonances of reference~\cite{Zuern2013}. 

\clearpage{}

\part{Dimensionality\label{part:Dimensionality}}

The dimensionality of space is crucial for the Efimov effect to occur.
For three identical bosons in $d$ dimensions, Esben Nielsen and
co-workers~\cite{Nielsen2001} have shown that the Efimov effect
can occur only when 
\[
2.3<d<3.8.
\]
As a result, for integral dimensions, only $d=3$ leads to the Efimov
effect.  In principle, the Efimov effect may also happen for fractional
dimensions in the allowed range, although this has not been demonstrated
explicitly yet.

\section{Situation in 1D and 2D\label{sec:Situation-in-2D}}

According to Nielsen's dimension criterion, there is no Efimov physics
in one and two dimensions~\cite{Bruch1979,Lim1980}. Still, one-
and two-dimensional three-body systems exhibit universal states when
their interaction is tuned near the appearance of two-body bound states.
However, because of the absence of Efimov attraction, no three-body
parameter needs introducing, and these universal states depend only
the scattering length, as the universal two-body states. As a result,
in the universal region, the energies of three-body and two-body states
are proportional to each other, since they both scale with the scattering
length, and there are no Borromean states. Moreover, because of the
absence of long-range Efimov attraction, the number of three-body
bound states is finite. This is consistent with the theorem by S.
A. Vugal'ter and G. M. Zhislin~\cite{Vugalter1983} stating that
few-body systems with finite-range interactions in one or two dimensions
can only support a finite number of bound states.

Even though Efimov physics does not occur, we briefly review the situation
for these dimensions to contrast them with the Efimov regime, and
understand the connection with Efimov physics in confined geometries.

\subsection{One dimension}

In one dimension, for a short-range pairwise interaction of range
$b$, the even-parity two-body wave function has the form 
\begin{equation}
\psi(x)\propto\vert x\vert-a_{\mbox{\tiny1D}}\label{eq:TwoBodyWaveFunction1D}
\end{equation}
for $b\ll\vert x\vert\ll k^{-1}$, where $a_{\mbox{\tiny1D}}$ is
the one-dimensional scattering length and $k$ is the relative wave
number. The zero-range theory reproducing equation~(\ref{eq:TwoBodyWaveFunction1D})
can be obtained by setting the following Bethe-Peierls boundary condition
at $x=0$,
\begin{equation}
-\frac{1}{\psi}\frac{d\psi}{dx}\xrightarrow[x\to0]{}\frac{1}{a_{\mbox{\tiny1D}}}\label{eq:BethePeierls1D}
\end{equation}
 or by replacing the interaction by a contact potential $\lambda\delta(x)$
of strength $\lambda=-\frac{\hbar^{2}}{\mu a_{\mbox{\tiny1D}}}$ where
and $\mu$ is the two-particle reduce mass. Unlike in 3D, no renormalisation
or regularisation of the delta function is needed here.

In the one-dimensional zero-range theory, there is one two-body bound
state of energy
\begin{equation}
E_{2}=-\frac{\hbar^{2}}{2\mu a_{\tiny\mbox{1D}}^{2}}\label{eq:1DdimerEnergy}
\end{equation}
for positive $a_{\mbox{\tiny1D}}$ (attractive interactions) and no
bound state for negative $a_{\mbox{\tiny1D}}$ (repulsive interactions).

\subsubsection{Identical bosons}

For identical bosons in one dimensions, the problem admits analytical
solutions~\cite{Lieb1963,McGuire1964}. For $a_{1D}>0$, a system
of $N$ particles admits exactly one $N$-body bound state of energy
$E_{N}^{(0)}=\frac{N(N^{2}-1)}{6}E_{2},$ therefore the three-body
bound state energy is
\begin{equation}
E_{3}^{(0)}=4\,E_{2}.\label{eq:1DtrimerEnergy}
\end{equation}

In fact, there is also a virtual three-body bound state at zero energy
($E_{3}^{(1)}=0$)~\cite{Kartavtsev2009}, which can be seen by the
fact that the particle-dimer scattering length is infinite. It can
also be seen by considering particles with different masses, which
can exhibit more than one bound state, and see that in the limit of
equal masses, the second bound state vanishes at the three-body threshold.
In this sense, the case of identical particles is a critical point
for the appearance of the second three-body bound state.

\subsubsection{2 particles + 1 particle}

The more general situation of two identical bosonic particles $A$
of mass $M$ and one extra particle $X$ of mass $m$ was extensively
studied by Oleg Kartavtsev and collaborators~\cite{Kartavtsev2009}.
If the coupling strength $\lambda_{AX}$ between two different particles
is positive (repulsive interaction) then there is no three-body bound
state, irrespective of the coupling $\lambda_{AA}$ between the identical
particles. For negative $\lambda_{AX}$ (attractive interaction between
different particles), the results depend on the mass ratio $M/m$
and the interaction strength $\lambda_{AA}$ between the two identical
particles. 

\paragraph{Effect of the mass ratio}

For mass ratios $M/m\le1$, there is at most one three-body bound
state. Larger mass ratios favor the existence of an increasingly larger
number of three-body bound states. 

\paragraph{Effect of the interaction between identical particles}

When the identical particle interaction is more attractive than the
attraction between different particles ($\lambda_{AA}\le\lambda_{AX}$),
there is a strongly bound AA dimer, and only one trimer with lower
energy, for any mass ratio. For similar attractions $\lambda_{AA}=\lambda_{AX}$
and equal masses $M=m$, we retrieve the case of three identical particles.
The trimer energy is given by equation~(\ref{eq:1DtrimerEnergy}),
where $E_{2}=E_{AX}=E_{AA}$. When $\lambda_{AA}$ is reduced to zero
(non-interacting identical particles), the trimer energy is reduced
to
\begin{equation}
E_{AAX}^{(0)}\simeq2.087719\,E_{AX}\label{eq:1DtrimerEnergyNoninteractingBosons}
\end{equation}

As one would expect, repulsion between identical particles ($\lambda_{AA}>0$)
tends to further reduce the trimer energy and the number of trimers.
In the limit of strong repulsion ($\lambda_{AA}\to\infty$), there
is no trimer for mass ratio $M/m\le1$. Interestingly, this situation
also describes the case of two non-interacting identical fermions
and one extra particle, due to the one-to-one correspondence in one
dimension between strongly interacting bosons and non-interacting
fermions established by Marvin D. Girardeau~\cite{Girardeau1960}.

\paragraph{Five-body Efimov effect in one-dimension }

So far our consideration has been restricted to two-body short-range
interactions. It is worthwhile to note that low-dimensional systems
with many-body interactions, although unlikely to be realised, may
in fact exhibit a many-body Efimov effect. In reference \cite{Nishida2010a},
Yusuke Nishida and Dam T. Son found that five identical bosons in
one dimension resonantly interacting through a four-body short-range
interaction, but without three- and two-body interactions, exhibit
a five-body Efimov effect with scaling factor 12.4. The effect survives,
although weakened, for four-component bosons, but disappears for fermionic
particles. 

\subsection{Two dimensions}

In two dimensions, the $s$-wave scattering length $a_{\mbox{\tiny2D}}$
is always positive. When it is much larger than the range of the interaction,
the interaction can be described by the following contact condition
on the two-body wave function $\psi(r)$:
\begin{equation}
\psi(r)\propto\ln\left(\frac{r}{a_{\mbox{\tiny2D}}}\right)+O(r),\label{eq:BethePeierls2D}
\end{equation}
which plays the role of the Bethe-Peierls condition~(\ref{eq:BethePeierls})
for two dimensions. Such zero-range model supports exactly one two-body
bound state of energy $E_{2}$ given by
\begin{equation}
E_{2}=-4e^{-2\gamma}\frac{\hbar^{2}}{2\mu a_{\mbox{\tiny2D}}^{2}}.\label{eq:2DdimerEnergy}
\end{equation}
where $\gamma$ is Euler's constant and $\mu$ is the reduced mass
between the two particles. Thus, unlike in 3D and 1D, there always
is a two-body bound state in two dimensions.

\subsubsection{Identical bosons}

For identical bosons in two dimensions with zero-range interactions,
one finds two three-body bound states, with energies~\cite{Bruch1979,Nielsen2001,Hammer2004,Kartavtsev2006}:
\begin{eqnarray}
E_{3}^{(0)} & = & 16.522688(1)\,E_{2},\label{eq:2Dtrimer0Energy}\\
E_{3}^{(1)} & = & 1.2704091(1)\,E_{2}.\label{eq:2Dtrimer1Energy}
\end{eqnarray}

The four-body spectrum has also been calculated~\cite{Platter2004a},
and it is found that there are two four-body bound states, with energies:
\begin{eqnarray}
E_{4}^{(0)} & = & 197.3(1)\,E_{2},\label{eq:2Dtetramer0Energy}\\
E_{4}^{(1)} & = & 25.5(1)\,E_{2}.\label{eq:2Dtetramer1Energy}
\end{eqnarray}

\subsubsection{2 particles + 1 particle}

The case of two non-interacting identical particles $A$ of mass $M$,
each interacting resonantly with one extra particle $X$ of mass $m$
was studied by Ludovic Pricoupenko and Paolo Pedri~\cite{Pricoupenko2010a},
as well as Filipe~F.~Belloti and co-workers~\cite{Bellotti2011}.
The interaction between particles $A$ and $X$ is described by the
scattering length $a_{\mbox{\tiny2D}}$, through the boundary condition~(\ref{eq:BethePeierls2D}).
Three-body bound states are characterised by a principal quantum number
$n$ corresponding to excitations of the hyper-radial motion, and
an internal angular momentum with a projection quantum number $\ell$.

When the two identical particles are bosons, the internal angular
momentum has an even projection quantum number $\ell=0,\,2,\,4,\dots$
etc. The ground-state trimer exists for any mass ratio and has the
quantum numbers $n=0$ and $\ell=0$. In the limit $M/m\to0$, its
energy is given by
\begin{equation}
E_{AAX}\simeq2\,E_{AX},\label{eq:2DAAXMm0}
\end{equation}
and for $M=m$, it is given by
\begin{equation}
E_{AAX}\simeq2.36\,E_{AX}.\label{eq:2DAAXMm1}
\end{equation}

Excited trimer states appear as the mass ratio is further increased,
and their number presumably grows indefinitely. The first trimers
appear at the following mass ratios: $M/m=1.770$ ($n=1,\,\ell=0$),
$M/m=8.341$ ($n=1,\,\ell=0$), $M/m=12.68$ ($n=0,\,\ell=2$), $M/m=18.27$
($n=3,\,\ell=0$), $M/m=23.76$ ($n=1,\,\ell=2$).

When the two identical particles are fermions, the internal angular
momentum has an odd projection quantum number $\ell=1,\,3,\,5,\dots$
etc. For a mass ratio $M/m$ smaller than $3.340$\footnote{The main text of reference~\cite{Pricoupenko2010a} reports the value
3.33, but the value 3.340 given in Table I is more accurate.}, there is no three-body bound state. For larger mass ratios, there
is a ground-state trimer with quantum numbers $n=0$ and $\ell=1$.
As in the bosonic case, the trimer energy and the number of trimer
states increase as the mass ratio is increased. Excited trimer states
appear at the following mass ratios: $M/m=10.41$ ($n=1,\,\ell=1$),
$M/m=20.85$ ($n=2,\,\ell=1$), $M/m=26.89$ ($n=0,\,\ell=3$), $M/m=34.59$
($n=3,\,\ell=1$), $M/m=41.98$ ($n=1,\,\ell=3$).

\begin{figure*}[!t]
\includegraphics[scale=0.4]{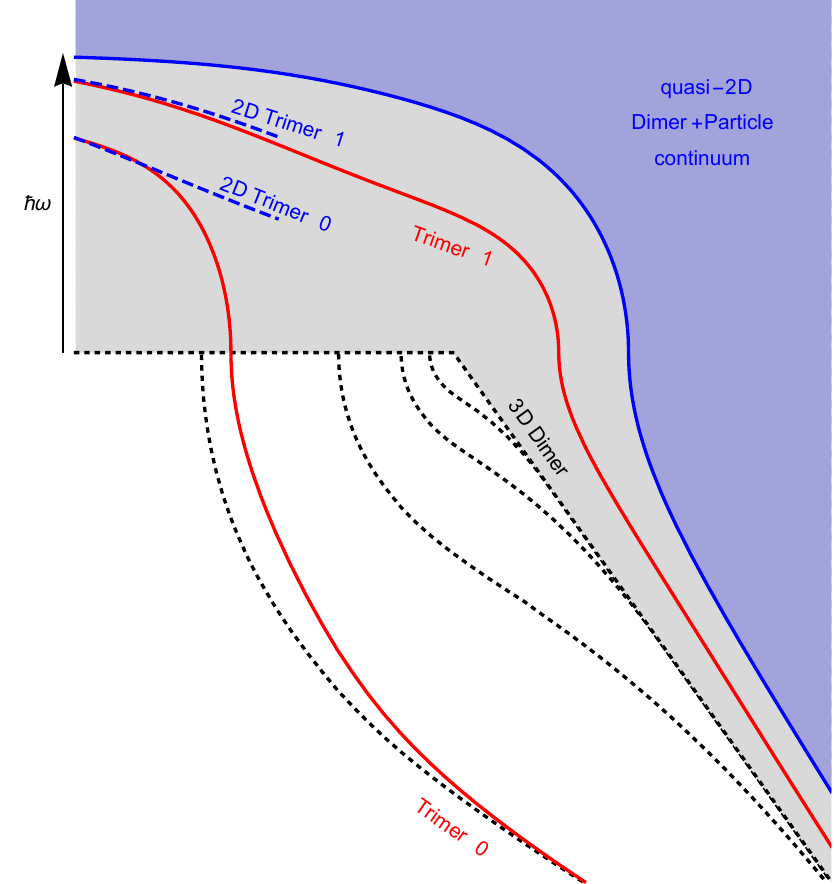}\includegraphics[scale=0.4]{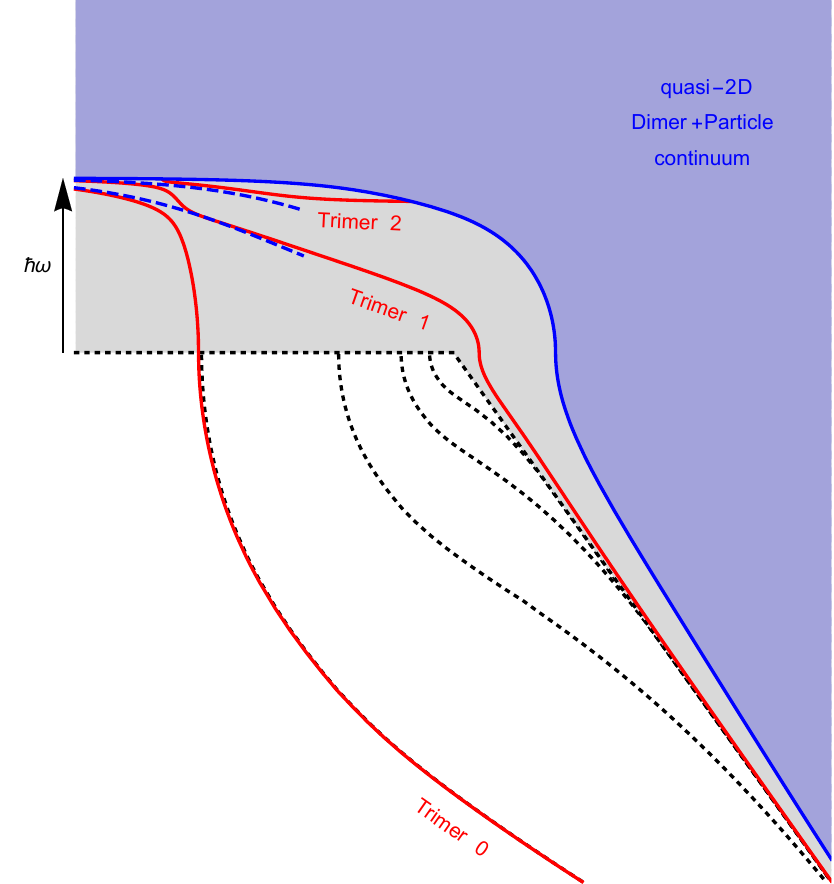}\includegraphics[scale=0.4]{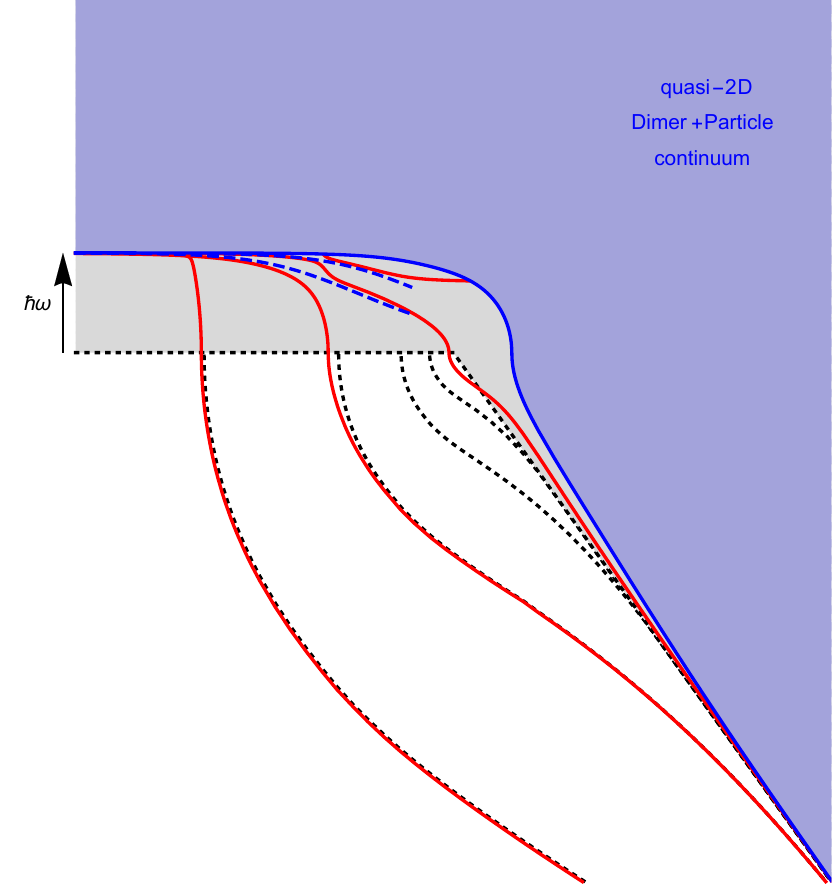}\caption{\label{fig:Quasi2D}Schematic three-body energy spectrum as a function
of inverse scattering length, under harmonic confinement in one direction.
The three-body bound-state energies are represented by red curves,
and the dimer-particle continuum is represented by the dark shaded
area. The threshold of this continuum is shifted by $\hbar\omega$
with respected to the free-space three-body threshold. The free-space
three-body threshold, along with the free-space dimer and the free-space
Efimov trimers are represented in dotted lines for reference - compare
with figure~\ref{fig:EfimovPlot}. The left panel corresponds to
strong confinement $l\sim a$. In this case there are only two trimers
that connect smoothly to the two trimers of the quasi-2D limit (shown
in dashed blue curves) for small binding energies. For gradually weaker
confinement (middle and right panels), new trimers appear from below
the continuum at around $a\sim a_{-}$, exhibiting an avoided crossing
structure between free-space states and quasi-2D states. In the limit
of weak confinement, one retrieves the free-space spectrum with many
Efimov trimers.}
\end{figure*}

\section{Effects of confinement\label{sec:Effects-of-confinement}}

In ultra-cold atom experiments, it is possible to create lower-dimensional
systems by confining the atoms in a narrow plane or tube by means
of laser light. The technique is called \emph{optical lattice}~\cite{Bloch2005}\emph{
}and consists in shining counter-propagating laser beams onto the
atomic cloud. The interference of the beams results in a sinusoidal
pattern of light, which creates a one-body potential for the atoms
proportional to the light intensity through the Stark effect. Atoms
are thus attracted to the nodes of the interference pattern, and repelled
by the maxima for a blue-detuned light. With sufficiently strong lasers
and appropriate layout of the beams, it is thus possible to confine
the atoms in quasi-2D or quasi-1D geometries. This also gives the
interesting possibility of going continuously from a three-dimensional
to a lower-dimensional systems. From a theoretical point of view,
since the Efimov effect occurs only in three dimensions, it is natural
to wonder how the infinity of Efimov states continously transform
into a finite number of trimer states at lower dimensions. This problem
is reminiscent of the crossover that happens in systems with two heavy
fermions and one light particle discussed in section~\ref{subsec:Kartavtsev-Malykh-universal-trimers}.

\subsection{From 3D to quasi-2D}

In reference~\cite{Levinsen2014}, Jesper Levinsen, Pietro Massignan,
and Meera Parish have looked theoretically into the case of confining
Efimov trimers in one direction, thus going from 3D to quasi-2D. They
modelled the interaction by a three-dimensional zero-range interaction
of scattering length $a$, and the confinement by a harmonic trap
of trapping frequency $\omega$. A harmonic trap is a good approximation
of a strongly confining optical lattice when the atoms reside at the
bottom of a well in the lattice. It also offers the theoretical advantage
of decoupling the centre of mass from the relative motion. Jesper
Levinsen and co-workers obtained a version of the Skorniakov and Ter-Martirosian
equation for this problem, which they solved numerically.

The two-body spectrum of such a system had already been investigated
by Dmitry~Petrov and Georgy~Shlyapnikov~\cite{Petrov2001} in 2000.
The 3D two-body bound state for positive scattering length $a$ is
unchanged as long as its extent, given by $a$, is smaller than the
confinement length $l=\sqrt{\hbar/m\omega}$. However, when $a\gtrsim l$,
the dimer acquires a two-dimensional character. Unlike the free space
case, it continues to exist even for negative scattering length $a$,
where it becomes a low-energy quasi-2D dimer, whose energy is given
by equation~(\ref{eq:2DdimerEnergy}) plus the zero-point energy
$\frac{1}{2}\hbar\omega$, which constitutes the energy of the two-body
threshold for particles whose motion along the confinement direction
is in the ground-state of the harmonic trap. Here, the two-dimensional
scattering length is given by:
\begin{equation}
a_{\mbox{\tiny2D}}=2\sqrt{\frac{\pi}{B}}e^{-\gamma}l\exp\left(-\sqrt{\pi/2}\frac{l}{a}\right),\label{eq:a2D}
\end{equation}
where $B\approx0.915$ is a numerical constant and $\gamma$ is Euler's
constant. Thus, unlike the free space case, a dimer exists for any
value of the 3D scattering length.

\begin{figure*}[t]
\includegraphics[scale=0.6]{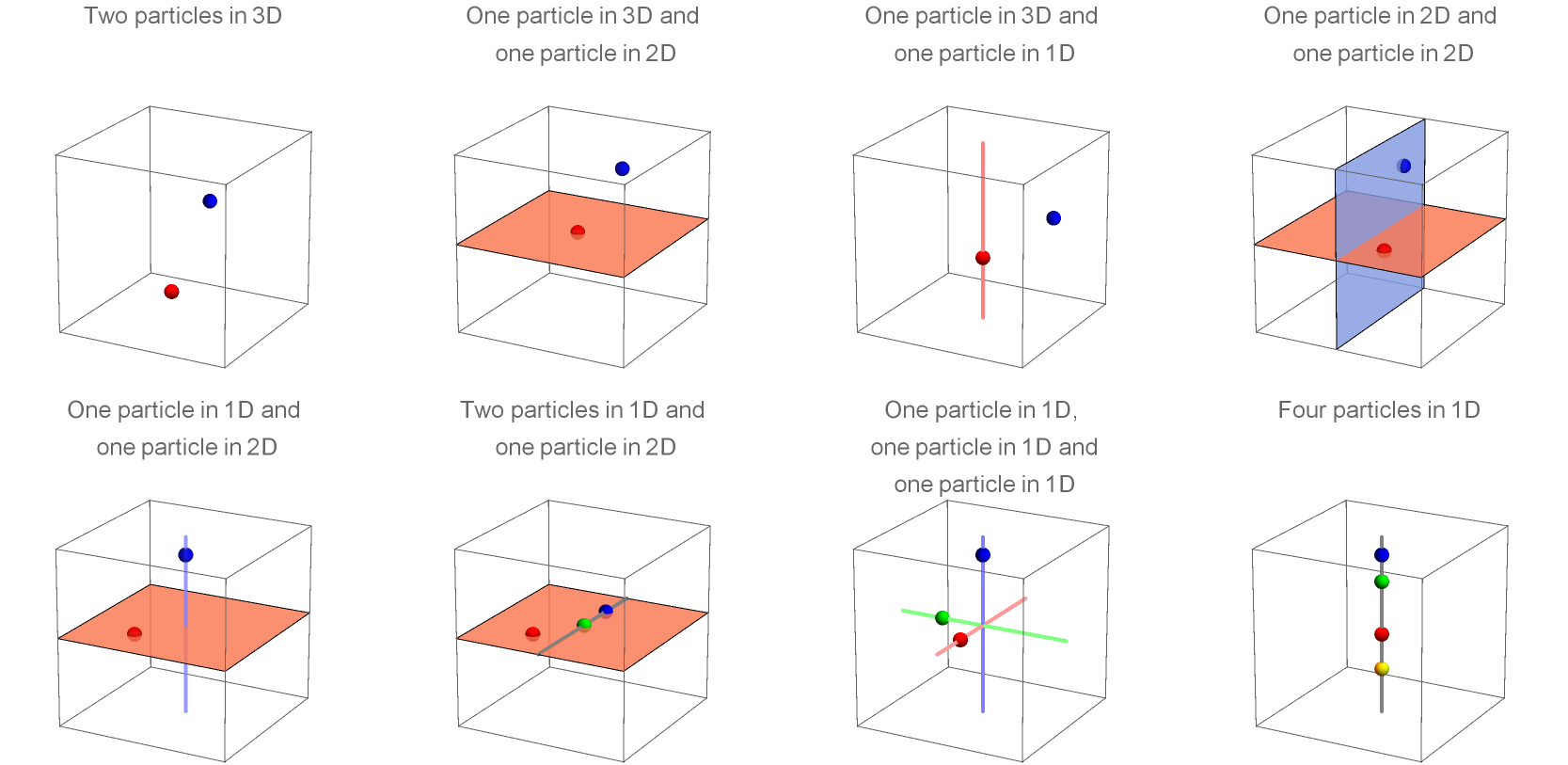}

\caption{\label{fig:Relative-Motion-Mixed-Dimensions}All possible cases of
$N$ scattering particles in mixed dimensions, whose relative motion
is described by exactly three coordinates. The number of coordinates
$\sum_{i}D_{i}-D_{\cap}$ is obtained by summing the dimensions $D_{i}$
of subspaces where each particle lives and subtracting the dimension
$D_{\cap}$ of the intersection of all subspaces, where the centre
of mass coordinates can be eliminated. The first case is the familiar
situation of two scattering particles in three dimensions, for which
$D_{1}=D_{2}=3$ and $D_{\cap}=3$. The cases involving $D_{i}=0$
are not represented, as they correspond to the trivial problems of
fixed scatterers.}
\end{figure*}

Similarly to the dimer case, it is clear that as soon as the system
is confined, the trimers' size cannot extend beyond the confinement
length $l$. Thus, among the infinity of Efimov trimers, only those
which can fit within the confinement region can exist. It would seem
natural that the last two Efimov trimers, which are the largest in
size, and thus the most sensitive to confinement, would turn into
2D trimers in the limit of strong confinement. However, the calculations
of Levinsen and collaborators show that the situation is slightly
more sophisticated. It is the ground-state and first-excited Efimov
states which continuously turn into quasi-2D trimers as their binding
energy is lowered by tuning the (3D) scattering length to negative
values. These states thus never dissociate into the three-body threshold,
unlike in free three-dimensional space (the $a_{-}$ point). This
happens, however, through avoided crossings between the 3D Efimov
trimers and 2D universal trimers, as shown in figure~\ref{fig:Quasi2D}.
The other trimer states are pushed up by these avoided crossings.
For strong enough confinement, they do not exist all, and there are
just two trimers. For weaker confinement, they appear from the dissociation
threshold at values of the scattering length that get closer to the
value of $a_{-}$ in 3D as the confinement is reduced. In the limit
of weak confinement, one retrieves the free-space Efimov spectrum
with many trimers.

The authors of reference~\cite{Levinsen2014} conjecture that a similar
picture should hold for systems confined in two directions, going
from 3D to quasi-1D. Similar crossover behavior may also be expected
for the mass-imbalanced two-component Fermi systems discussed in sections~\ref{subsec:2-fermions-1particle}
and \ref{subsec:Kartavtsev-Malykh-universal-trimers}~\cite{Levinsen2013}.

We also note that the crossover between three-dimensional bosonic
trimers and their two-dimensional counterparts was also studied in
two works~\cite{Yamashita2015,Lammers2016} in the case when one
of the three dimensions is compacted to a radius with periodic boundary
conditions. The resulting crossover is qualitatively similar to that
of~\cite{Levinsen2014}.

\section{Mixed dimensions\label{sec:Mixed-dimensions}}

\begin{table*}[!t]
\begin{tabular}{|c|c|c|c|c|c|}
\hline 
\multirow{5}{*}{\begin{turn}{90}
Particle B
\end{turn}} & \multicolumn{5}{c|}{}\tabularnewline
 & \multicolumn{5}{c|}{Two particles A}\tabularnewline
 & \multicolumn{5}{c|}{}\tabularnewline
\cline{2-6} \cline{3-6} \cline{4-6} \cline{5-6} \cline{6-6} 
 & \multirow{2}{*}{3D$^{2}$} & \multirow{2}{*}{2D$^{2}$} & \multirow{2}{*}{2D$\times$2D} & \multirow{2}{*}{1D$^{2}$} & \multirow{2}{*}{1D$\times$1D}\tabularnewline
 &  &  &  &  & \tabularnewline
\hline 
3D & \includegraphics[scale=0.4]{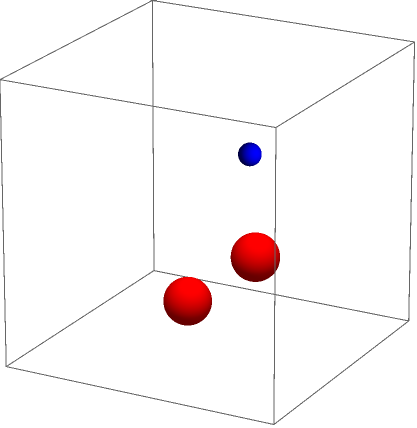} & \includegraphics[scale=0.4]{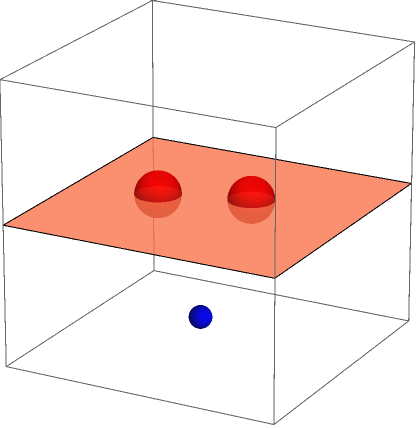} & \includegraphics[scale=0.4]{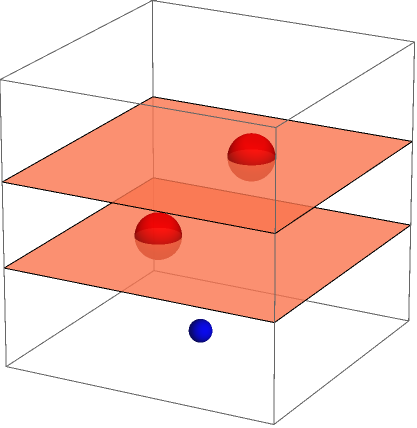} & \includegraphics[scale=0.4]{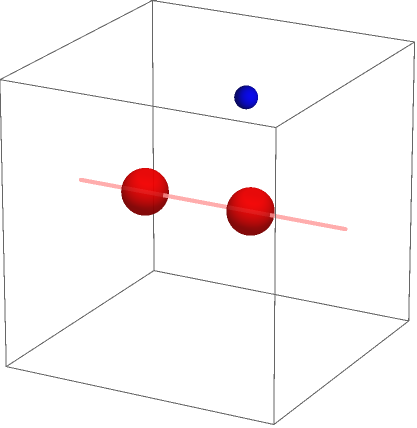} & \includegraphics[scale=0.4]{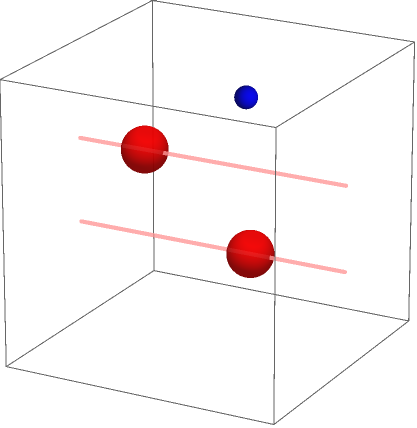}\tabularnewline
 & {\footnotesize{}Bosons: $\checkmark$} & {\footnotesize{}Bosons: $\checkmark$} & {\footnotesize{}Bosons: $\checkmark$} & {\footnotesize{}Bosons: $\checkmark$} & {\footnotesize{}Bosons: $\checkmark$}\tabularnewline
 & {\footnotesize{}Fermions: $\frac{m_{A}}{m_{B}}>13.6$} & {\footnotesize{}Fermions: $\frac{m_{A}}{m_{B}}>6.35$} & {\footnotesize{}Fermions: $\checkmark$} & {\footnotesize{}Fermions: $\frac{m_{A}}{m_{B}}>2.06$} & {\footnotesize{}Fermions: $\checkmark$}\tabularnewline
\hline 
2D & \includegraphics[scale=0.4]{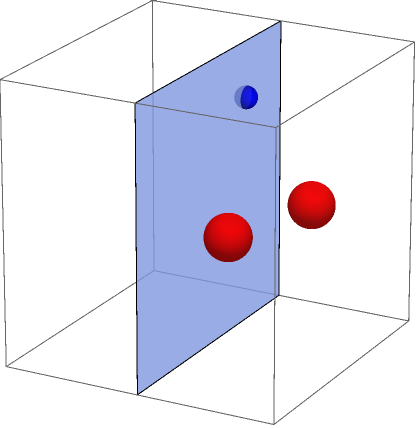} & \includegraphics[scale=0.4]{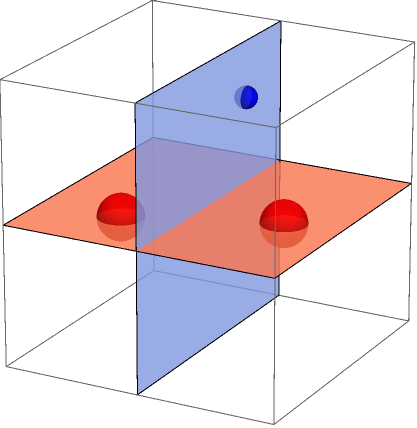} & \includegraphics[scale=0.4]{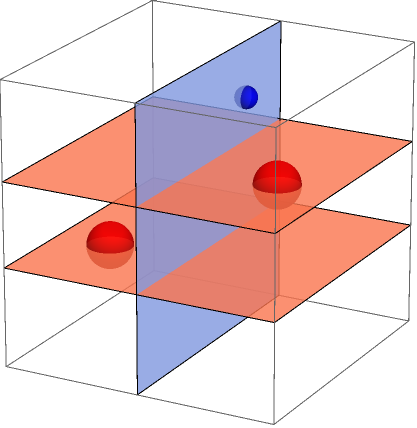} & \includegraphics[scale=0.4]{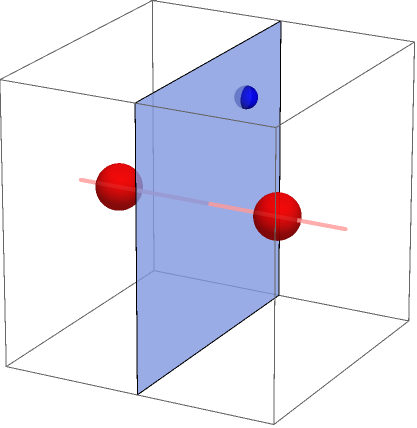} & \includegraphics[scale=0.4]{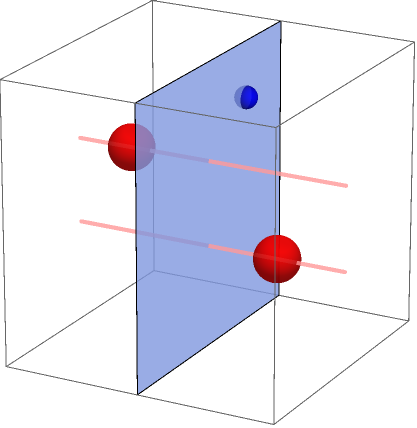}\tabularnewline
 & {\footnotesize{}Bosons: $\checkmark$} & {\footnotesize{}Bosons: $\checkmark$} & {\footnotesize{}Bosons: $\checkmark$} & {\footnotesize{}Bosons: $\checkmark$} & {\footnotesize{}Bosons: $\checkmark$}\tabularnewline
 & {\footnotesize{}Fermions: $\frac{m_{A}}{m_{B}}>28.5$} & {\footnotesize{}Fermions: $\frac{m_{A}}{m_{B}}>11.0$} & {\footnotesize{}Fermions: $\checkmark$} & {\footnotesize{}Fermions: \text{\sffamily X}} & {\footnotesize{}Fermions: $\checkmark$}\tabularnewline
\hline 
1D & \includegraphics[scale=0.4]{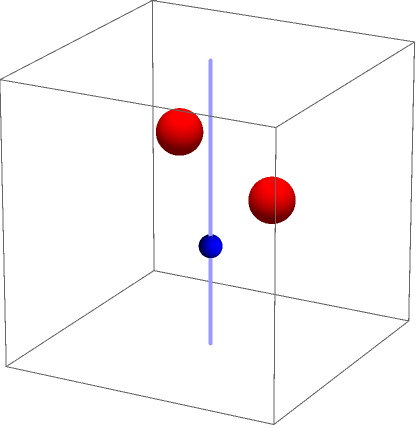} & \includegraphics[scale=0.4]{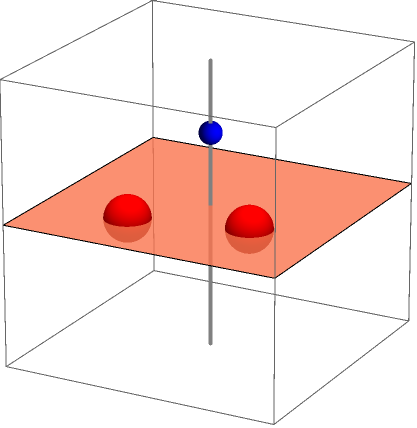} & \includegraphics[scale=0.4]{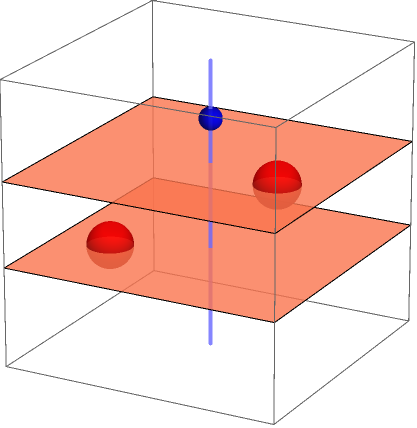} & \includegraphics[scale=0.4]{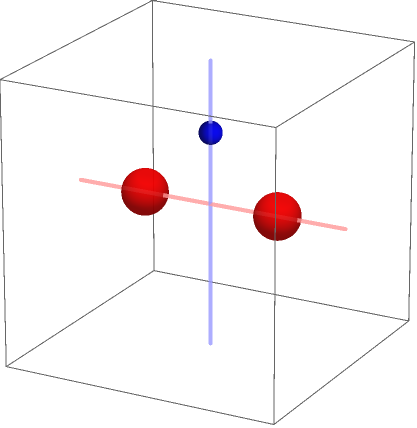} & \includegraphics[scale=0.4]{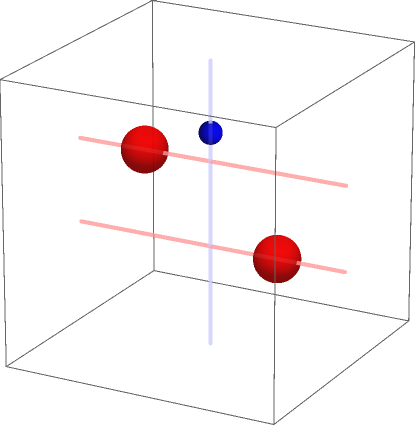}\tabularnewline
 & {\footnotesize{}Bosons: $\checkmark$} & {\footnotesize{}Bosons: $\checkmark$} & {\footnotesize{}Bosons: $\checkmark$} & {\footnotesize{}Bosons: $\checkmark$} & {\footnotesize{}Bosons: $\checkmark$}\tabularnewline
 & {\footnotesize{}Fermions: $\frac{m_{A}}{m_{B}}>155$} & {\footnotesize{}Fermions: \text{\sffamily X}} & {\footnotesize{}Fermions: $\checkmark$} & {\footnotesize{}Fermions: \text{\sffamily X}} & {\footnotesize{}Fermions: $\checkmark$}\tabularnewline
\hline 
\end{tabular}

\caption{\label{tab:Mixed-Dimensions-Systems}Various systems of two particles
A of mass $m_{A}$ resontantly interacting with one particle B of
mass $m_{B}$ in mixed dimensions, which are expected to exhibit the
Efimov effect. These systems are constructed by requiring that at
least two two-body subsystems correspond to a case listed in figure~\ref{fig:Relative-Motion-Mixed-Dimensions}.
The first case in the top left corner is the familiar situation where
all particles move in three-dimensional space. While the Efimov effect
always occurs if the two particles A are identical bosons or distinguishable
particles, it may depend on the mass ratio $m_{A}/m_{B}$ if the two
particles A are identical (polarised) fermions. For each case, it
is indicated whether the Efimov effect occurs ($\checkmark$) or not
({\footnotesize{}\text{\sffamily X}}), or the mass ratio $m_{A}/m_{B}$
beyond which the Efimov effect occurs. In the case of bilayer (3rd
column) and biwire (5th column) geometries, the two particles are
by construction distinguishable, and their statistics does not matter.}
\end{table*}

Although the Efimov effect is restricted to particles in three dimensions,
Yusuke Nishida and Shina Tan have found that Efimov physics can extend
to particles moving in subspaces of different dimensions, a situation
called \emph{mixed dimensions}~\cite{Nishida2008} that can be realised
by confining only certain particles instead of all three particles.
In an article entitled ``Liberating Efimov physics from three dimensions''~\cite{Nishida2011},
they explained the general arguments to identify such favourable situations.

\subsection{The specificity of $D=3$}

First of all, Yusuke Nishida and Shina Tan gave an intuitive interpretation
of the absence of Efimov physics in other dimensions than $D=3$.
They argued that a necessary condition for Efimov physics is that
the dimensionless two-body wave function at unitarity exhibits a scale-invariant
attraction. At short separations $r$, the two-body wave function
at unitarity is the singular solution of the Laplace equation. In
$D$-dimensions, it is thus
\begin{equation}
\psi(r)\propto\frac{1}{r^{D-2}}+O(r^{4-D},r^{2}).\label{eq:Unitary-Wave-Function-In-D-dimensions}
\end{equation}
For $D\ge4$, this wave function cannot be normalised due to the divergence
at the origin $r=0$, which means that unitarity does not exist in
the zero-range limit. Physically, this implies that particles form
tight dimers at separations on the order of the range $b$ of the
interaction~\cite{Nussinov2006}. This tight binding prevents the
binding of a third particle at large separations, and thus the Efimov
effect. On the other hand, for $D=2$, the unitarity wave function
of equation~(\ref{eq:Unitary-Wave-Function-In-D-dimensions}) is
constant near the origin (this can also be seen from equation~(\ref{eq:BethePeierls2D}):
when $a_{\mbox{\tiny2D}}\gg b$, the wave function is nearly constant
around $r\sim b$). This implies that the two particles are effectively
non-interacting, which prevents any binding of the particles - indeed,
the two-body bound state disappears for $a_{\mbox{\tiny2D}}\to\infty$.
For $D=1$, the unitarity wave function of equation~(\ref{eq:Unitary-Wave-Function-In-D-dimensions})
vanishes at the origin as $r$ (this can also be seen from equation~(\ref{eq:TwoBodyWaveFunction1D})
when $a_{\mbox{\tiny1D}}\to0$), which corresponds to a hardcore repulsion,
equivalent to the Pauli repulsion (node in the wave function) between
non-interacting fermions~\cite{Girardeau1960}. This repulsion also
prevents particles from binding. Therefore, only the case $D=3$ presents
a unitary two-body wave function that is scale-invariant and exhibits
an attractive effect.

\subsection{Interactions with three relative coordinates}

In a second step, Nishida and Tan point out that what matters in general
is not the dimension $D=3$ of space itself, but the dimensionality
of the relative motion. For two particles moving in the same space,
the dimensionality of the relative motion is equal to that of the
space, because the locations of the two particles are described by
two $D$-vectors, and after elimination of the centre of mass which
is also described by a $D$-vector, there remain exactly $D$ coordinates.
However, the situation is different for particles moving in subspaces
of different dimensions $D_{i}$. The dimensionality is given by the
sum of the dimensions of the subspaces in which each particle moves,
minus the dimension of the intersection of all sub-spaces, in which
the centre-of-mass coordinates can be separated from the relative
motion. Figure \ref{fig:Relative-Motion-Mixed-Dimensions} shows all
the possible combinations of scattering particles in mixed dimensions
such that the relative motion is described by exactly three coordinates.
In all cases, the equation of relative motion is therefore the same
Laplace equation, assuming that the particles interact through an
$N$-particle contact interaction specified at unitarity by the scale-invariant
boundary condition:
\begin{equation}
\psi(x_{1},x_{2},x_{3})\propto\frac{1}{r}+O(r)\mbox{ with }r=\sqrt{x_{1}^{2}+x_{2}^{2}+x_{3}^{2}}\label{eq:Three-coordinate-Boundary-condition}
\end{equation}

Hence, by analogy with the purely three-dimensional case, one can
expect that the addition of an extra particle in these systems will
lead the Efimov effect as well. According to Yusuke Nishida and Shina
Tan, this is indeed true, although they have postponed the actual
demonstration in the general case to a future publication. Nevertheless,
they give results for several cases involving two identical particles
A and a particle B. The cases are shown in Table~\ref{tab:Mixed-Dimensions-Systems}. 

\subsection{Confinement-induced Efimov effect}

For each case, Nishida and Tan have numerically calculated the scaling
strength $\vert s_{0}\vert$ as a function of the mass ratio $m_{A}/m_{B}$.
As in the pure 3D case (see section~\ref{subsec:Observations-in-nuclear-Multi-component}),
$\vert s_{0}\vert$ monotonically increases with the mass ratio, and
in the case of fermions cancels at a critical mass ratio, below which
the Efimov effect does not occur. The critical mass ratios are given
in Table~\ref{tab:Mixed-Dimensions-Systems}. Interestingly, the
strength $\vert s_{0}\vert$ also increases when the dimension of
space for one or two particles is reduced. Thus confining one or two
particles into lower-dimensional spaces makes it easier to observe
Efimov states. In particular, in the case of particles including identical
fermions for which the mass ratio is not enough to yield Efimov attraction,
the Efimov effect may occur by confining one or two particles~\cite{Nishida2009}.
Yusuke Nishida and Shina Tan call this effect the \emph{confinement-induced
Efimov effect.}

\subsection{Stable Efimov trimers in bilayer or biwire geometries}

Another remarkable situation is the case of particles separated in
disjoint subspaces, such as parallel layers~\cite{Nishida2010} or
wires, as shown in the third and fifth column of Table~\ref{tab:Mixed-Dimensions-Systems}.
Of course, for a two-body contact interaction to take place, these
disjoint subspaces must intersect the space of the third particle.
This way, both particles can interact with the third particle, and
thus the third particle can mediate interaction between the two particles.
Yet, the two particles are always spatially separated and cannot come
in contact. This has two major consequences. 

First, the statistics of the particles does not matter any more, because
they can be regarded as distinguishable particles, being with certainty
in different locations. Thus, the Pauli repulsion between identical
fermions that limits the Efimov effect to sufficiently large mass
ratios, does not play a role any more. In such mixed-dimensional settings,
fermions always exhibit the Efimov effect. 

Second, even though two of the three particles can come in contact,
the three particles can never come closer than the separation between
the two disjoint subspaces. In systems undergoing loss by three-recombination
(such as ultra-cold atoms), this fact can completely suppress the
loss, by preventing the three particles from coming all three at distances
where recombination occurs. It is thus a clever way to realise Efimov
states that are inherently stable, unlike their counterpart in free
space. 

Since the separation $d$ between the two subspaces constitutes the
smallest distance that the three particles can come to, it is the
length scale that breaks the discrete scale invariance and determines
the ground-state energy and three-body parameter of the trimers. In
particular, at unitarity,
\[
E^{(0)}=-\frac{\hbar^{2}\kappa_{*}^{(0)2}}{2\mu}\mbox{ and }E^{(n\gg1)}=-\frac{\hbar^{2}\kappa_{*}^{2}}{2\mu}e^{-2n\pi/\vert s_{0}\vert}
\]
\[
\mbox{with }\kappa_{*}^{(0)}=\frac{\alpha_{0}}{d}\mbox{ and }\kappa_{*}=\frac{\alpha}{d}
\]
The constants $\alpha_{0}$ and $\alpha$ have been calculated numerically
by Yusuke Nishida and Shina Tan for both the bilayer-free (2D$\times$2D$\times$3D)
and biwire-free (1D$\times$1D$\times$3D) geometries, for the two
mass ratios 40/6 and 6/40. The strength $\vert s_{0}\vert$ is the
same as for the single-layer (2D$^{2}\times$3D) and single-wire (1D$^{2}\times$3D)
geometries, because in the limit of weakly bound Efimov trimers, the
separation between the layers or wires is vanishly small compared
with the size of the trimers, and can be regarded as a single layer
or wire in the calculation of $s_{0}$.

Explicit calculations of the trimer energies as a function of the
scattering length for bilayer-free and biwire-free geometries were
performed by Tao Tin, Peng Zhang, and Wei Zhang~\cite{Yin2011},
using the Born-Oppenheimer approximation. The authors also calculated
the ground-state tetramer energy for the triwire-free geometry. They
found that the trimers and tetramers exist in a range of negative
and positive scattering lengths, as in the 3D case, and the binding
energy reaches a maximum when the scattering length is close to the
separation between the layers or wires.

\subsection{Observations with ultra-cold atoms}

The mixed dimension setting has been studied experimentally in only
one group so far. The group of Massimo Inguscio and Francesco Minardi
in Florence have realised a 2D-3D mixed-dimensional system by confining
potassium atoms in the 2D layers of an optical lattice, and let them
interact with rubidium atoms \cite{Lamporesi2010}. The interactions
between the two species can be changed by a magnetic Feshbach resonance.
This way, the researchers could explore the variation of the effective
mix-dimensional scattering length, and observed through recombination
loss spectroscopy the location of mix-dimensional two-body resonances
that are required for Efimov physics to set in \cite{Minardi2011}.
However, Efimov features predicted in the preceding sections have
not been revealed yet.

\clearpage{}

\part{The three-body parameter\label{part:The-three-body-parameter}}

\section{What is the three-body parameter?\label{sec:What-is-the-three-body-parameter}}

\subsection{In the zero-range theory\label{subsec:3BP-In-the-zero-range}}

As explained in section~\ref{subsec:Efimov-theory}, the three-body
parameter is a parameter that needs to be introduced to regularise
the zero-range theory of three particles. This parameter can be introduced
in several ways depending on the formalism used: a three-body short-range
boundary condition or a three-body short-range phase~\cite{Efimov1970a},
a two- or three-body momentum cut-off~\cite{Kharchenko1972}, or
a three-body contact interaction~\cite{Bedaque1999a}. Therefore,
there is not a single formal definition of this parameter, and it
can take the form of a length or an energy, etc.

In any case, it is associated with three-body observables, such as
the trimer energy or the particle-dimer scattering length. When the
three-body parameter is changed, the value of these observables is
rescaled accordingly. In the Efimov spectrum as a function of inverse
scattering length shown in figure~\ref{fig:EfimovPlot}, this corresponds
to a radial rescaling of the curves with respect to the accumulation
point at the centre. The following observables (shown in figure~\ref{fig:EfimovPlot})
are often taken as convenient references to characterise the three-body
parameter: 
\begin{enumerate}
\item the binding wave number $\kappa_{*}$ associated with the trimer energy
$-\hbar^{2}\kappa_{*}^{2}/m$ at unitarity (the scattering length
$a\to\pm\infty$).\label{enu:the-binding-wave-number}
\item the dissociation scattering length $a_{-}$ at which the trimer vanishes
in the three-body scattering threshold.\label{enu:the-dissociation-scattering-length}
\item the scattering length $a_{+}$ at which the three-body recombination
rate has a minimum (for lossy systems).\label{enu:the-scattering-length-a+}
\item the scattering length $a_{*}$ at which the trimer vanishes at the
particle + dimer scattering threshold.\label{enu:the-scattering-length-astar}
\end{enumerate}
These observables are related to each other by universal relations
in the zero-range theory. Namely, for identical bosons~\cite{Braaten2006,Gogolin2008},

\begin{equation}
a_{-}\simeq-1.50763\kappa_{*}^{-1}\label{eq:Universal_Relation_aminus_kappastar}
\end{equation}
\begin{equation}
a_{+}\simeq0.32\kappa_{*}^{-1}\label{eq:Universal_Relation_aplus_kappastar}
\end{equation}
\begin{equation}
a_{*}\simeq0.0707645086901\kappa_{*}^{-1}\label{eq:Universal_Relation_astar_kappastar}
\end{equation}
Note that because of the discrete scale invariance, these quantities
are defined up to a factor $e^{n\pi/\vert s_{0}\vert}$, with $n\in\mathbb{Z}$.
In other words, $\kappa_{*}$ and $\kappa_{*}e^{n\pi/\vert s_{0}\vert}$
represent the same three-body parameter.

\subsection{In systems with finite-range interactions\label{sub3BP-:In-systems-with-finite-range}}

The zero-range theory is of course an idealisation, since in reality
interactions do have a finite (i.e. non-zero) range. In reality, the
zero-range theory is applicable in the universal window of very large
scattering length and very small energy. In this window, the above
observables can be used to determine the three-body parameter. In
this sense, the three-body parameter, although originally a parameter
of the zero-range theory, can also be determined from finite-range
calculations or experimental data provided they can access the universal
window of highly-excited states or low-energy scattering. In general,
it might be better to talk about the \emph{three-body phase}, as in
equation~(\ref{eq:ThreeBodyPhase}), which is a physical quantity
independent of the model, and reserve the expression \emph{three-body
parameter} for the parameter of zero-range theories that fixes that
phase. Nevertheless, the expression three-body parameter is commonly
used in the sense of three-body phase, even in experimental contexts.

In a realistic system, it is often experimentally or computationally
difficult to access the universal window corresponding to highly excited
trimers. From the first few trimers of the Efimov series of a realistic
(finite-range) system, one may extract quantities $\kappa_{*}^{(i)}$
, $a_{-}^{(i)}$, $a_{+}^{(i)}$, $a_{*}^{(i)}$ for each trimer $i=0,\,1,\,2\dots$,
and consider them as approximate measures of the three-body parameter.
Strictly speaking, only in the large $i$ limit do they tend to values
representing the three-body parameter (up to a factor $e^{n\pi/\vert s_{0}\vert}$),
and the universal relations (\ref{eq:Universal_Relation_aminus_kappastar}-\ref{eq:Universal_Relation_astar_kappastar})
hold only approximately for $\kappa_{*}^{(i)}$ , $a_{-}^{(i)}$,
$a_{+}^{(i)}$, $a_{*}^{(i)}$ with small $i$. On the other hand,
the finite-range system quantities $\kappa_{*}^{(i)}$ , $a_{-}^{(i)}$,
$a_{+}^{(i)}$, $a_{*}^{(i)}$ can help remove the ambiguity over
the factor $e^{n\pi/\vert s_{0}\vert}$ in the definition of $\kappa_{*},\,a_{-},\,a_{+},\,a_{*}$.
For instance, one can restrict the definition of $\kappa_{*}$ such
that $\kappa_{*}\left(e^{\pi/\vert s_{0}\vert}\right)^{i}$ corresponds
approximately to the binding wave number $\kappa_{*}^{(i)}$ of the
$i$-th Efimov state. However, this definition still depends on what
is considered to be the first Efimov state. For simplicity, and in
view of the recent definitions of ``Efimov state'' (see section~\ref{subsec:What-is-an-Efimov-state}),
we will always label the ground-state trimer as the first Efimov state
$i=0$.

\subsection{In systems with loss\label{subsec:3BP-in-systems-with-loss}}

When the Efimov effect occurs in systems supporting several two-body
bound states, the Efimov states are trimer resonances that can decay
by recombining into a lower two-body bound state scattering off a
third particle. In the zero-range theory, such lower two-body bound
states are not present, yet Eric Braaten, Hans-Werner Hammer, and
Masaoki Kusunoki have shown that their combined effect can be described
in the zero-range theory by introducing an inelasiticity parameter
$\eta>0$~\cite{Braaten2003,Braaten2006,Werner2009}. Physically,
$e^{-4\eta}$ represents the probability of an incoming hyper-radial
flow to be reflected back to large hyper-radii, the rest being lost
at short distance by recombination. Formally, this amounts to giving
the three-body parameter a complex value $\kappa_{*}e^{i2\eta/\vert s_{0}\vert}$.
This makes the energy of the Efimov states complex, $E+\frac{i}{2}\Gamma$,
where the imaginary part corresponds to the energy width associated
with their finite lifetime. In the limit of small $\eta$, one finds
$\Gamma=\frac{4\eta}{\vert s_{0}\vert}E$.

\section{What sets the three-body parameter?\label{sec:What-sets-the-3BP}}

A natural question that arises is what in the microscopic details
of the finite-range interactions of real systems determines the three-body
parameter. At first glance, two different views can be proposed:
\begin{figure*}[t]
\includegraphics[scale=0.6]{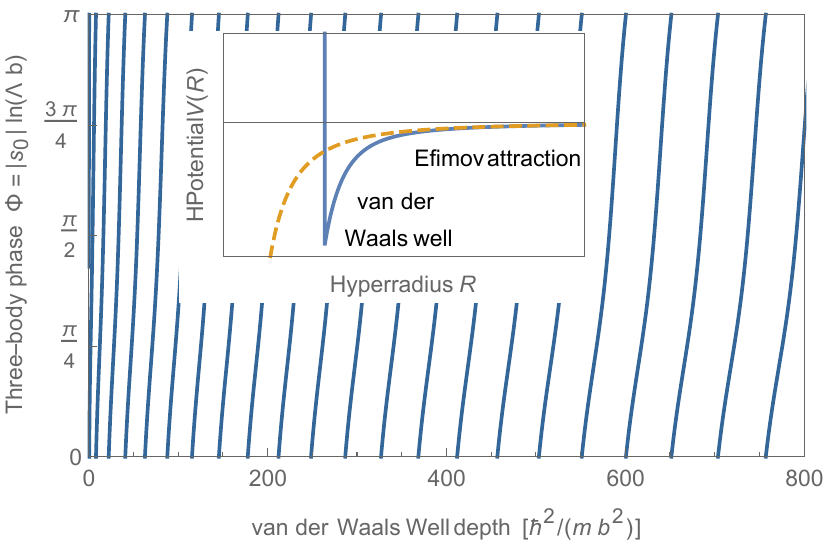}\hfill{}\includegraphics[scale=0.6]{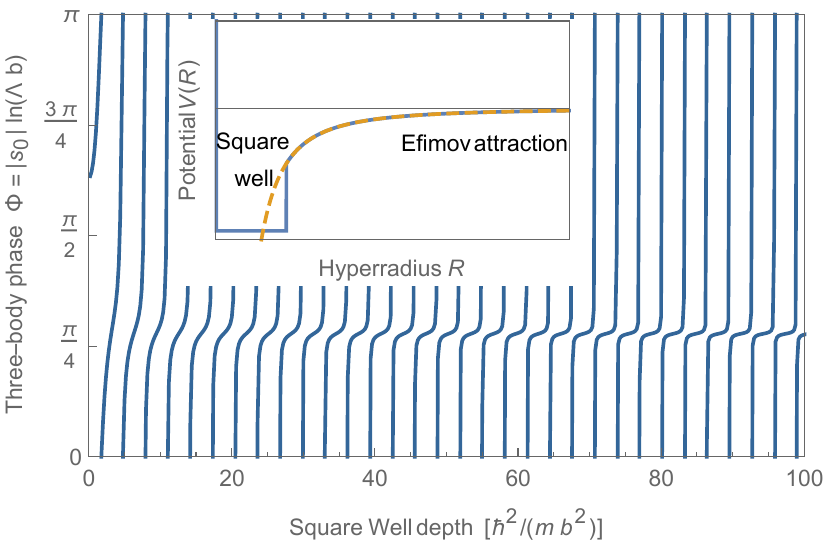}\caption{\label{fig:Three-body-phase}Three-body phase of equation~(\ref{eq:ThreeBodyPhase})
for the solution of the hyper-radial equation~(\ref{eq:HyperradialEquation2}),
where the hyper-radial potential $V(R)$ (represented in the insets)
consists of the Efimov long-range attraction $V_{0}$ of equation~(\ref{eq:EfimovAttraction})
at large distance, and an attractive well of range $b$ setting the
boundary condition at short distance. Left panel: the attractive well
is a van der Waals potential with a van der Waals length $b$, and
a repulsive wall at some variable hyper-radius $R_{0}$, i.e. $V(R)=V_{0}(R)-(2b)^{4}/R^{6}$
for $R>R_{0}$ and $V(R)=\infty$ for $R\le R_{0}$. Right panel:
the attractive well is a square-well potential of radius $b$ and
variable depth. The reference length scale $\Lambda_{0}^{-1}$ of
equation~(\ref{eq:ThreeBodyPhase}) has been set to $b$.}
\end{figure*}

\paragraph{1. The three-body parameter is roughly determined by the range of
interaction}

The first view is that the scale of the three-body parameter is roughly
determined by the range $b$ of the interaction, i.e. the length beyond
which the interaction can be taken as null or negligible. Indeed,
this range constitutes the length scale below which the zero-range
theory ceases to be valid. It is when the three particles approach
this range that the three-body boundary condition determining the
three-body parameter is set. It is therefore a natural guess that
the resulting three-body parameter, expressed as a length scale, is
on the order of this range.

This idea is supported by the early work of Llewellyn H. Thomas~\cite{Thomas1935},
although the Efimov effect was not known at the time. In this work,
it was found that the ground state of the three-body problem depends
on the range of the interaction. Namely, its energy becomes deeper
as the range of interaction is reduced; in the limit of zero-range
interaction, the energy goes to minus infinity, in other words the
spectrum is unbound from below, a phenomenon later known as the Thomas
collapse. Llewellyn Thomas used this finding to estimate the triton
binding energy from the range of nuclear forces.

\paragraph{2. The three-body parameter can take any value}

The second view is that the three-body parameter is a three-body boundary
condition for the free wave, in the same way as the two-body scattering
length being a two-body boundary condition for the free wave - see
the Bethe-Peierls condition~(\ref{eq:BethePeierls}). It is known
that the scattering length is in general difficult to predict from
the details of the two-body interaction. For a purely repulsive interaction,
the scattering length is on the order of the interaction range, but
for an attractive interaction, it can be very different. Although
on average (over many interaction potentials) the scattering length
is on the order of the interaction range, it may be much larger as
well as positive or negative~\cite{Flambaum1999}, and one is often
required to solve the two-body Schr\"odinger equation to obtain the
value of the scattering length for a given interaction potential.
Especially for interaction potentials with a deep well, the scattering
wave function accumulates a large phase inside the well, and a minute
change in the short-range details of the potential can completely
change the value of the scattering length.

By analogy, in the three-body problem at unitarity, one can view the
three-body body parameter as the result of the short-range boundary
condition for the hyper-radial problem with the Efimov attractive
potential of equation~(\ref{eq:EfimovAttraction}). The short-range
boundary condition arises from the complicated three-body dynamics
at short range where all hyper-angular channels are coupled. To get
some insight, though, one can model this dynamics by an effective
short-range hyper-radial potential.

As in the two-body case, if this short-range potential is purely repulsive
(such as a hard wall at $R=R_{0}$) the three-body parameter is set
by the range of the repulsion. On the other hand, if the short-range
potential is attractive, it can take very different values. For instance,
if an attractive $-1/R^{6}$ potential well (van der Waals type) is
used as a short-range boundary condition, one finds that the three-body
parameter can take any value with substantial probability - see the
left panel of figure~\ref{fig:Three-body-phase}. From these simple
calculations, one concludes that in general the three-body parameter,
like the two-body scattering lengh, can take on any value and is sensitive
to the short-range details of the interaction.\\

The above two views 1 and 2 can be conciled by saying that in general
the three-body parameter can take any value depending on the short-range
details of the interaction, but on average (over many interaction
potentials) it scales with the range of the interaction.

While it is true that the three-body parameter may in general take
any value, there are some notable cases presented in the following
sections for which it is simply related to a length scale of the two-body
interaction. 

\subsection{First calculations\label{subsec:First-calculations}}

There have been many calculations of the three-body problem with finite-range
interactions near unitarity, from which one can extract the Efimov
three-body parameter, at least approximately. For example, the work
of Moszkowski and co-workers~\cite{Moszkowski2000} has investigated,
for many shapes of two-body potential, the strength required to bind
three particles with respect to the strength required to bind two
particles - such a calculation is equivalent to the determination
of $a_{-}^{(0)}$, which is an approximate measure of the three-body
parameter. Their calculation indicates that the strength ratio could
vary from 2/3~to~1, although it is often close to~0.8, which suggests
a nominal value for the three-body parameter, supporting view~1.

Later, motivated by the experimental development of Efimov physics
in ultra-cold atomic gases, Jos\'e D'Incao and co-workers~\cite{DIncao2009}
investigated more specifically the variation of the three-body parameter
from one two-body resonance to another, as the strength of the same
two-body potential is increased (the Efimov states in this case are
resonances -see the end of section~\ref{subsec:What-is-an-Efimov-state}).
They concluded that the value of the three-body parameter changes
significantly, and is further modified by the presence of a three-body
force, supporting view~2.

\subsection{Van der Waals universality\label{subsec:Van-der-Waals-Universality}}

\subsubsection{Three identical bosons\label{subsec:VdW-Three-identical-bosons}}

\begin{figure}[t]
\hfill{}\includegraphics[scale=0.95]{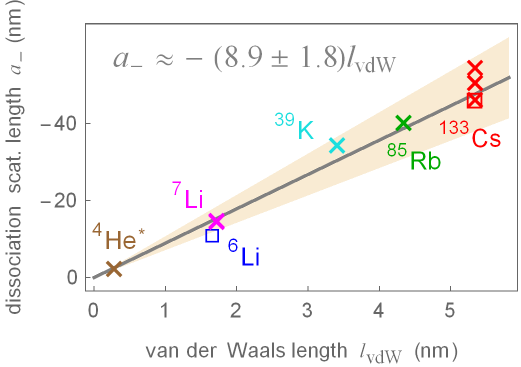}\hfill{}

\caption{\label{fig:van-der-Waals-Universality}Van der Waals universality
of the three-body parameter observed in ultra-cold atomic gases for
caesium-133~\cite{Berninger2011,Kraemer2006,Huang2014}, rubidium-85~\cite{Wild2012},
potassium-39~\cite{Zaccanti2009}, lithium-7~\cite{Gross2009,Pollack2009,Gross2010},
and metastable helium-4~\cite{Knoop2012}. For potassium-39, the
data corresponds to a point originally assigned to a four-body resonance~\cite{Roy2013}.
Also shown are the results~\cite{Huang2014a} of lithium-6 experiments~\cite{Huckans2009,Williams2009,Ottenstein2008,Lompe2010,Lompe2010a,Nakajima2010,Nakajima2011,Huang2014a},
which involve three distinguishable fermions behaving like identical
bosons, except that they have three different scattering lengths.
The value of $a_{-}$ for these experiments is estimated by the formula
$a_{-}^{(n)}e^{-n\pi/s_{0}}$. Results for $n=0$ (ground state) are
shown by crosses, whereas results for $n=1$ (excited state) are shown
by squares. A linear fit of the results is plotted, along with a shaded
area indicating two standard deviations. }
\end{figure}

Until Efimov states could be observed in ultra-cold atomic gases,
there was no experimental data directly related to the three-body
parameter. From 2006 onwards, quantities such as $\kappa_{*}$, $a_{-}$,
$a_{+}$, and $a_{*}$ could be measured around different resonances
in various atomic species (see section~\ref{subsec:Ultracold-atoms}),
at least for the ground Efimov state and in some cases the first excited
Efimov state, thus providing some experimental estimate of the three-body
parameter. Surprisingly, the obtained values~\cite{Gross2010,Berninger2011}
were not randomly distributed over an Efimov log-period $[1,\,e^{\pi/s_{0}}\approx22.7]$,
but instead showed a strong correlation with the range of the atomic
interaction, taken to be the van der Waals length\footnote{Not to be confused with the van der Waals radius $r_{W}$ that accounts
for the finite size of atoms and molecules in the equation of state
of gases.}, 
\[
\ell_{\tiny\mbox{vdW}}=\frac{1}{2}(2\mu C_{6}/\hbar^{2})^{1/4},
\]
associated with the van der Waals tail $-C_{6}/r^{6}$ of the interaction
betwen two neutral atoms of reduced mass $\mu$~\cite{Chin2010}.
The correlation of experimental results for $a_{-}$ for identical
bosons is shown in figure~\ref{fig:van-der-Waals-Universality}.
It gives on average:
\begin{equation}
a_{-}=-(8.9\pm1.8)\,\ell_{\tiny\mbox{vdW}}.\label{eq:ExperimentalVdWAminus}
\end{equation}
This average value is obtained from values of $a_{-}^{(n)}e^{-n\pi/s_{0}}$
for the ground-state ($n=0$) and first-excited ($n=1$) Efimov states.
The stated uncertainty is two standard deviations of the data; all
experimental data fall into this range.

The values obtained from the ground-state resonances are expected
to deviate (by up to 25\%) from the exact three-body parameter. These
values alone give: 
\begin{equation}
\boxed{a_{-}^{(0)}=-(9.1\pm1.5)\,\ell_{\tiny\mbox{vdW}}}.\label{eq:ExperimentalVdWAminus0}
\end{equation}
The first-excited-state resonances are expected to be closer to the
exact three-body parameter (within a few percent). There are currently
two experimental values for such resonances. One is for a resonance
in caesium-133 at a magnetic field around 800 G~\cite{Huang2014},
\begin{equation}
a_{-}^{(1)}e^{-\pi/\vert s_{0}\vert}=-8.8(4)\,\ell_{\tiny\mbox{vdW}}.\label{eq:ExperimentalVdWAminus1Cs}
\end{equation}
The other comes from a resonance at around 900 G in lithium-6 experiments~\cite{Williams2009,Lompe2010,Lompe2010a,Nakajima2010,Nakajima2011},
which involve three distinguishable fermions behaving like identical
bosons. A recent and thorough analysis of the data~\cite{Huang2014a},
including thermal corrections, gives the value 
\begin{equation}
a_{-}^{(1)}e^{-\pi/\vert s_{0}\vert}=-7.11(6)\,\ell_{\tiny\mbox{vdW}}.\label{eq:ExperimentalVdWAminus1Li6}
\end{equation}

All these results indicate that within an error of 20\%, the three-body
parameter is universally determined by the van der Waals length, with
$a_{-}\approx-9\,\ell_{\mbox{\tiny\mbox{vdW}}}$.

This remarkable experimental finding triggered theoretical activity
to understand its origin. The first proposed explanation \cite{Chin2011}
invoked quantum reflection of the three-body wave function in the
region where atoms undergo van der Waals attraction. The argument
supporting this idea is based on the calculation of the three-body
phase from a hyper-radial potential which consists of an Efimov attraction
at large distance and a deep square well potential at short distance
(approximating the effects of the two-body van der Waals attraction).
It is represented in the right panel of figure~\ref{fig:Three-body-phase}.
For deep wells, the three-body phase appears to stabilise around a
well-defined value, close to $\pi/4$. However, the stability of the
three-body phase turns out to be a peculiarity of the square well
approximation. As we discussed previously, when the square well in
the hyper-radius potential is replaced by a van der Waals attraction,
the three-body phase can take different values with significant probability
- see the left panel of figure~\ref{fig:Three-body-phase}. Nevertheless,
the idea of quantum reflection looked compelling and prompted theorists
to check thoroughly the physics of three particles interacting via
deeply attractive van der Waals potentials.

The work of Jia Wang and co-workers~\cite{Wang2012} presented the
first exact calculation, using the adiabatic hyper-spherical representation.
In this method, the three-body wave function is expressed as a linear
combination of products of hyper-radial and hyper-angular wave functions
$\Psi=\sum_{n}F_{n}(R)\Phi_{n}(\Omega;R)$, where the hyper-radius
$R$ characterises the size of the three-body system and is defined\footnote{Note that the authors of reference~\cite{Wang2012} use a different
convention for the definition of the hyper-radius, namely $R^{2}=\frac{1}{\sqrt{3}}(r_{12}^{2}+r_{23}^{2}+r_{31}^{2})$.
Here, we use a definition that is consistent with the one used in
section \ref{subsec:Efimov-theory} - see equation~(\ref{eq:Hyper-radius}).} as $R^{2}=\frac{2}{3}(r_{12}^{2}+r_{23}^{2}+r_{31}^{2})$, and $\Omega$
denotes the set of remaining coordinates describing the geometry of
the three-body system, which are called hyper-angles. Note that this
form becomes separable only for the zero-range interaction at unitarity,
see equation~(\ref{eq:SeparableSolution}). For finite-range interactions
such as the van der Waals type, the hyper-radius $R$ is coupled to
the hyper-angles $\Omega$ at short distance. Hyper-angular wave functions
$\Phi_{n}$ are thus calculated to obtain hyper-radial potentials
$W_{n}(R)$, as well as coupling terms $W_{nn^{\prime}}(R)$, which
lead to a set of coupled equations for the hyper-radial motion:
\begin{equation}
\left(-\frac{d^{2}}{dR^{2}}+W_{n}(R)-E\right)F_{n}(R)+\sum_{n^{\prime}\ne n}W_{n,n^{\prime}}(R)F_{n^{\prime}}(R)=0,\label{eq:CoupledHyperradialEquations}
\end{equation}
where $R$ is expressed in units of van der Waals length $\ell_{\mbox{\tiny vdW}}$
and the potentials $W_{n}$, $W_{n,n^{\prime}}$ is expressed in units
of van der Waals energy $E_{\tiny\mbox{vdW}}=\frac{\hbar^{2}}{m\ell_{\mbox{\tiny vdW}}^{2}}$.
These hyper-radial equations uncouple at large hyper-radius, and one
of them features a hyper-radial potential $W_{0}(R)$ that asymptotes
to the Efimov attraction $V_{0}(R)$ of equation~(\ref{eq:EfimovAttraction}).
Solving the coupled equations~(\ref{eq:CoupledHyperradialEquations})
for various two-body interactions with a van der Waals tail, Jia Wang
and co-workers found that the three-body parameter is essentially
determined by the van der Waals tail and relatively insensitive to
other details. In the limit of deep van der Waals interactions, they
found: 
\begin{equation}
\boxed{\kappa_{*}^{(0)}=(0.21\pm0.01)/\ell_{\tiny\mbox{vdW}}}\label{eq:KappaForVdW}
\end{equation}
\begin{equation}
\boxed{a_{-}^{(0)}=-(10.70\pm0.35)\,\ell_{\mbox{\tiny vdW}}}\label{eq:AminusForVdW}
\end{equation}
which is close to ultra-cold atom observations given in figure~\ref{fig:van-der-Waals-Universality}
and equation~(\ref{eq:ExperimentalVdWAminus0}). For shallower potentials
for which the van der Waals character is less pronounced, they found
slightly larger values of $\kappa_{*}^{(0)}$ and $a_{-}^{(0)}$.
These values were recently confirmed in Refs.~\cite{Hiyama2014,Blume2015}
for shallow van der Waals potentials with a repulsive core and supporting
only one two-body bound state, giving: 
\begin{eqnarray}
\kappa_{*}^{(0)} & = & (0.226\pm0.004)/\ell_{\tiny\mbox{vdW}}\label{eq:KappaForShallowVdW}\\
a_{-}^{(0)} & = & -(9.69\pm0.20)\,\ell_{\mbox{\tiny vdW}}\label{eq:AminusForShallowVdW}
\end{eqnarray}

For all van der Waals potentials investigated so far, deep or shallow,
with or without a repulsive core, the values of $a_{-}^{(0)}/\ell_{\tiny\mbox{vdW}}$
are comprised in the range $[-11.2,-8.3]$, i.e. $-9.73\pm15\%$.
This theoretical value is often cited in the literature, as it agrees
well with the experimental value of equation~(\ref{eq:ExperimentalVdWAminus0}).
However it is more relevant to shallow potentials, and the value in
equation~(\ref{eq:AminusForVdW}) should be taken as the universal
value relevant to interactions dominated by their van der Waals tail.
The more important deviation between this value and the experimental
value of equation~(\ref{eq:ExperimentalVdWAminus0}) suggests that
other physics beyond the van der Waals universality may contribute
to the three-body parameter of atoms close to a broad magnetic Feshbach
resonance~\cite{Huang2014a}.

\begin{figure*}[!t]
\hfill{}\includegraphics[viewport=0bp 0bp 400bp 263bp,clip,width=12cm]{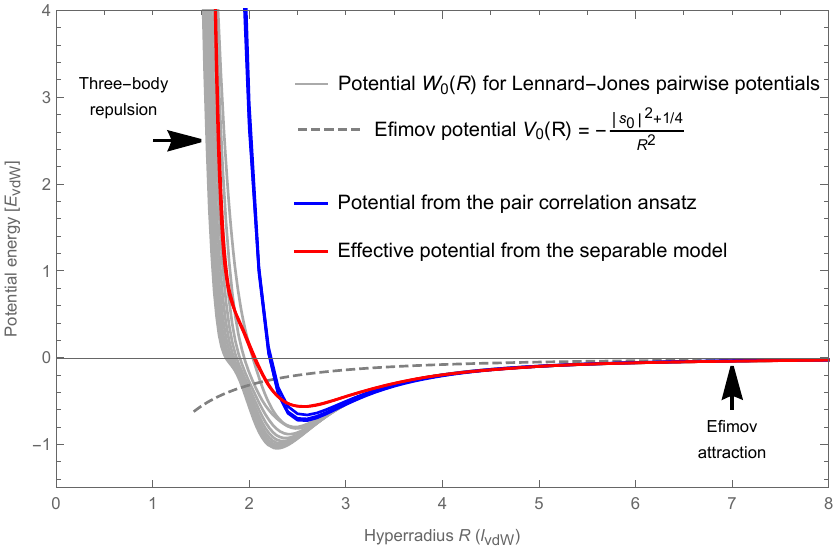}\hfill{}

\hfill{}~~\includegraphics[width=12cm]{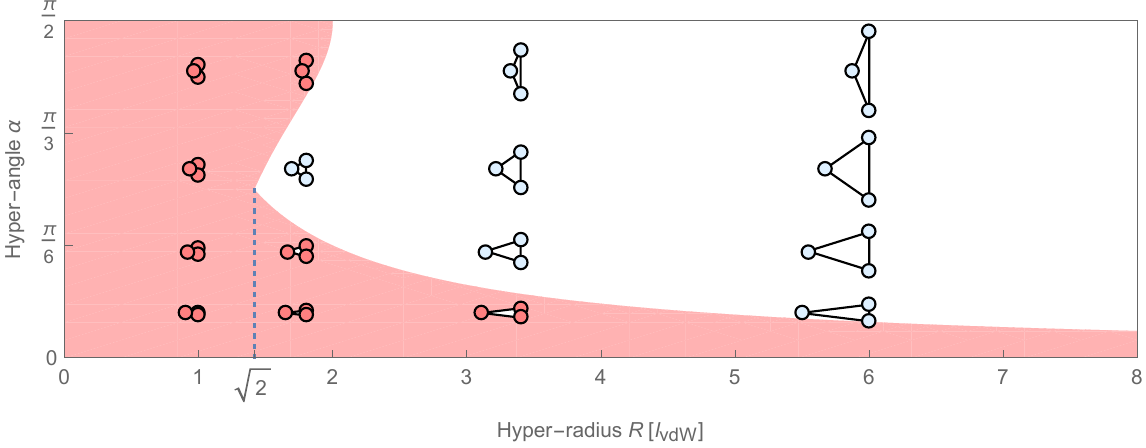}\hfill{}

\caption{\label{fig:Three-body-repulsion}\textbf{Top}: Potential energy for
three bosons resonantly interacting via two-body van der Waals interactions
as a function of their hyper-radius. The grey curves represent the
diagonal potential $W_{0}(R)$ obtained in reference~\cite{Wang2012}
for the channel that asymptotes to the Efimov attraction $V_{0}(R)$
(shown in dashes). The different curves correspond to calculations
for two-body potentials (6-12 Lennard-Jones type~\cite{Jones1924})
of different depths, supporting from 2 to 10 $s$-wave two-body bound
states (from top to bottom). The green curves correspond to the effective
potential~(\ref{eq:EffectiveHyperradialPotential}) obtained from
the ground and first-excited helium trimer wave function calculated
in reference~\cite{Hiyama2014} with various helium potentials scaled
to unitarity. They overlap within their thickness. The blue curves
represent the potential obtained from the pair ansatz~(\ref{eq:Hyper-angular-ansatz})
of reference~\cite{Naidon2014}. They correspond to the zero-energy
two-body wave functions $\varphi(r)$ for Lennard-Jones potentials
supporting 2 to 5 $s$-wave bound states. The red curves represent
the effective hyper-radial potential~(\ref{eq:EffectiveHyperradialPotential})
obtained from the separable model of reference~\cite{Naidon2014}
given by equation~(\ref{eq:SeparablePotential}). They correspond
to the zero-energy two-body wave functions $\varphi(r)$ for Lennard-Jones
potentials supporting 3 and 7 $s$-wave bound states.\protect \\
\textbf{Bottom}: Configuration space of the three-boson system as
a function of the hyper-radius and one hyper-angle, for a right angle
between the two Jacobi vectors.. A few configurations are shown, where
bosons are represented by circles with a diameter equal to the van
der Waals length $\ell_{\text{\tiny vdW}}$. The region of configurations
for which at least two atoms are within the distance $\ell_{\text{\tiny vdW}}$
from each other (\textit{i.e.} circles touching each other) is shaded
in pink. These configurations have near-zero probability due to the
two-body suppression induced by the van der Waals attraction - see
figure~\ref{fig:Two-body-radial-density}. Below the hyper-radius
$R=2\ell_{\text{\tiny vdW}}$, configurations therefore have to deform
to a nearly equilateral configuration. The kinetic energy cost associated
with this deformation results in the repulsive barrier of the top
panel. }
\end{figure*}

\paragraph{Three-body repulsion\label{par:Three-body-repulsion}}

Jia Wang and co-workers looked for a simple reason for this universality
in their calculated hyper-radial potentials. They found that the potential
$W_{0}(R)$ exhibits a steep repulsive barrier at a hyper-radius $R\approx2\ell_{\mbox{\tiny vdW}}$,
instead of a strong attraction as one would na\"ively expect from
the van der Waals attraction. Furthermore, they found that this repulsive
barrier is always present, \emph{irrespective of the presence or absence
of a repulsive core}\footnote{Shortly after the findings of Jia Wang and co-workers, a work~\cite{Sorensen2012}
proposed that the three-body repulsion originates from the repulsive
core in the two-body potential. However, this statement is not consistent
with the results of Jia Wang and co-workers. } in the two-body potential, and in the limit of deeply attractive
potentials has a seemingly universal form that depends only on the
van der Waals tail of the potential - see the grey curves in the top
panel of Fig\@.~\ref{fig:Three-body-repulsion}. This observation
alone cannot explain their results, because the potential $W_{0}(R)$
is strongly coupled to other channels in the region of the repulsive
barrier. Solving the coupled equations, Jia Wang and co-workers found
that this repulsive barrier is actually enhanced by the hyper-radial
couplings $W_{nn\prime}$ and effectively suppresses the probability
of finding the three particles within a hyper-radius $R<2\ell_{\mbox{\tiny vdW}}$.
The authors confirmed that imposing such a steep barrier by hand in
the potential $W_{0}(R)$ alone sets the three-body phase to a value
in good agreement with their exact results~(\ref{eq:KappaForVdW}-\ref{eq:AminusForVdW}).
The repulsive barrier also explains why the results do not depend
much on the interatomic three-body forces, such as the Axilrod-Teller
potential~\cite{Axilrod1943}: the repulsive barrier prevents the
three atoms from coming to the short distances where they would experience
the three-body force.

The three-body repulsive barrier was confirmed in a different manner
by Emiko Hiyama and Masayasu Kamimura~\cite{Hiyama2014} from exact
three-body calculations with various helium potentials. Using the
Gaussian expansion method to solve the three-body problem, they obtained
the energies $E$ and the three-body wave functions of trimers at
unitarity. Integrating the squared wave function over hyper-angles
to obtain the integrated hyper-radial density $\rho(R)$ gives the
effective hyper-radial wave function $F(R)=R\rho(R)^{1/2}$. This
can be converted into the effective hyper-radial potential
\begin{equation}
U(R)=\frac{1}{F(R)}\frac{d^{2}}{dR^{2}}F(R)+E,\label{eq:EffectiveHyperradialPotential}
\end{equation}
assuming that $F(R)$ satisfies a single Schr\"odinger equation akin
to equation~(\ref{eq:CoupledHyperradialEquations}). Hiyama and Kamimura
found that for all the helium pairwise potentials, both the ground
and first-excited trimers give the same potential $U(R)$ in the van
der Waals region and it exhibits the universal repulsive barrier -
see the green curve in the top panel of figure~\ref{fig:Three-body-repulsion}.
This results in $a_{-}^{(0)}=-9.78(1)\ell_{\mbox{\tiny vdW}}$ for
all the helium potentials, in agreement with the calculations of Refs.~\cite{Wang2012,Blume2015}
for shallow Lennard-Jones potentials.

The calculations of Refs.~\cite{Wang2012,Hiyama2014} thus indicate
that the origin of the van der Waals universality of the three-body
parameter is a \emph{three-body repulsion} rather than quantum reflection.
Jia Wang and co-workers noted that this repulsion originates from
the suppression of probability to find two atoms at short separation.
This suppression of probability is due, as in classical mechanics,
to the acceleration of the relative motion by the attractive potential,
which makes the two atoms spend little time at short separation. This
suppression squeezes the three-body wave function, which results in
an increase of kinetic energy that is responsible for the repulsive
barrier.

\paragraph{Connection with two-body physics\label{par:Connection-with-two-body-physics}}

The work of reference~\cite{Naidon2014a} confirmed the interpretation
of Jia Wang and co-workers~\cite{Wang2012}, and showed that this
increase of kinetic energy is physically related to a deformation
of the three-body system from its configuration at large separations
to shorter hyper-radii. At large hyper-radius, the three-body configuration
is independent of the hyper-radius, and distributed according to the
Efimov hyper-angular wave function, 
\begin{equation}
\Phi_{0}^{(ZR)}(\Omega)=\sum_{i=1}^{3}\frac{\phi_{0}(\alpha_{i})}{\sin2\alpha_{i}},\label{eq:EfimovHyperangularWavefunction}
\end{equation}
where $\phi_{0}$ is given by equation~(\ref{eq:HyperangularSolution})
and $\alpha_{i}$ denote the hyper-angles $\alpha$ for the three
possible Jacobi sets of coordinates given by equations~(\ref{eq:Jacobi-r}-\ref{eq:Jacobi-rho}).
As mentioned in section~\ref{subsec:EfimovStructure}, the most probable
configurations have an elongated-triangle geometry: two particles
are close and one is farther away. However, at hyper-radii on the
order of the van der Waals length, the suppression of two-body probability
forces the system to adopt an equilateral configuration - see the
bottom panel of figure~\ref{fig:Three-body-repulsion}. This change
of configuration can be seen in the hyper-radius dependence of the
hyper-angular wave function $\Phi_{0}$, and the kinetic energy associated
with this deformation is described by the non-adiabatic term 
\begin{equation}
Q_{00}(R)=\int d\Omega\left|\frac{d\Phi_{0}}{dR}\right|^{2}\label{eq:NonadiabaticDiagonalTerm}
\end{equation}
contained in $W_{0}(R)$. As the three-body system probes shorter
hyper-radii, the deformational energy due to the two-body suppression
arises as a repulsive barrier in $W_{0}(R)$. To support this interpretation,
the authors of reference~\cite{Naidon2014} presented two models
built on the van der Waals two-body suppression, that lead to the
universal three-body repulsion.

The first model consists in making the following ansatz for the hyper-angular
wave function, 
\begin{equation}
\Phi_{0}(\Omega;R)=\Phi_{0}^{(ZR)}(\Omega)\times\varphi(r_{12})\varphi(r_{23})\varphi(r_{31}),\label{eq:Hyper-angular-ansatz}
\end{equation}
i.e. the hyper-angular wavefunction $\Phi_{0}^{(ZR)}$ from the zero-range
theory is multiplied by a two-body correlation function $\varphi$
for all three pairs. This two-body correlation is taken to be the
two-body radial wave function $\varphi(r)=r\psi(r)$ at zero-energy
for a potential at unitarity with a van der Waals tail. This pair
correlation ansatz describes the suppression of two-body probability
in a simple fashion, and can be used to calculate the hyper-radial
potential $W_{0}(R)$.

\begin{figure}[t]
\hfill{}\includegraphics[scale=0.8]{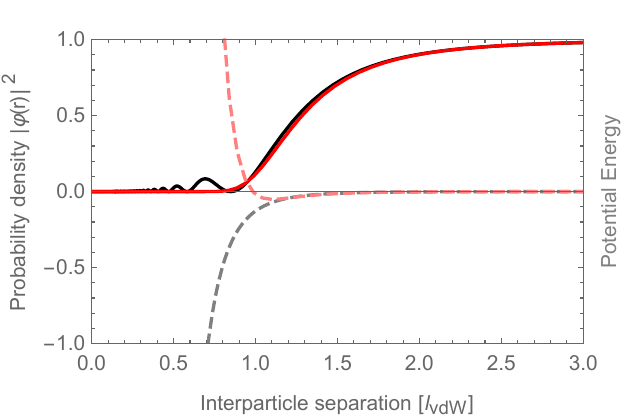}\hfill{}

\caption{\label{fig:Two-body-radial-density}Two-body radial probability density
$\vert\varphi(r)\vert^{2}$ at unitarity ($a\approx\infty$) as a
function of interparticle separation $r$, for a deep van der Waals
potential (black, obtained from equation~(\ref{eq:VanDerWaalsTwoBodyWaveFunction}))
and for a shallow van der Waals potential with a repulsive core (red).
The corresponding potentials are shown in dashed curves. In both cases,
the probability density is suppressed for $r\lesssim\ell_{\text{\tiny vdW}}$.
In the first case, this is due the acceleration in the attractive
van der Waals potential. In the second case, this is due to the repulsive
core. The density profile is nearly the same for $r\gtrsim\ell_{\text{\tiny vdW}}$
since both potentials have the van der Waals form in this region.}
\end{figure}

The two-body radial wave function $\varphi(r)$ is known to have a
universal form in the van der Waals region:
\begin{equation}
\varphi(r)\underset{r\gtrsim\ell_{\mbox{\tiny vdW}}}{=}\Gamma(5/4)\sqrt{x}J_{\frac{1}{4}}(2x^{-2})\label{eq:VanDerWaalsTwoBodyWaveFunction}
\end{equation}
where $x=r/\ell_{\mbox{\tiny vdW}}$ and $\Gamma$ and $J_{\alpha}$
denote the gamma and Bessel functions. At shorter distances $r\ll\ell_{\mbox{\tiny\mbox{vdW}}}$
, $\varphi(r)$ has a potential-dependent form and vanishes at $r=0$.
The number of nodes in $\varphi(r)$ corresponds to the number of
$s$-wave two-body bound states supported by the potential. The probability
density $\vert\varphi(r)\vert^{2}$ is plotted as a solid black curve
in figure~\ref{fig:Two-body-radial-density}.

The authors of reference~\cite{Naidon2014a} found that the resulting
hyper-radial potential $W_{0}(R)$ exhibits a repulsive barrier similar
to that of reference~\cite{Wang2012} - see the blue curves in the
top panel of figure~\ref{fig:Three-body-repulsion}. The barrier
is due to the large value of the non-adiabatic term $Q_{00}$ of equation~(\ref{eq:NonadiabaticDiagonalTerm}),
confirming the deformation scenario. The repulsive barrier does not
depend much on the short-range form of the potential nor its number
of bound states, as long as the two-body probability is sufficiently
suppressed at short-distance, i.e. $\varphi(r)$ has most of its amplitude
in the van der Waals region where it has the universal form~(\ref{eq:VanDerWaalsTwoBodyWaveFunction}).
This occurs for most physical potentials, which feature either a strongly
repulsive core or a deep well that reduces the short-distance probability
- see figure~\ref{fig:Two-body-radial-density}.

In the second model of reference~\cite{Naidon2014a}, the two-body
suppression is introduced through a separable representation of two-body
van der Waals potentials. Any local potential $V$ can be represented
as a superposition of non-local separable potentials, a representation
introduced by Ernst, Shakin and Thaler (EST)~\cite{Ernst1973}. Truncating
this representation to a single separable potential $\hat{V}$ gives
an approximation of the original potential $V$ that reproduces one
of its eigenstates $\vert\psi\rangle$ exactly at a chosen energy,
and other eigenstates approximately around that energy. The separable
potential $\hat{V}$ is explicitly constructed from that eigenstate
as
\begin{equation}
\hat{V}=\frac{1}{\langle\psi\vert V\vert\psi\rangle}V\vert\psi\rangle\langle\psi\vert V.\label{eq:SeparablePotential}
\end{equation}
From this expression, one can easily check that the action of $\hat{V}$
and $V$ onto $\vert\psi\rangle$ is the same. Choosing again $\vert\psi\rangle$
to be the zero-energy scattering state, the resulting separable potential
reproduces that state exactly by construction, and therefore the two-body
suppression at low energy. Solving the three-body problem with this
separable potential (see Appendix for details), constructed from a
$\varphi(r)$ that is dominated by the van der Waals form~(\ref{eq:VanDerWaalsTwoBodyWaveFunction}),
gives a three-body parameter that is consistent with the results of
Jia Wang and co-workers~\cite{Wang2012}. In the limit of a deep
van der Waals potential $V$, i.e. when $\varphi(r)$ tends to the
van der Waals form~(\ref{eq:VanDerWaalsTwoBodyWaveFunction}), the
authors find 
\begin{equation}
\kappa_{*}^{(0)}=0.187(1)\ell_{\tiny\mbox{vdW}}^{-1}\label{eq:KappaForVdWSeparable}
\end{equation}
\begin{equation}
a_{-}^{(0)}=-10.86(1)\ell_{\mbox{\tiny vdW}},\label{eq:AminusFOrVdWSeparable}
\end{equation}
in fair agreement with equations~(\ref{eq:KappaForVdW}-\ref{eq:AminusForVdW}).
Furthermore, the authors integrated the obtained three-body probability
density over the hyper-angles to obtain a hyper-radial wave function,
and converted that wave function into an effective hyper-radial potential
through equation~(\ref{eq:EffectiveHyperradialPotential}). The resulting
potential is shown by the red curves in the top panel of figure~\ref{fig:Three-body-repulsion}
and compares well with the potentials of reference~\cite{Wang2012}~and~\cite{Hiyama2014}.

These results show that the van der Waals universality of the three-body
parameter is a consequence of the van der Waals two-body universality
given by equation~(\ref{eq:VanDerWaalsTwoBodyWaveFunction}).

\subsubsection{2 bosons + 1 particle\label{subsec:VdW-2-bosons+1particle}}

The van der Waals universality extends to systems of different particles,
such as heteronuclear atomic systems. The physics is richer because
the particles may have different masses, different quantum statistics,
and the interactions may have different van der Waals lengths in addition
to different scattering lengths. The case of two identical bosons
$A$ of mass $m_{A}$ and one particle $X$ of mass $m_{X}$ was theoretically
studied by Yujun Wang and co-workers in reference~\cite{Wang2012a}
using 6-12 Lennard-Jones potentials~\cite{Jones1924} $V_{AA}$ and
$V_{AX}$. In this case, as discussed in section~\ref{sec:Multi-component-systems},
there are two scattering lengths $a_{AA}$ and $a_{AX}$ and two limits
for which the Efimov effect occurs: $a_{AX}\to\pm\infty$ with $a_{AA}$
finite, and $a_{AX},a_{AA}\to\pm\infty$. Accordingly, for these two
Efimov regimes, there are two geometric scaling strengths, respectively
$s_{0}$ and $s_{0}^{*}$ (solutions of equations (\ref{eq:TwoResonantPairs})
and (\ref{eq:ThreeResonantPairs})), and two three-body parameters,
respectively $\kappa$ and $\kappa^{*}$. The authors of reference~\cite{Wang2012a}
found that, for a given mass ratio $m_{A}/m_{X}$ between $A$ and
$X$, these three-body parameters are determined solely from the van
der Waals lengths $\ell_{\mbox{\tiny vdW,AA}}$ and $\ell_{\mbox{\tiny vdW,AX}}$
(as well as $a_{AA}$ for the first Efimov regime), and do not depend
upon other short-range details, as in the case of three identical
bosons. Nevertheless, they found that this van der Waals universality
is explained differently for the Efimov-favoured limit (the limit
of large mass ratio $m_{A}/m_{X}$, i.e. two heavy bosons and one
light particle) and the Efimov-unfavoured limit (the limit of small
mass ratio $m_{A}/m_{X}$, i.e. two light bosons and one heavy particle).

In the Efimov-unfavoured limit, the two Efimov regimes are well separated.
The first Efimov regime corresponds to a family of bound states in
a hyper-radial potential (that asymptotes to the dissociation threshold
$AX+A$ for $a_{AX}>0$), whereas the second Efimov regime corresponds
to a family of bound states in another hyper-radial potential (that
asymptotes to the dissociation threshold $AA+X$ for $a_{AA}>0$).
The authors of reference~\cite{Wang2012a} found that the hyper-radial
potential for the first Efimov regime has a universal form for a fixed
$\ell_{\mbox{\tiny vdW,AA}}$ and $\ell_{\mbox{\tiny vdW,AX}}$ when
it is expressed in units of $a_{AA}$, while the hyper-radial potential
for the second Efimov regime has a universal form for a fixed $\ell_{\mbox{\tiny vdW,AA}}$
and $\ell_{\mbox{\tiny vdW,AX}}$, which is independent of $a_{AA}$.
This form exhibits the Efimov attraction at large distance, and a
short-range repulsion that is similar to that of three identical bosons.
This three-body repulsion presumably follows from the same two-body
induced deformation mechanism explained in the previous section.

In the Efimov-favoured limit, on the other hand, the hyper-radial
repulsion picture becomes inadequate to understand the three-body
parameter: the repulsive barrier in the hyper-radial potential featuring
the Efimov attraction goes to shorter and shorter distances and becomes
irrelevant, while the couplings to other channels become increasingly
strong. The authors of reference~\cite{Wang2012a} found that the
Born-Oppenheimer picture is better suited for this situation. This
picture consists in considering the two heavy bosons as slow, and
take their relative coordinate $r$ as the adiabatic variable instead
of the hyper-radius $R$. As the two bosons slowly move, the light
particle is assumed to adiabatically follow a wave function $\Phi(\vec{\rho};r)$
that is calculated at each separation $r$ of the two bosons. The
three-body wave function is therefore
\begin{equation}
\Psi_{BO}(r,\vec{\rho})=F(r)\Phi(\vec{\rho};r)\label{eq:Born-Oppenheimer-Wave-Function}
\end{equation}
The determination of $\Phi(\vec{\rho};r)$ yields eigenstates $\Phi_{n}(\vec{\rho};r)$
and eigenvalues $U_{n}(r)$. One of these eigenvalues corresponds
to a potential $U(r)$ exhibiting the Efimov attraction at large $r$
(when $a_{AX}\gg\ell_{\mbox{\tiny vdW,AX}}$). The relative wave function
$F(r)$ for the two bosons is then simply given by the following Schr\"odinger
equation:
\begin{equation}
\left(-\frac{\hbar^{2}}{m_{A}}\nabla_{r}^{2}+V_{AA}(r)+U(r)-E\right)F(r)=0\label{eq:Born-Oppenheimer-Relative-Motion}
\end{equation}

Note that in this Efimov-favoured limit, the two Efimov regimes are
nearly the same ($s_{0}\approx s_{0}^{*}$), and appear in the Born-Oppenheimer
approximation as a single Efimov effect described by a single channel
(corresponding to the potential $V_{AA}+U$) and a single three-body
parameter approximating the two three-body parameters. This three-body
parameter is then determined by the form of the potential $V_{AA}(r)+U(r)$.
According to reference~\cite{Wang2012a}, $U(r)$ turns out to be
negligible in the short-range region, and therefore the short-range
phase of $F(r)$ is set by the van der Waals potential $V_{AA}$ only.
As is known from the theory of van der Waals potentials~\cite{Gribakin1993,Gao1998},
that phase depends only on the van der Waals tail of this potential
and its scattering length $a_{AA}$. As a result, the three-body parameter
depends only on the van der Waals length $\ell_{\mbox{\tiny vdW,AA}}$
and the scattering length $a_{AA}$. If $\ell_{\mbox{\tiny vdW,AX}}>\ell_{\mbox{\tiny vdW,AA}}$,
it may also depend on $\ell_{\mbox{\tiny vdW,AX}}$ since the potential
$V_{AA}+U$ may depend on the van der Waals tail of $V_{AX}$ at intermediate
distances. The conclusion is therefore the same as for the Efimov-unfavoured
limit: the three-body parameters depend only on the van der Waals
lengths $\ell_{\mbox{\tiny vdW,AA}}$ and $\ell_{\mbox{\tiny vdW,AX}}$
, and the scattering length $a_{AA}$. However, the origin of this
universality is different from that of the Efimov-unfavoured limit
and the case of three identical bosons. Instead of a three-body repulsion,
it is the van der Waals potential $V_{AA}$ between the two bosons
that sets the three-body parameters, the same way it sets the scattering
length between the two bosons.

The authors of reference~\cite{Wang2012a} have checked that this
conclusion, obtained from the Born-Oppenheimer picture, is validated
by exact calculations using the fully-coupled hyper-radial equations
for mass ratios ranging from 14 to 29. For these large mass ratios,
they obtained a very good agreement between the Born-Oppenheimer and
exact calculations for the wave functions and energies, and demonstrated
that the results are nearly insensitive to the number of bound states
in the two-body Lennard-Jones potentials $V_{AA}$ and $V_{AX}$.

\subsection{Other types of short-range interactions\label{subsec:Other-types-of-interactions}}

The discovery of the van der Waals universality for three particles
rekindles the question of which conditions lead to a simple determination
of the three-body parameter. The van der Waals universality for two
particles is a particular case of the universality of power-law tail
potentials, which decay as $1/r^{n}$, with $n>3$. One can therefore
expect that a similar universality exists for three particles interacting
via power-law potentials. As for other finite-range interactions,
reference~\cite{Wang2012} indicates that van der Waals universality
likely extends to any finite-range interaction that sufficiently suppresses
the two-body probability at short distance. Their conclusion is drawn
from the similitude between the three-body parameter for deep P\"oschl-Teller
potentials~\cite{Poeschl1933,Fluegge1947} and that of deep van der
Waals potentials, once both are expressed in units of effective range.
The relevant length scale for the three-body parameter would therefore
be the effective range $r_{e}$, which in the case of van der Waals
interactions is simply related to the van der Waals length~\cite{Gao1998,Flambaum1999}.

\subsubsection{Two-body correlation and effective range}

Extending the two-body analysis of section~\ref{par:Connection-with-two-body-physics}
to arbitrary potentials qualitatively corroborates this point: if
the interaction suppresses the two-body probability within some range
$r_{0}$, the restriction of three-body configurations at small sizes
imposes the three-body system to deform, and the kinetic energy of
that deformation creates a three-body repulsive barrier at a hyper-radius
comparable with the two-body suppression range $r_{0}$. The effective
range $r_{e}$ provides a good estimate of the suppression range.
At unitarity, the effective range $r_{e}$ is defined by~\cite{Bethe1949}
\begin{equation}
\frac{1}{2}r_{e}=\int_{0}^{\infty}dr\left[\bar{\varphi}(r)^{2}-\varphi(r)^{2}\right],\label{eq:EffectiveRange}
\end{equation}
where $\bar{\varphi}(r)=1$ is the asymptotic form of $\varphi(r)$
that is a solution of the free two-body problem. By construction,
if the amplitude of $\varphi$ is suppressed within some range $r_{0}$
with respect to that of $\bar{\varphi}$, then the effective range
is positive, and $\frac{1}{2}r_{e}$ is a good estimate of $r_{0}$.
Therefore the location of the three-body repulsive barrier, and thus
the three-body parameter, should be given by the effective range.

Of course, the precise value of the three-body parameter should depend
upon the precise location and shape of the repulsive barrier. These
in turn should depend upon the precise shape of the two-body wave
function $\varphi(r)$. For interactions with a van der Waals tail,
we know that $\varphi(r)$ has a universal form in the region of suppression
given by equation~(\ref{eq:VanDerWaalsTwoBodyWaveFunction}), and
thus the three-body parameter is universally related to $\ell_{\mbox{\tiny vdW}}$.
More generally, for interactions with a power-law tail decaying as
$-C_{n}/r^{n}$ ($n>3$), the two-body wave function at zero energy
has the following universal form at unitarity: 
\begin{equation}
\varphi(r)\underset{r\gtrsim\ell_{n}}{=}\Gamma\left(\frac{n-1}{n-2}\right)\sqrt{x}J_{\frac{1}{n-2}}(2x^{-(n-2)/2})\label{eq:PowerLawTwoBodyWaveFunction}
\end{equation}
where $x=r/\ell_{n}$, and $\ell_{n}=\left[\frac{1}{n-2}\frac{\sqrt{mC_{n}}}{\hbar}\right]^{\frac{2}{n-2}}$.
With this notation, $\ell_{6}=\ell_{\mbox{\tiny vdW}}$. For each
power $n$, there should thus be a universal relation between the
three-body parameter and $\ell_{n}$. In contrast, for interactions
decaying faster than power laws, such as exponentially decaying potentials,
there is no such universality of the two-body wave function. Only
when such potentials are very deep does the two-body wave function
show a very abrupt depletion of probability that approaches a step
function~\cite{Naidon2014}:
\begin{equation}
\varphi(r)=\begin{cases}
0 & \mbox{ for }r<\frac{1}{2}r_{e}\\
1 & \mbox{ for }r\ge\frac{1}{2}r_{e}
\end{cases}.\label{eq:StepFunctionWaveFunction}
\end{equation}

The work of reference~\cite{Naidon2014} investigates these ideas
quantitatively using the separable potential approximation described
in section~\ref{par:Connection-with-two-body-physics}. The authors
checked that for various finite-range interaction potentials near
unitarity (Gaussian, exponential~\cite{Fluegge1947}, P\"oschl-Teller~\cite{Poeschl1933,Fluegge1947},
Yukawa~\cite{Yukawa1955,Fluegge1947}, Morse~\cite{Morse1929,Fluegge1947},
6-12 Lennard-Jones~\cite{Jones1924}), the separable potential~(\ref{eq:SeparablePotential})
built with the corresponding zero-energy two-body wave function $\varphi$
does reproduce within a few percent the three-body parameter extracted
from exact calculations for three identical bosons~\cite{Moszkowski2000}.
This confirms that, for pairwise potentials inducing a pronounced
suppression of two-body probability, the three-body parameter is essentially
governed by the zero-energy two-body wave function. 

As a result, it follows that the three-body parameter for such interactions
roughly scales with the effective range. The numerical values of the
three-body parameters for all the potentials considered in~\cite{Naidon2014},
when expressed in units of the effective range of the potential, differ
by just a factor of two. Namely,
\begin{equation}
\kappa_{*}^{(0)}\in[0.2,\,0.4]\times(\frac{_{1}}{^{2}}r_{e})^{-1}.\label{eq:ApproximateUniversality}
\end{equation}
Although this does not constitute a universal result, it is a significant
reduction of variance with respect to the Efimov log-period $[1,\,e^{\pi/s_{0}}\approx22.7]$
allowed by the zero-range theory. This fact may also be seen in the
context of functional renormalisation group (FRG)~\cite{Horinouchi2015}.
In this framework, one applies a regulator that cuts off momenta smaller
than $k$, and looks at the flow of quantities, such as the effective
three-body coupling constant, as a function of $k$. The occurrence
of the Efimov effect at zero energy and the unitary limit implies
a limit cycle, i.e. a flow that is log-periodic in $k$, instead of
a fixed-point limit. This can be intuitively understood by considering
the log-periodic three-body wavefunction at zero energy in equation~(\ref{eq:F0bis})
confined in a box of size $1/k$. The phase of the log-periodic oscillations
is related to the three-body parameter. The authors of reference~\cite{Horinouchi2015}
checked that plotting the flow as a function of $kr_{e}$ for different
separable potentials (similar to those of reference~\cite{Naidon2014})
results in phase differences that are small compared to $2\pi$. This
confirms that the three-body parameter for these potentials is roughly
determined by the effective range, in accordance with equation~(\ref{eq:ApproximateUniversality}).

\subsubsection{Deep-potential limits}

The exact calculations of reference~\cite{Moszkowski2000} only feature
potentials near unitarity with at most one $s$-wave two-body bound
state. The authors of reference~\cite{Naidon2014} then extended
their separable potential calculations to the case of deeper potentials
near unitarity. Within the separable potential approximation, this
amounts to constructing the separable potential from a two-body wave
function with a larger number of nodes corresponding to the number
of $s$-wave bound states in the original potential.

\subsubsection*{Power-law potentials}

For potentials with a power-law tail $r^{-n}$, because of the universality
of the two-body wave function given in equation~(\ref{eq:PowerLawTwoBodyWaveFunction}),
changing the depth of the potential results in little change in the
three-body parameter. In the limit of deep potentials, the authors
numerically found the following three-body parameters:
\begin{eqnarray}
\kappa_{*} & = & 0.364(1)/(\frac{_{1}}{^{2}}r_{e})\mbox{\;\ for }n=4\label{eq:3BP-for-C4}\\
\kappa_{*} & = & 0.2614(1)/(\frac{_{1}}{^{2}}r_{e})\mbox{\;\ for }n=6.\label{eq:3BP-for-C6}
\end{eqnarray}
Here, the effective range $r_{e}$ is given by
\begin{eqnarray}
\frac{_{1}}{^{2}}r_{e} & \approx & \frac{\pi}{\sin(\frac{\pi}{n-2})}\frac{2^{\frac{2}{n-2}}}{n-2}\frac{\Gamma(\frac{n/2+1}{n-2})}{\Gamma(\frac{n/2}{n-2})\Gamma(\frac{n+1}{n-2})}\,\ell_{n}\qquad\label{eq:EffectiveRangeForPowerLaw}\\
 & \sim & \left(\frac{n-2}{n-1}\right)^{\frac{1}{n-2}}\frac{2(n-2)^{2}}{(n-3)(2n-5)}\,\ell_{n},\qquad\label{eq:ApproximateEffectiveRangeForPowerLaw}
\end{eqnarray}
which can be obtained from equations~(\ref{eq:EffectiveRange}) and
(\ref{eq:PowerLawTwoBodyWaveFunction}). Note that for $n=4$ and
$n=6$, equation~(\ref{eq:EffectiveRangeForPowerLaw}) reduces to
$\frac{1}{2}r_{e}/\ell_{4}=\frac{2\pi}{3}\simeq2.0944$ and $\frac{1}{2}r_{e}/\ell_{6}=\frac{16\Gamma\left(\frac{5}{4}\right)^{2}}{3\pi}\simeq1.39473$,
respectively, and $\frac{1}{2}r_{e}\approx\ell_{n}$ for large $n$.

\subsubsection*{Faster-than-power-law potentials}

For potentials decaying faster than a power law, as the depth of the
potential is increased, the two-body wave function slowly converges
to the step function given by equation~(\ref{eq:StepFunctionWaveFunction}).
As a result, the three-body parameter slowly converges to a universal
limit, which reference~\cite{Naidon2014} found to be numerically:
\begin{equation}
\kappa_{*}=0.2190(1)/(\frac{_{1}}{^{2}}r_{e}).\label{eq:3BP-for-step-function}
\end{equation}

\subsubsection{Classes of universality}

The authors of reference~\cite{Naidon2014} concluded that there
are two classes of universality for the three-body parameter: the
class of potentials decaying as power law, which exhibit a robust
universality, and the class of potentials decaying faster than power
laws, which give a universal parameter only in the limit of very deep
potentials. The deep-potential limits are continuously connected,
because the two-body wave function~(\ref{eq:PowerLawTwoBodyWaveFunction})
for power-law potentials also tends to the step function~(\ref{eq:StepFunctionWaveFunction})
for very large $n$. Hence, as the power $n$ is continuously increased,
the universal value of the three-body parameter continuously decreases
from equations~(\ref{eq:3BP-for-C4}-\ref{eq:3BP-for-C6}) to equation~(\ref{eq:3BP-for-step-function}),
but requires deeper and deeper potentials to be reached.

It is important to note that these results are obtained within the
separable potential approximation; they need to be confirmed and refined
by exact calculations.

Finally, let us remark that the square-well potential
\begin{equation}
V(r)=\begin{cases}
-V_{0} & \mbox{ for }r<r_{0}\\
0 & \mbox{ for }r\ge r_{0}
\end{cases},\label{eq:SquareWellPotential}
\end{equation}
which is often used in model calculations, stands out as a particular
case. Its two-body wave function at unitarity does not show a suppression
of probability at short distance, even when the depth $V_{0}$ is
increased, and does not converge to equation~(\ref{eq:StepFunctionWaveFunction}).
Therefore, it is not expected to exhibit the deformation-induced three-body
repulsion, nor the universal three-body parameter of equation~(\ref{eq:3BP-for-step-function}).
The problem stems from the absence of tail: there is only an abrupt
variation of the potential that precludes any acceleration, even classically.
If the abrupt variation is smoothed a bit, the three-body parameter
should eventually reach the universal value of equation~(\ref{eq:3BP-for-step-function})
in the limit of deep wells.

\subsection{Coupled-channel interactions\label{subsec:Coupled-channel-interactions}}

The preceding discussions only considered interactions described by
a single potential. In many physical systems with resonant interactions,
the reality is more complex as the interaction involves the coupling
of potentials from different channels, corresponding to different
internal states of the colliding particles. For instance, as explained
in section~\ref{subsec:Observations-with-atoms}, the two-body interaction
between ultra-cold atoms is made resonant by using Feshbach resonances~\cite{Tiesinga1993,Chin2010},
which result from the coupling between two or more hyperfine channels.
In such situations, the value of three-body parameter may not follow
the results presented in the previous sections, and depends on the
characteristics of the coupled channels. Here, we consider the case
of isolated Feshbach resonances.

\subsubsection{Feshbach resonances}

In the neighbourhood of a Feshbach resonance, two scattering particles
at relative energy $E=\frac{\hbar^{2}k^{2}}{m}$ approach each other
in some entrance channel corresponding to their internal states, but
during their collision can couple to a bound state of energy $E_{c}$
in a closed channel (i.e. whose potential dissociates above the energy
$E$) corresponding to different internal states. The $s$-wave scattering
phase shift is then the sum of two contributions: 
\begin{equation}
\delta_{0}(k)=\delta_{0,\mbox{\tiny bg}}(k)+\delta_{0,\mbox{\tiny res}}(k)\label{eq:TotalSwaveShift}
\end{equation}
The first contribution is the background phase shift $\delta_{0,\mbox{\tiny bg}}$
corresponding to the entrance channel, and the second contribution
is a resonant phase shift $\delta_{0,\mbox{\tiny res}}$ induced by
the coupling to the bound state. It has a Breit-Wigner form~\cite{Mott1965,Chin2010}:
\begin{equation}
\tan\delta_{0,\mbox{\tiny res}}(k)=-\frac{\frac{1}{2}\Gamma(k)}{E-E_{C}-\Delta(k)},\label{eq:ResonantSWaveShift}
\end{equation}
where $\Gamma$ and $\Delta$ are the width and shift of the resonance.

At small scattering energy, $\Gamma(k)\approx\alpha k$, and thus
$\delta_{0,\mbox{\tiny res}}$ can be expanded as:
\begin{equation}
\frac{k}{\tan\delta_{0,\mbox{\tiny res}}(k)}=-\frac{1}{a_{\mbox{\tiny res}}}-R_{*}k^{2}+o(k^{2})\label{eq:phaseShiftNarrowResonance}
\end{equation}
where $a_{\mbox{\tiny res}}=-\frac{\frac{1}{2}\alpha}{E_{c}+\Delta(0)}$
and $R_{*}=\frac{2\hbar^{2}}{m\alpha}$. It follows that the scattering
length $a$ is
\begin{equation}
a=a_{\mbox{\tiny bg}}+a_{\mbox{\tiny res}},\label{eq:TotalScatteringLength}
\end{equation}
where $a_{\mbox{\tiny}bg}=-\lim_{k\to0}\tan\delta_{0,\mbox{\tiny}bg}(k)/k$
is the background scattering length. The resonance condition is met
when the energy of the bound state $E_{c}$ is tuned to compensate
the shift $\Delta(0)$, making $a_{\mbox{\tiny res}}$, and thus $a$,
divergent.

The strength of the resonance is characterised by the length $R_{*}$.
When $R_{*}$ is much smaller than the range $b$ of the interaction
(the van der Waals length $\ell_{\mbox{\tiny vdW}}$ in the case of
atoms), the resonance is strong and dominated by the entrance channel;
the particles are most likely to be found in the entrance channel.
Such entrance-channel dominated resonances usually (although not necessarily)
occur over a broad range of the tuning parameter (such as an applied
magnetic field), and thus are also called broad resonances. On the
other hand, when $R_{*}$ is much larger than the range $b$, the
resonance is weak and dominated by the closed channel. Such resonances
are usually (although not necessarily) observed as narrow resonances.
The strength of a resonance is thus conveniently characterised by
the dimensionless ratio $s_{\mbox{\tiny res}}=b/R_{*}$.

\subsubsection{Broad resonances\label{subsec:Broad-Feshbach-resonances}}

Broad resonances are dominated by their entrance channel and can be
effectively described by a single potential~\cite{Chin2010}. The
results of sections~\ref{subsec:Van-der-Waals-Universality} and
\ref{subsec:Other-types-of-interactions} based on single interaction
potentials can therefore apply to the case of such resonances. In
particular, for broad atomic Feshbach resonances, the van der Waals
tail of the open-channel potential is the main feature that determines
the three-body parameter. 

\subsubsection{Narrow resonances\label{subsec:Narrow-Feshbach-resonances}}

In 2004, Dmitry Petrov~\cite{Petrov2004} investigated the physics
of three bosons near a narrow Feshbach resonance. For narrow resonances,
$R_{*}$ is much larger than the range $b$ of inter-particle forces,
which induces the strong energy dependence of equation~(\ref{eq:phaseShiftNarrowResonance}).
Close to the resonance condition, one can neglect the background contribution
$\delta_{s,\mbox{\tiny bg}}$, and comparing equation~(\ref{eq:phaseShiftNarrowResonance})
with equation~(\ref{eq:LowEnergyPhaseShift}), we see that such resonant
interactions have a large and negative effective range 
\begin{equation}
r_{e}\simeq-2R_{*}\ll-b.\label{eq:EffectiveRangeForNarrowResonances}
\end{equation}
This confers to these systems some sort of long-range property, which
has the effect of providing a three-body boundary condition for the
Efimov-attracted particles at distances on the order of $R_{*}$,
at much larger distance than the range of inter-particle forces. 

Formally though, these systems can still be treated in the zero-range
theory, on the basis of equation~(\ref{eq:phaseShiftNarrowResonance}),
implying that the scattering length $a$ in equations~(\ref{eq:LeeHuangYangPotential})
or (\ref{eq:BethePeierls}) is to be replaced by the energy-dependent
scattering length $a(k)$ given by
\begin{equation}
\frac{1}{a(k)}=\frac{1}{a}+R_{*}k^{2},\label{eq:EnergyDependentScatteringLength}
\end{equation}
where $k$ is the relative wave number between two particles. The
work of reference~\cite{Petrov2004} shows that unlike the original
zero-range theory, the resulting three-body equations are well-behaved.
Indeed, the presence of the term $R_{*}k^{2}$ in equation~(\ref{eq:EnergyDependentScatteringLength})
turns the Efimov attraction~(\ref{eq:EfimovAttraction}) at short
hyper-radius $R$ into a Coulomb-type attraction $-1/(R_{*}R)$, which
does not necessitate the introduction of a three-body parameter. Another
way to put it is that the three-body phase is set by $R_{*}$. Namely,
the three-body phase as measured by $\kappa_{*},\,a_{*},\,a_{+}$
was numerically found to be (up to a factor $e^{n\pi/s_{0}}$) \cite{Petrov2004,Braaten2006}:
\begin{eqnarray}
\kappa_{*} & \simeq & 0.11/R_{*}\simeq-0.22/r_{e},\label{eq:3BPforNarrowResonances}
\end{eqnarray}
\begin{equation}
a_{+}\simeq2.9\,R_{*}\simeq-1.45\,r_{e},\label{eq:aplusforNarrowResonances}
\end{equation}
\begin{equation}
a_{*}\simeq0.64\,R_{*}\simeq-0.32\,r_{e},\label{eq:astarforNarrowResonances}
\end{equation}
which predicts accurately the quantities $a_{+}^{(i)}\approx a_{+}\left(e^{\pi/\vert s_{0}\vert}\right)^{i}$
and $a_{*}^{(i)}\approx a_{*}\left(e^{\pi/\vert s_{0}\vert}\right)^{i}$,
except for $i=0$, for which $a_{+}^{(0)}\simeq3.3\,R_{*}$ and $a_{*}^{(0)}\simeq0.45\,R_{*}$.

In 2008, Alexander Gogolin and co-workers~\cite{Gogolin2008} found
an analytical solution to the three-boson problem near a narrow Feshbach
resonance. Instead of the single-channel zero-range model based on
equation~(\ref{eq:EnergyDependentScatteringLength}), they used a
two-channel model with no interaction in the entrance channel and
a zero-range coupling between the entrance and closed channel. Integrating
the corresponding Schr\"odinger equation, they showed that the resulting
integral equation can be mapped to a single-particle Schr\"odinger-like
equation. From this equation, the three-body parameter can be expressed
analytically, leading to an accurate value:
\begin{eqnarray}
\kappa_{*} & \simeq & 0.11691/R_{*}\simeq-0.23381/r_{e},\label{eq:3BPforNarrowResonancesAccurate}
\end{eqnarray}
\begin{equation}
a_{-}\simeq-12.895\,R_{*}\simeq6.448\,r_{e}.\label{eq:aminusForNarrowResonancesAccurate}
\end{equation}
It is from these calculations that the universal relation between
$a_{-}$ and $\kappa_{*}$ given in Eq\@.~(\ref{eq:Universal_Relation_aminus_kappastar})
was calculated accurately. For the ground-state trimer, the theory
gives~\cite{Nishida2012a} 
\begin{eqnarray}
a_{-}^{(0)} & \simeq & -10.90216\,R_{*},\label{eq:aminusForNarrowResonancesGroundState}\\
a_{*}^{(0)} & \simeq & 0.458398\,R_{*}.\label{eq:astarForNarrowResonancesGroundState}
\end{eqnarray}

This calculation was generalised to the case of 2+1 fermions~\cite{Castin2011}
and to the case of 2+1 bosons~\cite{Endo2016a}. The authors have
found that the three-body parameter in these systems is universally
described by the effective range and the mass ratio. In the limit
of large mass imbalance, in particular, Efimov trimers' energies at
unitarity are analytically found as 

{\footnotesize{}\medskip{}
\[
\vert E^{(\!n\!)}\vert\!\!=\!\!\frac{\hbar^{2}}{2\mu R_{*}^{2}}e^{-\frac{2\pi}{\vert s_{0}\vert}(1\!+\!n)}\!\times\!\left\{ \begin{array}{ll}
\!\!\!\!4e^{-\pi/2} & {\scriptstyle \!\!\!\!(\mathrm{2\ light\ bosons\ \!\!+\!\!\ 1\ heavy\ particle})}\\
\!\!\!\!4e^{2J_{0}} & {\scriptstyle \!\!\!\!(\mathrm{2\ heavy\ bosons\ \!\!+\!\!\ 1\ light\ particle})}\\
\!\!\!\!4e^{2J_{0}} & {\scriptstyle \!\!\!\!(\mathrm{2\ heavy\ fermions\ \!\!+\!\!\ 1\ light\ particle})}
\end{array}\right.
\]
}{\footnotesize\par}

{\footnotesize{}\bigskip{}
}where $n=0,1,2....$, $J_{0}=0.505560...$, and $s_{0}$ is the imaginary
solution of Eq.~(\ref{eq:TwoResonantPairs}). The energies of the
2 heavy bosons + 1 light particle system and 2 heavy fermions + 1
light particle system converge to the same values since the repulsion
originating from the antisymmetrisation becomes negligible in the
large mass imbalance limit.

\subsubsection{Intermediate resonances\label{subsec:Intermediate-resonances}}

It is remarkable that both the limit of very broad resonances and
that of very narrow resonances lead to a universal three-body parameter
in terms of the effective range of the interaction. In the case of
broad resonances, the interaction can be described by a single potential,
which lead to universal three-body parameters such as~(\ref{eq:3BP-for-C4})
or (\ref{eq:3BP-for-C6}). In the case of narrow resonances, the large
and negative effective range leads to a universal three-body parameter
given by equation~(\ref{eq:3BPforNarrowResonancesAccurate}). An
intriguing question is how these two limits are connected for intermediate
resonances which are neither very broad nor very narrow. In particular,
one may wonder whether it can still be universally expressed in terms
of the effective range, as the two limits suggest.

The work of Richard Schmidt and co-workers~\cite{Schmidt2012} investigates
this point using a separable potential model to describe the coupled
channels. This model treats the coupling between the entrance and
closed channels, but does not include any interaction in the entrance
channel. As a result, it does not reproduce the deformation and three-body
repulsion discussed in section~\ref{par:Connection-with-two-body-physics}
that lead to the universal three-body parameter for broad resonances.
The authors therefore adjusted the arbitrary form of of their inter-channel
coupling (an exponential function) so that the three-body parameter
in the limit of a broad resonance ($s_{\mbox{\tiny res}}\gg1$) coincides
with the known value for van der Waals interactions,~given by equation~(\ref{eq:KappaForVdW}).
They could then calculate the three-body parameter as the strength
of the resonance $s_{\mbox{\tiny res}}$ is decreased, using functional
renormalisation group techniques. In the limit of narrow resonances
($s_{\mbox{\tiny res}}\ll1$), their calculation give $a_{-}\simeq-12.9\,R_{*}$
and $\kappa_{*}R_{*}\simeq0.117$, reproducing the results~(\ref{eq:3BPforNarrowResonancesAccurate}-\ref{eq:aminusForNarrowResonancesAccurate}).
For intermediate resonances ($s_{\mbox{\tiny res}}\sim1$), they obtain
a smooth and continuous crossover connecting the two limits\footnote{An alternative crossover was proposed in reference~\cite{Sorensen2012},
but the model used in that work appears to rely on some invalid assumptions,
such as the universal three-body parameter being set by the repulsive
core of the two-body interaction potential, which is not the case,
as noted in section~\ref{par:Three-body-repulsion}.}. This is in contrast with the effective range, which changes sign
as the strength of the resonance reduces: for broad resonances, the
effective range is positive and on the order of the true range $b$
of the interaction (see equation~(\ref{eq:EffectiveRangeForPowerLaw}))
and for narrow resonance, the effective range is negative and given
by the length $R_{*}$ (see equation~(\ref{eq:EffectiveRangeForNarrowResonances})).
This indicates that in this crossover region of intermediate resonances,
the three-body parameter is not simply given by the effective range.
This is to be expected because the universal mechanisms relating the
three-body parameter to the effective range are different for the
two limits: in the case of broad resonances, universality stems from
the two-body short-range correlation given by equation~(\ref{eq:VanDerWaalsTwoBodyWaveFunction}),
which is an off-the-energy-shell property, whereas in the case of
narrow resonance, universality stems from the energy-dependence of
the scattering phase shift~(\ref{eq:phaseShiftNarrowResonance}),
which is an on-the-energy-shell property.

The work of Yujun Wang and Paul S. Julienne~\cite{Wang2014} presents
the most complete model so far. The two-body interaction is described
by a set of two or three channels corresponding to the spin states
involved in the Feshbach resonance, and the potential in each channel
is modelled by 6-12 Lennard-Jones potential. By incorporating both
the resonance and van der Waals physics, this kind of models is known
to describe the two-body physics very accurately over an energy range
comparable with the van der Waals energy $E_{\text{\mbox{\tiny vdW}}}$,
in the same spirit as multi-channel quantum-defect theory~\cite{Gao2005,Gao2011}.
The authors of reference~\cite{Wang2014} argue that their model
should equally provide an accurate description of the three-body physics,
in particular the three-body parameter. Indeed, the Feshbach resonance
is decribed properly, both in the limit of narrow resonances and the
limit of broad resonances where the van der Waals tail of the potential
determines the three-body parameter. Moreover, the Lennard-Jones potentials
support more than one bound state, allowing the description of recombination
and relaxation processes to these bound states. Solving the three-body
problem numerically with such models, the authors of reference~\cite{Wang2014}
could indeed reproduce the experimentally observed loss by three-body
recombination or atom-dimer relaxation around several resonances:
a broad resonance in caesium-133 ($s_{\mbox{\tiny res}}\approx200$),
a double resonance in caesium-133 ($s_{\mbox{\tiny res}}\approx200$
and $s_{\mbox{\tiny res}}\approx200$), and a broad resonance in rubidium-85
($s_{\mbox{\tiny res}}\approx200$). Not only could they reproduce
the location of Efimov peaks giving the three-body parameter, but
also the loss rate for scattering lengths outside the window of zero-range
universality. This remarkable agreement is a further evidence of the
three-body van der Waals universality.

The experimental Efimov features considered in reference~\cite{Wang2014}
are driven by broad resonances, and therefore satisfy the van der
Waals universal value~(\ref{eq:AminusForVdW}) of the three-body
parameter. In principle, the model could also be used to investigate
the case of intermediate and narrow resonances, although this was
not detailed in reference~\cite{Wang2014}. The authors point out
that for these resonances the results should depend not only on the
van der Waals length $\ell_{\mbox{\tiny vdW}}$ and the resonance
strength $s_{\tiny\mbox{res}}$ but also on other parameters such
as the background scattering length $a_{\mbox{\tiny bg}}$ (a fact
missing by construction in the work of Richard Schmidt and co-workers~\cite{Schmidt2012}).
Even for the broad resonances, their calculations indicate that while
$a_{-}^{(0)}$ remains largely insensitive to $a_{\mbox{\tiny bg}}$,
the value of $a_{+}^{(0)}$ may depends on $a_{\mbox{\tiny bg}}$.
A full mapping of the parameter space for Feshbach resonances remains
to be done.

\subsubsection{Experimental observations}

\paragraph{Broad resonances}

In ultra-cold atom experiments, it is usually easier to deal with
open-channel dominated resonances because they correspond to broad
resonances in terms of the applied magnetic field.  Most of the measurements
reported in Fig\@.~\ref{fig:van-der-Waals-Universality} are obtained
from relatively broad resonances. As mentioned in section~\ref{subsec:Broad-Feshbach-resonances},
broad resonances are expected to be well described by single-channel
two-body potentials. Indeed, the theoretical results of section~\ref{subsec:Van-der-Waals-Universality}
for single-channel van der Waals potentials agree with the observations
within 20\%. This agreement has confirmed the van der Waals universality
of the three-body parameter for these resonances. 

However, reducing the tolerance reveals some discrepancies that seem
significant. First of all, the theoretical calculation of $a_{-}^{(0)}/\ell_{\mbox{\tiny vdW}}$
for single-channel deep van der Waals potentials gives a value close
to $-11$,~see equations~(\ref{eq:AminusForVdW}) and (\ref{eq:AminusFOrVdWSeparable}),
whereas most experimental values are above $-9$. Another issue pointed
out in reference~\cite{Huang2014a} is that the presumably most precise
experimental determinations of the three-body parameter, based on
first-excited-state resonances, give the values~(\ref{eq:ExperimentalVdWAminus1Cs})
and (\ref{eq:ExperimentalVdWAminus1Li6}) that differ significantly
by 20\%. In the absence of systematic errors, one has to conclude
that while van der Waals physics is the main ingredient determining
the three-body parameter, it is not the only one. Possible candidates
are the deviation from the van der Waals tail ($C_{8}$ coefficient)
in the single-channel potential and coupled-channel effects. According
to the calculations discussed in section~\ref{subsec:Intermediate-resonances},
the coupled-channel effects tend to increase the value of $a_{-}$,
which may make it closer to the observed values. Yet, more work is
needed to refine our understanding. From the current theoretical and
experimental results, one may only say that the value of $a_{-}$
for the broad atomic resonances is about $-9\,\ell_{\mbox{\tiny vdW}}$,
with an uncertainty of 20\%.

\paragraph{Intermediate and narrow resonances}

Closed-channel dominated resonances, on the other hand, are most often
narrow and necessitate a fine tuning of the magnetic field, which
in turn requires a high stabilisation of the intensity of the current
in the coil creating the magnetic field. There are therefore much
less experimental observations for narrow resonances. The experimental
group of Giovanni Modugno in Florence~\cite{Roy2013} have reported
the measurement of the three-body parameter for seven different Feshbach
resonances in potassium-39. These resonances correspond to a magnetic
field of 58.92, 60.1, 65.67, and 471.0 gauss for atoms polarised in
the hyperfine state with projection $m_{F}=0$ along the magnetic
field, 33.64, 162.35, and 560.7 gauss for $m_{F}=-1$, and 402.6 gauss
for $m_{F}=+1$. The strength $s_{\mbox{\tiny res}}$ of these resonances
has been calculated to range from 0.11 to 2.8, and the background
scattering $a_{\mbox{\tiny bg}}$ ranges from -1.54 nm to -0.95 nm.
The group found that the measured value of $a_{-}^{(0)}$ does not
vary significantly from its expected value~(\ref{eq:ExperimentalVdWAminus0})
for broad resonances. Although the resonances are not narrow enough
to be fully in the regime described theoretically in section~\ref{subsec:Narrow-Feshbach-resonances},
one would have expected that the observations would show some indication
of the crossover between the narrow and broad limits, as suggested
by the models of \cite{Schmidt2012,Wang2014}. As of now, these observations
remain to be interpreted theoretically.

\clearpage{}

\part{More than three particles\label{part:More-than-three-particles}}

After the Efimov effect was discovered, a natural question was whether
the same effect could apply to a larger number of particles. In 1973,
Amado and Greenwood~\cite{Amado1973} already concluded that there
is no Efimov effect for four identical bosons or more, in the sense
that for $N\ge4$, there is not an infinite number of $N$-body bound
states near the appearance of an $(N-1)$-body bound state, as is
the case for $N=3$. In spite of this early negative result, it was
later found that not only a $4$-body Efimov effect can occur for
mass-imbalanced fermions, but a variety of universal $N$-body bound
states were found in the region where the $3$-body Efimov effect
occurs. Here, we review the situation for bosons and mass-imbalanced
Fermi mixtures.

\section{Bosons\label{sec:Many-Identical-bosons}}

\subsection{Tetramers tied to Efimov trimers\label{subsec:4-body-states-associated}}

\subsubsection{Four identical bosons}

\begin{figure}[t]
\hfill{}\includegraphics[scale=0.34]{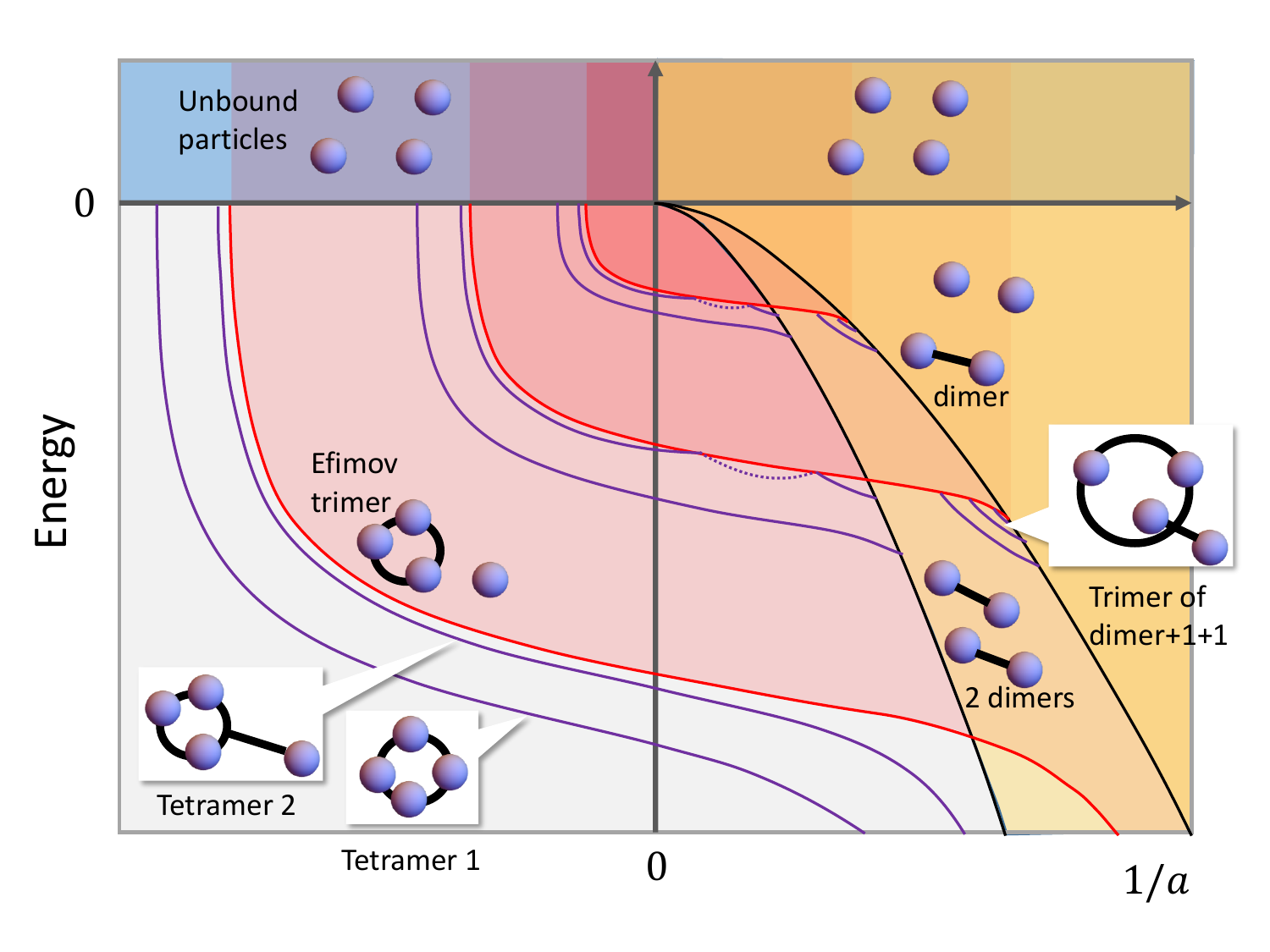}\hfill{}

\caption{\label{fig:Four-bosons}Schematic four-boson energy spectrum as a
function of the inverse scattering length $1/a$ between each pair
of bosons (adapted from~\cite{Ferlaino2009} and~\cite{Deltuva2013}).
The black curves represent the dimer+boson+boson and dimer+dimer thresholds.
The red curves represent the trimer+boson thresholds (compare with
the trimer energies of figure~\ref{fig:EfimovPlot}). The solid purple
curves represent the tetramer energies, and the dashed purple curves
correspond to inelastic virtual tetramer states. Here, the two tetramers
associated with the ground-state trimer do not follow exactly the
universal pattern exhibited by the tetramers tied to excited trimers,
although this depends on the microscopic model. }
\end{figure}

Although Amado and Greenwood \cite{Amado1973} found from the trace
of the four-body kernel of the four-boson integral equation that no
four-body Efimov effect occurs, they also acknowledged that near a
two-body resonance where the three-body Efimov effect occurs, there
should also be four-body bound states. However, one fundamental question
was whether in the limit of zero-range interactions a four-body parameter
is required to set the four-body energy, just as a three-body parameter
is required to set the three-body energy. Amado and Greenwood's result
suggests that, in the limit $\vert a\vert\gg b$, no four-body parameter
is required. As a further evidence, J. A. Tjon found by solving numerically
the four-body problem for different models that the four-body energy
is correlated with the three-body energy, a correlation referred to
as the ``Tjon line''~\cite{Tjon1975,Tjon1981}.

On the other hand, the works of Sadhan K. Adhikari, Tobias Frederico,
I. D. Goldman, and Yamashita~\cite{Adhikari1995} on the perturbative
renormalisation of the few-body problem with delta function potentials,
and subsequent works~\cite{Yamashita2006,Hadizadeh2011}, advocated
the introduction of an $(N+1)$-parameter for each particle added
to the $N$-body system. This question was also addressed by Hans-Werner
Hammer, Lucas Platter, and Ulf-G. Mei{\ss}ner~\cite{Platter2004}\cite{Hammer2007},
by solving the Yakubovsky equations~\cite{Yakubovsky1967} (a generalisation
to four bodies of the Faddeev equations) with a Gaussian separable
two-body and three-body potentials. The strengths of these potentials
are adjusted (renormalised) such that the two-body and three-body
energies are independent of the Gaussian cutoff $\Lambda$ of the
potentials. The authors found that there are two tetramer states below
the ground-state trimer and that their energy is relatively independent
of the cutoff $\Lambda$, suggesting that no four-body renormalisation,
and thus no four-body parameter is required. This led to some controversy
on the necessity of a four-body parameter.

The work of Hans-Werner Hammer and Lucas Platter \cite{Hammer2007}
also suggested that the presence of two tetramer states below the
ground-state trimer is a universal feature that also occurs below
each of the excited trimer states. This point was confirmed by Javier
von Stecher, Jos\'e P. D'Incao and Chris H. Greene in their study~\cite{Stecher2009}
where they solved the four-boson problem in hyperspherical coordinates.
The use of four-body hyper-radial potentials reveals the presence
of a well below each trimer-boson scattering threshold, following
the Efimov geometric scaling of these thresholds. Each potential well
can support two bound states, one of which being just below the trimer-boson
threshold, that is to say a trimer weakly bound to a boson. The existence
of these states was confirmed by solving the coupled hyper-radial
equations using the correlated Gaussian basis set expansion. This
infinite set of tetramer states have been referred to as ``universal
tetramers tied to (or associated with) Efimov trimers'', to avoid
the designation ``Efimov tetramers'' which would suggest a four-body
Efimov effect ruled out in reference~\cite{Amado1973}.

The work of von Stecher and co-workers also proposed a solution to
the controversy on the necessity of a four-body parameter. Indeed,
the location of the well in the four-body hyper-radial potential moves
to a larger hyper-radius by a factor of 22.7 at each new trimer-boson
threshold, making the corresponding tetramer states essentially insensitive
to any short-range four-body force, as advocated in \cite{Platter2004,Hammer2007}.
On the other hand, the well associated with the ground-state trimer-boson
threshold is located at shorter hyper-radii comparable with the range
of interaction, thus making the corresponding tetramers sensitive
to a short-range four-body force. This sensitivity makes to some extent
the ground-state tetramer energies independent of the trimer energy,
thus requiring in zero-range models the introduction of the four-body
parameter advocated in \cite{Yamashita2006,Hadizadeh2011}. 

The precise four-body spectrum for excited states was calculated by
Arnoldas Deltuva in a series of papers~\cite{Deltuva2010a,Deltuva2010,Deltuva2011,Deltuva2011a,Deltuva2011b,Deltuva2012,Deltuva2012a,Deltuva2012b,Deltuva2012c,Deltuva2013}.
This calculation had been challenging because the excited tetramers
are resonant states embedded in trimer-boson continua. Deltuva obtained
the binding energies $B_{n,i}$ and widths $\Gamma_{n,i}/2$ of these
states by calculating four-body scattering properties using the Alt-Grassberger-Sandhas
(AGS) equations~\cite{Grassberger1967,Alt1972}. Here, $n$ refers
to the $n$-th trimer of binding energy $b_{n}$, which the two tetramers
$i=1,2$ are associated with. At the unitary limit, and for large
$n$, the following universal relations were found:
\begin{eqnarray*}
B_{1}+i\Gamma_{1}/2 & = & (4.610(1)+i0.01483(1))b_{n}\\
B_{2}+i\Gamma_{2}/2 & = & (1.00227(1)+i0.000238(1))b_{n}
\end{eqnarray*}

The whole four-body spectrum is shown in figure~\ref{fig:Four-bosons}.
Generally speaking, each pair of tetramers follows the trimer energy.
The tetramer states thus form a superposition of two geometrical series
with the Efimov scaling ratio $e^{\pi/s_{0}}\approx22.7$. The two
tetramers appear from the four-body threshold before the appearance
of the associated trimer and dissociate in the dimer-dimer scattering
threshold. However, Deltuva found that in the case of the tetramer
resonances, the most weakly bound state first dissociates in the trimer-boson
threshold, surviving as a inelastic virtual state, before reappearing
from that threshold and eventually dissociating in the dimer-dimer
threshold. There is an additional family of tetramer states which
appear near the crossing of a trimer and the dimer-boson threshold.
These tetramer states are simply a consequence of the Efimov effect
occurring for the system of dimer and two bosons, since the scattering
length between the dimer and boson is resonant near this crossing.
These particular features may not occur for the ground state tetramers,
depending on the details of the short-range interaction.

\subsubsection{3 bosons + 1 particle}

\begin{figure}
\hfill{}\includegraphics[scale=0.34]{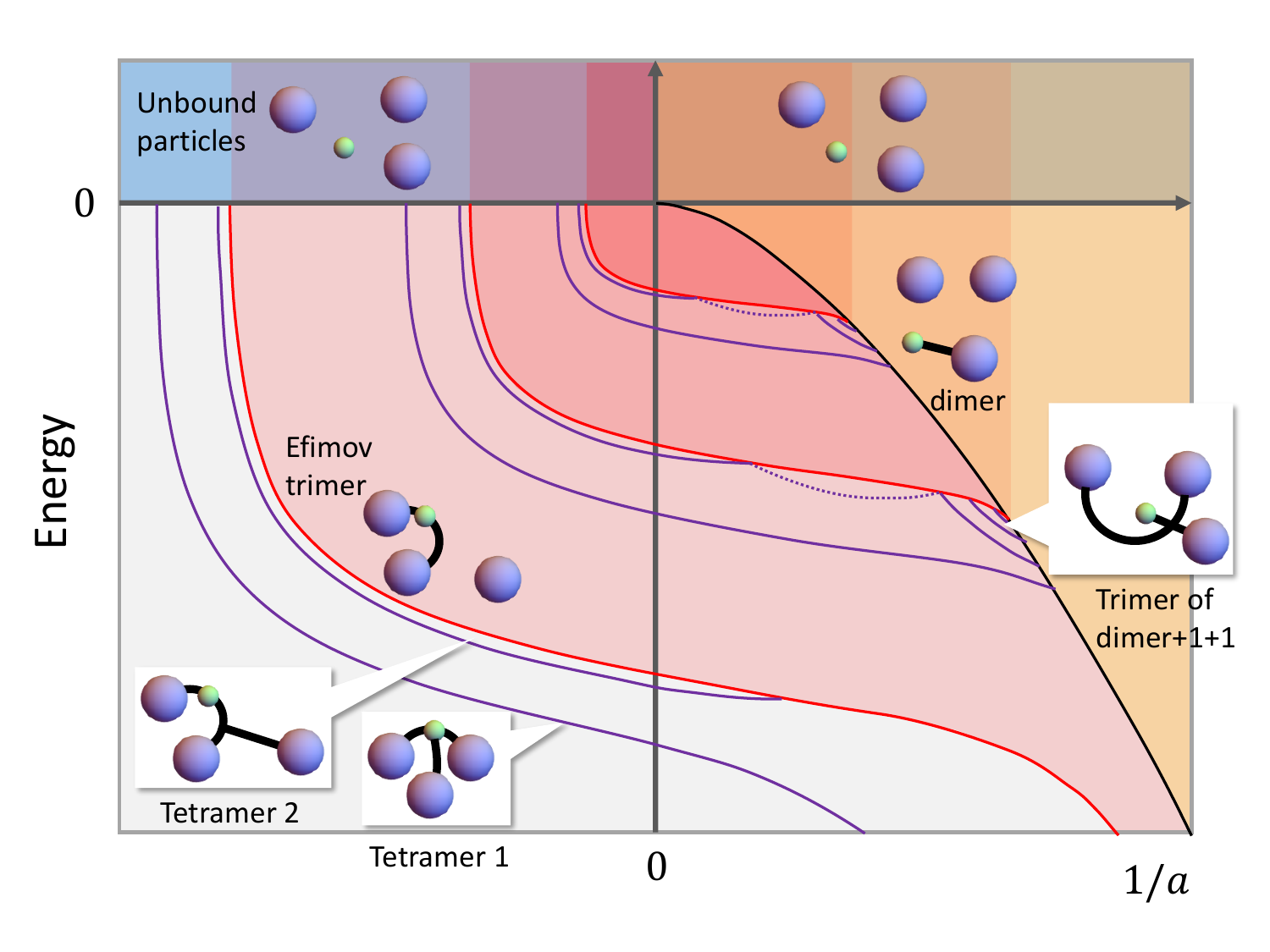}\hfill{}

\caption{\label{fig:Three+One}Schematic energy spectrum of three heavy bosons
and one light particle as a function of the inverse scattering length
$1/a$ between the heavy boson and light particle (adapted from Refs.~\cite{Wang2012b,Blume2014b}).
The same conventions as those of figure~\ref{fig:Four-bosons} are
used. It is assumed that the mass ratio between heavy and light is
larger than $\sim$13. For mass ratios smaller than $\sim$13, the
second tetramer state would disappear at a negative value of the scattering
length, instead of positive. The virtual states indicated by the dashed
curves are a conjecture made by analogy with figure~\ref{fig:Four-bosons}.}
\end{figure}

The four-body Efimov spectrum was investigated in the case of heavy
and light bosons mixtures by Yujun Wang and co-workers~\cite{Wang2012b},
and subsequently by Doerte Blume and Yangqian Yan~\cite{Blume2014b}.
As in the case of identical bosons, there appears to be no four-body
Efimov effect in these mixtures, despite earlier claims based on the
Born-Oppenheimer approximation~\cite{Naus1987}. On the other hand,
the works of Refs.~\cite{Wang2012b,Blume2014b} show that, as in
the case of identical bosons, universal tetramer states consisting
of three heavy and one light bosons are tied to the heavy+heavy+light
Efimov trimers (see section~\ref{subsec:2-bosons+1particle} for
a discussion of these trimers). The schematic four-body spectrum is
given in figure~\ref{fig:Three+One}. Tetramer states tied to excited
trimers were evidenced in reference~\cite{Wang2012b} from the calculated
four-body recombination rate, while the authors of reference~\cite{Blume2014b}
only calculated the tetramers tied to the ground-state trimer but
conjectured that similar tetramers exist for each excited trimer.
In this latter work, two tetramers were found below the ground-state
Efimov trimer. They appear at negative scattering lengths, but the
excited tetramer disappears in the trimer+heavy threshold for the
mass ratio $\kappa\lesssim13$ before reaching the unitary limit,
while it persists to some positive scattering length for $\kappa\gtrsim13$.
On the other hand, in the work of reference~\cite{Wang2012b}, only
one tetramer state was found at mass ratios $\kappa=30$ and $\kappa=50$.
It is possible that the second tetramer was missed in that study because
of its very weak binding energy.

In reference~\cite{Blume2014b}, the ground-state pentamer and hexamer
were also calculated. They were found to follow the energy of the
ground-state tetramer, as in the case of identical bosons.

\subsection{Universal clusters\label{subsec:Universal-clusters}}

The previous results on universal tetramers tied to Efimov trimers
naturally raise the question of the universality of larger cluster
of bosonic particles. Although there is an extensive theoretical literature
on bosonic clusters, motivated in particular by helium droplets, recent
works have focused on the connection to Efimov physics, and in particular
whether bosonic clusters with resonant interactions can be universally
described by a scattering length and a three-body parameter, as in
the case of universal tetramers.

This question has not been conclusively answered yet, because it requires
the daunting task of calculating excited $N$-body cluster resonant
states embedded in scattering continua of sub-clusters. So far, most
studies \cite{Stecher2010,Stecher2011,Yamashita2011,Horne2014,Gattobigio2014,Yan2015}
have focused on the $N$-body bound states below the ground-state
trimer, although one study~\cite{Stecher2011} was able to find $N$-body
resonances up to $N=6$ below the first-excited trimer. These studies
give us some idea of the properties of the conjectured universal clusters
associated with excited Efimov trimers. 

\subsubsection{Clusters below the ground-state trimer}

\begin{figure}
\hfill{}\includegraphics[scale=0.35]{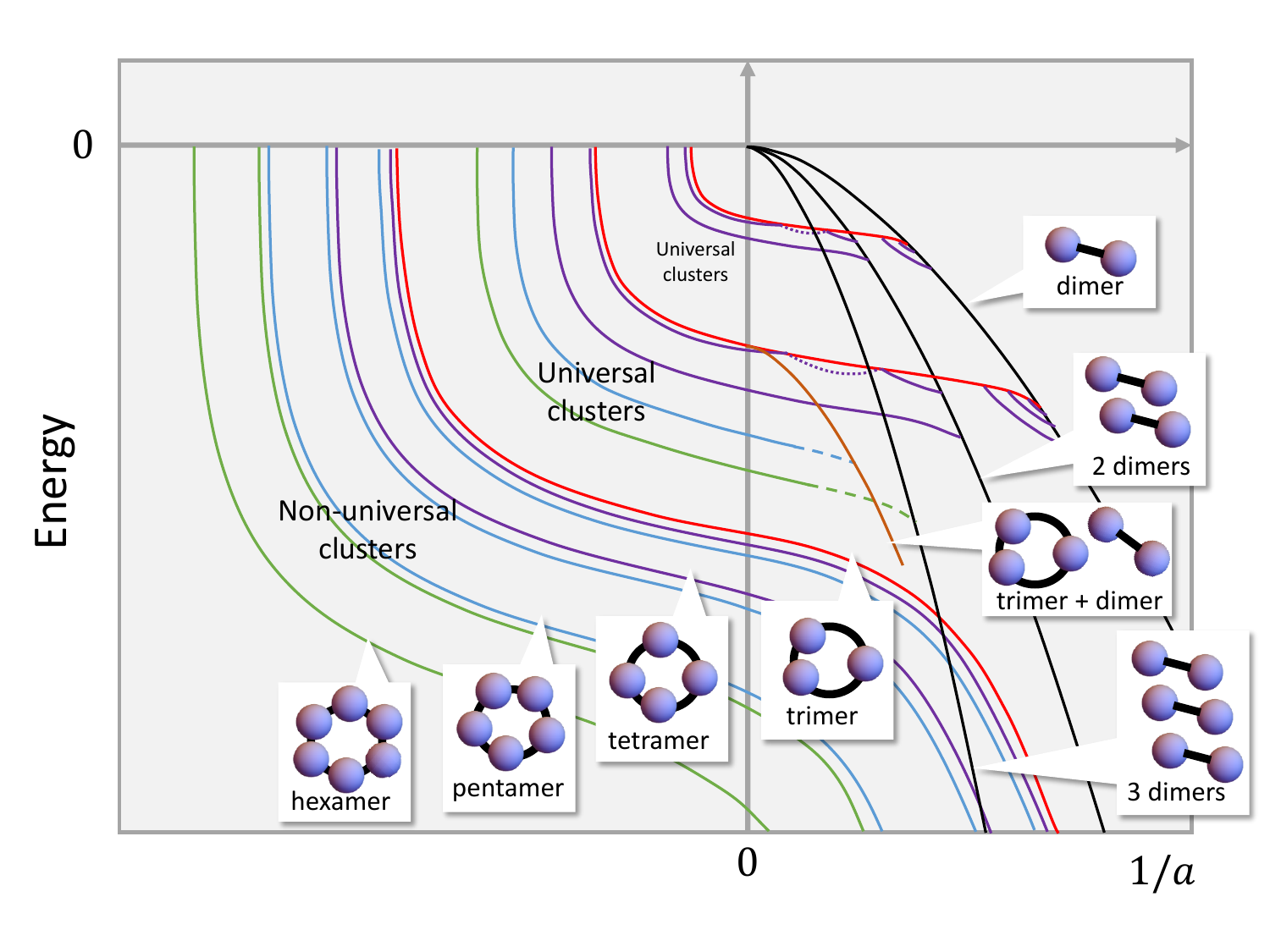}\hfill{}

\caption{\label{fig:Nbosons}Schematic spectra of the $N$-boson clusters up
to $N=6$, as a function of the inverse scattering length $1/a$ between
the bosons. This figure is based on the qualitative results of Refs.~\cite{Stecher2010,Stecher2011,Gattobigio2014}.
Trimers and tetramers correspond to the red and purple curves, as
in figure~\ref{fig:Four-bosons}, and pentamers and hexamers are
shown in blue and green, respectively. Clusters below the ground-state
trimer are non-universal, as they depend on the details of the interactions,
whereas clusters below the excited trimers are believed to approach
a universal pattern. This pattern is not well known: only a a few
states have been calculated in reference~\cite{Stecher2011} for
negative scattering lengths. The region of positive scattering lengths
offers many possible crossings and is virtually unknown.}
\end{figure}

The earliest attempt at calculating the $N$-body clusters tied to
the ground-state Efimov trimer is the work of Javier von Stecher~\cite{Stecher2010},
solving by the Diffusion Monte-Carlo method a model consisting of
two-body square-well interactions and three-body repulsive hard-core
interactions. Von Stecher found that the $N$-boson ground-state cluster
systematically appears at a weaker two-body attraction than the $N-1$-boson
cluster, and remains at a lower energy. This Borromean binding property
can be easily understood from the kinetic and interaction energy counting
argument given in section~\ref{sec:What-is-Efimov}. In this study,
only the ground bound state has been calculated. In a subsequent study~\cite{Stecher2011}
using different potential models and numerical techniques, von Stecher
identified three bound states for $N=5$, and two bound states for
$N=6$. These results are qualitatively summarised in figure~\ref{fig:Nbosons}. 

In the work of Mario Gattobigio and Alejandro Kievsky~\cite{Gattobigio2014},
the same problem was investigated for systems of up to $N=6$ bosons
interacting either through two-body Gaussian potentials, or two-body
P\"oschl-Teller potentials. In each case, the first and second $N$-body
bound state were calculated. The results with Gaussian two-body potentials
were confirmed by Yangqian Yan and Doerte Blume~\cite{Yan2014,Yan2015},
who also obtained results for two-body Lennard-Jones potentials, adjusted
two-body Helium-4 potentials, and two-body zero-range interactions
with power-law $C_{p}R^{-p}$ three-body repulsive interactions. 

\begin{figure}
\hfill{}\includegraphics[scale=0.65]{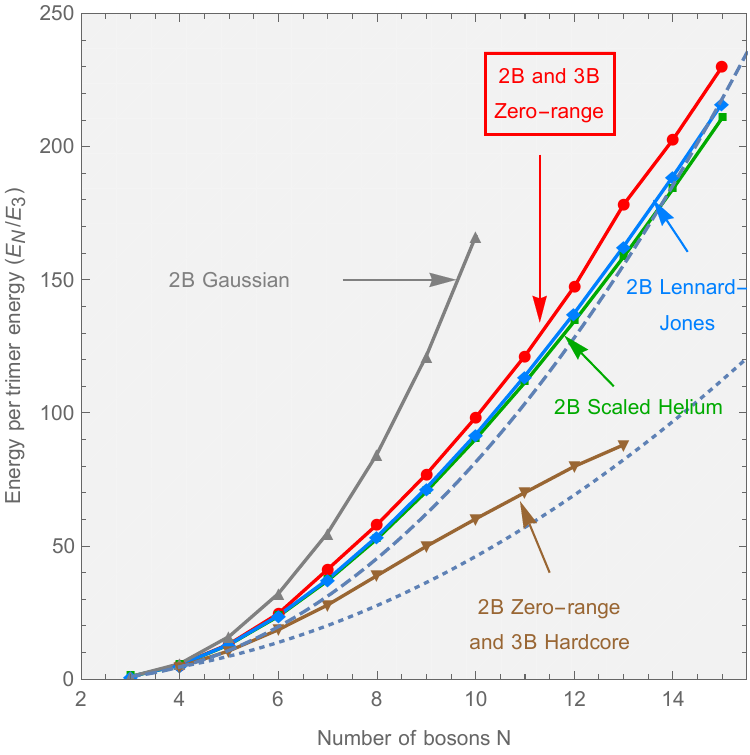}\hfill{}

\caption{\label{fig:NBosonClusters}Bosons at unitarity (adapted from Refs.~\cite{Stecher2010,Gattobigio2014,Yan2015}):
Energy of the ground-state bosonic cluster (normalised by the ground-state
trimer energy) as a function of the number of bosons, for different
models of interactions at unitarity ($1/a=0$) indicated by the arrows.
The dotted and dashed curves are analytical predictions from Refs.~\cite{Nicholson2012}
and \cite{Gattobigio2014}, respectively. }

\end{figure}

Although all these results agree qualitatively with figure~\ref{fig:Nbosons},
they quantitatively disagree. Figure~\ref{fig:NBosonClusters} shows
the energy of the ground state cluster for some of these models. These
models deviate as the number of particles $N$ is increased. This
is not unexpected: as we have learnt from the preceding sections,
the ground-state properties often deviate significantly from the universal
pattern exhibited by excited states because the spatial extent of
the ground states is comparable to the range of the interactions.
Interestingly, though, Gattobigio and Kievsky showed in reference~\cite{Gattobigio2014}
that these non-universal deviations (for their own results and those
of reference~\cite{Stecher2010}) could be mapped back for all $N$
(up to $N=6$) to a single universal curve corresponding to the Efimov
trimer curve given by equation~(\ref{eq:UniversalFormula}), generalising
the modified universal formula~(\ref{eq:ModifiedUniversalFormula})
for $N>3$. This observation suggests that there is a yet to be understood
relationship between the non-universal ground-state clusters and the
universal Efimov trimer structure.

\subsubsection{Universal $N$-body clusters }

Much less is known about the clusters tied to excited Efimov trimers,
which are resonant states presumed to have a universal structure.
In reference~\cite{Stecher2011}, Javier von Stecher could identify
some of these states up to $N=6$. They are schematically shown in
figure~\ref{fig:Nbosons}, and qualitatively follow the pattern of
the clusters associated with the ground-state Efimov trimer for negative
scattering lengths. This indicates that the ground-state clusters
give us some idea of the spectrum of the excited clusters. The earlier
work of reference~\cite{Stecher2010} attempted to make these ground-state
clusters close to the universal clusters by using a model consisting
of two-body square-well interactions and three-body hard core repulsive
interactions. The resulting ground-state pentamer and hexamer were
indeed found to agree relatively well with the excited pentamer and
hexamer resonances found in reference~\cite{Stecher2011}, much more
so than the ground-state clusters of other models. Nevertheless, it
is unclear how much universal this model is, especially for large
$N$.

To improve this situation, Yan and Blume considered an alternative
model in an attempt to approach a purely zero-range model, which supposedly
yields universal results for both the excited and ground clusters
(in fact, a purely zero-range model has no ground state). Although
two-body zero-range interactions can be implemented in their numerical
method (based on the path integral Monte Carlo method), three-body
zero-range interactions cannot be treated. The authors thus resorted
to a three-body repulsive interaction of the form $C_{p}R^{-p}$,
where $R$ is the three-body hyper-radius - see equation~(\ref{eq:Hyper-radius}),
and carefully checked that when $p$ is increased, the three-body
ground-state observables convincingly converge to that of the universal
(zero-range) theory. It is thus plausible that the ground-state clusters
of this model converge to the universal clusters for large enough
$p$. Although the cluster energies are found to be rather insensitive
to $p$ for small $N$, for larger $N$ their energies become strongly
dependent on $p$. Nevertheless, we noticed that the dependence on
$1/p$ appears to be linear, which makes it easy to extrapolate to
$p\to\infty$. In figure~\ref{fig:NBosonClusters}, we have represented
the result of this extrapolation by the red curve. At present, this
curve represents the most plausible variation of the energy with $N$
for universal clusters at the unitarity limit. Interestingly, it turns
out to be close to the curves obtained for ground-state clusters for
Lennard-Jones and scaled helium potentials, although these curves
are expected to represent a different class of universality associated
with van der Waals potentials. 

There remain many open questions about the universal clusters. The
region of positive scattering lengths have not been addressed yet,
and promises to reveal a rich pattern of clusters dissociating into
subclusters. There is also an intriguing possibility that the energy
of clusters of a very large number of particles, if they exist, may
leave the scaling window of the trimer they are originally associated
with. 

\subsection{Observation with atoms}

In the theoretical study reference~\cite{Stecher2009} on tetramers
tied to Efimov trimers, it was realised that experimental evidence
of these tetramers could be seen through the enhancement of the four-body
recombination loss rate at the scattering lengths where the tetramers
appear at the four-body scattering threshold, something which had
in fact already been observed in experimental data from ultra-cold
caesium atom experiments at the University of Innsbruck. In reference~\cite{Ferlaino2009},
the experimentalists in Innsbruck confirmed that the locations of
these loss peaks at $a=0.47a_{-}$ and $a=0.84a_{-}$ (where $a_{-}$
designates the scattering length at which a trimer appears, seen experimentally
by a three-body loss peak) are consistent with the expected locations
$a_{-\mbox{\tiny Tetra1}}\approx0.43a_{-}$ and $a_{-\mbox{\tiny Tetra2}}\approx0.9a_{-}$.
A few years later, the same group reported the observation in the
same system of a weak five-body loss peak at the scattering length
where a ground-state pentamer is expected to appear~\cite{Zenesini2013}.

Two other groups, the group of Randall G. Hulet at Rice University~\cite{Pollack2009}
and the group of Giovanni Modugno and Massimo Inguscio in Florence~\cite{Zaccanti2009},
also reported in 2009 the observation of the four-body loss peaks
at the scattering lengths consistent with the above values, but it
turned out later that their results were plagued by misassignments
of the loss peaks and recalibration of the scattering length with
respect to the magnetic field (see section~\ref{subsec:Ultracold-atoms}),
and their results were modified in 2013 \cite{Roy2013,Dyke2013}.
The former group originally found two four-body loss peaks~\cite{Pollack2009},
but later in reference~\cite{Dyke2013}, they could only observe
a single peak at $a=0.37a_{-}$ which is likely to correspond to one
of the four-body bound states, while the other one is not observed.
The latter group originally reported the observation of a single four-body
loss peak at $a=0.43a_{-}$~\cite{Zaccanti2009}, but after reassignment
of the peaks, no four-body loss peak was found \cite{Roy2013}.

\section{Mass-Imbalanced Fermi mixtures\label{sec:Imbalanced-Fermi-mixtures}}

Resonantly interacting four-body systems with identical fermions have
also been studied recently. The first treatment of these systems was
the exact calculation by Dmitry Petrov and co-workers~\cite{Petrov2005,Petrov2004a}
of the dimer-dimer scattering length for a four-body system of two
identical (i.e., spin-polarised) fermions plus two other identical
fermions (i.e. in a different spin state). The four-body problem for
fermions, while more challenging than that for identical bosons, is
still tractable analytically and numerically. Systems with identical
fermions are subject to the repulsion originating from the Pauli exclusion
principle and are, generally speaking, less likely to support bound
states than bosonic systems. However, this repulsion can be overcome
by increasing the mass imbalance between the two fermionic species,
leading to the formation of a four-body bound state, as in the case
of mass-imbalanced three-body system with fermions (see section \ref{subsec:Kartavtsev-Malykh-universal-trimers}).
In this section, we review the recent theoretical progress on two
classes of four-body systems which consist of resonantly-interacting
identical fermions of two species: a system of two identical (i.e.
spin-polarised) fermions plus two other identical fermions in section~\ref{subsec:2-fermions-+2-fermions},
and a system of three identical fermions plus another distinguishable
particle in section~\ref{subsec:3-fermions-+1particle}. We review
and discuss prospects for systems of more than four particles in section~\ref{subsec:Five-bodies-and-beyond}.

\subsection{2 fermions + 2 fermions\label{subsec:2-fermions-+2-fermions}}

The earliest works on the four-body problem in a two-component Fermi
mixture were motivated by the studies of the BEC-BCS crossover~\cite{Leggett1980,Nozieres1985,Bourdel2004,SadeMelo1993,Haussmann1993,Ohashi2005,Diehl2010,Levinsen2006}.
The nature and stability of the Bose-Einstein condensate phase, consisting
of dimers made of different fermions, is sensitive to scattering properties
of the dimers. In some works~\cite{SadeMelo1993,Haussmann1993,Ohashi2005,Diehl2010,Pieri2000,Birse2011},
the dimer-dimer $s$-wave scattering length has been estimated with
some approximations. Exact calculations have been performed in Refs.~\cite{Petrov2005,Petrov2004a,Levinsen2006,Brodsky2006,DIncao2009b},
and the dimer-dimer scattering length at equal mass has been found
to be $a_{dd}/a=0.59...$ ($a$ is the $s$-wave scattering length
between two non-identical fermions), which is significantly smaller
than the mean-field (i.e., the Born approximation) value $a_{dd}/a=2.0$
due to the Pauli exclusion principle. The exact calculation of the
elastic dimer-dimer $s$-wave scattering length has also been extended
to the mass-imbalanced case in Refs.~\cite{Petrov2005a,Levinsen2011,Iskin2008},
and the asymptotic behavior at large mass ratio $a_{dd}\approx a_{ad}/2$
($a_{ad}$ is the fermion-dimer $s$-wave scattering length) has also
been found analytically~\cite{Alzetto2013}.

For a four-body system made of 2 identical fermions + 2 identical
fermions, there necessarily exists a repulsion between each identical
fermions due to the Pauli exclusion principle, preventing the formation
of a universal four-body bound state~\cite{Petrov2005a,Blume2012,Endo2015}.
Indeed, for any parity and angular momentum channel, there is no universal
four-body bound state when the mass ratio is smaller than the three-body
Efimov critical mass ratio $\kappa_{c}^{(3)}=13.6069657...$~\cite{Blume2010a,Blume2012,Endo2015}:
there is neither a four-body Efimov effect at the unitary limit~\cite{Blume2010a,Endo2015,Michelangeli2016}
nor a universal four-body bound state of Kartavtsev-Malykh character
for a positive scattering length~\cite{Blume2012}. This is in marked
contrast to the 3+1 Fermi system discussed in the next section, where
four-body bound states can appear below the three-body Efimov critical
mass ratio. 

On the other hand, if the mass ratio exceeds the three-body Efimov
critical mass ratio $\kappa_{c}^{(3)}=13.6069657...$, the three-body
Efimov attraction arises and may bind the four particles. The resulting
four-body bound states, however, are not expected to be four-body
Efimov states. Since the introduction of the three-body parameter
is known to regularise the four-body problem for a bosonic four-body
system~\cite{Amado1973,Platter2004,Hammer2007}, it is expected that
the four-body bound states of the 2+2 Fermi system, if they exist,
should also be characterized by the three-body parameter only, and
would not require the introduction of a four-body parameter. They
would therefore be tied to the three-body Efimov states, in an analogous
manner to the four-body bound states of a Bose system (see section~\ref{sec:Many-Identical-bosons}
and Refs.~\cite{Platter2004,Hammer2007,Stecher2009,Deltuva2011,Wang2012b}),
and show the discrete-scaling symmetry with the same scaling factor
as the Efimov trimers.

\subsection{3 fermions + 1 particle\label{subsec:3-fermions-+1particle}}

In contrast to the 2 + 2 Fermi system, the system of 3 identical fermions
of mass $M$ plus a distinguishable particle of mass $m$ exhibits
richer physics. In particular, the repulsion between the identical
fermions can be overcome by the attraction mediated by the other particle
when the mass of the identical fermions is much larger than that of
the other particle. This situation is similar to the 2+1 system (see
section~\ref{subsec:Kartavtsev-Malykh-universal-trimers}). As a
result, physics similar to that of the 2+1 Fermi system has been found
to occur in the 3+1 Fermi systems, as discussed below.

\subsubsection{Four-body Efimov effect\label{subsec:Four-body-Efimov-effect}}

Yvan Castin and co-workers have studied the 3+1 system at the unitary
limit, and have found that a four-body Efimov effect occurs, i.e.
the existence of a four-body attraction leading to an infinite number
of four-body bound states in the absence of three-body bound states.
This effect occurs in the $L^{\Pi}=1^{+}$ channel, where $L$ is
the total angular momentum and $\Pi$ is the parity of the four-body
state, and for a mass ratio $M/m$ above the critical ratio $\kappa_{c}^{(4)}=13.384...$~\cite{Castin2010},
which is below the critical mass ratio $\kappa_{c}^{(3)}=13.606...$
for the occurence of three-body Efimov effect for 2+1 systems. The
same result has also been obtained more recently by Betzalel Bazak
and Dmitry Petrov~\cite{Bazak2017}. Since there is no trimer at
unitarity in the range of mass ratio $\kappa_{c}^{(4)}<M/m<\kappa_{c}^{(3)}$,
the binding of the 3+1 states is a genuine four-body phenomenon. These
four-body Efimov states exhibit discrete scale invariance with a scaling
factor $e^{\pi/s}$, where the exponent $s$ in the scaling factor
is obtained by numerically solving the four-body Schr\"odinger equation
at the unitary limit. As in the three-body Efimov effect, this leads
to a transcendental equation that determines the value of $s$. It
is worthwhile to note that this is the first and so far only known
example of the four-body Efimov effect in any physical system: for
a four-body system of bosons, the four-body Efimov effect does not
occur~\cite{Amado1973}, and there exists only a finite number of
four-body bound states tied to each Efimov trimers~\cite{Platter2004,Hammer2007,Stecher2009,Deltuva2011}
- see section~\ref{subsec:4-body-states-associated}. It is also
interesting to note that some of the four-body Efimov states are likely
to persist beyond the three-body critical mass ratio $\kappa_{c}^{(3)}=13.606...$,
while some others are expected to mix with the trimer+fermion scattering
continua.

\subsubsection{Universal four-body bound state\label{subsec:Universal-four-body-bound}}

Below the four-body Efimov critical mass ratio $\kappa_{c}^{(4)}=13.384...$,
there is no universal four-body bound state at the unitary limit.
However, at positive scattering length for a mass ratio $M/m\gtrsim9.5$
, a universal four-body bound state was found numerically by Doerte
Blume in reference~\cite{Blume2012} in the same $L^{\Pi}=1^{+}$
channel as the four-body Efimov states. This four-body bound state
is universally characterised by the $s$-wave scattering length, and
is independent of other short-range parameters. It can therefore be
interpreted as a four-body analogue of the universal three-body bound
states found by Kartavtsev and Malykh for mass ratio $M/m>8.17...$
and $M/m>12.91...$ (see section~\ref{subsec:Kartavtsev-Malykh-universal-trimers}).
Just as the Kartavtsev-Malykh three-body bound states are remnants
of the three-body Efimov states appearing above the critical mass
ratio $\kappa_{c}^{(3)}=13.606...$, the universal four-body bound
state may be regarded as a remnant of the four-body Efimov states
appearing above the critical mass ratio $\kappa_{c}^{(4)}=13.384...$~\cite{Castin2010}.
We note that the critical mass ratio for this universal four-body
state has recently been refined to $M/m=8.862(1)$~\cite{Bazak2017}.

Note that, as in the case of the 2 + 1 Fermi system (see section~\ref{subsec:Kartavtsev-Malykh-universal-trimers}
and Refs.~\cite{Nishida2008b,Blume2010a,Blume2010,Safavi-Naini2013}),
other kinds of four-body bound states may exist for the 3 + 1 Fermi
system at unitarity even below the critical mass ratio, but they are
non-universal~\cite{Blume2010a,Blume2012,Blume2010} in the sense
that they depend on a short-range boundary condition that is not set
by the scattering length but other short-range details of the interactions.

\subsubsection{Five bodies and beyond\label{subsec:Five-bodies-and-beyond}}

It is rather challenging to study the universal and Efimov states
of five or more particles since it requires a highly accurate calculation
of the extremely small binding energy of these clusters at the critical
mass in the presence of the fermionic sign problem. Nevertheless,
Betzalel Bazak and Dmitry Petrov have recently succeeded in studying
the mass-imbalanced 4+1 Fermi system by using the diffusion Monte
Carlo method to solve the Skorniakov - Ter-Martirosian equation for
this system~\cite{Bazak2017}. They found the critical mass ratios
for the Efimov pentamers and universal pentamers in the $L^{\Pi}=0^{-}$
channel to be 13.279(2) and 9.672(6), respectively. 

From the critical mass ratios for the 2+1, 3+1, and 4+1 systems, it
is tempting to make an extrapolation to $N+1$ system with $N\rightarrow\infty$.
The extrapolation suggests that the critical mass ratios for the Efimov
and universal clusters approach $\sim13.0(1)$ and $\sim12(2)$ in
that limit. But this naive extrapolation should not be taken too seriously
since it neglects shell-closure-oscillation which may be significant
around the 4+1 system corresponding to the shell closure of the $p$-wave
orbital in the shell model picture~\cite{Bazak2017}. The authors
of reference~\cite{Bazak2017} expect that 5+1 Efimov hexamers or
universal hexamers, if they do exist, show qualitatively different
behaviours from 2+1, 3+1, and 4+1 clusters since to form these hexamers
a heavy fermion needs to occupy a higher angular momentum or radial
quantum number state in the shell model picture.

\clearpage{}

\part{Many-body systems\label{part:Many-body-systems}}

So far, our focus has been on isolated systems with a relatively small
number of particles. This approach is relevant to the description
of some systems such as light nuclei or isolated molecules. However,
many physical systems are composed of a very large number of particles.
It is thus natural to ask oneself about the implications of Efimov
physics in this context. Several studies have recently started to
address this question. The first studies have looked into the influence
of a many-body background onto Efimov states, while more recent studies
deal with the consequences of few-body Efimov physics at the many-body
level.

\section{Many-body background\label{sec:Many-body-background}}

The observation of Efimov physics in ultra-cold atom experiments involve
the existence of triatomic molecules that can be formed from an extremely
dilute cloud of atoms. Even though the density of the cloud, typically
$10^{12}$ atoms per cubic centimetre, is small enough to regard the
triatomic molecules as isolated, it is natural to wonder how the properties
of such molecules are changed under the influence of a surrounding
medium. It is particularly interesting that the medium may be composed
of the same constituents as the those of the molecule, as this consideration
constitutes a first step towards many-body problems. Depending on
the quantum statistics of the constituents, there are thus two possible
media, a Fermi sea or a Bose gas.

\subsection{Efimov states in a Fermi sea}

Let us consider an Efimov trimer containing one or several fermionic
particles, and surrounded by one or several seas of fermions that
are identical to those of the Efimov trimer. The presence of these
many additional fermions constitutes a non-trivial many-body problem,
but it can be treated in a first approximation as a static Fermi sea.
In this case, the main effect of the Fermi sea is to prevent the fermions
in the Efimov trimer from occupying the states already occupied by
the other fermions, owing to the Pauli exclusion between identical
fermions. This idea has been explored in the works of Refs.~\cite{MacNeill2011,Niemann2012,Nygaard2014}.
Generally speaking, the Pauli exclusion tends to reduce the binding
of the Efimov trimer. For a large enough density of the Fermi seas,
the Efimov trimer eventually disappears. The successive disappearances
of the different excited Efimov states as the density is increased
follow a scaling law with the universal scaling ratio $e^{\pi/\vert s_{0}\vert}$
\cite{MacNeill2011,Nygaard2014}. Interestingly, the trimers can survive
at positive energy, as a ``Cooper triples'', in analogy with the
Cooper pairs~\cite{Cooper1956}, a fact emphasised in reference~\cite{Niemann2012}.

\subsubsection{Two bosons and a fermion in a Fermi sea}

The first study on the influence of a Fermi sea on Efimov states is
that of Davis James MacNeill and Fei Zhou~\cite{MacNeill2011}. They
considered two heavy bosons of mass $M$ immerged in a sea of light
fermions of mass $m$, assuming a resonant pairwise interaction between
the bosons and the fermions. If the sea was composed of a single fermion,
we know from section~\ref{subsec:2-bosons+1particle} that Efimov
trimers exist and can be interpreted in the Born-Oppenheimer picture
(see section~\ref{subsec:The-Born-Oppenheimer-picture}) as the binding
of the two bosons by the light fermion. The authors thus used the
Born-Oppenheimer approximation, along with a semi-classical approximation,
to calculate the three-body spectrum with the constraint the fermion
cannot occupy states of momenta smaller than the Fermi momentum $k_{F}$
of the Fermi sea. Since this constraint breaks the translational invariance,
the total momentum of the system does not simply shift the energy
of the spectrum. For simplicity, the authors restricted their consideration
to states of zero total momentum. For $k_{F}=0$, they retrieve the
vacuum Efimov states of section~\ref{subsec:The-Born-Oppenheimer-picture}.
Increasing $k_{F}$ results in an increase of the trimer energies
(a decrease of their binding energy). As a result, the infinite tower
of Efimov states is gradually pushed up, until the ground state disappears
in the scattering threshold. The successive disappearances of Efimov
states follow a scaling law with the universal scaling ratio $e^{\pi/\vert s_{0}\vert}$.
 The required $k_{F}$ to observe this density dependence, however,
seems to be too large for cold-atom experiments with a large mass
imbalance~\cite{Zinner2015}.

\subsubsection{Three fermions in a Fermi sea}

A similar treatment for three kinds of fermions of equal mass was
done in the works of Nicolai Gayle Nygaard and Nikolaj Thomas Zinner~\cite{Nygaard2014},
and Patrick Niemann and Hans-Werner Hammer~\cite{Niemann2012}. The
work of reference~\cite{Nygaard2014} considered explicitly one Fermi
sea associated with one of the three fermions, while the work of reference~\cite{Niemann2012}
considered three Fermi seas of equal Fermi momentum $k_{F}$. In both
works, the problem was solved using the Skorniakov~-~Ter-Martirosian
three-body integral equation with the Pauli exclusion constraint from
the static Fermi seas. The results are qualitatively the same as those
of MacNeill and Zhou~\cite{MacNeill2011}, although the work of Niemann
and Hammer~\cite{Niemann2012} emphasises the existence of solutions
at positive energies, which they call ``Cooper triples'', in analogy
with the Cooper pairs. 

A limitation of the previous results is that the Fermi seas are treated
as static. In reality, the back action of the particles on the Fermi
seas can create particle-hole excitations near the surface of the
Fermi sea, an effect called polarisation of the sea. The authors of
reference~\cite{MacNeill2011} estimated that the polarisation of
the sea would reduce even further the binding of the Efimov trimers.
On the other hand, Charles J. M. Mathy and co-workers ~\cite{Mathy2011}
found in a somewhat different system that the polarisation of the
sea may on the contrary enhance the stability of trimers. This work
considers the case of a trimer made of a light fermion and two heavy
fermions (known as Kartavtsev-Malykh universal trimer, see section~\ref{subsec:Kartavtsev-Malykh-universal-trimers}),
surrounded by a sea of identical heavy fermions. The authors used
a variational ansatz involving both the trimer and the particle-hole
excitation and found that the stability of the trimer is enhanced
by a moderate density of the Fermi sea.

\subsection{Efimov states in a condensate}

\subsubsection{Two impurities and a boson from a BEC}

The case of Efimov trimers in a BEC (Bose-Einstein Condensate) has
been considered by Nikolaj Thomas Zinner in reference~\cite{Zinner2013a}.
In that study, two heavy particles (impurities) resonantly interact
with bosons from a surrounding condensate of bosons. At low density,
it is known that one of the bosons can bind to two heavy particles
to form an Efimov trimer, which can be easily interpreted in the Born-Oppenheimer
approximation, as detailed in section~\ref{subsec:The-Born-Oppenheimer-picture}.
The author of reference~\cite{Zinner2013a} thus used the Born-Oppenheimer
approximation to calculate the effective interaction between the two
heavy particles at fixed separations. To treat the effect of the surrounding
condensate, the author used the Bogoliubov approach~\cite{Bogoliubov1947,Pethick2002}
to describe the bosons as a ground-state condensate with quasiparticle
excitations. The author then retained only the coupling to quasi-particles,
resulting in a three-body problem for two heavy impurities and a quasi-particle.
Although the validity of this approximation is not clear, it leads
to results that are similar to those of reference~\cite{MacNeill2011}.
The resulting Born-Oppenheimer potential between the two heavy impurities
is influenced by the medium at large separations $R$, on the order
of the coherence length $\xi$ of the condensate. In this region,
the bonding potential is reduced with respect to the bonding potential
in the absence of condensate (see section~\ref{subsec:The-Born-Oppenheimer-picture}
and figure~\ref{fig:Bonding-potential}), resulting in a faster decay.
This in turn reduces the binding of the Efimov states, as in the case
of a Fermi sea. However, in this theory, the potential appears to
be not defined beyond a certain separation $R_{0}\approx0.5\xi$,
so that these preliminary conclusions are to be confirmed by a more
consistent theory.

\subsubsection{One impurity and two bosons from a BEC}

Systems of a single impurity atom resonantly interacting with a Bose
condensate have recently been realised by two independent groups;
one with potassium-39 atoms in two different hyperfine states by the
group of Jan Arlt at Aarhus University~\cite{Jorgensen2016}, and
the other one with a mixture of rubidium-87 and potassium-40 by the
group of Eric Cornell and Deborah Jin at JILA~\cite{Hu2016}. Efimov
states may appear in these systems since two identical bosons can
bind with the impurity to form an Efimov trimer, as discussed in section~\ref{subsec:2-bosons+1particle}.
This possibility was explicitly taken into account for the impurity-BEC
system in the work of Jesper Levinsen, Meera M. Parish, and Georg
M. Bruun~\cite{Levinsen2015}, by constructing a variational ansatz
that includes up to two Bogoliubov excitations of the BEC. This allows
to take into account the three-body correlations of the impurity with
two bosons of the condensate. If one includes only two-body correlations,
the resulting variational ground-state is a ``Bose polaron'', i.e.
the impurity is dressed by the surrounding condensate and, as the
impurity-boson interaction is increased, progressively binds with
one of the bosons to form a dimer surrounded by the remaining bosons.
By contrast, taking into account three-body correlations, the Bose
polaron instead turns into a trimer composed of the impurity bound
to two bosons and surrounded by the remaining bosons. The authors
indicate that this state may be seen as the avoided crossing between
the two-body correlated polaron with the vacuum ground-state Efimov
trimer composed of the impurity and two bosons. In the experiments
\cite{Jorgensen2016,Hu2016}, such an avoided crossing could not be
observed since the Efimov trimers are much larger than the mean atomic
distance, but it would appear if one can prepare a gas with much smaller
density. Interestingly, the energy of the trimer dressed by the medium
is lower than that of the vacuum trimer, showing that in this case
the surrounding condensate stabilises the Efimov trimer.

Given these results, it would be tempting to think that by taking
a larger number of Bogoliubov excitations, one could couple the polaron
to universal clusters similar to those discussed in section~\ref{subsec:3-fermions-+1particle},
such as a tetramer formed of three bosons and the impurity. While
this is certainly to be expected, the authors of reference~\cite{Levinsen2015}
argue that the couplings to these clusters would be smaller than that
to the trimer, and possibly negligible. Their argument is that the
repulsive interaction between the bosons tend to reduce the coupling
between the polaron and universal clusters. They checked that the
coupling to the trimer is indeed reduced when the scattering length
$a_{B}$ between the bosons is increased, relative to the scattering
length $a_{-}$ between a boson and the impurity at which the trimer
appears. Since universal clusters are expected to appear at scattering
lengths between a boson and the impurity that are even smaller than
$\vert a_{-}\vert$, they expect that the reduction of the coupling
is comparatively stronger for larger clusters. Luis Aldemar Peña Ardila
and Stefano Giorgini have also arrived at a similar conclusion with
a quantum Monte Carlo calculation \cite{Ardila2015}. 

\section{Many-body phases\label{sec:Many-body-phases}}

\subsection{Identical bosons\label{subsec:Many-body-Identical-bosons}}

Since it became possible to adjust the scattering length of ultra-cold
atoms close to unitarity thanks to magnetic Feshbach resonances, experimentalists
have hoped to realise the unitary Bose gas, an intriguing strongly-correlated
system which theorists have been excited about for many years~\cite{Cowell2002,Ho2004,Lee2010,Diederix2011,Li2012,Borzov2012,Jiang2014,Piatecki2014}.
In particular, some theorists have wondered about the role of Efimov
physics in such a gas, and how it can be described by the three-body
parameter of the atoms, unlike the unitary Fermi gas which has no
microscopic length scale. Unfortunately, when the scattering length
is tuned to large values, ultra-cold bosonic gases are found to be
unstable because of enhanced loss by recombination of the atoms into
the many bound states that exist for atomic interactions. It is for
this reason that Efimov trimers in ultra-cold gases have been evidenced
essentially through features in these losses, rather than more direct
obervations. Nevertheless a metastable unitary gas of density $\rho$
and temperature $T$ (or thermal de Broglie wavelength $\lambda=h/\sqrt{2\pi mk_{B}T}$)
can be prepared in the non-degenerate regime $\rho\lambda^{3}\ll1$~\cite{Fletcher2013},
and it is possible to observe the gas dynamics in the degenerate regime
$\rho\lambda^{3}\gtrsim1$~\cite{Makotyn2014}. Here we review some
theoretical and experimental works that attempt to reveal the role
of Efimov physics in an atomic Bose gas.

\subsubsection{Three-body contact in a Bose gas}

Shina Tan~\cite{Tan2008,Tan2008a,Tan2008b} has introduced a set
of universal relations for the two-component Fermi gas that relates
its properties to the scattering length $a$ between two different
fermions through the strength of two-particle short-range correlations,
characterised by an extensive quantity called the ``contact'' $C_{2}$.
This quantity could be directly measured through radio-frequency (rf)
spectroscopy in a number of experiments with ultra-cold atomic Fermi
gases. The notion of contact was generalised for bosons by several
authors~\cite{Combescot2009,Werner2010,Schakel2010,Braaten2011,Smith2014}.
It appears that, in addition to the two-body contact, that gives the
variation of energy $E$ of the Bose gas with respect to the scattering
length $a$,
\[
\frac{dE}{da}=\frac{\hbar^{2}}{8\pi ma^{2}}C_{2},
\]
 the occurrence of Efimov physics for bosons requires to introduce
a three-body contact $C_{3}$, that gives the variation of energy
with respect to the three-body parameter $\kappa_{*}$,
\[
\frac{dE}{d\kappa_{*}}=\frac{\hbar^{2}\kappa_{*}^{2}}{m}C_{3}.
\]

\paragraph{Contacts of the dilute Bose gas}

The rf response of a Bose-Einstein condensate of rubidium-85 was recently
investigated experimentally at JILA~\cite{Wild2012}. The three-body
contact is expected to introduce in the rf response a frequency dependence
that is proportional to $C_{3}$ and has a log-periodicity that characterises
Efimov physics~\cite{Braaten2011}. Although the contribution from
the two-body contact was clearly seen in the experiment, no log-periodic
contribution from the three-body contact could be evidenced. This
was explained by subsequent calculations by D. Hudson Smith and co-workers
in reference~\cite{Smith2014} which determined that for the dilute
Bose gas of volume $V$:
\begin{eqnarray*}
C_{2}/V & \approx & 16\pi^{2}a^{2}\rho^{2}\\
C_{3}/V & \approx & \frac{16\pi^{2}(4\pi-3\sqrt{3})s_{0}\cosh(\pi s_{0})}{3\sinh^{3}(\pi s_{0})}a^{4}\rho^{3}.
\end{eqnarray*}
From this expression, the authors found that the value of $C_{3}$
is indeed too small to be observed in the experiment.

\paragraph{Contact of the unitary Bose gas}

In a subsequent experiment, the experimentalists at JILA were able
to measure the momentum distribution in their Bose-Einstein condensate
of rubidium at unitarity, before the three-body losses set in and
deplete the gas~\cite{Makotyn2014}. This distribution was found
to saturate to a seemingly universal distribution. The tail of this
momentum distribution is known to be related to the two-body and three-body
contacts~\cite{Braaten2011,Werner2010}. As in the rf spectroscopy
case, the three-body contact introduces a contribution which is proportional
to $C_{3}$ and has the Efimov log-periodicity in momentum. From dimensional
analysis, D. Hudson Smith and co-workers~\cite{Smith2014} found
that for the unitary Bose gas,
\begin{eqnarray*}
C_{2}/V & \approx & \alpha\rho^{4/3}\\
C_{3}/V & \approx & \beta\rho^{5/3}
\end{eqnarray*}

By fitting the experimental measurement of the momentum distribution,
assuming that the observed variations correspond to the log-periodic
prediction for the tail, the authors of reference~\cite{Smith2014}
obtained $\alpha=22(1)$ and $\beta=2.1(1)$. Other theoretical works
have made predictions for the value of $\alpha$, namely $\alpha=10.3$~\cite{Diederix2011},
$\alpha=12$~\cite{Sykes2014}, and $\alpha=9.02$~\cite{Rossi2014}.
The two-body contact of the unitary Bose gas has also been calculated
at finite temperature using a three-body-cutoff model~\cite{Comparin2016}
but the small values of the three-body contact and $\beta$ have proved
difficult to obtain. 

\subsubsection{The non-degenerate unitary Bose gas}

In the non-degenerate regime $\rho\lambda^{3}\ll1$, the equation
of state of the unitary Bose gas can be treated by the so-called virial
expansion~\cite{Huang1987,Pais1959}:
\[
\frac{P\lambda^{3}}{k_{B}T}=\sum_{n\ge1}a_{n}(\rho\lambda^{3})^{n}
\]
where $a_{1}=1$, $a_{2},\,a_{3}$... are the virial coefficients.
Because the third coefficient is related to three-body correlations,
it is expected to depend on Efimov physics. Yvan Castin and F\'elix
Werner have studied the virial coefficients at unitarity $1/a=0$~\cite{Castin2013}
and found an analytical expression for the first virial coefficients
in the zero-range limit:

\begin{eqnarray*}
a_{2} & = & -\sqrt{2}\frac{9}{8}\\
a_{3} & = & \frac{81}{8}-6\sqrt{3}f(\kappa_{*}\lambda)
\end{eqnarray*}
where $\kappa_{*}$ is the three-body parameter defined from the ground-state
trimer energy $E_{0}=-\hbar^{2}\kappa_{*}^{2}/m$ - see section~\ref{subsec:3BP-In-the-zero-range},
and the function $f$ admits the low- and high-temperature limits\footnote{The calculation of the third virial coefficient has also been generalised
to the case of 2+1 fermions~\cite{Gao2015} and to the case of 2+1
bosons~\cite{Endo2016a}.}:
\[
f(\kappa_{*}\lambda)\simeq e^{E_{0}/k_{B}T}\qquad\mbox{for }k_{B}T\ll E_{0}
\]

\[
f(\kappa_{*}\lambda)\simeq\frac{\vert s_{0}\vert}{2\pi}\ln(e^{\gamma+2\pi C/\vert s_{0}\vert}E_{0}/k_{B}T)\qquad\mbox{for }k_{B}T\gg E_{0},
\]
where $C\simeq0.648...$ and $\gamma\simeq0.577...$ is Euler's constant.
At high temperature, the gas therefore exhibits a dependence on temperature
that has the log-periodicity associated with Efimov physics. The authors
however noted that the validity of the zero-range model may break
down in this limit.

Away from unitarity, the second and third virial coefficients were
recently calculated numerically in reference~\cite{Barth2015}, as
a function of the scattering length $a$ and three-body parameter
$\kappa_{*}$. These numerical results are consistent with the above
analytical predictions at unitarity.

\subsubsection{The Efimov liquid phase}

\begin{figure}
\hfill{}\includegraphics[scale=0.35]{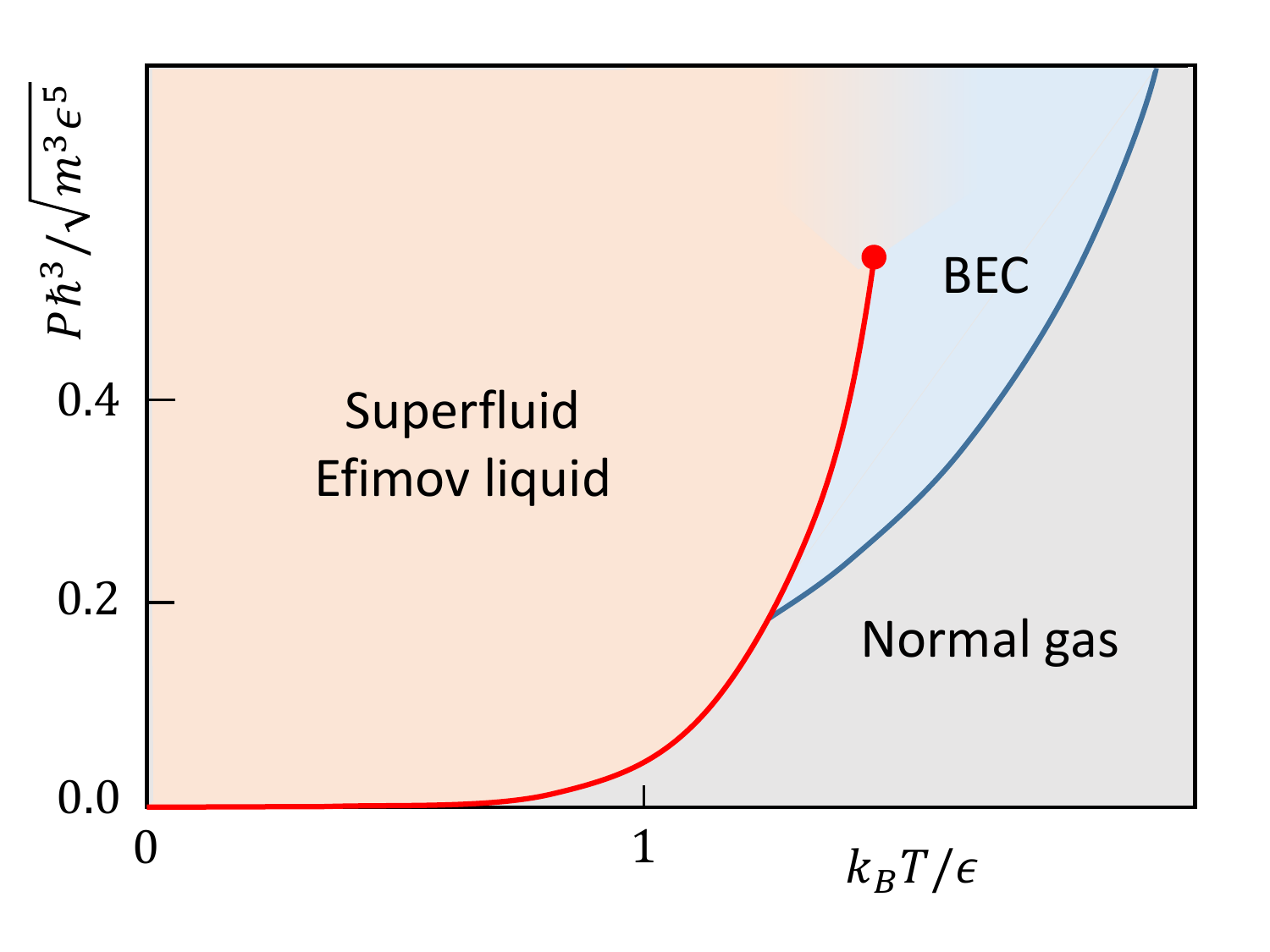}\hfill{}

\caption{\label{fig:EfimovPhase}Schematic phase diagram of unitary bosons
as a function of pressure and temperature normalised by the binding
energy per particle $\epsilon$ in the Efimov liquid phase (adapted
from the path-integral Monte Carlo results of reference~\cite{Piatecki2014}).
The value of $\epsilon$ is related to the energy $E_{0}$ of the
ground-state Efimov trimer at unitarity, although it appears to be
model-dependent (see figure~\ref{fig:NBosonClusters}). In the calculation
of reference~\cite{Piatecki2014}, $\epsilon\approx8\vert E_{0}\vert$.}

\end{figure}

The degenerate regime $\rho\lambda^{3}\gtrsim1$ of the unitary Bose
gas is more speculative since strong losses occur in experiments with
atoms. Nevertheless, the recent experimental achievement of reference~\cite{Makotyn2014}
has raised hope for the investigation of the degenerate unitary Bose
gas.

The recent theoretical developments in few-body Efimov physics indicate
that bosons may form universal excited $N$-body clusters governed
by Efimov physics - see section~\ref{subsec:Universal-clusters}.
Similar clusters exist below the ground-state Efimov trimer, although
they are not necessarily universal. The tendency of bosons to cluster
near unitarity indicates that the system as a gas is metastable. This
raises questions about the ground state of the system. The results
of numerical investigations of $N$-body clusters shown in figure~\ref{fig:NBosonClusters}
suggest that the energy per particle tends to a negative constant
for large $N$, although this constant depends strongly on the model.
A negative constant energy per particle would imply that the system
becomes a liquid for large $N$. Based on this idea, Swann Piatecki
and Werner Krauth~\cite{Piatecki2014} investigated the possibility
of a liquid phase in the unitary Bose system, using the path-integral
Monte Carlo method to numerically solve the same model as that of
reference~\cite{Stecher2010}, \emph{i.e.} bosons with two-body zero-range
interactions and three-body hard repulsive core. They obtain the phase
diagram shown in figure~\ref{fig:EfimovPhase} for a homogeneous
system. They found indeed that for sufficiently low temperature and
pressure, the system becomes a liquid, which they call the ``superfluid
Efimov liquid''. As expected, the density of this liquid is fixed
by the trimer energy, \emph{i.e.} the three-body parameter. We should
note that the model is quantitatively different from that of zero-range
interactions or shallow van der Waals interactions, as can be seen
in figure~\ref{fig:NBosonClusters}. The obtained phase boundaries
are therefore likely to be neither universal nor quantitative for
a realistic atomic system. Moreover, it is yet unclear how stable
the Efimov liquid phase is in an realistic atomic system, which interacts
through deep van der Waals potentials, since further decay would occur
to more deeply bound states.

Yet, the idea of an Efimov liquid, essentially bound by the Efimov
attraction is quite compelling. A particularly intriguing point is
whether the discrete scale invariance still holds in this many-body
system, implying that excited metastable liquid phases of densities
smaller by factors of $22.7^{3}\approx10^{4}$ could exist, at least
in principle.

\subsection{Trimer phases in Fermi mixtures}

As discussed in Section~\ref{sec:Multi-component-systems}, the Efimov
effect occurs in multi-component Fermi systems, i.e. mixtures of fermions
of different kinds or with different spins that interact resonantly.
Efimov trimers or related universal trimers may therefore be formed
in these systems. A striking difference with bosonic systems, however,
is that it is more difficult for more than three fermions to bind,
due to the Pauli exclusion between identical fermions. As a result,
a relatively stable phase of trimers may appear under certain conditions
in these systems. Recent studies have started to investigate this
possibility.

\subsubsection{Three-component Fermi mixtures}

For fermions of equal masses, three kinds of fermions are necessary
to exhibit the Efimov effect. In the work of Paulo Bedaque and Jos\'e
D'Incao~\cite{Bedaque2009}, the zero-temperature phase diagram of
the equal-mass three-component Fermi gas is sketched out qualitatively
as a function of the scattering lengths $a_{ij}$ between the different
components 1, 2, 3. Their reasonning is based on how energetically
favourable it is for certain components to pair, and how the condensation
of these pairs at zero temperature gives different symmetries of the
order parameter, corresponding to different phases. However, they
note that, beyond pairing, Efimov three-body physics should also be
taken into account, in particular the resonant enhancement of fermion-pair
scattering at certain values of the scattering lengths $a_{i}$. According
to the authors, this would result in additional phases where the pairs
and fermions are spatially separated due this enhanced repulsion,
instead of forming a mixture.

\begin{figure}
\hfill{}\includegraphics[scale=0.34]{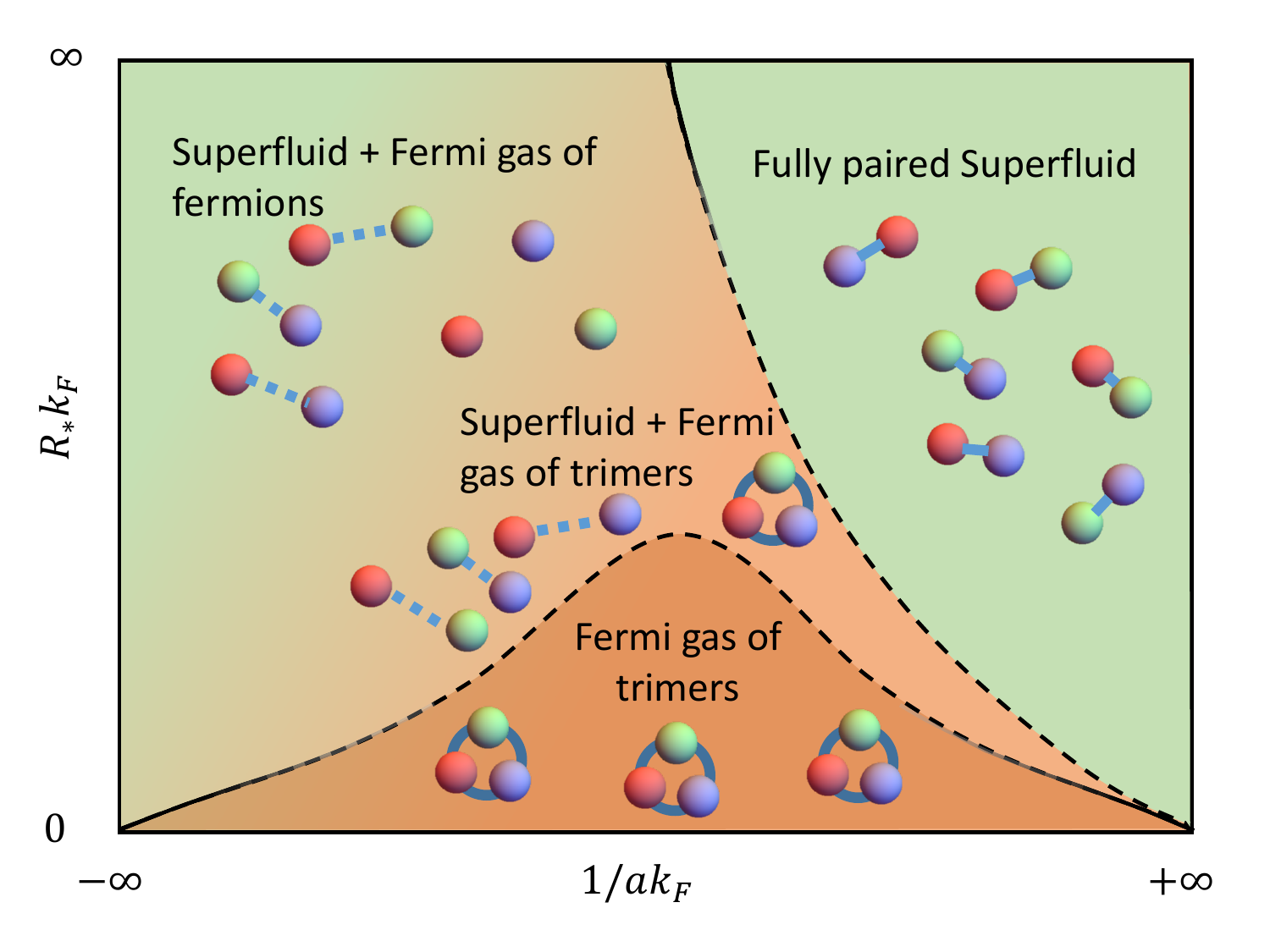}\hfill{}

\caption{\label{fig:Three-component-phase-diagram}Conjectured phase diagram
of the three-component Fermi mixture with Fermi momentum $k_{F}$,
for the same narrow scattering resonance, parameterised by $a$ and
$R_{*}$, between all pairs of fermions (adapted from reference~\cite{Nishida2012a}). }

\end{figure}

In his work~\cite{Nishida2012a}, Yusuke Nishida goes one step further
by explicitly considering the possibility of three fermions to form
a ground-state Efimov trimer to miminise their energy. The zero temperature
phase diagram, as a function of the scattering length $a$ (assumed
to be the same for all pairs of components) and the three-body parameter,
consists of three phases: a fully paired supefluid, a partially paired
superfluid with remaining unpaired fermions, and a trimer phase, which
is assumed to be a Fermi gas of ground-state trimers. Nishida calculates
the boundaries of these phases in limits where they can be calculated
exactly using a narrow-resonance model parameterised by $a$ and $R_{*}$
(see section~\ref{subsec:Coupled-channel-interactions}):
\begin{itemize}
\item the high-density limit $k_{F}R_{*}\to\infty$ of the transition between
the fully paired superfluid and partially paired superfluid,
\item the dilute limit $k_{F}a\to0^{-}$ of the transition between the superfluid
and the trimer phase at negative scattering length $a<0$, which is
given by equation~(\ref{eq:aminusForNarrowResonancesGroundState})
indicating the appearance of a ground-state trimer
\item the dilute limit $k_{F}a\to0^{+}$ of the transition between the superfluid
and trimer phase at positive scattering length $a>0$, which is obtained
by comparing the energies of a Fermi gas of trimers and a Bose-Einstein
condensate of dimers.
\end{itemize}
Here, $k_{F}$ denotes the Fermi momentum of the system. The schematic
phase diagram is represented in figure~\ref{fig:Three-component-phase-diagram}.
The author confirmed the qualitative aspects of that diagram by solving
a simple mean-field model including the trimers as a non-interacting
Fermi gas. The author also pointed out that within the partially paired
superfluid phase, there occurs a ``fermion-trimer\textquotedbl{}
continuity: the unpaired fermions gradually turn into trimers by binding
with pairs as $k_{F}R_{*}$ is decreased. This effect is anologous
to the quark-hadron continuity in nuclear matter. A qualitatively
similar phase diagram is expected for broad resonances, where in general
the three-body parameter $\kappa_{*}^{(0)}$ plays the role of $R_{*}^{-1}$. 

We note that in a previous work~\cite{Nishida2010} Yusuke Nishida
also conjectured the existence of a trimer phase in a two-component
Fermi gas where one component is free and the other is confined in
two separate layers, making the system resemble a three-component
system since the fermions confined in the different layers are distinguishable.
As seen in section~\ref{sec:Mixed-dimensions}, this situation leads
to the Efimov effect for one free particle resonantly interacting
with two particles confined in different layers. Near this resonance,
it is therefore expected that the ground-state Efimov trimer leads
to a trimer phase at sufficiently low density, as in the three-component
Fermi gas.

\subsubsection{Two-component Fermi mixtures}

In mass-imbalanced two-component Fermi systems, Efimov trimers exist
when the mass ratio between heavy and light fermions exceeds $\kappa_{c}^{(3)}=13.606...$,
as seen in sections~\ref{subsec:2-fermions-1particle} and~\ref{subsec:Kartavtsev-Malykh-universal-trimers}.
However, it is challenging to observe many-body physics induced by
these Efimov trimers in ultra-cold atom experiments, because such
Efimov trimers, when made of atoms, are unstable against recombination
into an atom plus a tightly bound dimer (see section~\ref{subsec:Many-body-Identical-bosons}).
On the other hand, the universal trimers, which are remnants of the
Efimov trimers in the range of mass ratios $\kappa_{1}=8.17260...<M/m<\kappa_{c}^{(3)}$,
are stable against three-body recombination thanks to the repulsion
created by the Pauli exclusion principle (see section~\ref{subsec:Kartavtsev-Malykh-universal-trimers}).
Therefore, a gas composed of the universal trimers seems to be a promising
candidate to observe a stable Efimov-induced many-body phase in ultra-cold
atoms. 

Such a possibility has been studied in Refs.~\cite{Endo2016} and~\cite{Naidon2016}.
Considering the mass ratio $\kappa_{1}<M/m<8.862...$ to avoid the
formation of universal tetramers (see Ref ~\cite{Blume2012,Bazak2017}
and section~\ref{subsec:Universal-four-body-bound}), a stable many-body
phase of the universal trimers has been predicted to exist in a range
of scattering lengths and population imbalance between heavy and light
fermions. When the scattering length is small and positive so that
the binding energy of the universal trimer is large enough, the trimer
can be regarded as a point-like composite fermion, having three internal
degrees of freedom originating from its angular momentum $L=1$. The
trimer gas thus becomes a three-component Fermi gas, each component
corresponding to one of the three rotational states, $m=-1,0,1$.
Since the $s$-wave interaction between the universal trimers would
be the dominant interaction of the trimer phase at low energy, the
authors have found that its low-energy effective Hamiltonian has an
SU(3) symmetry. Furthermore, in reference~\cite{Naidon2016}, the
Resonating Group Method was used to obtain an estimate of the interaction
between the two universal trimers. The interaction was found to be
of the soft-core repulsion type, leading to a positive $s$-wave scattering
length for the trimer-trimer scattering. This would imply that the
corresponding trimer phase, in the limit of low density and low temperature,
is not superfluid, but a three-component, SU(3) Fermi liquid~\cite{Endo2016}.

We should note that the existence and nature of this trimer phase
await further confirmation, since the crucial assumption about the
absence of larger clusters (e.g. pentamers, hexamers) has not been
precisely validated, and the method used to determine the trimer-trimer
interaction is only approximate. While the mass ratio window $8.17...<M/m<8.862$
seems rather restrictive, an ultra-cold mixture of $^{53}\mathrm{Cr}$
and $^{6}\mathrm{Li}$ atoms falls into this window ($M/m=8.80..$)
and is a promising candidate to confirm the existence and investigate
the properties of this trimer phase. We should note however that several
points may affect the existence of a trimer phase in this mixture.
First of all, the mass ratio of $^{53}$Cr and $^{6}$Li atoms happens
to be very close to the critical mass ratio for the appearance of
a universal four-body bound state. This suggests that the $p$-wave
atom-trimer scattering volume is likely to be strongly enhanced in
this system, similarly to the enhanced atom-dimer $p$-wave scattering
observed in a fermionic $^{40}$K-$^{6}$Li mixture~\cite{Endo2011,Levinsen2011,Jag2014}.
Second, below a certain scattering length, effective-range corrections
and magnetic dipole-dipole interactions between $^{53}$Cr atoms may
not be negligible~\cite{Endo2016}.

\clearpage{}

\part{Conclusion\label{part:Conclusion}}

The Efimov effect, in its most restrictive definition (see section~\ref{subsec:What-is-an-Efimov-state}),
could be simply regarded as an oddity in the energy spectrum of three
particles with short-range interactions. On the other hand, it could
be argued that the Efimov effect is on the contrary a central concept
around which a wide range of strongly-interacting systems may be described.
Such is the case of long-studied systems such as the tritium nucleus
or two-neutron halo nuclei in nuclear physics~(see sections~\ref{subsec:Observations-in-nuclear}
and \ref{subsec:Observations-in-nuclear-Multi-component}), or the
triatomic molecule of helium-4 in atomic physics (see section~\ref{subsec:Observations-with-atoms}).
Even though these systems may not be recognised as Efimov states in
the strictest sense, the Efimov effect provides a simple framework
for the existence of such compounds. The recent experimental observations
with ultra-cold atoms and theoretical developments have now opened
an even richer variety of systems related to Efimov physics, from
$N$-body universal clusters tied to Efimov states, to the super-Efimov
effect, or mixed-dimensional Efimov states. Perhaps one of the key
points of Efimov physics is to shift the paradigm of two-body correlations
in pairwise interacting systems to three-body correlations due, or
partly due, to the Efimov attraction. In this regard, what we have
learnt from Efimov physics appears to be quite promising for the study
of many-body systems where such three-body correlations may play an
important role.

\subsection*{Acknowledgments}

This work has greatly benefitted from discussions with Doerte Blume,
Takumi Doi, Yasuro Funaki, Chao Gao, Chris Greene, Hui Hu, Jesper
Levinsen, Haozhao Liang, Xia-Ji Liu, Sergej Moroz, Meera Parish, Dmitry
Petrov, Lucas Platter, Ludovic Pricoupenko, Jean-Marc Richard, and
Zheyu Shi. It was partially supported by the RIKEN iTHES project and
the RIKEN Incentive Research Project.\\

\begin{strip}
\begin{center}
\begin{minipage}[t]{0.965\textwidth}%
\begin{shaded}%
\vspace{0.3cm}

\textbf{}%
\begin{tabular*}{1\textwidth}{@{\extracolsep{\fill}}>{\raggedright}m{2cm}>{\raggedright}m{14cm}}
\includegraphics[scale=0.06]{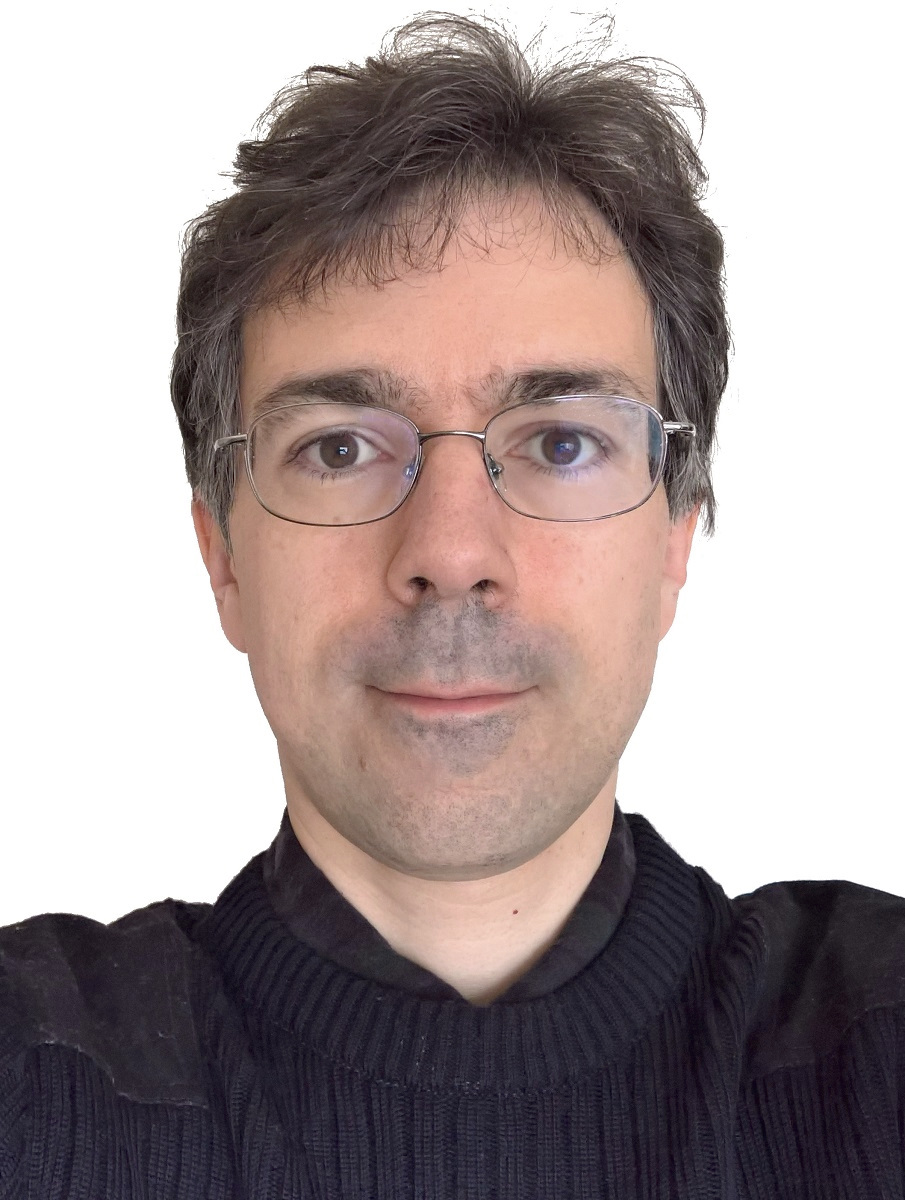} & \textbf{Pascal Naidon} is a senior research scientist at the Nishina
Center for Accelerator-Based Science of RIKEN. He received his PhD
in 2005 from the University of Paris 6, after working at Laboratoire
Aimé Cotton in Orsay. He worked as a guest researcher at NIST until
2008 and as a postdoctoral researcher at the University of Tokyo until
2011. He has been working on the theory of ultra-cold atoms, in particular
universal few-body and many-body phenomena occurring for resonant
interactions.

\vspace{0.3cm}
\tabularnewline
\includegraphics[scale=0.06]{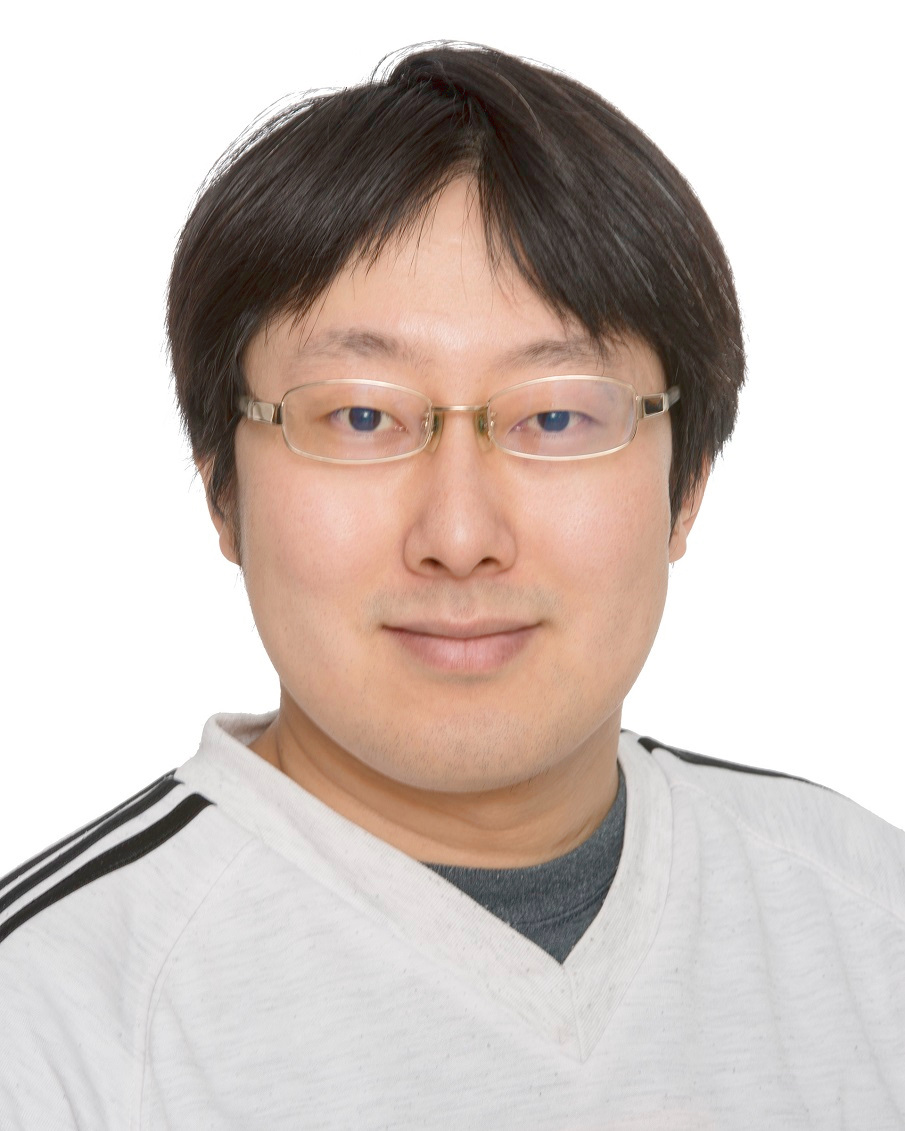} & \textbf{Shimpei Endo} is a research fellow at School of Physics and
Astronomy, Monash University. He received his PhD in 2014 at University
of Tokyo. He worked as a postdoctoral researcher at Laboratoire Kastler
Brossel, École Normale Supérieure, and then started working at Monash
University from 2016. His research interests are directed towards
understanding universal few-body and many-body physics in strongly
interacting quantum systems, in particular ultra-cold quantum gases.

\vspace{0.3cm}
\tabularnewline
\end{tabular*}

\vspace{0.3cm}
\end{shaded}%
\end{minipage}
\par\end{center}

\end{strip}

\rule[0.5ex]{0.95\textwidth}{1pt}

\cleardoublepage{}

\part*{Appendix: the Skorniakov - Ter-Martirosian equation}

In section~\ref{subsec:Efimov-theory}, the problem of three identical
bosons interacting via zero-range forces was solved using the Faddeev
equations in hyper-spherical coordinates. That approach has the advantage
of revealing the Efimov effect in a transparent manner, and provides
some analytic results. For the purpose of solving the problem numerically,
however, it is often preferred to use integral equations in momentum
space. These equations were first derived by G.~V.~Skorniakov and
Karen~A~Ter-Martirosian~\cite{Skorniakov1957}, and take advantage
of the contact nature of the interaction to reduce the dimensionality
of the problem. In general, the three-body problem requires $3d$
coordinates in $d$ dimensions. For translationally invariant systems,
one can eliminate $d$ coordinates associated with the centre of mass.
With contact interactions, the number of coordinates can be further
reduced by $d$. Additional rotational invariance may further reduce
the remaining $d$ coordinates to just one, making the problem easy
to solve numerically.

It turns out that contact interactions are not necessary to obtain
such simplification. It is sufficient to have a separable interaction,
which enables to treat finite-range effects. In this appendix, we
derive the Skorniakov - Ter-Martirosian equation for three identical
bosons interacting via a separable interaction; the zero-range equation
can be obtained by considering the limit when the separable interaction
becomes a contact interaction.

\subsubsection*{Separable interaction}

A separable interaction~\cite{Yamaguchi1954} is represented by an
operator of the form:
\begin{equation}
\hat{V}=\xi\vert\phi\rangle\langle\phi\vert,\label{eq:SeparablePotential-Definition}
\end{equation}
which is a projector onto a ``state\textquotedbl{} $\vert\phi\rangle$
multiplied by a scalar $\xi$. When applied to a two-body state $\vert\psi\rangle$,
it gives in momentum representation:
\begin{equation}
\langle\vec{p}\vert\hat{V}\vert\psi\rangle=\xi\langle\vec{p}\vert\phi\rangle\langle\phi\vert\psi\rangle=\xi\phi(\vec{p})\int\!\!\frac{d^{3}\vec{p}}{(2\pi)^{3}}\phi^{*}(\vec{p})\psi(\vec{p})\label{eq:SeparablePotential-Action}
\end{equation}
where $\vec{p}$ denotes the relative wave vector between two particles.
Note that for later convenience, we use the same notation $\vec{p}$
for the integration variable: it should be understood that all occurences
of $\vec{p}$ inside the integral refer to the integration variable.
To make the choice of $\xi$ unique for a given $\hat{V}$, the function
$\phi$ is normalised such that $\phi(\vec{0})=1$.

\subsubsection*{Two-body problem}

The two-body Schr\"odinger equation in momentum space reads:
\begin{equation}
\left(\frac{\hbar^{2}p^{2}}{m}-E\right)\psi(\vec{p})+\xi\phi(\vec{p})\langle\phi\vert\psi\rangle=0\label{eq:SeparablePotential-TwoBodyEquation}
\end{equation}

For scattering states $\psi_{\vec{k}}$ with an incoming wave vector
$\vec{k}$, the previous equation can be written as 
\begin{equation}
\psi_{\vec{k}}(\vec{p})=(2\pi)^{2}\delta^{3}(\vec{p}-\vec{k})-\frac{\frac{m}{\hbar^{2}}\xi\phi(\vec{p})}{p^{2}-k^{2}+i\varepsilon}\langle\phi\vert\psi_{\vec{k}}\rangle\label{eq:SeparablePotential-ScatteringState}
\end{equation}
with $E=\hbar^{2}k^{2}/m$ and $\varepsilon\to0^{+}$. In this equation,
one can recognise the $T$-matrix element:
\begin{equation}
T(\vec{k},\vec{p})=\langle\vec{p}\vert V\vert\psi_{\vec{k}}\rangle=\xi\phi(\vec{p})\langle\phi\vert\psi_{\vec{k}}\rangle\label{eq:SeparablePotential-TwoBodyTMatrix}
\end{equation}
Projecting equation~(\ref{eq:SeparablePotential-ScatteringState})
onto $\phi$, one gets a closed equation for $\langle\phi\vert\psi_{\vec{k}}\rangle$
which gives the straightforward solution,
\begin{equation}
\langle\phi\vert\psi_{\vec{k}}\rangle=\left(1+\frac{m}{\hbar^{2}}\xi\int\!\!\frac{d^{3}\vec{p}}{(2\pi)^{3}}\frac{\vert\phi(\vec{p})\vert^{2}}{p^{2}-k^{2}+i\varepsilon}\right)^{-1}\phi^{*}(\vec{k}),\label{eq:SeparablePotential-Overlap}
\end{equation}
from which one obtains $T(\vec{k},\vec{p})$,
\begin{equation}
T(\vec{k},\vec{p})=\left(\frac{1}{\xi}+\frac{m}{\hbar^{2}}\int\!\!\frac{d^{3}\vec{p}^{\prime}}{(2\pi)^{3}}\frac{\vert\phi(\vec{p}^{\prime})\vert^{2}}{p^{\prime2}-k^{2}+i\varepsilon}\right)^{-1}\phi(\vec{p})\phi^{*}(\vec{k})\label{eq:SeparablePotential-TwoBodyTMatrix-Expression}
\end{equation}
This relation can be used to express the parameters $\xi$ and $\phi$
of the separable potential in terms of physical quantities such as
the scattering length $a$:
\begin{equation}
\frac{4\pi\hbar^{2}}{m}a=T(\vec{0},\vec{0})=\left(\frac{1}{\xi}+\frac{m}{\hbar^{2}}\int\!\!\frac{d^{3}\vec{p}^{\prime}}{(2\pi)^{3}}\frac{\vert\phi(\vec{p}^{\prime})\vert^{2}}{p^{\prime2}}\right)^{-1}\label{eq:SeparablePotential-ScatteringLength}
\end{equation}

\subsubsection*{Three-body problem}

The three-body problem is expressed in a particular Jacobi wave vector
set $(\vec{P},\vec{p})$ chosen among the three possible sets $(\vec{P}_{k},\vec{p}_{k})$,

\begin{align}
\vec{P}_{k} & =\frac{2}{3}\left(\vec{k}_{k}-\frac{\vec{k}_{i}+\vec{k}_{j}}{2}\right)\label{eq:JacobiMomentum1}\\
\vec{p}_{k} & =\frac{1}{2}(\vec{k}_{j}-\vec{k}_{i})\label{eq:JacobiMomentum2}
\end{align}
where $(i,j,k)$ is a cyclic permutation of $(1,2,3)$ and $\vec{k}_{i}$
is the wave vector of the $i$th particle. In these coordinates, the
three-body Schr\"odinger equation reads:
\begin{align}
\left(\frac{3}{4}\frac{\hbar^{2}}{m}P^{2}+\frac{\hbar^{2}}{m}p^{2}-E\right)\Psi(\vec{P},\vec{p})\nonumber \\
+\sum_{i=1,2,3}\xi\phi(\vec{p}_{i})\int\!\!\frac{d^{3}\vec{p}_{i}}{(2\pi)^{3}}\phi^{*}(\vec{p}_{i})\Psi(\vec{P},\vec{p}) & =0,\label{eq:AppendixB1}
\end{align}
where the first term is associated with the kinetic energy, and the
second term is the sum of the action of the separable interaction
on the wave function $\Psi$ over the three pairs 12, 23, and 31.
Because of the bosonic exchange symmetry, $\Psi(\vec{P},\vec{p})$
can be replaced by $\Psi(\vec{P}_{i},\vec{p}_{i})$ inside the integral
of equation~(\ref{eq:AppendixB1}). As in equation~(\ref{eq:SeparablePotential-Action}),
it should be understood that $\vec{p}_{i}$ inside this integral refers
to the integration variable, such that the integral term, once integrated,
depends only the remaining Jacobi wave vector $\vec{P}_{i}$. One
can thus write: 
\begin{equation}
\left(\frac{3}{4}P^{2}+p^{2}-\frac{m}{\hbar^{2}}E\right)\Psi(\vec{P},\vec{p})+\sum_{i=1,2,3}F(\vec{P}_{i})\phi(\vec{p}_{i})=0,\label{eq:AppendixB2}
\end{equation}
where 
\begin{equation}
F(\vec{P})=\frac{m}{\hbar^{2}}\xi\int\!\!\frac{d^{3}\vec{p}}{(2\pi)^{3}}\phi^{*}(\vec{p})\Psi(\vec{P},\vec{p}).\label{eq:AppendixB3}
\end{equation}

Equation~(\ref{eq:AppendixB2}) can be inverted as
\begin{equation}
\Psi(\vec{P},\vec{p})=\Psi_{0}(\vec{P},\vec{p})-\sum_{i=1,2,3}\frac{F(\vec{P}_{i})\phi(\vec{p_{i}})}{\frac{3}{4}P^{2}+p^{2}-\frac{m}{\hbar^{2}}E+i\varepsilon},\label{eq:AppendixB4}
\end{equation}
where $\varepsilon\to0^{+}$. For three-body scattering states, $E\ge0$
and $\Psi_{0}$ is a solution of the non-interacting problem for three
particles at that energy, providing the asymptotic boundary condition.
For states with at least two bound particles, one has $\Psi_{0}=0$.
In the remainder, we will restrict our consideration to the latter
case. Inserting equation~(\ref{eq:AppendixB4}) into equation~(\ref{eq:AppendixB3})
gives:
\begin{equation}
\frac{\hbar^{2}}{m\xi}F(\vec{P})=-\sum_{i=1,2,3}\int\!\!\frac{d^{3}\vec{p}}{(2\pi)^{3}}\phi^{*}(p)\frac{F(\vec{P}_{i})\phi(\vec{p}_{i})}{\frac{3}{4}P^{2}+p^{2}-\frac{m}{\hbar^{2}}E+i\varepsilon}.\label{eq:AppendixB5}
\end{equation}
Making the choice $(\vec{P},\vec{p})=(\vec{P}_{3},\vec{p}_{3})$,
one can factorise one of the terms in the sum with the left-hand side
of equation~(\ref{eq:AppendixB5}) as follows:
\begin{align}
\left(\frac{\hbar^{2}}{m\xi}+\int\!\!\frac{d^{3}\vec{p}}{(2\pi)^{3}}\frac{\vert\phi(\vec{p})\vert^{2}}{\frac{3}{4}P^{2}+p^{2}-\frac{m}{\hbar^{2}}E+i\varepsilon}\right)F(\vec{P})\nonumber \\
+\sum_{i=1,2}\int\!\!\frac{d^{3}\vec{p}}{(2\pi)^{3}}\phi^{*}(\vec{p})\frac{F(\vec{P}_{i})\phi(\vec{p}_{i})}{\frac{3}{4}P^{2}+p^{2}-\frac{m}{\hbar^{2}}E+i\varepsilon} & =0.\label{eq:AppendixB6}
\end{align}
Expressing the Jacobi coordinate sets $(\vec{P}_{1},\vec{p}_{1})$
and ($\vec{P}_{2},\vec{p}_{2}$) in terms of $(\vec{P},\vec{p})$,
one finds 

\begin{align}
\vec{P}_{1} & =-\vec{p}-\frac{1}{2}\vec{P}\label{eq:JacobiMomentumTransform1}\\
\vec{p}_{1} & =-\frac{1}{2}\vec{p}+\frac{3}{4}\vec{P}\label{eq:JacobiMomentumTransform2}
\end{align}
and
\begin{align}
\vec{P}_{2} & =\vec{p}-\frac{1}{2}\vec{P}\label{eq:JacobiMomentumTransform3}\\
\vec{p}_{2} & =-\frac{1}{2}\vec{p}-\frac{3}{4}\vec{P}\label{eq:JacobiMomentumTransform4}
\end{align}
so that $\vec{p}=-\vec{P}_{1}-\frac{1}{2}\vec{P}=\vec{P}_{2}+\frac{1}{2}\vec{P}$
and therefore $\vec{p}_{1}=\frac{1}{2}\vec{P}_{1}+\vec{P}$ and $\vec{p}_{2}=-\frac{1}{2}\vec{P}_{2}-\vec{P}$.
Performing a change of integration variable $\vec{p}\to\vec{P}_{1}$
and $\vec{p}\to\vec{P}_{2}$ in the first and second integrals of
the sum, and relabelling the integration variable as $\vec{Q}$ in
both integrals, one finally arrives at the integral equation for $F$:
{\small{}
\begin{align}
\frac{\hbar^{2}}{m}\frac{\vert\phi(\vec{k})\vert^{2}}{T(\vec{k},\vec{k})}F(\vec{P})\;\;+\qquad\qquad\qquad\qquad\qquad\qquad\qquad\qquad\qquad\nonumber \\
\int\!\!\frac{d^{3}\vec{Q}}{(2\pi)^{3}}\,\frac{\phi^{*}(-\vec{Q}\!-\!\frac{\vec{P}}{2})\phi(\frac{\vec{Q}}{2}\!+\!\vec{P})+\phi^{*}(\vec{Q}\!+\!\frac{\vec{P}}{2})\phi(-\frac{\vec{Q}}{2}\!-\!\vec{P})}{P^{2}+Q^{2}+\vec{Q}\cdot\vec{P}-\frac{mE}{\hbar^{2}}}F(\vec{Q})\nonumber \\
=0\label{eq:Skorniakov-TerMartirosyan}
\end{align}
}where the wave vector $\vec{k}$ corresponds to the energy $\frac{\hbar^{2}k^{2}}{m}=E-\frac{3}{4}\frac{\hbar^{2}P^{2}}{m}$.
This integral equation on $F$ constitutes the Skorniakov - Ter-Martirosian
equation for three identical bosons interacting via a separable interaction.
The remarkable point of this equation is that it replaces the original
three-body Schr\"odinger equation (\ref{eq:AppendixB1}) for the
unknown function $\Psi(\vec{P},\vec{p})$ of two three-dimensional
variables by an equation on a function of only one three-dimensional
variable, which is not possible for a general interaction. For a rotationally
invariant system, the equation can be reduced to independent equations
for each partial wave $F_{\ell}(P)$ that depends only on the one-dimensional
variable $P=\vert\vec{P}\vert$.  In order to solve equation~(\ref{eq:Skorniakov-TerMartirosyan})
numerically to obtain the three-body bound states, one writes the
left-hand side of equation~(\ref{eq:Skorniakov-TerMartirosyan})
as a matrix acting on $F$ (through a discretisation scheme or spectral
method) and looks for the energies $E<0$ that make one of its eigenvalues
equal to zero, in order to satisfy the right-hand side of equation~(\ref{eq:Skorniakov-TerMartirosyan}).
The corresponding eigenvectors $F$ give the three-body wave functions
$\Psi$ through equation~(\ref{eq:AppendixB4}).

\subsubsection*{Generalisations}

The Skorniakov - Ter-Martirosian equation can be generalised to distinguishable,
fermionic or any mixture of particles. In general, there are three
functions $F_{i}$
\begin{equation}
F_{i}(\vec{P}_{i})=\frac{m}{\hbar^{2}}\xi\int\!\!\frac{d^{3}\vec{p}_{i}}{(2\pi)^{3}}\phi^{*}(\vec{p}_{i})\Psi(\vec{P},\vec{p})\label{eq:GeneralisedF}
\end{equation}
to describe the three pairs of particles, and they are solutions of
three coupled integral equations~\cite{Braaten2010}. The interaction
can also be generalised to multi-channel interaction~\cite{Lee2007,Jona-Lasinio2010,Naidon2011}.
There have also been generalisations to the relativistic case - see
section~\ref{subsec:Relativistic-case}.

\subsubsection*{Zero-range limit}

The zero-range limit can be obtained by setting $\phi(\vec{p})$ to
a constant $\phi(\vec{p})=1$. In this limit, the separable potential
is just a constant $\xi$ in momentum space, corresponding to a delta
function in real space. However, this limit gives rise to an ultraviolet
divergence of the integrals. These divergences can be cured by imposing
a momentum cutoff $\Lambda$ and renormalising $\xi$ in terms of
the physical scattering length $a$. Formally, this is equivalent
to setting $\phi(\vec{p})=\theta(\Lambda-p)$ in the formulas, where
$\theta$ denotes the unit step function. The integral equation~(\ref{eq:Skorniakov-TerMartirosyan})
thus becomes,

\begin{align}
\frac{\hbar^{2}}{m}\frac{1}{T(\vec{k},\vec{k})}F(\vec{P})\qquad\qquad\qquad\qquad\qquad\qquad\qquad\nonumber \\
+2\int_{\begin{array}{c}
{\scriptstyle \vert\vec{Q}+\frac{\vec{P}}{2}\vert<\Lambda}\\
{\scriptstyle \vert\vec{P}+\frac{\vec{Q}}{2}\vert<\Lambda}
\end{array}}\frac{d^{3}\vec{Q}}{(2\pi)^{3}}\frac{F(\vec{Q})}{P^{2}+Q^{2}+\vec{Q}\cdot\vec{P}-\frac{mE}{\hbar^{2}}} & =0.\label{eq:STM-ZeroRange1}
\end{align}

The equation (\ref{eq:SeparablePotential-ScatteringLength}) gives
the renormalisation relation:
\begin{equation}
\frac{1}{a}=\frac{4\pi\hbar^{2}}{m}\frac{1}{\xi}+\frac{2}{\pi}\Lambda.\label{eq:TwoBody-Renormalisation-Relation}
\end{equation}
From this relation, one can express the $T$-matrix elements:{\small{}
\begin{equation}
T(\vec{k},\vec{p})=\frac{m}{4\pi\hbar^{2}}\!\!\left(\frac{1}{a}\!-\!\frac{2}{\pi}\sqrt{\!-\!k^{2}}\arctan\frac{\Lambda}{\sqrt{\!-\!k^{2}}}\right)^{\!\!-\!1}\!\!\theta(\Lambda-k)\theta(\Lambda-p)\label{eq:ZeroRange-TMatrix}
\end{equation}
}For sufficiently large $\Lambda$, the on-shell two- body $T$-matrix
element approaches
\begin{equation}
T(\vec{k},\vec{k})\approx\frac{m}{4\pi\hbar^{2}}\left(\frac{1}{a}-\sqrt{-k^{2}}\right)^{-1},\label{eq:ZeroRange-TMatrix-Approximation}
\end{equation}
and the restriction $\vert\vec{Q}+\frac{\vec{P}}{2}\vert<\Lambda$
and $\vert\vec{P}+\frac{\vec{Q}}{2}\vert<\Lambda$ over the integration
volume can be approximated by $Q<\Lambda$. This results in the simplified
equation,
\begin{align}
\left(\frac{1}{a}-\sqrt{\frac{3}{4}P^{2}-\frac{m}{\hbar^{2}}E}\right)F(\vec{P})\qquad\qquad\qquad\qquad\qquad\nonumber \\
+8\pi\int_{Q<\Lambda}\frac{d^{3}\vec{Q}}{(2\pi)^{3}}\frac{F(\vec{Q})}{P^{2}+Q^{2}+\vec{Q}\cdot\vec{P}-\frac{mE}{\hbar^{2}}}=0.\label{eq:STM-ZeroRange}
\end{align}
which is the original equation derived by G.~V.~Skorniakov and K.~A.~Ter-Martirosian~\cite{Skorniakov1957}.
A remarkable point of this equation is that the two-body physics enters
only through the on-shell $T$-matrix elements $T(\vec{k},\vec{k})$
and the cutoff $\Lambda$. In general, for non-separable interactions,
the off-shell $T$-matrix elements $T(\vec{k},\vec{p})$ would be
required. Here, all the off-shell information is captured by the cutoff
$\Lambda$, as can be checked from the expression of $T(\vec{k},\vec{p})$
in equation~(\ref{eq:ZeroRange-TMatrix}). This information is however
essential for the three-body problem, and $\Lambda$ cannot be set
to infinity in the above integral, as the results would not converge
but exhibit logarithmic oscillations with $\Lambda$. It is the large
but finite value of $\Lambda$ that sets the three-body parameter
of the Efimov states. Because of the discrete scale invariance discussed
in the Efimov theory (see section~\ref{subsec:Efimov-theory}), scaling
the value of $\Lambda$ by powers of $e^{\pi/\vert s_{0}\vert}$ gives
the same three-body parameters and the same observables.

It should be noted that the impossibility to take $\Lambda$ to infinity
is due to the fact that only two-body interactions have been assumed.
The three-body parameter may also be fixed by a three-body force.
Adding a $\Lambda$-dependent zero-range three-body interaction can
cancel the dependence of the observables on $\Lambda$, making it
possible to take the limit $\Lambda\to\infty$~\cite{Bedaque1999,Bedaque1999a}.
The renormalised observables then depends on the two-body scattering
length $a$ and a three-body parameter introduced by the three-body
interaction.

\subsubsection*{Finite-range effects}

The choice of the parameters $\xi$ and $\phi(\vec{p})$ of the separable
potential in equation~(\ref{eq:SeparablePotential-Definition}) depends
on the system and observables of interest. In the windows of universality
shown in figure~\ref{fig:EfimovPlotFiniteRange}, the observables
depend only the two-body scattering length $a$ and the three-body
parameter $\kappa_{*}$. To calculate these universal observables,
any choice of $\xi$ and $\phi(\vec{p})$ leading to a desired $a$
and $\kappa_{*}$ is possible. The most common choices are the zero-range
limit $\phi(\vec{p})=\theta(\Lambda-p)$ described above, and the
Gaussian separable potential with $\phi(\vec{p})=\exp(-\Lambda^{2}p^{2})$~\cite{Lee2007,Jona-Lasinio2010,Naidon2011},
since they lead to analytical simplifications and simple relations
between the physical quantities $(a,\kappa_{*})$ and the parameters
$(\xi,\Lambda)$.

On the other hand, to calculate observables outside the universal
region, that are affected by finite-range corrections, a more precise
choice of the parameters is needed. As discussed in section~\ref{subsec:Finite-range-interactions},
the interest of separable potential models over zero-range models
is that they can account for the finite-range effects non-perturbatively.
Indeed, separable potential models can be parameterised to reproduce
exactly the two-body scattering length and effective range, i.e. reproduce
exactly the low-energy two-body observables. Even when the scattering
length and effective range are fixed, there remains some freedom to
parameterise the separable potential, and the precise choice of the
separable potential can change the value of the three-body parameter.
As discussed in section~\ref{subsec:Other-types-of-interactions},
the three-body parameter is to a great extent (although not completely)
set by the effective range, so that many separable potentials (such
as Gaussian) having the correct scattering length and effective range
already give a rough estimate of the three-body energies beyond the
zero-range limit.

If one has further information on the two-body interaction, such as
off-shell $T$-matrix elements or a model potential for the interaction\footnote{which are strictly speaking not two-body observables, but may be known
to model accurately the interaction of certain particles.}, one may construct an even more precise separable potential by fully
exploiting its parameter space. Namely, a separable potential $\hat{V}$
can be designed to reproduce exactly an eigenstate $\vert\psi\rangle$
at a given energy $E$ of a given local potential $V$. It has the
form~\cite{Ernst1973}:
\begin{equation}
\hat{V}=\frac{1}{\langle\psi\vert V\vert\psi\rangle}V\vert\psi\rangle\langle\psi\vert V.\label{eq:SeparablePotential-VParameterisation}
\end{equation}
which shows that the action of $\hat{V}$ onto $\vert\psi\rangle$
is indeed the same as $V$, thus ensuring that $\vert\psi\rangle$
is also an eigenstate of $\hat{V}$ at the same energy. If $\vert\psi\rangle$
is chosen to be the zero-energy scattering eigenstate, the separable
potential has by construction the correct scattering length and correct
effective range. Indeed, the scattering length is given by the asymptotic
form of the zero-energy radial wave function $\varphi(r)=r\psi(\vec{r})$
(see equation~\ref{eq:TwoBodyWaveFunction}):
\[
\varphi(r)\xrightarrow[r\to\infty]{}\bar{\varphi}(r)=1-\frac{r}{a}
\]
and the effective range $r_{e}$ is obtained from $\varphi$ through
the formula~\cite{Bethe1949},
\[
\frac{1}{2}r_{e}=\int_{0}^{\infty}dr\left[\bar{\varphi}(r)^{2}-\varphi(r)^{2}\right].
\]

In this case, the separable potential of equation~(\ref{eq:SeparablePotential-VParameterisation})
has the explicit parameterisation~\cite{Naidon2014a,Naidon2014}:
\begin{equation}
\phi(p)=1-p\int_{0}^{\infty}dr\left(\bar{\varphi}(r)-\varphi(r)\right)\sin pr\label{eq:SeparablePotential-ZeroEnergyParameterisation1}
\end{equation}
and $\xi$ is given by equation~(\ref{eq:SeparablePotential-ScatteringLength}),
\begin{equation}
\xi=\frac{4\pi\hbar^{2}}{m}\left(\frac{1}{a}-\frac{2}{\pi}\int_{0}^{\infty}\!\!dp\vert\phi(p)\vert^{2}\right)^{-1}\label{eq:SeparablePotential-ZeroEnergyParameterisation2}
\end{equation}

This separable representation was shown to reproduce approximately
the three-body energies of the original potential~\cite{Naidon2014},
and in particular the three-body parameters of van der Waals potentials~\cite{Naidon2014a}
- see section~\ref{subsec:VdW-Three-identical-bosons}.

As an example, we have constructed such a separable potential for
helium-4, using the zero-energy eigenstate obtained from the scaled
LM2M2 potential of reference~\cite{Aziz1991}, as solved for the
ground-state energy as a function of scattering length (varied by
scaling the LM2M2 potential). The obtained energy is shown as a dotted
curve in figure~\ref{fig:Helium1}.

Note that the representation of a given potential in terms of a separable
potential was generalised by Ernst, Shakin and Thaler (EST)~\cite{Ernst1973},
who have shown that any local potential $V$ can be represented exactly
as a superposition of non-local separable potentials. 

\subsubsection*{Integral equations for the triton}

We present here the separable model of the triton, first introduced
by Kharchenko~\cite{Kharchenko1967} and used to calculate the surfaces
and curves of figure~\ref{fig:Triton}. This model describes the
triplet and singlet interactions of nucleons by separable potentials
of the form,
\begin{align*}
\hat{V}_{t} & =\xi_{t}\vert\phi_{t}\rangle\langle\phi_{t}\vert\\
\hat{V}_{s} & =\xi_{s}\vert\phi_{s}\rangle\langle\phi_{s}\vert
\end{align*}
leading to the following integral equations: 
\begin{align}
\left[\frac{1}{a_{t}}+\frac{2}{\pi}\int_{0}^{\infty}\!\!dp\vert\phi_{t}(\vec{p})\vert^{2}\left(\frac{p^{2}}{\frac{3}{4}P^{2}+p^{2}-\frac{m}{\hbar^{2}}E}-1\right)\right]F_{t}(k)\nonumber \\
+4\pi\int\frac{d^{3}\vec{Q}}{(2\pi)^{3}}\frac{\frac{1}{2}I_{\vec{k},\vec{k}^{\prime}}^{tt}F_{t}(Q)+\frac{3}{2}I_{\vec{k},\vec{k}^{\prime}}^{ts}F_{s}(Q)}{P^{2}+Q^{2}+\vec{P}\cdot\vec{Q}-\frac{mE}{\hbar^{2}}}=0\label{eq:tritonSTM1}
\end{align}
\begin{align}
\left[\frac{1}{a_{s}}+\frac{2}{\pi}\int_{0}^{\infty}\!\!dp\vert\phi_{s}(\vec{p})\vert^{2}\left(\frac{p^{2}}{\frac{3}{4}P^{2}+p^{2}-\frac{m}{\hbar^{2}}E}-1\right)\right]F_{s}(k)\nonumber \\
+4\pi\int\frac{d^{3}\vec{Q}}{(2\pi)^{3}}\frac{\frac{3}{2}I_{\vec{k},\vec{k}^{\prime}}^{st}F_{t}(Q)+\frac{1}{2}I_{\vec{k},\vec{k}^{\prime}}^{ss}F_{s}(Q)}{P^{2}+Q^{2}+\vec{P}\cdot\vec{Q}-\frac{mE}{\hbar^{2}}}=0\label{eq:tritonSTM2}
\end{align}
where
\[
I_{\vec{k},\vec{k}^{\prime}}^{ij}=\phi_{i}^{*}(\vert\vec{Q}+\frac{1}{2}\vec{P}\vert)\phi_{j}(\vert\vec{P}+\frac{1}{2}\vec{Q}\vert).
\]
The functions $\phi_{t}$ and $\phi_{s}$ are obtained from the zero-energy
parameterisation of equation~(\ref{eq:SeparablePotential-ZeroEnergyParameterisation1})
where $\varphi(r)$ is the two-body radial wave function,
\[
\varphi(r)=\frac{r_{0}}{a}\left(\frac{Q_{\lambda}(0)}{P_{\lambda}(0)}P_{\lambda}(\tanh r/r_{0})-Q_{\lambda}(\tanh r/r_{0})\right)
\]
of the zero-energy scattering eigenstate for the P\"oschl-Teller
potential,
\[
V(r)=-\frac{\hbar^{2}}{m}\lambda(\lambda+1)\text{sech}^{2}(r/r_{0}),
\]
with the scattering length 
\[
a=r_{0}\left(\frac{Q_{\lambda}(0)}{P_{\lambda}(0)}+H_{\lambda}\right),
\]
where $P_{\lambda}(x)$ and $Q_{\lambda}(x)$ designate the Legendre
polynomials and $H_{\lambda}=\int_{0}^{1}\frac{1-t^{\lambda}}{1-t}dt$
is the harmonic number, which is equal to $\sum_{n=1}^{\lambda}\frac{1}{n}$
for integral values of $\lambda$.

For $\phi_{t}$ (respectively $\phi_{s}$), the parameters $r_{0}$
and $\lambda$ are chosen to reproduce the triplet scattering length
$a_{t}=5.4112$~fm (respectively the singlet scattering length $a_{s}=-23.7148$~fm)
and the triplet effective range $r_{e,t}=1.7436$~fm (respectively
the singlet effective range $r_{e,s}=2.750$~fm~). The values are
taken from reference~~\cite{Hackenburg2006}. To obtain the surfaces
and curves of figure~\ref{fig:Triton}, both $a_{t}$ and $a_{s}$
were varied away from their physical values.\pagebreak{}

\hypersetup{urlcolor=myred}\bibliographystyle{IEEEtran2}
\bibliography{paper24}

\begin{strip}

\part*{List of changes}
\begin{itemize}
\item Introduction: A short history of Efimov physics: the word ``length\textquotedbl{}
was added after ``van der Waals\textquotedbl .
\item Section ``4.5.2 Structure\textquotedbl{} {[}Section 2.1.5.2 in the
published version{]}, end of third paragraph: the following incorrect
sentence was removed: ``For the full wave function of equation (2.16),
one obtains the average value of $\alpha$ to be close to $\pi/6$
, corresponding to an elongated triangle\textquotedbl . See https://doi.org/10.1103/PhysRevA.104.059903.
\item Section ``4.7.1 Helium-4\textquotedbl{} {[}Section 2.1.7.1 in the
published version{]}: ``The range of this potential can be characterised
by its van der Waals length, which is about 0.54 nm\textquotedbl ,
the value was changed to 0.27 nm to be consistent with the definition
given in the van der Waals universality section.
\item Section ``5.2 The super-Efimov effect\textquotedbl , after Eq. (5.5)
{[}Section 2.2.2 after Eq. (2.57) in the published version{]}: the
symbol $\log$ was replaced by $\ln$.
\item Figure 11.3 {[}Figure 22 in the published version{]}: the top panel
was corrected, due to a wrongly calculted curve in the original figure.
See https://doi.org/10.1103/PhysRevA.104.059903.
\item Figure 11.3 {[}Figure 22 in the published version{]}: the words ``for
a right angle between the Jacobi vectors\textquotedbl{} were added
to the caption.
\item Formula after Eq. (11.42) {[}Eq. (4.45) on page 54 in the published
version{]}: a minus sign was added before the factor $2\pi$ in the
exponential and absolute value bars were added to E(n).
\end{itemize}
\end{strip}
\end{document}